\documentclass[prx,twocolumn,showpacs,groupedaddress,superscriptaddress,nofootinbib,floatfix,preprintnumbers,longbibliography]{revtex4-2}

\usepackage{amssymb,amsmath,graphicx,color}
\usepackage{bm}
\usepackage{bbm}
\usepackage[tight]{subfigure}
\usepackage[export]{adjustbox}
\usepackage{enumerate}
\usepackage{braket}
\usepackage{appendix}
\usepackage[
  bookmarks=true,
  colorlinks,
  linkcolor=black,
  urlcolor=black,
  citecolor=black,
  plainpages=false,
  pdfpagelabels,
  final,
  breaklinks=true
]{hyperref}

\usepackage{multirow}
\usepackage{rotating,booktabs}
\usepackage[verbose]{placeins}
\usepackage{physics}
\usepackage[dvipsnames]{xcolor}

\newcommand{\PS}[1]{ {\color{magenta} #1 }}
\newcommand{\JRD}[1]{ {\color{violet} #1 }}

\begin{document}
\title{Quantum electrodynamics of intense laser-matter interactions: \textcolor{black}{A tool for quantum state engineering}}

\author{Philipp Stammer}
\email{philipp.stammer@icfo.eu}
\affiliation{ICFO-Institut de Ciencies Fotoniques, The Barcelona Institute of Science and Technology, Castelldefels (Barcelona) 08860, Spain.}
\author{Javier Rivera-Dean}
\affiliation{ICFO-Institut de Ciencies Fotoniques, The Barcelona Institute of Science and Technology, Castelldefels (Barcelona) 08860, Spain.}
\author{Andrew~Maxwell}
\affiliation{ICFO-Institut de Ciencies Fotoniques, The Barcelona Institute of Science and Technology, Castelldefels (Barcelona) 08860, Spain.}
\affiliation{Department of Physics and Astronomy, Aarhus University, DK-8000 Aarhus C, Denmark}
\author{Theocharis Lamprou}
\affiliation{Foundation for Research and Technology-Hellas, Institute of Electronic Structure \& Laser, GR-
70013 Heraklion (Crete), Greece.}
\affiliation{Department of Physics, University of Crete, P.O. Box 2208, GR-71003 Heraklion (Crete), Greece.}
\author{Andrés Ordóñez}
\affiliation{ICFO-Institut de Ciencies Fotoniques, The Barcelona Institute of Science and Technology, Castelldefels (Barcelona) 08860, Spain.}
\author{Marcelo~F.~Ciappina}
\affiliation{Physics Program, Guangdong Technion - Israel Institute of Technology, Shantou, Guangdong
515063, China.}
\affiliation{Technion - Israel Institute of Technology, Haifa, 32000, Israel.}
\affiliation{Guangdong Provincial Key Laboratory of Materials and Technologies for Energy Conversion, Guangdong Technion – Israel Institute of Technology, Shantou, Guangdong 515063, China}
\author{Paraskevas Tzallas}
\affiliation{Foundation for Research and Technology-Hellas, Institute of Electronic Structure \& Laser, GR-
70013 Heraklion (Crete), Greece.}
\affiliation{ELI-ALPS, ELI-Hu Non-Profit Ltd., Dugonics tér 13, H-6720 Szeged, Hungary}
\author{Maciej Lewenstein }
\affiliation{ICFO-Institut de Ciencies Fotoniques, The Barcelona Institute of Science and Technology, Castelldefels (Barcelona) 08860, Spain.}
\affiliation{ICREA, Pg.~Lluís Companys 23, 08010 Barcelona, Spain.}

\begin{abstract}

Intense laser-matter interactions are at the center of interest in research and technology since the development of high power lasers. They have been widely used for fundamental studies in atomic, molecular, and optical physics, and they are at the core of attosecond physics and ultrafast opto-electronics. Although the majority of these studies have been successfully described using classical electromagnetic fields, recent investigations based on fully quantized approaches have shown that intense laser-atom interactions can be used for the generation of controllable high-photon-number entangled coherent states and coherent state superpositions. In this tutorial, we provide a comprehensive fully quantized description of intense laser-atom interactions. We elaborate on the processes of high harmonic generation, above-threshold-ionization, and we discuss new phenomena that cannot be revealed within the context of semi-classical theories. We provide the description for conditioning the light field on different electronic processes, and their consequences for quantum state engineering of light. Finally, we discuss the extension of the approach to more complex materials, and the impact to quantum technologies for a new photonic platform composed by the symbiosis of attosecond physics and quantum information science.

\end{abstract}

\date{\today}

\maketitle
\tableofcontents

\section{Introduction}

\subsection{Quantum optics and super intense laser atom interaction physics} There were two instances that gave rise to the birth of quantum optics: the invention of the laser (and earlier maser) (cf\@. \cite{Siegman-book} and references therein), and the formulation of the quantum theory of optical coherence by Roy Glauber and George Sudarshan (cf\@. \cite{glauber1963coherent, glauber1963quantum, sudarshan1963equivalence}). At the time of its infancy, and also much later, quantum optics was considered to be “an applied Quantum Electrodynamics (QED)” (see for instance \cite{Eberlybook}). From the perspective of atomic-molecular and optical (AMO) physics, QED concerns itself with very advanced, and precise perturbation theory calculations of various quantities relevant for spectroscopic measurements (cf\@. \cite{Pachucki1,Pachucki2,Drake}). 
Quantum optics, on the other hand, focused for many years on single or few body phenomena, trying to understand non-perturbative aspects of atom-light interaction (for a recent overview see \cite{Booklarson}). 

One of the areas of quantum optics that aimed at exploring non-perturbative effects was the so-called strong laser fields physics, sometimes termed as super intense laser atom interaction physics (SILAP). The history of this area is well illustrated in the collection of essays on the relevant subjects in the Handbook \cite{Drake-handbook}. At the beginning there was light, but then there were photons. The lasers were strong, but not strong enough, so they were mainly used to generate multi-photon resonant, and then non-resonant processes such as absorption, ionization or very sophisticated high order perturbation theory.

With the advent of stronger and stronger lasers, came the revolution. The multiphoton processes became non-accessible to even very high-order perturbation theory. 
The key strong-field non-perturbative processes are depicted in Fig.~\ref{fig:SF_Processes}, which includes above-threshold ionization (ATI), also termed strong field ionization; high harmonic generation (HHG), where the photoelectron laser driven recombination leads to the emission of a high energy photon; and non-sequential double ionization (NSDI), in-which the photoelectron laser-driven recollision leads to the ionization of a second electron.
The history of these events is well described in the recent review on the Strong Field Approximation (SFA) \cite{amini2019symphony}, and in several reviews on the subject, which started with the observations of non-perturbative plateaus in ATI \cite{Agostini1979,Kruit1983,Becker2002}, together with non-perturbative plateaus and cutoff in HHG \cite{McPherson1987,Ferray1988,AnneML2}, or multi-electron ionization, in particular NSDI (cf\@. \cite{Anne1983a,Corkum1993,Walker1994,Feuerstein2001,Eberly2012}). Nowadays, this physics extends from atomic targets \cite{AnneML2}, to molecules (cf\@. \cite{Bandraukbook,Calegari1,Calegari2,Chinesebook}), atomic clusters (cf\@. \cite{Lezius,Brabec,Rost}), solids (cf\@. \cite{vampa_merge_2017}), 2D (topological) materials (cf\@. \cite{Ivanov,Chacon}), and much more. The theoretical methods of this era forgot about QED, and describe the strong laser fields entirely classically. Independently of the process of interest, various theoretical approaches may be used, but all of them assume only the classical nature of the driving laser fields and neglect their quantum fluctuations\footnote{Classical random fluctuation of the laser fields, such as the phase or amplitude fluctuations, obviously, were already considered both in multiphoton physics and in SILAP \cite{Drake-handbook,Wodkiewicz1,Wodkiewicz2}.}.

\subsection{SILAP - theoretical approaches} 
\begin{figure*}
    \centering
    \includegraphics[width=\linewidth]{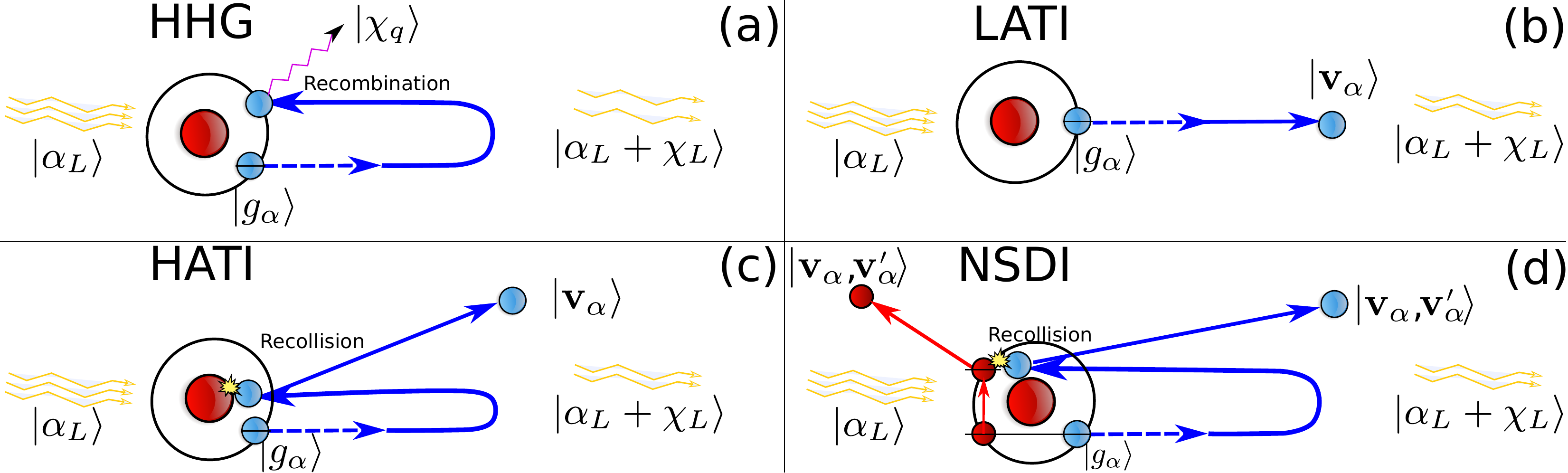}
    \caption{Depiction of the key quasi-classical strong field processes, along with the quantum states of light and matter. (a) High harmonic generation (HHG), (b)  \textcolor{black}{low-order above-threshold ionization (LATI) without rescattering events}, (c) high-order above-threshold ionization (HATI) \textcolor{black}{including rescattering events}, (d) non-sequential double ionization (NSDI). The states of the driving field mode before and after interaction are given by $\ket{\alpha_{L}}$ and $\ket{\alpha_{L}+\chi_L}$, respectively. High-order harmonic modes, due to HHG, are given by $\ket{\chi_q}$. The initial and final electronic states are given by $\ket{g_{\alpha}}$ and $\ket{\mathbf{v}_{\alpha}}$. 
    \textcolor{black}{To emphasize the electrons in NSDI may be entangled \cite{maxwell2021entanglement}, we have written them as a two-particle state. The quantum optical degrees of freedom may also be entangled with each other and the electronic states, however, for clearer labelling we have written these separately.}
    }
    \label{fig:SF_Processes}
\end{figure*}

Below we list the most important theory approaches to the three most relevant processes of SILAP, each of which is pictorically represented in Fig.~\ref{fig:SF_Processes}:    

\begin{itemize}

    \item {\it High Harmonic Generation - simple atoms.} For simple atoms, that is, for instance, hydrogen or noble gases, the single atom response may be calculated using the time-dependent Schrödinger equation in the single active electron (SAE) approximation; it  works usually perfectly. Going beyond SAE is rarely necessary, but may be achieved using the Time-Dependent Hartree-Fock (TDHF) approach \cite{Cederbaum1,Cederbaum2}, the Time-Dependent Density Functional method \cite{Gross1,Gross2}, or quantum chemistry methods such as the Time-Dependent configuration-interaction method \cite{Moszynski1,Moszynski2}. Unfortunately, the full description of the HHG process, requires also taking into account propagation effects (cf\@. \cite{AnneML2}). That makes often the use of the Time-Dependent Schrödinger Equation (TDSE) too demanding, leaving as the reasonable alternative the SFA \cite{amini2019symphony}. The SFA, combined with propagation codes solving classical Maxwell equations, gives usually very satisfactory results. Of course, there are more approaches one can use at the single atom level: i) for longer pulses one can use the Floquet or $R$-matrix theory (cf\@. \cite{Joachain-book}); ii) for short, intense pulses the ultimate  description belongs to the TDSE; iii) in some regimes of parameters, even more ``brutal'' than the SFA, classical (truncated Wigner) methods can be useful (cf\@. \cite{Percival1,Percival2,Grochmalicki}). 
    
    \item {\it High Harmonic Generation - complex targets.} For more complex atoms (with non-negligible electronic correlations) or  molecules, the SFA seems to be the only method, although Floquet methods and $R$-matrix work reasonable well for long laser pulses. The situation changes when we go to solids and truly many body systems. In the weakly correlated cases one can apply the semiconductor Bloch equations \cite{vampa_theoretical_2014,vampa_merge_2017}, or a little more flexible Wannier-Bloch approach \cite{osika_wannier-bloch_2017}. But, in the case of strongly correlated systems, more sophisticated approaches must be used: from Density Functional Theory (DFT), through exact diagonalization for small systems (cf\@. \cite{Bauer1,Bauer2,Baldelli}), to more advanced mean field theories like the generalized Time-Dependent Pairing Theory (TDPT) (cf\@. \cite{Jensandus}),
    Dynamical Mean Field Theory (DMFT) \cite{DMFT1,DMFT2,Vollhardt},  or slave bosons theories \cite{PALee} in a time dependent version.
    
    \item {\it Above Threshold ionization - simple atoms.} For single targets, the same methods as mentioned above can be used \cite{ATI-review1,ATI-review2}. While propagation effects are not relevant here as the output in this case are photoelectrons, ponderomotive effects and averaging over the spatial intensity profile of the laser pulse are, however, extremely relevant.
    
    \item {\it Above Threshold ionization - complex targets.} Obviously, this is particularly challenging, even if the multielectron effects play no role. The high energy parts of ATI spectra come from rescattering processes which are possible, but technically hard, to describe with the SFA \cite{Noslen1,Noslen2,Noslen3,NOslen4,Noslen5}. Again, the ponderomotive effect and averaging over spatial intensity profile of the laser pulse should be taken into account \cite{Schuricke2011}.
        
    \item {\it Sequential and non-sequential double ionization.} For simple two electron systems (Helium \cite{Taylor1,Taylor2,moore_two_2001,Muller1}), 1D two-electron models \cite{Ebery-Grobe,eberly-corr,Becker-rmp,demorissonfaria_electron_2011}, or quasi-3D models \cite{sacha1,sacha2}), one can use the TDSE approach in various ``flavours''. The quasi-3D approach can be even generalized to 3-electron systems \cite{Dmitry}. Otherwise, we can only apply approximate methods, such as the SFA, see e.g. Ref.~\cite{maxwell_controlling_2016}.
        
\end{itemize}

\subsection{SILAP and QED}
        
All of these approaches are amazingly impressive and have led to a tremendous amount of important results, but all of them neglect the genuine QED nature of light. Recently, this orthodox paradigm seem to be more and more often questioned. 
For instance,  some of the early  papers with  ``QED" flavour, treated laser field in a QED manner to develop a frequency domain theory for HHG, ATI and NSDI, which later they used to describe X-ray pulse in assisted HHG/NSDI \cite{guo_quantum_1988,fu_interrelation_2001,wang_frequencydomain_2007,wang_frequencydomain_2012}. 

More recently, a new line of research was initiated by P. Tzallas and his collaborators, combining quantum optics methods with SILAP. In their two pioneering papers, they observed quantum optical signatures in
strong-field laser physics by looking at the infrared
photon counting in HHG \cite{Paris-srep}. In the subsequent paper \cite{Paris-ncomm}, high-order harmonics were measured by the photon
statistics of the infrared driving-field exiting the
atomic medium. These ideas stimulated other groups to look on QED aspects of SILAP in a more detailed way. I. Kaminer with his collaborators discussed the quantum-optical nature of HHG \cite{Ido-ncomm} (see also \cite{Ido-nphysrev}). They present a fully-quantum theory of extreme nonlinear optics, predicting quantum effects that alter both the spectrum and photon statistics of HHG. In particular, they predict the emission of shifted frequency combs and identify spectral features arising from the breakdown of the dipole approximation for the emission. Each frequency component of HHG can be bunched and squeezed, and each emitted photon is a superposition of all frequencies in the spectrum, i.e. each photon is a comb. G. Paulus {\it et al.} looked at the photon counting of extreme ultraviolet high harmonics using
a superconducting nanowire single‑photon detector \cite{Gerhard-aphysB}. 
\textcolor{black}{Triggered by recent developments on radiation sources of high power pulsed squeezed light \cite{spasibko2017multiphoton, manceau2019indefinite}, the authors in Ref.~\cite{gorlach2022high} initiated investigations towards the generation of high harmonics driven by intense squeezed light sources.}
In a series of papers, S. Varro's group studied the quantum optical aspects of HHG \cite{Sandor-phot1}, developing quantum optical models \cite{Sandor-phot2}, and a quantum-optical description of photon statistics and cross correlation \cite{Sandor-pra}. 
        
Our approach followed the earlier ideas of the Crete group, and focused on the experiments in which quantum electrodynamical properties of  the driving laser field were observed upon a) conditioning on  HHG, and b) coherent diminution of the amplitudes of the light field, allowing  for the final reconstruction of the Wigner function of the laser mode using quantum tomography. Our results, and their possible implications for applications in quantum information (QI) and quantum technologies (QT), were described in a series of papers \cite{lewenstein2021generation,stammer2022high,rivera2022strong,stammer2022theory,Lamprou_QS_Photonics2021,rivera2021jcompelec}. The culmination of this approach was the observation of a massive optical Schrödinger ``cat'' state of the fundamental laser mode, conditioned on the HHG process. 

\subsection{Motivation and goals of this paper}

The main motivation and goals of the present manuscript are: 
\begin{itemize}

\item We will provide a full detailed theoretical quantum electrodynamical (QED) formulation of the quantum optics of intense laser-atom interactions. In particular, we describe the processes of High Harmonic Generation (HHG) and Above Threshold Ionization (ATI), and characterize their back-action on the coherent state of the driving laser field. Furthermore, we show how quantum operations (conditioning measurements)  on HHG and ATI can lead to coherent state superpositions with controllable quantum features. Such Schrödinger ``cat-like'' or ``kitten-like'' states offer, in principle, fascinating possibilities for applications in QTs.

\item  We will discuss what are the experimental conditions needed to control the quantum features of the light state after HHG and ATI. In particular, we show how it is possible to switch the quantum character of the generated  Schrödinger optical “cat-like” states to “kitten-like” states,  and how new observables, e.g. the photon number, inherit information about the quantum nature of the laser-matter interaction processes. Finally, we show the conditions under which the generation of high-photon number optical “cat” states is possible.  

\end{itemize}

All of this material will be presented in a didactic form.  We believe that our tutorial will contribute to the more complete understanding of the strong-field matter processes from a full quantum electrodynamical viewpoint, and to the development of new methods that naturally lead to the creation of massive superpositions and massive entangled states, such as high photon number coherent state superpositions with controllable quantum features. Generating, certifying, and applying such states  could greatly advance  quantum technology.

We attempt, {\it toutes proportiones gardées}, to shape our tutorial inspired by the best of the best: C. Cohen-Tannoudji's books \cite{tannoudji1992atom,cohen1997photons}, or the reference paper on quantum optics of dielectric media \cite{Roy1991pra}. Therefore, we first formulate a complete QED theory, including: general description of dynamics, starting from the minimal coupling $\bf p\cdot A$ (``velocity gauge") Hamiltonian, and discussing cutoff issues. We then discuss carefully going to the $\bf d\cdot E$ (``length gauge") description, considering also the divergent role of the ${\bf x}^2$ term, and introducing the dipole approximation for one atom.

We formulate the problem by assuming as initial state the coherent state of the modes, contributing to the fundamental laser pulse,  and vacuum otherwise. We employ various unitary transformations by going to the interaction picture with respect to the free-field Hamiltonian, shifting away the initial coherent state of the driving laser, and finally moving to the interaction picture with respect to the semi-classical Hamiltonian. We pay a lot of attention to the conditioning idea and its “advantages”: conditioning is here the basic, natural tool that allows generation of quantum correlated states.

The presentation is therefore quite technical, but we hope it will be indeed useful for newcomers to the area: it will allow them to understand the basics of the theory and  the known results, and, more importantly, the essence of the future challenges and questions to be studied and tackled.

The plan of this tutorial is thus as follows
\begin{itemize}
    \item Section I is an introduction, discussing the state of the art, motivations and objectives.
    \item Section II contains the preliminaries, and in particular provides a brief introduction to QED.
    \item Section III focuses on laser-matter interactions, and describes the basic Hamiltonian of a single active electron in a laser field, both in the ``velocity'' and ``length'' gauges. We discuss then various approximations used (the dipole approximation, in particular), and the dynamics of the systems.
    \item Section IV is devoted to the discussion of the conditioning methods, which provide our basic tool for the engineering of massively quantum correlated states. 
    \item Section V discusses the experimental aspects of the proposed quantum engineering approach based on conditioning. 
    \item Finally, Sections VI and VII contain a discussion of the results, and the outlook for future investigations, respectively. 
\end{itemize}
The paper also contains 7 appendices with more technical aspects of the derivations, calculations, and numerical simulations. 

\section{Preliminaries}

\subsection{Electromagnetic field operator and Hamiltonian}

To derive  the microscopic response  of the matter to the optical field, and the respective change of the electromagnetic (EM) field quantities, we consider the following Hamiltonian \cite{tannoudji1992atom, cohen1997photons, vogel2006quantum}
\begin{align}
\label{eq:hamiltonian}
H = H_f + H_{A+f},
\end{align}
where the free field Hamiltonian $H_f$ consists of the transverse part of the electromagnetic field
\begin{align}
\label{eq:H_free_field}
H_f = \frac{1}{2} \int \dd^3r \left\{ \frac{1}{\epsilon_0} \vb{\Pi}^2(\vb{r}) + \frac{1}{\mu_0} \left[ \nabla \times \vb{A}(\vb{r})\right]^2 \right\}.
\end{align}

The dynamical field variables are given by the vector potential $\vb{A}(\vb{r})$, and the canonical conjugate field momentum $\vb{\Pi}(\vb{r}) = \epsilon_0 \pdv{t} \vb{A}(\vb{r},t)$, with the corresponding commutation relation given by $\left[ A_m(\vb{r}), \Pi_n(\vb{r}^\prime) \right] = i \hbar \delta^\perp_{mn}(\vb{r} - \vb{r}^\prime) $, where $\delta_{mn}^\perp(\vb{r} - \vb{r}')$ is the transverse $\delta$-function (see Appendix \ref{App:quantization:EM} for details). 
We will use the common procedure of expressing the dynamical variables in a plane wave expansion 
\begin{align}
\label{eq:vector_potential}
\mathbf{A} (\mathbf{r}) = \sum_\mu \int \dd^3k \left[ \sqrt{\frac{\hbar}{2 \epsilon_0 c k (2 \pi)^3}} g(k) \mathbf{\epsilon}_{\mathbf{k} \mu} a_{\mathbf{k}\mu} e^{i \mathbf{k } \cdot \mathbf{r}} + \operatorname{h.c.} \right],
\end{align}
where $\mathbf{\epsilon}_{\bf{k} \mu}$ is the polarization vector for the field mode $\vb{k} = (k_x,k_y,k_z)^T$ in the polarization mode $\mu$. We introduced here a regularizing factor $g(k)$, describing the cutoff of the matter-field coupling at high frequencies/high momenta (see Remark 1 below).  The bosonic creation- and annihilation operators $a_{\bf{k} \mu}^{(\dagger)}$ obey the commutation relation
\begin{align}
\left[ a_{\bf{k} \mu}, a^{\dagger}_{\bf{k}^\prime \nu} \right] = \delta_{\mu \nu} \delta(\bf{k} - \bf{k}^\prime).
\end{align}

Note that we have used the Coulomb gauge $\nabla \cdot \vb{A} = 0$ in which $\bf{\epsilon}_{\bf{k} \mu} \perp \bf{k}$. The canonical conjugate momentum field is given by  

\begin{align}\label{eq:field:momentum}
\vb{\Pi} (\vb{r}) = - i \sum_\mu \int \dd^3k  \sqrt{\frac{\hbar \epsilon_0 c k}{2(2\pi)^3}} g(k) \vb{\epsilon}_{\vb{k} \mu} a_{\vb{k}\mu} e^{i \vb{k} \cdot \vb{r}}  + \operatorname{h.c.} ,
\end{align}
such that we can express the free field Hamiltonian \eqref{eq:H_free_field} in the form 
\begin{align}
H_f =  \sum_\mu \int \dd^3k \, \hbar \omega_k \left(  a^\dag_{\bf{k} \mu} a_{\bf{k} \mu} + \frac{1}{2} \right),
\end{align}
with frequency $\omega_k = c |{\bf{k}}|$ of mode $\bf{k}$.

\subsection{\label{intro_coherent_state}Coherent states of light and their superposition}
From a classical perspective, the EM field generated by a laser source is often described as a \textcolor{black}{single mode} wave with a well-defined amplitude and phase. 
However, in the quantum theory of radiation the amplitude and the phase are conjugate variables, and can therefore not be determined with arbitrary accuracy in the same experiment. The state for which the product of the variances of those two quantities reaches the lower limit, and therefore provides the optimal description of the classical field, are given by \emph{coherent states of light}
\begin{align}
    \ket{\alpha} = e^{-\abs{\alpha}^2 /2} \sum_{n=0}^\infty \frac{\alpha^n}{\sqrt{n!}} \ket{n}, 
\end{align}
which is a coherent superposition of Fock states $\ket{n}$ \textcolor{black}{of a single mode}, with $\alpha \in \mathbbm{C}$ the coherent state amplitude. Coherent states of light are said to be classical states of light, and their properties can be described by classical EM theory.

In order to provide physical intuition on how coherent states are generated, we consider the most basic example of a light source which is that of a \textit{classical} charge described by the current $\vb{J}(\vb{r},t)\in \mathbbm{R}^3$ \cite{ScullyBook}. The charge current couples to the EM field \textcolor{black}{(now for the general multi-mode case)} via the vector potential, given by the operator $\vb{A}(\vb{r},t)$ in \eqref{eq:vector_potential}. The Schrödinger equation describing the dynamics of the state of the EM field is
\begin{equation}
    i \hbar \dv{t}\ket{\phi(t)}
        = \int \dd^3 r \  
            \vb{J}(\vb{r},t) \cdot \vb{A}(\vb{r},t) \ket{\phi(t)},
\end{equation}
with the solution given by
\begin{equation}
    \begin{aligned}
    \ket{\phi(t)}
        &= \mathcal{T}
            \exp{-\dfrac{i}{\hbar}
                \int^t_{t_0}\dd t' \int \dd^3 r
                    \vb{J}(\vb{r},t^\prime) \cdot \vb{A}(\vb{r},t^\prime)}
                        \ket{\phi(t_0)}
        \\ &= \prod_{\vb{k}\mu} 
                e^{i\varphi_{\vb{k}\mu}}
                \exp[\alpha_{\vb{k},\mu}(t)a_{\vb{k}\mu}^\dagger
                    -\alpha^*_{\vb{k},\mu}(t)a_{\vb{k}\mu}]
                \ket{\phi(t_0)},
    \end{aligned}
\end{equation}
\textcolor{black}{where we have used the definition of the vector potential in \eqref{eq:vector_potential}, and} the initial state of the EM field $\ket{\phi(t_0)}$.
Thus, the coupling of the charge current to the EM field induces a shift of the initial state via the displacement operator
\begin{equation}
    D(\alpha_{\vb{k}\mu})
        = \exp[\alpha_{\vb{k}\mu} a_{\vb{k}\mu}^\dagger
                -\alpha^*_{\vb{k}\mu}a_{\vb{k}\mu}].
\end{equation}

This operator shifts each mode of the initial state by a quantity $\alpha_{\vb{k},\mu}(t)$ given by
\begin{equation}
    \alpha_{\vb{k},\mu}(t) 
        = \dfrac{i}{\hbar}\Tilde{g}(k)
            \int^t_{t_0} \dd t' \int \dd^3 r \
                \vb{J}(\vb{r},t) \cdot \vb{\epsilon}_{\vb{k},\mu} e^{-i\omega_k t + i \vb{k}\cdot \vb{r}},
\end{equation}
which corresponds to the Fourier transform of the classical charge current. Thus, up to a global phase, we can define the coherent state of light in the mode $\{\vb{k},\mu\}$ when the displacement operator is acting on the initial vacuum state 
\begin{equation}
    \ket{\alpha_{\vb{k}\mu}}
        = \exp[\alpha_{\vb{k},\mu}(t)a_{\vb{k}\mu}^\dagger
                    -\alpha^*_{\vb{k},\mu}(t)a_{\vb{k}\mu}] \ket{0_{\vb{k}\mu}},
\end{equation}
which are eigenstates of the annihilation operator $a_{\vb{k}\mu} \ket{\alpha_{\vb{k}\mu}} = \alpha_{\vb{k}\mu} \ket{\alpha_{\vb{k}\mu}}$. Hence, the oscillation of a classical charge current generates states of light that satisfy the lower bound in the uncertainty relation between the field amplitude and phase.
Furthermore, this uncertainty keeps saturated during their free-field evolution, that is, these states \textcolor{black}{remain coherent states} when they freely propagate.

For the reasons stated before, coherent states are said to be \emph{classical states}, since their properties can be recovered by the classical EM theory. 
However, despite the fact that coherent states are classical states of light, the superposition of two different coherent states can exhibit genuine quantum features without a classical counterpart \cite{Haroche-book}. The generic form of a coherent state superposition (CSS) in a single field mode reads
\begin{equation}\label{eq:coh:sup:gen}
    \ket{\phi} = \sum_i a_i \ket{\alpha_i}.
\end{equation}

One of the most well-known examples of CSS are the so-called Schrödinger cat states \cite{Haroche-book}, which consist of the superposition of two macroscopically distinguishable coherent states.  
The superposition of coherent states can lead to highly non-classical features which can be witnessed in different ways, e.g. by means of negativities in their Wigner function representation \cite{hudson_when_1974,kenfack_negativity_2004}. Furthermore, these states can be combined with linear optical elements such as beam splitters \cite{asboth_coherent-state_2004} to generate entangled coherent states \cite{sanders_entangled_1992}, which are useful in many different applications of quantum information science \cite{gilchrist_schrodinger_2004} such as quantum communications \cite{jouguet_experimental_2013}, quantum computation \cite{lloyd_quantum_1999, ralph_quantum_2003} and quantum metrology \cite{joo_quantum_2011}. 

\begin{figure}
    \centering
    \includegraphics[width = 1.\columnwidth]{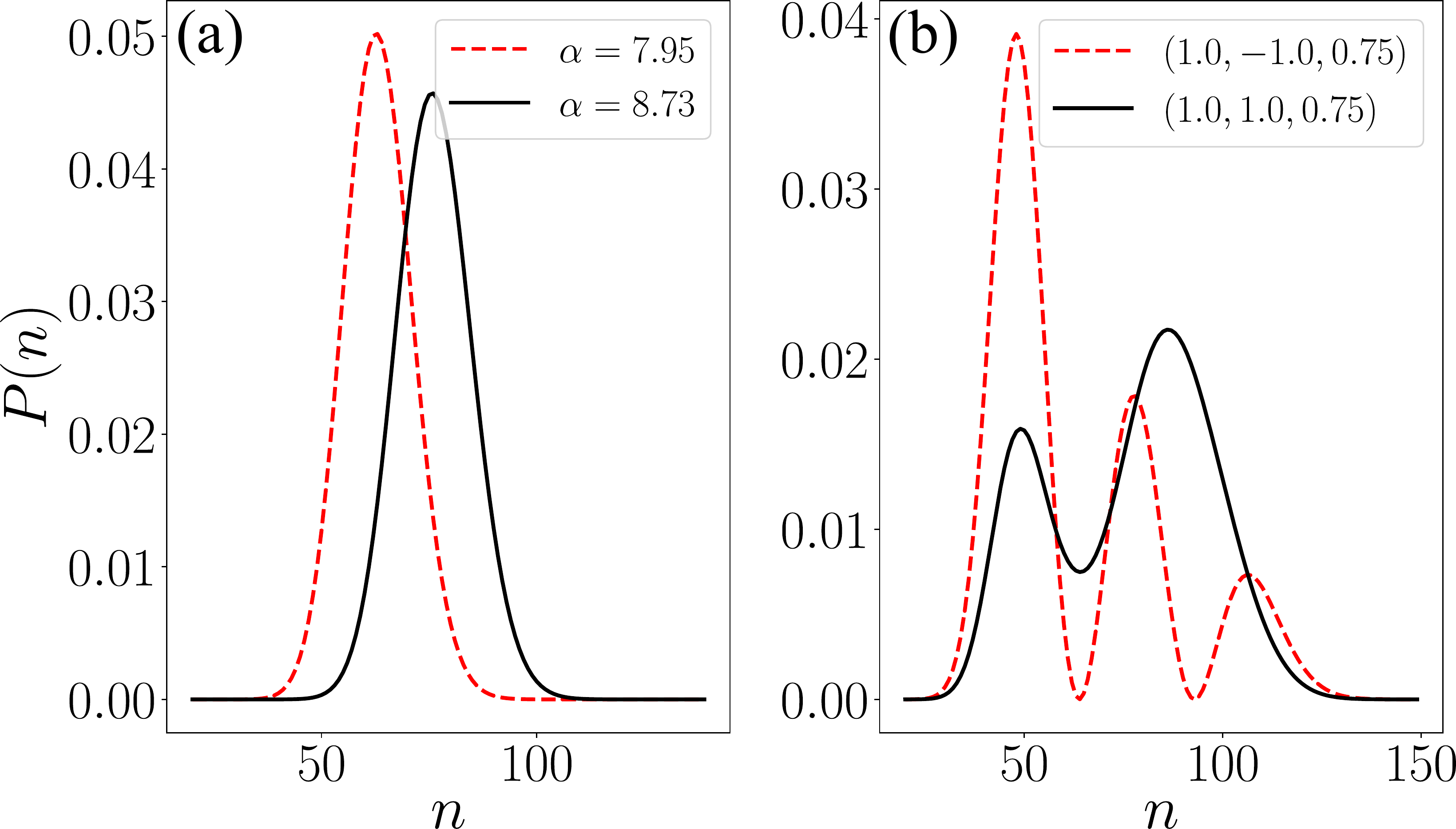}
    \caption{Photon number probability distribution for (a) two coherent states with $\alpha = 7.95$ (red dashed curve) and $\alpha = 8.73$ (black solid curve), and (b) for a superposition of three coherent states of the form $\ket{\phi} = a_1 \ket{7} + a_2\ket{9} + a_3 \ket{10}$, with $(a_1,a_2,a_3)=(1.0,-1.0,0.75)$ (red dashed curve) and $(a_1,a_2,a_3)=(1.0,1.0,0.75)$ (black solid curve). The coefficients in the superposition are shown up to a normalization which,  nonetheless, has been taken into account when doing the plots. For the red dashed curve probability distribution in plots (a) and (b), the mean photon number is $\langle n \rangle \approx 63.2$ while for the black solid line is $\langle n \rangle \approx 76.2$.}
    \label{Fig:coherent:superp:prob}
\end{figure}

In the context of the present paper, we are mainly interested in generating non-classical field states in the fundamental driving laser mode, and therefore compute the backaction on this mode due to the interaction with a gas medium. To measure the change of the driving laser mode, we will consider the photon number probability distribution of coherent states and their superposition. The probability distribution of finding $n$ photons in a \textcolor{black}{single mode} coherent state with amplitude $\alpha$ is Poissonian
\begin{equation}
    P(n)
        = \lvert \braket{n}{\alpha}\rvert^2
        = e^{-\lvert \alpha \rvert^2}
            \dfrac{\lvert \alpha \rvert^{2n}}{n!},
\end{equation}
where the mean $\expval{n}=\bra{\alpha} a^\dagger a\ket{\alpha} = \abs{\alpha}^2$ and the variance $(\Delta n)^2 = \abs{\alpha}^2$ are equal. 
A photon distribution is shown in Fig.~\ref{Fig:coherent:superp:prob}(a) for $\alpha = 7.95$ (red dashed curve) and $8.73$ (black solid curve). Both distributions show a maximum peak at the mean photon value of the distribution, $\langle n \rangle \approx 63.2$ and $\langle n \rangle \approx 76.2$, respectively.
On the other hand, for the coherent state superposition presented in \eqref{eq:coh:sup:gen} the photon number probability distribution is given by
\begin{equation}
    P(n) = \Bigg\lvert 
                \sum_{i} a_i e^{-\lvert\alpha_i\rvert^2/2} 
                \dfrac{\big(\alpha_i\big)^n}{\sqrt{n!}}
            \Bigg\rvert^2,
\end{equation}
which is not a Poissonian distribution anymore due to the interference between the different coherent states $\ket{\alpha_i}$. In Fig.~\ref{Fig:coherent:superp:prob}(b), we show the obtained probability distribution of a superposition of three coherent states of the form $\ket{\phi} = a_1 \ket{7} + a_2\ket{9}+ a_3\ket{10}$, when considering different relative phases of the corresponding probability amplitudes $a_i$. In particular, the mean photon number of the two distributions that are shown coincide with the ones considered for the coherent states in Fig.~\ref{Fig:coherent:superp:prob}(a). Yet, the distributions are very different as in for the coherent state superposition, and we observe a multipeak structure. The main differences between the black solid line and the red dashed curve in (b) lies in the relative phase of the $\ket{9}$ coherent state as it differs in $\pi$ from one case to the other. Depending on this relative phase, we either get a multipeak probability distribution, or a situation where two of the peaks are very close to each other and are indistinguishable.

\section{Intense laser-matter interaction}\label{Sec:las:mat:int}

\subsection{Minimal coupling Hamiltonian - the velocity gauge}

The primary goal of the present tutorial is to describe the dynamics of many particles in gas phase in strong laser fields, taking into account the full quantum electrodynamic nature of the intense driving laser field. We shall consider $N$ atoms/molecules located  at $\vb{R}_\alpha$, $\alpha=1, \ldots, N$ in the single active electron (SAE) approximation. 
\textcolor{black}{Readers familiar with fully quantized light-matter interaction in the $\vb{p} \cdot \vb{A}$ and $\vb{r}\cdot \vb{E}$ gauge can skip the next subsections, and start with section \ref{sec_dynamics_HHG} directly.}

The Hamiltonian $H_{A+f}$ in \eqref{eq:hamiltonian} describes the coupling of the collection of charges in the SAE approximation with the EM field. It is given by the minimal-coupling Hamiltonian describing the interaction of an electron of mass  $m$ and charge $q = - e$ with the EM field \cite{tannoudji1992atom, cohen1997photons, vogel2006quantum}
\begin{align}
H_{A+f} = \sum_\alpha \frac{[\vb{p_\alpha} - q \vb{A}(\vb{r}_\alpha)]^2}{2m} + V_{\rm at}(\vb{r}_\alpha,\vb{R}_\alpha),
\end{align}
where $\vb{p}_\alpha = m \dot{ \vb{r}}_\alpha + q \vb{A}(\vb{r}_\alpha)$ is the canonical conjugate momentum of the electron with coordinate $\vb{r}_\alpha$, which obey $[r_{i\alpha}, p_{j \alpha^\prime} ] = i \hbar \delta_{\alpha \alpha^\prime} \delta_{i j}$, and $V_{\rm at}(\vb{r}_\alpha,\vb{R}_\alpha)$ is the  effective potential felt by the single active electron (see Remark 2 below). 

The total Hamiltonian is then of the following form
\begin{equation}\label{eq:H_pAdip}
    \begin{aligned}
    H &= \frac{1}{2} \int \dd^3r \left\{ \frac{1}{\epsilon_0} \vb{\Pi}^2(\vb{r}) + \frac{1}{\mu_0} \left[ \nabla \times \vb{A}(\vb{r})\right]^2 \right\}\\
    &\quad+ \sum_\alpha \left[\frac{[\vb{p_\alpha} + e\vb{A}(\vb{r}_\alpha)]^2}{2m} + V_{\rm at}(\vb{r}_\alpha,\vb{R}_\alpha)\right],
    \end{aligned}
\end{equation}
and will be used for the transformation to the length gauge. Furthermore, at this point we present some remarks that need to be taken into account in the present description:
\\

\noindent{\it Remark 1.} Note that we are considering here a non-relativistic theory of electrons interacting with the quantum EM field. Moreover, we consider the dipole approximation. Obviously,  the considered model is ultraviolet divergent. This divergence has no physical meaning, and is related partially to the incorrect treatment of large wave-vectors in the dipole approximation. It cannot be removed  by a satisfactory renormalization procedure, which leads to non-physical run-away solutions \cite{aichelburg_exactly_1977,aichelburg_model_1977}. Instead, we introduce a form factor tempering the coupling for high frequencies, just as the $e^{i \vb{k} \cdot \vb{r}}$
term  does. Following \cite{rzazewski_initial_1976,rzazewski_initial_1980}, we take the form factor
\begin{align}
\label{eq:form_factor}
g(k)= \frac{\Gamma}{\sqrt{\Gamma^2 + k^2}},    
\end{align}
where the cut-off parameter $\Gamma$ is much larger than the laser frequency, and is of the order of $\Gamma\simeq d^{-1}$, where $d$ is equal to the characteristic amplitude of the electronic oscillations. With such a choice, the atomic lifetime remains finite, while the frequency (Lamb) shift is linearly divergent as $\Gamma\to\infty$.\\

\noindent{\it Remark 2.} 
The  effective potential $V_{\rm at}(\vb{r}_\alpha,\vb{R}_\alpha)$  felt by the single active electron in a given atom/molecule centered at $\vb{R}_\alpha$, and for neutral atoms, has a long range tail corresponding to the Coulomb interaction between the charges, related to the parallel component of the electric field. For any specific atom/molecule, with exception of atomic hydrogen,  it has to be calculated carefully using advanced Hartree-Fock methods \cite{tong_density-functional_1997,tong_empirical_2005}.\\

\subsection{Transformation to the length gauge}

In the following we are interested in localized systems, e.g., atoms or molecules, which are located at a position $\vb{R}_\alpha$. Moreover, for the moment we only consider systems within the SAE approximation. Since we are dealing with localized atomic or molecular systems, in which a single active  bound electron is driven by the intense laser field,  we further use that the induced polarization is due to the displacement of electrons with charge $e$, and we can define the polarization via 
\begin{align}
\label{eq:def_polarization}
\nabla \cdot \vb{P}(\vb{r}) = e \sum_\alpha \delta(\vb{r} - \vb{r}_\alpha).
\end{align}

To describe the interaction of intense laser fields with matter, we transform the Hamiltonian \eqref{eq:H_pAdip} into a more convenient form, in which the dipole moment of the electrons
\begin{align}
    \mathbf{d} = \sum_\alpha q_\alpha \mathbf{r}_\alpha = -e \sum_\alpha \vb{r}_\alpha 
\end{align}
is coupled to the electric field $\vb{E}(\vb{r})$. We will therefore first obtain a multipolar-coupling Hamiltonian by performing a unitary transformation followed by the dipole approximation. 

\subsubsection{Power-Zienau-Woolley transformation}

The unitary transformation which turns the minimal-coupling Hamiltonian \eqref{eq:H_pAdip} into the multipolar-coupling is given by the Power-Zienau-Woolley (PZW) transformation \cite{tannoudji1992atom, cohen1997photons, vogel2006quantum}
\begin{align}
\label{eq:PZW}
T \equiv \exp{\frac{i }{\hbar} \int \dd^3 r \mathbf{P}(\bf{r}) \cdot \mathbf{A}(\mathbf{r})},
\end{align}
where the atomic polarization $\mathbf{P}(\mathbf{r})$ given by
\begin{equation}
    \begin{aligned}
    \mathbf{P}(\mathbf{r}) &= -e \sum_\alpha \int_0^1 \dd s \left( \mathbf{r}_\alpha - \mathbf{R}_\alpha \right)
    \\
    & \hspace{2.5cm}\times
    \delta \left( \mathbf{r} - \mathbf{R}_\alpha - s \left( \mathbf{r}_\alpha - \mathbf{R}_\alpha \right) \right),
    \end{aligned}
\end{equation}
solves \eqref{eq:def_polarization}. 
The PZW transformation obviously leaves both the vector potential $\vb{A}^\prime(\vb{r}) = T \vb{A}(\vb{r}) T^\dagger = \vb{A}(\vb{r})$ and the electron coordinates $\vb{r}^\prime_\alpha = T \vb{r}_\alpha T^\dagger= \vb{r}_\alpha$ invariant. However, the canonical conjugate field momentum $\vb{\Pi}(\vb{r})$, and the electron canonical momentum $\vb{p}_\alpha$ transform according to 
\begin{equation}
\label{eq:new_field_momentum}
    \vb{\Pi}^\prime(\vb{r}) = T \vb{\Pi}(\vb{r}) T^\dagger = \vb{\Pi}(\vb{r}) - \vb{P}_\perp(\vb{r}),
\end{equation}
\begin{equation}\label{eq:new_electron_momentum}
    \begin{aligned}
    \mathbf{p}^\prime_\alpha  &= T \mathbf{p}_\alpha T^\dagger
     = \mathbf{p}_\alpha + e \mathbf{A}(\mathbf{r}_\alpha)
    \\
    & \quad + e \int_0^1 \dd s \, s (\mathbf{r}_\alpha - \mathbf{R}_\alpha) \\
    & \quad \times \left\{ \nabla \times \mathbf{A}(\mathbf{R}_\alpha + s (\mathbf{r}_\alpha - \mathbf{R}_\alpha))  \right\},
    \end{aligned}
\end{equation}
where $\vb{P}_\perp(\vb{r}) $ is the transverse part of the polarization $\vb{P}(\vb{r})$. We can now express the minimal-coupling Hamiltonian \eqref{eq:H_pAdip} in terms of the new dynamical variables from \eqref{eq:new_field_momentum}, \eqref{eq:new_electron_momentum}, and we obtain the multipolar-coupling Hamiltonian $H' = T H T^\dagger$ (see Appendix \ref{App:PWZ:transf}). 
In this multipolar Hamiltonian we will only consider the term in which the transverse displacement field $\vb{\Pi}^\prime (\vb{r}) = -  [\epsilon_0 \vb{E}_\perp (\vb{r}) + \vb{P}_\perp (\vb{r})]$ is coupled to the atomic polarization $\vb{P} (\vb{r})$.
The terms describing the interaction of the paramagnetic magnetization with the magnetic induction field $\vb{B}(\vb{r}) = \nabla \times \vb{A}(\vb{r})$, and the diamagnetic energy of the system, which is quadratic in the magnetic induction field, respectively,  will be neglected in the following when performing the dipole approximation.

\subsubsection{Dipole approximation}

Since we are considering electrons which are bound by the Coulomb potential $V_C$ to the atomic core at position $\vb{R}_\alpha$, we can expand the field variables in powers of $\vb{r}_\alpha - \vb{R}_\alpha$. Within the electric dipole approximation, we only keep the zeroth-order term, such that the vector potential $\vb{A}(\vb{R_\alpha})$ and the canonical conjugate momentum field $\vb{\Pi}(\vb{R}_\alpha)$ are evaluated at the atomic position $\vb{R}_\alpha$. 
Therefore, the interaction Hamiltonian $H_{int}^\prime$ in the multipolar form simplifies significantly. Since the vector potential loses its spatial dependence in the electric dipole approximation, the last two terms in \eqref{eq:H_multipolar} vanish due to $\nabla_{\vb{r}_\alpha} \times \vb{A}(\vb{R}_\alpha) = 0$. Furthermore, the canonical momentum field is evaluated at the atomic position $\vb{R}_\alpha$, and we have
\begin{align}
H_{int}^\prime = - \vb{d} \cdot \vb{E}_\perp = e \sum_\alpha \vb{r}_\alpha \cdot \vb{E}_\perp(\vb{R}_\alpha), 
\end{align}
where $\vb{d} = - e \sum_\alpha  \vb{r}_\alpha $ is the electric dipole moment for global neutral systems. Now, the total Hamiltonian after the unitary PZW transformation within the dipole approximation has the following form
\begin{align}
\label{eq:hamiltonian_dip1}
\nonumber
H^\prime = &  \sum_\mu \int \dd^3k \, \hbar \omega_k \left( a^\dag_{\bf{k} \mu} a_{\bf{k} \mu} + \frac{1}{2} \right) \\
& + \sum_\alpha \frac{\mathbf{p}^2_\alpha}{2 m} + V_{\rm at}(\vb{r}_\alpha, \vb{R}_\alpha)  \\
& + e \sum_\alpha \vb{r}_\alpha \cdot \vb{E}_\perp(\vb{R}_\alpha) + \frac{1}{2\epsilon_0} \int d^3r \mathbf{P}_\perp^2 (\mathbf{r})\nonumber .
\end{align}

\subsubsection{Renormalization of polarization self-energy}

The last term of \eqref{eq:hamiltonian_dip1} is, in our non-relativistic theory, and within the dipole approximation, strictly speaking infinite. If we use the large momentum cutoff, it becomes regularized, but gives rise to non-physical quadratic contribution to the electronic potential. We thus ``renormalize'' this term as above, including it into the effective potential $\tilde V_{\rm at}(\vb{r}_\alpha, \vb{R}_\alpha)$. The final  Hamiltonian in the dipole approximation (sometimes known as length gauge) reads 
\begin{align}
\label{eq:hamiltonian_dip_final}
H_{\rm dip} = &  \sum_\mu \int \dd^3k \, \hbar \omega_k \left( a^\dag_{\bf{k} \mu} a_{\bf{k} \mu} + \frac{1}{2} \right) \\
& \nonumber + \sum_\alpha \left[\frac{\mathbf{p}^2_\alpha}{2 m} + \tilde V_{\rm at}(\vb{r}_\alpha, \vb{R}_\alpha) + e\vb{r}_\alpha \cdot \vb{E}_\perp (\vb{R}_\alpha)\right].
\end{align}
This Hamiltonian will provide the starting point for the further discussion.

\subsection{\label{sec_dynamics_HHG}Dynamical evolution}

In this section we obtain the dynamical evolution of the EM field. We therefore solve the time-dependent Schrödinger equation (TDSE) \cite{lewenstein2021generation, rivera2022strong}
\begin{align}
\label{eq:TDSE_1}
i \hbar \dv{}{t} |{\psi(t)}\rangle = H_{\rm dip} |{\psi(t)}\rangle,
\end{align}
with the Hamiltonian of \eqref{eq:hamiltonian_dip_final}. To describe intense laser-matter interactions, we assume that the electrons are initially in the ground state $\otimes_\alpha \ket{g_\alpha}$, and that the EM field of the laser pulse is described by \textcolor{black}{multimode} coherent states $\ket{\alpha_{\vb{k} \mu}}$. The spectral profile, and wavevector of the laser pulse are centered around $\omega_L$ and $\mathbf{k}_{L}$, respectively, and the field is in the polarization mode $\mu$. All higher frequency modes (i.e.~in particular high harmonic modes) are assumed to be in the vacuum state $\ket{0_{\vb{k}\mu}}$ such that the initial condition is given by 
\begin{align}\label{eq:init:state}
\ket{\psi(t_0)} = \bigotimes_\alpha\ket{g_\alpha} \bigotimes_{\vb{k},\mu \simeq \mathbf{k}_{L \mu}} \ket{\alpha_{\vb{k}\mu}} \bigotimes_{\vb{k} \gg \vb{k}_L}  \ket{0_{\vb{k} \mu}}.
\end{align}

\begin{figure}
    \centering
    \includegraphics[width = 1.\columnwidth]{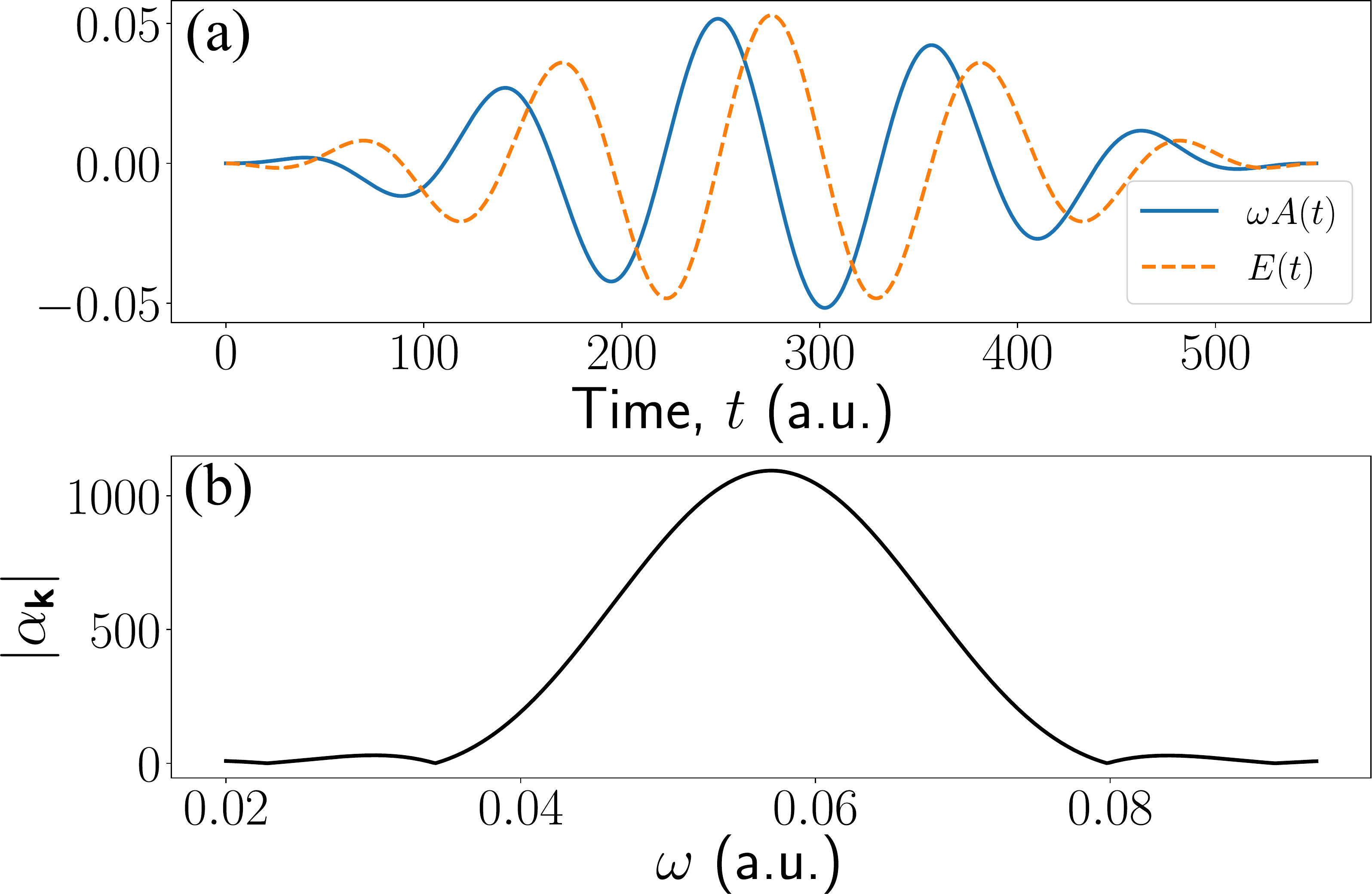}
    \caption{In (a) we consider a linearly polarized electromagnetic field with a sinusoidal squared envelope, central frequency $\omega = 0.057$ a.u., field amplitude $E=0.053$ a.u. and $n_{cyc} = 5$ number of cycles. The orange dashed curve shows the electric field while the blue solid curve the vector potential. In (b) we show the dependence of $\lvert\alpha_{\vb{k},\mu}\rvert$ on the frequency mode.}
    \label{Fig:behaviour:alpha}
\end{figure}

We want to emphasize that the coherent state amplitudes $\alpha_{\vb{k},\mu}$ of the driving laser source are a function of the field momentum $\vb{k}$ and polarization $\mu$, encoding the information of the spectral characteristics of the driving laser. For each frequency mode $\omega_{\vb{k},\mu}$, with given polarization $\mu$, the respective amplitude and phase varies. This allows to consider arbitrary driving field configurations including different frequency or spatial modes, and to include complex polarization states \cite{Haroche-book}. 
The magnitude of each mode $\abs{\alpha_{\vb{k},\mu}}$ is proportional to the frequency spectrum of the considered EM driving field, and is centered around the driving laser frequency $\omega_{\vb{k}_L}$. In Fig.~\ref{Fig:behaviour:alpha} we illustrate the magnitude $\abs{\alpha_{\vb{k},\mu}}$ of the spectral decomposition for a driving laser with a sinusoidal squared envelope with $n_{cyc}=5$ cycles of duration. The spectral amplitude is given by $\alpha_{\vb{k},\mu} \propto \int \dd t \epsilon_{\vb{k},\mu}\cdot \vb{A}(t) e^{-i\omega_k t}$.

We shall now solve the TDSE \eqref{eq:TDSE_1} by transforming into the interaction picture with respect to the free-field Hamiltonian $H_f$ such that $|{\psi^\prime (t)}\rangle = \exp[i H_f t / \hbar] \ket{\psi(t)}$ solves the following Schrödinger equation
\begin{equation}
\begin{aligned}
\label{eq:TDSE_trafo1}
    i \hbar \dv{}{t} |{\psi^\prime(t)}\rangle &= \sum_{\alpha} \bigg( \frac{\vb{p}^2_\alpha}{2 m_\alpha} +  \tilde V_{\rm at}(\vb{r}_\alpha, \vb{R}_\alpha)
    \\& \hspace{1.5cm}
        + e\vb{r}_\alpha\cdot \vb{E}_Q(t,\vb{R}_\alpha)\bigg) |{\psi^\prime(t)}\rangle ,
\end{aligned}
\end{equation}
where the time-dependent electric field operator is now given by 
\begin{equation}
\begin{aligned}
\label{eq:E_quantum}
\vb{E}_Q(t, \vb{R}_\alpha) &=  i \sum_\mu \int \dd^3k \tilde g(k) \epsilon_{k \mu} \big[ a_{k \mu} e^{-i \omega_k t +i \mathbf{k } \cdot \mathbf{R}_\alpha} \\
&  \quad - a^\dagger_{k \mu} e^{i \omega_k t-i \mathbf{k } \cdot \mathbf{R}_\alpha} \big].
\end{aligned}
\end{equation}
To separate the contribution of the classical electric field from the quantum corrections associated to it, we perform a unitary transformation that shifts the initial coherent state of the field, $\bigotimes_{\vb{k},\mu \simeq \vb{k}_{L,\mu}}\ket{\alpha_{\vb{k} \mu}}$, to the vacuum state. This is performed by applying the displacement operators $D^\dagger(\alpha_{k}) = D(- \alpha_{k})$, such that the new state $\ket{{\tilde \psi(t)}} = \bigotimes_{\vb{k},\mu \simeq \vb{k}_{L,\mu}} D^\dagger(\alpha_{\vb{k}, \mu}) \ket{\psi^\prime(t)}$ obeys the Schrödinger equation
\begin{equation}
\begin{aligned}
\label{eq:TDSE_shifted}
i \hbar \dv{}{t} |{\tilde \psi(t)}\rangle = & \sum_{\alpha} \bigg[ \frac{\vb{p}^2_\alpha}{2 m_\alpha} + \tilde V_{\rm at}(\vb{r}_\alpha, \vb{R}_\alpha) \\
&  - \vb{d} \cdot \vb{E}_{cl}(t, \vb{R}_\alpha) - \vb{d} \cdot \vb{E}_Q(t, \vb{R}_\alpha) \bigg] |{\tilde \psi(t)}\rangle,
\end{aligned}
\end{equation}
with the new initial condition given by 
\begin{align}
    \ket{\tilde \psi (t_0)} & = D^\dagger(\alpha_{k_L}) \ket{\psi(t_0)} \nonumber \\
    & = \bigotimes_\alpha\ket{g_\alpha} \bigotimes_{\vb{k},\mu \simeq \mathbf{k}_{L \mu}} \ket{0_{\vb{k}\mu}} \bigotimes_{\vb{k} \gg \vb{k}_L}  \ket{0_{\vb{k} \mu}}    
\end{align}

The classical part of the electric field is given by
\begin{equation}
\begin{aligned}
\vb{E}_{cl}(t, \vb{R}_\alpha) &=   i \sum_\mu \int \dd^3 k \tilde g(k) \epsilon_{k \mu} \big[ \alpha_{k, \mu}(k) e^{-i \omega_k t +i \mathbf{k } \cdot \mathbf{R}_\alpha} \\
&\quad - \alpha_{k, \mu}^*(k) e^{i \omega_k t-i \mathbf{k }\cdot\mathbf{R}_\alpha} \big] ,
\end{aligned}
\end{equation}
where $ \tilde g(k) = g(k) \sqrt{\hbar c k/(2 \epsilon_0 (2 \pi)^3)} $ with $g(k)$ from \eqref{eq:form_factor}, and the quantum part $\vb{E}_Q(t)$ is given by Eq. \eqref{eq:E_quantum}. The dependence of the amplitude $\alpha_{\vb{
k} \mu}(k)$ of mode $\{\vb{k}, \mu \}$ on the field momentum takes into account the spectral distribution of the driving laser field as shown in Fig. \ref{Fig:behaviour:alpha} (b).
By virtue of the separation of the classical and quantum part of the electric field, the semi-classical strong field Hamiltonian appears in \eqref{eq:TDSE_shifted}, i.e.
\begin{align}
H_{sc}(t) = \sum_{\alpha} \left[ \frac{\vb{p}_\alpha^2}{2 m}  + \tilde V(\vb{r}_\alpha, \vb{R}_\alpha) + e \vb{r}_\alpha \cdot \vb{E}_{cl}(t,\vb{R}_\alpha) \right].
\end{align}

We therefore transform the atomic variables into the interaction picture with respect to the semi-classical Hamiltonian $H_{sc}(t)$ by applying the unitary transformation $U_{sc}(t) = \mathcal{T} \exp[- i  \int_{t_0}^t H_{sc}(t^\prime) dt^\prime]$, with the time-ordering operator $\mathcal{T}$, such that the state $\ket{\Psi(t)} \equiv U_{sc}^\dagger(t) \ket{\tilde \psi(t)}$ solves the Schrödinger equation
\begin{align}
\label{eq:TDSE_final}
i \hbar \dv{}{t}|{\Psi(t)}\rangle = e \sum_\alpha \vb{r}_\alpha(t) \cdot \vb{E}_Q(t, \vb{R}_\alpha) \ket{\Psi(t)},
\end{align}
where $\vb{r}_\alpha(t) = U_{sc}^\dagger(t) \vb{r}_\alpha U_{sc}(t)$ is the dipole moment operator in the interaction picture with respect to the semi-classical Hamiltonian.
Note that this expression is still exact and no approximations have been performed so far (based on the approximate Hamiltonian in \eqref{eq:hamiltonian_dip_final}). To describe the different processes in strong field physics, such as high harmonic generation (HHG) or above-threshold ionization (ATI), we will use the Schrödinger equation of Eq.~\eqref{eq:TDSE_final} as the starting point for the respective discussions, and approximations compatible with the strong field processes will be applied from here on. This will serve as the origin for the light engineering protocols in intense laser-matter interactions, which will be introduced in the subsequent sections.

\section{Quantum state engineering using intense laser fields}

The description of the dynamical evolution of the total system, formed by many electrons, with the EM field has, thus far, been exact under the Hamiltonian within the dipole approximation.
This interaction will entangle the electronic and EM field states \cite{stammer2022theory}, and many different processes can occur (see Introduction). To engineer the photonic degree of freedom for generating non-classical states of light, we can now perform specific operations on the total state. For instance, we can condition the total evolution on certain electronic states, such as the ground state or a continuum state, which corresponds to the process of HHG and ATI, respectively. The conditional evolution of the EM field modes then allows to apply approximations based on assumptions in agreement with intense laser-matter interaction. 

For instance, it was shown that the EM field state is in an entangled state between all the field modes \cite{stammer2022high, stammer2022theory}, which then further allows us to perform a second stage of conditioning by measuring particular field modes leading to a quantum operation acting on the remaining modes \cite{stammer2022theory}. This will, for instance, generate high photon number optical ``cat-like'' states in the IR \cite{lewenstein2021generation, rivera2022strong} or in the XUV regime \cite{stammer2022high}. Those measurements provide access to the non-classical character of the entangled state. 

In the following we will introduce these aspects with the prospect to engineer the quantum state of the EM field. 

\subsection{Conditioning on electronic ground state: HHG}

To describe the process of high harmonic generation in atomic systems, we consider only those cases in which the electron is found in its ground state, leading to the emission of radiation of high-order harmonics of the driving frequency. We therefore project the TDSE \eqref{eq:TDSE_final} on the electronic ground state of all atoms $\otimes_\alpha \ket{g_\alpha}$ \textcolor{black}{in the original laboratory frame}, such that 
\begin{align}
\label{eq:TDSE_projected_GS}
    i \hbar \dv{}{t} \ket{\phi(t)} = e \sum_\alpha \vb{E}_Q(t, \vb{R}_\alpha )\cdot \bra{g...g} \vb{r}_\alpha (t) \ket{\Psi(t)},
\end{align}
where we have defined the EM field state conditioned on the electronic ground state $\ket{\phi(t)} = \bra{g...g} \ket{\Psi(t)}$. To simplify the right-hand side we introduce the identity in the spirit of the SFA by neglecting excited electronic bound states
\begin{align}
    \label{eq:idendity}
    I = \bigotimes_\alpha \left( \dyad{g_\alpha} + \int \dd^3 v_\alpha \dyad{\vb{v}_\alpha} \right),
\end{align}
where we have denoted the continuum states as $\ket{\vb{v}_\alpha}$ corresponding to an electron with kinetic momentum $m\vb{v}$. We then obtain 
\begin{equation}
\begin{aligned}
\label{eq:TDSE_projection_GS}
    i \hbar \dv{}{t} \ket{\phi(t)} &=  e \sum_\alpha \expval{\vb{r}_\alpha(t)}\cdot \vb{E}_Q(t, \vb{R}_\alpha) \ket{\phi(t)} \\
    &\quad + e \sum_\alpha \int d^3 v_\alpha \bra{g_\alpha} \vb{r}_\alpha(t) \ket{\vb{v}_\alpha}\cdot \vb{E}_Q(t, \vb{R}_\alpha)
    \\& \hspace{4cm}
        \times\ket{\phi_\alpha(\vb{v}_\alpha,t)},
\end{aligned}
\end{equation}
where $\expval{\vb{r}_\alpha(t)} = \bra{g} \vb{r}_\alpha(t) \ket{g}$ is the time-dependent dipole moment expectation value in the electronic ground state, and we have defined the state of the EM field with one electron conditioned on a continuum state $\ket{\phi_\alpha(\vb{v}_\alpha,t)} = \bra{\vb{v}_\alpha g...g} \ket{\Psi(t)}$.
For the process of HHG we neglect the second term corresponding to the projection of the total state onto an electronic continuum state, which are hardly occupied at the end of the pulse since the electron recombines to its ground state during the process of HHG. However, note that this approximation neglects the continuum population at all times, and not only at the end of the pulse, though this contribution is assumed to be small compared to the ground state amplitude \cite{lewenstein1994theory}. We can thus proceed and solve
\begin{equation}
\begin{aligned}
\label{eq:TDSE_classical_charge}
    i \hbar \dv{}{t} \ket{\phi(t)} =  e \sum_\alpha \expval{\vb{r}_\alpha(t)}\cdot \vb{E}_Q (t, \vb{R}_\alpha) \ket{\phi(t)}.
\end{aligned}
\end{equation}

Since the field operator is linear in the creation and annihilation operator we can solve the TDSE exactly and obtain for the propagator
\begin{equation}
\begin{aligned}
    U(t,t_0) & = \mathcal{T} \exp{ -\frac{ie}{\hbar} \int_{t_0}^t \dd t^\prime  \sum_\alpha \expval{\vb{r}_\alpha(t^\prime)} \cdot \vb{E}_Q(t^\prime, \vb{R}_\alpha) } \\
    & = \prod_{\vb{k} \mu} D[\chi_{\vb{k},\mu}(t) ] \, e^{i \varphi_{\vb{k},\mu}(t)},
\end{aligned}
\end{equation}
which is equivalent to a multimode displacement operator, where 
\begin{equation}
\begin{aligned}
    \varphi_{\vb{k},\mu}(t) = &  \frac{e^2}{\hbar^2} \tilde g^2(k) \sum_{\alpha, \alpha^\prime}  \int_{t_0}^t \dd t_1 \int_{t_0}^{t_1} \dd t_2 \, \epsilon_{\vb{k} \mu} \cdot \expval{\vb{r}_\alpha (t_1)}\\
    & \times \epsilon_{\vb{k} \mu} \cdot \expval{\vb{r}_{\alpha^\prime} (t_2)} \sin[\omega_k (t_1 - t_2)].
\end{aligned}
\end{equation}

Thus, the initial state of the field, up to a phase factor, after the interaction is given by 
\begin{equation}
\begin{aligned}
    \ket{\phi(t)} &= \prod_{\vb{k}, \mu} D[\chi_{\vb{k},\mu}(t) ] \ket{\phi(t_0)}
    \\
    &= \mathcal{D}[\boldsymbol{\chi}(t)]
        \ket{\phi(t_0)},
\end{aligned}
\end{equation}
where $\mathcal{D}[\boldsymbol{\chi}(t)]$ is a shorthand notation for the product of all displacement operators $D[\chi_{\vb{k},\mu} (t) ]$ on the field modes, each of which shifts the initial state of the respective mode by an amplitude 
\begin{align}
\label{eq:chi_displacement}
    \chi_{\vb{k} ,\mu}(t) = - e \sum_\alpha \tilde g(k) e^{- i \vb{k} \cdot \vb{R}_\alpha} \int_{t_0}^t d \tau \, \vb{\epsilon}_{\vb{k} \mu} \cdot  \expval{\vb{r}_\alpha (\tau)} e^{i \omega_k \tau}.
\end{align}

To obtain the field state in the original laboratory frame, we have to undo the transformations in \eqref{eq:TDSE_trafo1} and \eqref{eq:TDSE_shifted}
\begin{align}
    \ket{\Phi(t)} =  \prod_{\vb{k}, \mu \simeq \vb{k}_{L,\mu}} D(\alpha_{\vb{k},\mu}) e^{-\frac{i}{\hbar} H_f t }  \ket{\phi(t)},
\end{align}
and we obtain 
\begin{align}\label{eq:state:condit:ground}
    \ket{\Phi(t)} = &  \bigotimes_{\vb{k}, \mu \simeq \vb{k}_{L,\mu}} e^{i \varphi_{\vb{k},\mu}(t)} \ket{[\alpha_{\vb{k},\mu} + \chi_{\vb{k},\mu}(t) ]e^{- i \omega_{\vb{k}} t }} \\
    & \bigotimes_{\vb{k}, \mu \gg \vb{k}_{L,\mu}} \ket{\chi_{\vb{k},\mu}(t) e^{-i \omega_{\vb{k}} t} } \nonumber,
\end{align}
where 
\begin{align}
    \varphi_{\vb{k},\mu} (t) = \operatorname{Im} \left[ \alpha_{\vb{k},\mu} \chi^*_{\vb{k},\mu}(t) \right].
\end{align}

To obtain the spectrum of the scattered light, i.e. the HHG spectrum, we note that in our treatment the interatomic correlations of the dipole moment are neglected \cite{sundaram1990high, stammer2022theory}, and that the spectrum is solely governed by the coherent part due to the classical charge current of the dipole moment expectation value (see physical explanation of the generation of coherent states from classical charge currents in section \ref{intro_coherent_state}). 
\textcolor{black}{It should be noted that due to this approximation, e.g. neglecting the continuum contribution in \eqref{eq:TDSE_projection_GS} which is only valid for small depletion of the ground state, the final state of the total EM field is given by coherent product states in \eqref{eq:state:condit:ground}. This can also be seen from the point of view that only the dipole moment expectation value is coupled to the field (see \eqref{eq:TDSE_classical_charge}), acting as a classical charge current, and all dipole moment correlations are neglected \cite{sundaram1990high, stammer2022theory}. Terms including higher orders of $\vb{E}_Q(t)$ would lead, for instance, to squeezing in the field modes.}
The coherent contribution to the HHG spectrum is proportional to (see Appendix \ref{App:HHG:Spectrum})
\begin{align}
\label{eq:HHG_spectrum_main}
    S(\omega_{\vb{k},\mu}) \propto \lim_{t \to \infty} \abs{\chi_{\vb{k},\mu}(t)}^2 = \tilde g(k)^2 N^2  \abs{\vb{\epsilon}_{\vb{k},\mu} \cdot \expval{\vb{d}(\omega_{\vb{k}})}}^2.
\end{align}

\begin{figure}
    \centering
    \includegraphics[width = 1.\columnwidth]{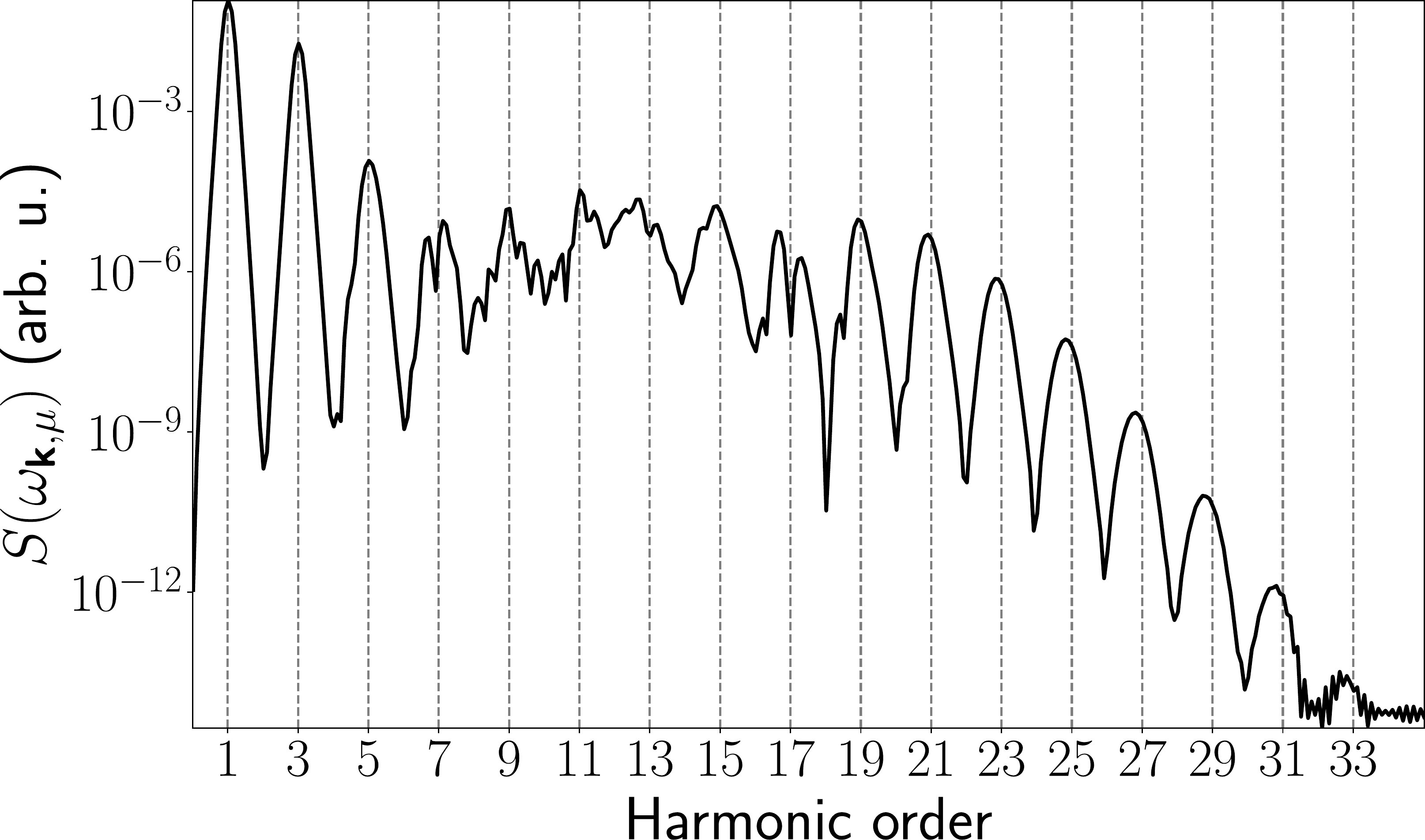}
    \caption{High harmonic generation spectrum obtained from the evaluation of Eq.~\eqref{eq:HHG_spectrum_main} \textcolor{black}{using the \textsc{Qprop} software \cite{Qprop}}, when a hydrogen atom in the ground state ($I_p=0.5$ a.u.) interacts with a linearly polarized electromagnetic field with a sinusoidal squared envelope that has 10 cycles, driving frequency $\omega = 0.057$ a.u., and electric field amplitude $E = 0.053$ a.u. The harmonic yield extends until the 21st harmonic, which constitutes the cutoff frequency of the spectrum, that is, the maximum energy that can be gained by the electron when accelerating in the continuum.}
    \label{fig:HHG:spectrum}
\end{figure}

We observe that the HHG spectrum is proportional to the Fourier components of the time-dependent dipole moment expectation value, which is in agreement with the results obtained when neglecting the correlations of the dipole moment operator \cite{sundaram1990high}. In Fig.~\ref{fig:HHG:spectrum} we show the HHG spectrum given by Eq.~\eqref{eq:HHG_spectrum_main}, for a linearly polarized driving field with a sinusoidal squared envelope and $n_{cyc} =10$ cycles. We observe the usual features of a HHG spectrum from linearly polarized, single color driving fields, namely that the peaks of the spectrum are located at the odd harmonics of the fundamental driving frequency, and that the spectrum exhibits a plateau structure that lasts until the cutoff region, here located around the 21st harmonic order. This cutoff frequency corresponds to the maximum kinetic energy that can be gained by the electron within the field~\cite{lewenstein1994theory}. We note that, by controlling the properties of the employed laser field (intensity, polarization, wavelength, etc.) as well as the atomic species used in the interaction region, the extension of the harmonic plateau can be further increased. 

\subsection{Conditioning on harmonic field modes}\label{Sec:Conditioning:Theory}

Thus far, we have presented the first step of conditioning, in which the post-selection was performed on the electronic degree of freedom, i.e. the electronic ground state, corresponding to the process of HHG.
In the next conditioning step for quantum state engineering of light, we perform measurements on the EM field itself. In particular, we perform a conditioning measurement on the harmonic field modes, leading to optical ``cat-like" states in the fundamental driving laser mode.

As a consequence of the non-linear interaction with the electron, the fundamental and harmonic modes get shifted by a quantity $\chi_{\vb{k},\mu}$, which depends on the time-dependent dipole moment of the electron and correlates the shift obtained in each of the modes. In fact, the mode that gets excited during the HHG process which takes into account all these correlations, is given by the corresponding number states $\ket{\Tilde{n}}$. Thus, the HHG process corresponds to the case in which these wavepacket modes get excited, i.e. whenever $\Tilde{n} \neq 0$. This allows us to introduce a set of positive-operator valued measure (POVM) \cite{NielsenandChuang} considering these two events \cite{stammer2022theory}
\begin{equation}
    \mathcal{A}_{\Tilde{n}}
        = \{\Pi_{\Tilde{0}},\Pi_{{\Tilde{n}\neq 0}}\},
\end{equation}
where $\Pi_{\Tilde{0}} = \dyad{\Tilde{0}}$ considers the case where no harmonic radiation is emitted, and $\Pi_{\Tilde{n}\neq0} = \sum_{\Tilde{n}\neq0} \dyad{\Tilde{n}}$ considers the case where the wavepacket mode is excited, i.e. harmonic radiation is generated. \textcolor{black}{Note that, the states $\ket{\tilde n}$ are the number states of the HHG wavepacket mode, and thus by definition all possible excitation sum up to the identity, i.e. $\Pi_{\Tilde{0}} + \Pi_{\Tilde{n}\neq0} = \mathbbm{1}$.} Having in mind that \textcolor{black}{the vacuum state of this wavepacket mode} $\ket{\Tilde{0}}$ state coincides with the state of the field prior to the interaction, i.e. the quantum state of the field modes in Eq.~\eqref{eq:init:state}, the conditioning on the case where harmonics are generated can be written as $\Pi_{\Tilde{n}\neq 0} = \mathbbm{1}-\Pi_{\Tilde{0}}$. Applying this operation to the state in \eqref{eq:state:condit:ground}, we get
\begin{equation}\label{eq:state:after:harm:cond}
    \begin{aligned}
    \ket{\Tilde{\Phi}(t)}
        &= \bigotimes_{\vb{k},\mu\simeq \vb{k}_{L,\mu}}
            e^{i\varphi_{\vb{k},\mu}(t)}
            \ket{[\alpha_{\vb{k},\mu} + \chi_{\vb{k},\mu}(t)]e^{-i\omega_k t}}
            \\
            &\quad\quad\quad\bigotimes_{\vb{k},\mu\gg \vb{k}_{L,\mu}}
            \ket{\chi_{\vb{k},\mu}(t)e^{-i\omega_k t}}
            \\
            &\quad-
                \bigotimes_{\vb{k},\mu\simeq \vb{k}_{L,\mu}}
            \xi_\text{IR}e^{i\varphi_{\vb{k},\mu}(t)}
            \ket{\alpha_{\vb{k},\mu}e^{-i\omega_k t}}
            \\
            &\quad\quad\quad\bigotimes_{\vb{k},\mu\gg \vb{k}_{L,\mu}}
            \xi_\text{HH}\ket{0_{\vb{k},\mu}},
    \end{aligned}
\end{equation}
where $\xi_\text{IR}$ and $\xi_\text{HH}$ are the overlaps between the initial state and the state we condition on, and are given by
\begin{equation}
\label{eq:xi_overlaps}
    \xi_\text{IR} 
        = \prod_{\vb{k},\mu \simeq \vb{k}_L,\mu} 
            \braket{\alpha_{\vb{k},\mu}e^{-i\omega_k t}}{[\alpha_{\vb{k},\mu} + \chi_{\vb{k},\mu}(t)]e^{-i\omega_k t}},
\end{equation}
\begin{equation}
    \xi_\text{HH} 
        =\prod_{\vb{k},\mu \gg \vb{k}_L,\mu} 
            \braket{0_{\vb{k},\mu}}{ \chi_{\vb{k},\mu}(t)e^{-i\omega_k t}}.
\end{equation}

The final state obtained in Eq.~\eqref{eq:state:after:harm:cond} is an entangled state, heralded by the emission of harmonic radiation, between all the field modes that get excited during the HHG process, and thus includes all the harmonic modes up to the cutoff frequency given by the harmonic spectrum. The harmonic cutoff depends on the driving laser field parameters, and the atomic species used for the HHG process \cite{krause_high-order_1992}. Furthermore, the amount of entanglement in the obtained state depends on the shift $\chi_{\vb{k},\mu}$ \cite{stammer2022high}, which can be controlled by means of the gas density in the interaction region \cite{rivera2022strong}.
In \cite{lewenstein2021generation,rivera2022strong}, the states in Eq.~\eqref{eq:state:after:harm:cond} were used to generate a coherent state superposition in the driving laser mode. The implemented scheme introduced a measurement where the radiation obtained in the harmonic modes was anticorrelated with the depletion obtained in the fundamental modes (see experimental section \ref{sec:experiment} for further details). In the present analysis, this measurement corresponds to a projective operation onto the harmonic modes, i.e. $\ket{\Tilde{\Phi}_{\vb{k}_{L,\mu}}} = \prod_{\vb{k},\mu\gg \vb{k}_{L,\mu}} \braket{\chi_{\vb{k},\mu}}{\Tilde{\Phi}(t)}$, such that the state of the fundamental mode is given by
\begin{equation} \label{eq:Paris_IR_Projection_HHG}
    \begin{aligned}
    \ket{\Tilde{\Phi}_{\vb{k}_{L,\mu}}}
        &= \bigotimes_{\vb{k},\mu\simeq \vb{k}_{L,\mu}}
                \ket{[\alpha_{\vb{k},\mu} + \chi_{\vb{k},\mu}(t)]e^{-i\omega_k t}}
                \\
                & \quad 
                    - \bigotimes_{\vb{k},\mu\simeq \vb{k}_{L,\mu}}
                    \xi_\text{IR}
                        \lvert\xi_\text{HH}\rvert^2
                        \ket{\alpha_{\vb{k},\mu}e^{-i\omega_k t}},
    \end{aligned}
\end{equation}
which represents a coherent state superposition \textcolor{black}{(CSS)} between two overlapping coherent states. \textcolor{black}{This superposition state has the particular form of two overlapping coherent states, i.e. two coherent states which are not too much separated in phase space, and can change from optical ``cat-like'' to ``kitten-like'' states depending on the parameter $\chi_{\vb{k}_L}$ \cite{rivera2022strong}. It is important to note that the distance between the two states in the superposition can not be too large since for increasing $\chi_{\vb{k}_L}$ the overlap $\xi_{IR}$ would decrease (see \eqref{eq:xi_overlaps}), and thus the second term in the superposition would vanish. However, this also brings an advantage for practical purposes, since traditional cat states are much more fragile against decoherence such as losses for increasing separation of the two states \cite{van2001entangled}. A fate the CSS generated from strong laser fields does not experience. Furthermore, this particular property allows this cat-like state to grow in size by increasing the amplitude $\alpha_{\vb{k}_L}$, leading to high photon number coherent state superpositions orders of magnitude higher than current schemes \cite{stammer2022high, rivera2022strong, stammer2022theory}. Those CSS show prominent non-classical features in their respective Wigner function (see Section \ref{Sec:Quantum:characterization}) showing the close analogy to optical cat states. An analysis of the properties, such as photon statistics and different non-classicality measures, of a CSS of this form can be found in \cite{othman2018quantum}.}

Similarly, one can instead perform the projective operation over the driving field mode, and all the harmonic modes except the one in mode $(\vb{k},\mu)$. This operation allows one to obtain coherent state superpositions that belong to the XUV region \cite{stammer2022high}
\begin{equation} \label{eq:Paris_XUV_Projection_HHG}
    \ket{\Tilde{\Phi}_{\vb{k},\mu}}
        = \ket{\xi_{\vb{k},\mu}e^{-i\omega_kt}}
        - \xi_{\vb{k},\mu} 
            \Big(
                \prod_{\vb{k}',\mu'\neq \vb{k},\mu}
                \lvert \xi_{\vb{k}',\mu'} \rvert^2
            \Big)
            \ket{0_{\vb{k},\mu}},
\end{equation}
where $\xi_{\vb{k}',\mu'} = \braket{\alpha_{\vb{k}',\mu'}e^{-i\omega_{k'} t}}{[\alpha_{\vb{k}',\mu'} + \chi_{\vb{k}',\mu'}(t)]e^{-i\omega_{k'} t}}$ if $\vb{k}',\mu' \simeq \vb{k}_{L, \mu}$, and equal to $\braket{0_{\vb{k}',\mu'}}{ \chi_{\vb{k}',\mu'}(t)e^{-i\omega_{k'} t}}$ otherwise.
\textcolor{black}{To relate the formal approach of this subsection to the actual experiment in which an optical ``cat-like'' CSS of the form of \eqref{eq:Paris_IR_Projection_HHG} was measured \cite{lewenstein2021generation}, we shall summarize the conditioning procedure introduced in this section. 
We first projected the evolution of the total system in \eqref{eq:TDSE_projected_GS} on the electronic ground state for taking into account the process of HHG. We then conditioned the shifted optical field state on the HHG wavepacket modes via projecting the state \eqref{eq:state:condit:ground} on $\Pi_{\Tilde{n}\neq 0} = \mathbbm{1}-\Pi_{\Tilde{0}}$. This measurement operation leads to the entangled field state, e.g. wavefunction ``collapse'', into the state \eqref{eq:state:after:harm:cond}. This does not affect the HHG dynamics itself, but formally conditions the field state onto the process of HHG. Finally, the harmonic field modes are detected, and thus the total entangled state is projected on the respective coherent states of all harmonic modes which are shifted by the respective harmonic amplitudes $\chi_{\vb{k}.\mu}$. This leads to the CSS as written in \eqref{eq:Paris_IR_Projection_HHG}, and which was reconstructed in \cite{lewenstein2021generation}. The actual experimental conditioning on HHG has its formal description by the sequence of these three measurements, and is derived in terms of a quantum theory of measurement via POVM in Ref.~\cite{stammer2022theory}. In simple terms, the conditioning leads to a projection on everything that was not in the initial field state $\ket{\alpha_{\vb{k}_L}}$, i.e. $\mathbbm{1}- \dyad{\alpha_{\vb{k}_L}}$ the identity subtracted by the initial state.}     

\subsection{Conditioning on electronic continuum states: ATI}

\subsubsection{Optical field state conditioned on ATI}

In the following we are interested in describing the process of above-threshold ionization (ATI) in which the electron is found in the continuum after the end of the pulse. We therefore project the TDSE \eqref{eq:TDSE_final} on the electronic state where one electron is found in a continuum state $\ket{\vb{v}_\alpha}$, and the other atoms are in the ground state 
\begin{equation}
\begin{aligned}
    i \hbar \dv{}{t} \ket{\phi_\alpha (\vb{v}_\alpha,t)} &=
    \\ &\hspace{-0.5cm}
    e \sum_{\alpha^\prime} \vb{E}_Q(t, \vb{R}_{\alpha^\prime})  \cdot\bra{\vb{v}_\alpha g ...g } \vb{r}_{\alpha^\prime}(t) \ket{\Psi(t)},
\end{aligned}
\end{equation}
where we have defined $\ket{\phi_\alpha (\vb{v}_\alpha,t)} = \bra{\vb{v}_\alpha g...g} \ket{\Psi(t)}$.
We again use the identity \eqref{eq:idendity} to obtain
\begin{equation}
\begin{aligned}
\label{eq:ManyAtoms:ATI:Sch_identity}
     i \hbar &  \dv{}{t} \ket{\phi_\alpha (\vb{v}_\alpha,t)} =  e \vb{E}_Q(t, \vb{R}_\alpha) \cdot \bra{\vb{v}_\alpha} \vb{r}_\alpha(t) \ket{g} \ket{\phi(t)} \\
     & + e \vb{E}_Q(t, \vb{R}_\alpha)\cdot \int d^3 v_\alpha^\prime \bra{\vb{v}_\alpha} \vb{r}_\alpha(t) \ket{\vb{v}_\alpha^\prime} \ket{\phi_\alpha (\vb{v}_\alpha^\prime, t)} \\
     & + e \sum_{\alpha^\prime \neq \alpha} \vb{E}_Q(t, \vb{R}_{\alpha^\prime})\cdot \bigg[ \expval{\vb{r}_{\alpha^\prime}(t)} \ket{\phi_\alpha (\vb{v}_\alpha, t)} \\
     & \hspace{1.1cm} + \int d^3 v_{\alpha^\prime}^\prime \bra{g} \vb{r}_{\alpha^\prime}(t) \ket{\vb{v}_{\alpha^\prime}^\prime} \ket{\phi_{\alpha, \alpha^\prime}(\vb{v}_\alpha, \vb{v}_{\alpha^\prime}^\prime, t) }\bigg].
\end{aligned}
\end{equation}
Here, we have defined the field state conditioned on ionization of two atoms $\ket{\phi_{\alpha, \alpha^\prime}(\vb{v}_\alpha, \vb{v}^\prime_{\alpha^\prime}, t) }  = \bra{\vb{v}_\alpha \vb{v}^\prime_{\alpha^\prime}g...g} \ket{\Psi(t)}$. 
In the following, we only take into account the single atom response such that we can neglect all terms in which $\alpha^\prime \neq \alpha$, i.e. we neglect the last two terms in \eqref{eq:ManyAtoms:ATI:Sch_identity}. 
Note that if one would include the sum over $\alpha^\prime$ of the remaining $N-1$ atoms, this gives rise to a strongly correlated many-body system, and should be considered when many-body phenomena are investigated. 
We are first interested in the process of direct ATI such that we will neglect the contributions from scattering events at the ionic potential after ionization. The time-dependent continuum-continuum (C-C) transition matrix elements $\bra{\vb{v}_\alpha} \vb{r}_\alpha(t) \ket{\vb{v}_\alpha^\prime}$ of the electronic degrees of freedom can be written as (see Appendix \ref{sec:app_cc_elements})
\begin{equation}\label{eq:C-C:approx}
    \begin{aligned}
	\mel{\vb{v}}{\vb{r}(t)}{\vb{v}^\prime} &\simeq
		 \int \dd^3 v^{\prime \prime} \int \dd^3 v^{\prime \prime\prime}
				b^*_{\vb{v}}(\vb{v}^{\prime \prime}, t)
				\\
				&\hspace{2.2cm} \times b_{\vb{v}'}(\vb{v}^{\prime \prime\prime}, t)
					\mel{\vb{v}^{\prime \prime}}{\vb{r}}{\vb{v}^{\prime \prime\prime}},
	\end{aligned}
\end{equation}
where (see Appendix \ref{sec:app_amplitudes}) 
\begin{align}
    b_{\vb{v}}(\vb{v}^\prime,t) = \exp{- \frac{i}{\hbar} S(\vb{v'},t,t_0)} \delta(\vb{v} - \vb{v}^\prime),
\end{align}
is the amplitude of a continuum state with momentum $\vb{v}^\prime$ when initially having momentum $\vb{v}$, and where we have defined the action
\begin{align}
    S(\vb{v}',t,t_0) = \int_{t_0}^t \dd t^\prime \left[ \frac{1}{2m} \left( m\vb{v}^\prime - \frac{e}{c} \vb{A}(t) + \frac{e}{c} \vb{A}(t^\prime) \right)^2+ I_p \right].
\end{align}
The transition matrix element in \eqref{eq:C-C:approx} is given by 
\begin{equation}\label{eq:dipCC:el}
	e\mel{\vb{v}^{\prime \prime}}{\vb{r}}{\vb{v}^{\prime \prime\prime}}
		= ie\hbar \nabla_{\vb{v}^{\prime \prime}} \delta(\vb{v}^{\prime \prime} - \vb{v}^{\prime \prime\prime})
			- \hbar \vb{g}(\vb{v}^{\prime \prime},\vb{v}^{\prime \prime\prime}),
\end{equation}
where the first and second term contribute to direct and re-scattering ATI, respectively.
In the following we only consider the first term which is the contribution not influenced by the scattering center, and neglect the re-scattering transition matrix element $\vb{g}(\vb{v}^{\prime \prime},\vb{v}^{\prime \prime\prime})$. 
It therefore remains to solve
\begin{equation}\label{eq:TDSE_coupled_HHG_ATI1}
\begin{aligned}
    i \hbar \dv{}{t} \ket{\phi_\alpha(\vb{v}_\alpha, t)} &=
       e \vb{E}_Q(t,\vb{R}_\alpha) \cdot \Delta \vb{r}_\alpha(t, \vb{v}_\alpha) \ket{\phi_\alpha (\vb{v}_\alpha,t)} \\
      &\quad + e \vb{E}_Q(t, \vb{R}_\alpha) \cdot\bra{\vb{v}_\alpha} \vb{r}_\alpha(t) \ket{g} \ket{\phi(t)},
\end{aligned}
\end{equation}
with the total displacement of the electron in terms of the canonical momentum $\vb{p}_\alpha = m\vb{v}_\alpha - \frac{e}{c}\vb{A}(t) $, given by
\begin{align}\label{eq:electr:displac}
    \Delta \vb{r}_\alpha(t,\vb{v}_\alpha) =  \frac{1}{m} \int_{t_0}^t \dd t^\prime  \left[\vb{p}_\alpha + \frac{e}{c}\vb{A}(t^\prime)\right].
\end{align}

The right-hand side of \eqref{eq:TDSE_coupled_HHG_ATI1} is decomposed into two terms. The first term is the homogeneous part of the differential equation in which the electric field operator is coupled to the total displacement of the electron during its propagation in the continuum $\Delta \vb{r}_\alpha(t, \vb{v}_\alpha)$. This takes into account the backaction of the electron's motion in the continuum over the EM field. In the second term, the electric field is coupled to the transition matrix element from the ground to the continuum state, which takes into account the effect of ionization. 
The solution of \eqref{eq:TDSE_coupled_HHG_ATI1} is given by 
\begin{equation}\label{eq:ATI:purestate}
\begin{aligned}
    \ket{\phi_\alpha(\vb{v}_\alpha, t)} = &   \frac{-i e}{\hbar}  \int_{t_0}^t \dd t_1 \mathcal{T} e^{-\frac{ie}{\hbar} \int_{t_1}^{t} \dd t^\prime\vb{E}_Q(t^\prime, \vb{R}_\alpha) \Delta \vb{r}_\alpha(t^\prime,\vb{v}_\alpha)} \\
    & \times \vb{E}_Q(t_1, \vb{R}_\alpha) \bra{\vb{v}_\alpha} \vb{r}_\alpha(t_1) \ket{g}  \ket{\phi(t_1)},
\end{aligned}
\end{equation}
where we have used the initial condition that the electron is initially in the ground state, i.e. $\ket{\phi_\alpha(\vb{v}_\alpha, t_0)} = \bra{\vb{v}_\alpha g...g} \ket{\Psi(t_0)} = 0$. However, the homogeneous part of the differential equation \eqref{eq:ATI:purestate} adds an important contribution to the total solution, namely, the displacement of the field due to the electron's propagation in the continuum.
The exponential term in \eqref{eq:ATI:purestate} can be solved analytically since it is linear in the field operators, and give rise to a multimode displacement operator 
\begin{equation}
\begin{aligned}
    \mathcal{T} & \exp{-\frac{ie}{\hbar} \int_{t_1}^t \dd t^\prime \, \Delta \vb{r}_\alpha(t^\prime,\vb{v}_\alpha) \cdot  \vb{E}_Q(t^\prime,\vb{R}_\alpha) } \\
    & = \prod_{\vb{k}, \mu} D[\delta_\alpha (t, t_1, \omega_k,\vb{v}_\alpha)] e^{i \varphi_{\vb{k} \mu}(t, t_1, \vb{v}_\alpha)}\\
    &= \mathcal{D}[\boldsymbol{\delta}(t,t_1,\vb{v}_\alpha)],
\end{aligned}
\end{equation}
where $\mathcal{D}[\boldsymbol{\delta}(t,t_1,\vb{v}_\alpha)]$ is a shorthand notation for the product over all the modes of the phase $\varphi_{\vb{k},\mu}$ and the displacement operators $D[\delta_\alpha (t, t_1, \omega_k,\vb{v}_\alpha)]$ for which we have  
\begin{equation}\label{eq:Delta}
    \delta_\alpha(t, t_1, \omega_k,\vb{v}_\alpha)  = - \frac{e}{\hbar} \tilde g(k) e^{- i \vb{k}\cdot \vb{R}_\alpha} \epsilon_{\vb{k}\mu} \cdot \Delta \tilde{\vb{r}}^*_\alpha(t, t_1, \omega_k,\vb{v}_\alpha)
\end{equation}
\begin{equation}\label{eq:Phase}
    \begin{aligned}
    \varphi_{\vb{k} \mu}(t, t_1, \vb{v}_\alpha) & = \frac{e^2}{\hbar^2} \tilde g^2(k) \int_{t_1}^t \dd\tau_1 \int_{t_1}^{\tau_1} \dd\tau_2  \epsilon_{\vb{k} \mu} \cdot \Delta \vb{r}_\alpha (\tau_1, \vb{v}_\alpha) \\
    & \quad \times \epsilon_{\vb{k} \mu} \cdot \Delta \vb{r}_\alpha (\tau_2, \vb{v}_\alpha) \sin[\omega_k (\tau_1 - \tau_2)],
    \end{aligned}
\end{equation}
with the Fourier transform of the electron displacement 
\begin{align}
\label{eq:FT_electron_displacement}
    \Delta \tilde{\vb{r}}_\alpha(t,t_1, \omega_k,\vb{v}_\alpha) = \int_{t_1}^t \dd t^\prime e^{i \omega_k t^\prime} \Delta \vb{r}_\alpha(t^\prime, \vb{v}_\alpha).
\end{align}

Finally, the state of the EM field conditioned on ATI is given by
\begin{equation}
    \begin{aligned}\label{eq:DI:term}
    \ket{\phi_\alpha(\vb{v}_\alpha,t)} &=  \frac{-ie}{\hbar} \int_{t_0}^t \dd t_1 
    \mathcal{D}[\boldsymbol{\delta}(t,t_1,\vb{v}_\alpha)] \\
    & \hspace{1.5cm}\times \vb{E}_Q(t_1,\vb{R}_\alpha) \cdot \bra{\vb{v}_\alpha} \vb{r}_\alpha(t_1) \ket{g}
    \\
    &\hspace{1.5cm} \times \mathcal{D}[\boldsymbol{\chi}(t_1)] \ket{\phi(t_0)}.
    \end{aligned}
\end{equation}

The time-dependent transition matrix element from the ground to the continuum state is given by (see Appendix \ref{sec:app_bc_element})
\begin{equation}\label{eq:transition_gs_continuum_aprox}
    \begin{aligned}
    \bra{\vb{v}_\alpha} \vb{r}_\alpha(t_1) \ket{g} &\simeq  \bra{\vb{p}_\alpha + e/c \vb{A}(t_1) } \vb{r}_\alpha \ket{g}
    \\
    &\hspace{-1cm} \times
    \exp{\frac{i}{\hbar}
    \int_{t_0}^{t_1} \dd t^\prime \left[ \frac{1}{2m} \left( \vb{p}_\alpha + \frac{e}{c} \vb{A}_{cl}(t^\prime)  \right)^2 + I_p \right]}.
    \end{aligned}
\end{equation}

\begin{figure}
    \centering
    \includegraphics[width = 1.\columnwidth]{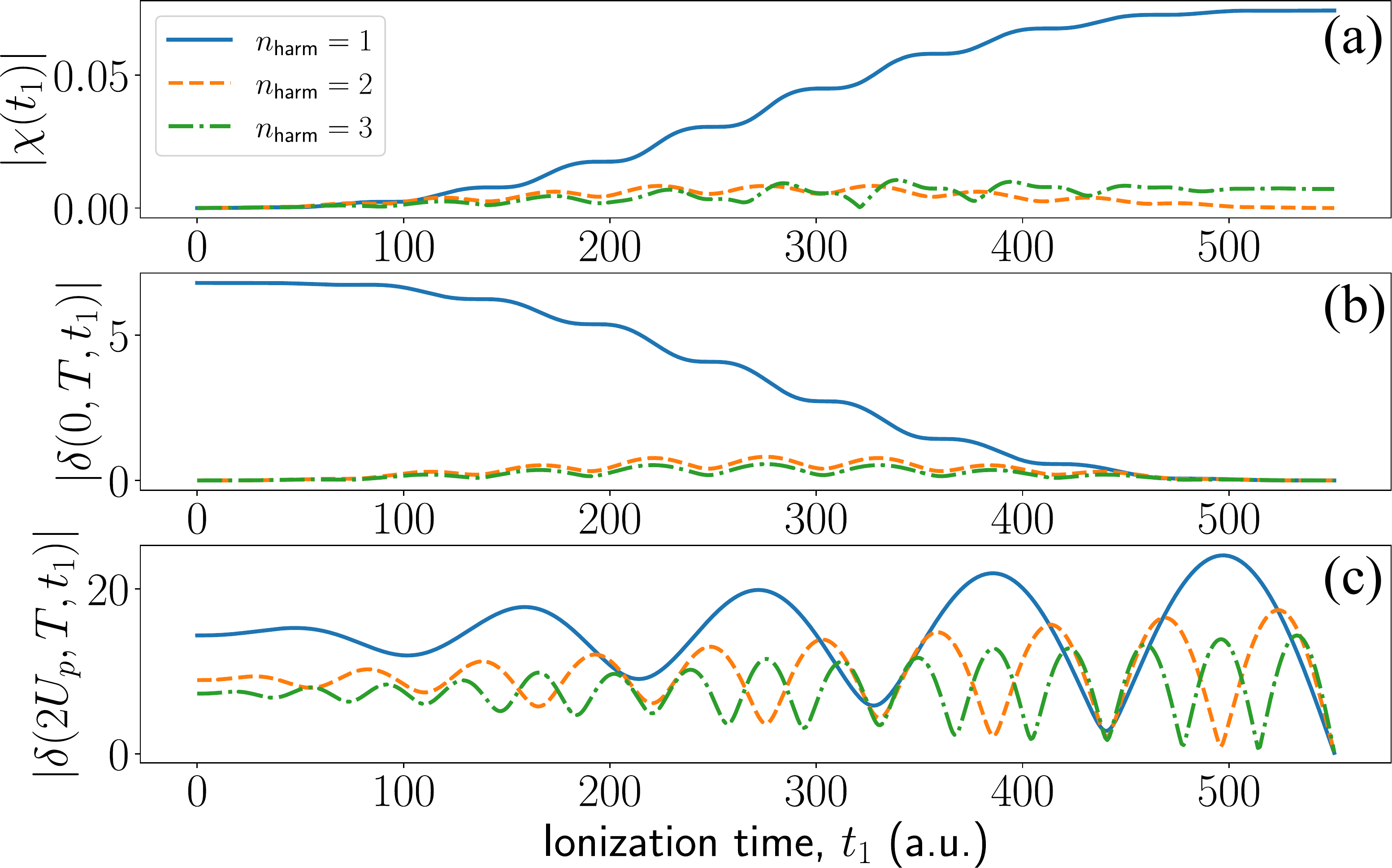}
    \caption{Behaviour of the shifts $\abs{\chi(t_1)}$ (in (a)) and $\abs{\delta(T, t_1,\omega_k,\vb{p})}$ ((b) $p=0$ and (c) \textcolor{black}{$p=0.93\sqrt{U_p}$}) over ionization time $t_1$. The different curves in each of the plots represent the contribution to different harmonic modes ($n_\text{harm}$). In particular, the blue solid curve corresponds to the fundamental mode $n = 1$, the orange dashed curve to the second harmonic $n=2$ and the green dash-dotted curve to the third harmonic $n=3$. In general, the displacement from $\abs{\chi(t_1)}$ is, at least, two orders of magnitude smaller than the one coming from $\abs{\delta(T, t_1,\omega_k,\vb{p})}$, which furthermore gets enhanced for increasing values of the canonical momentum $\vb{p}$.}
    \label{Fig:delta:xi}
\end{figure}

The state of the EM field \eqref{eq:DI:term}, conditioned on ATI, has an illustrative interpretation in terms of the backaction on the field due to the electron dynamics. The initial state of the field $\ket{\phi(t_0)}$ is displaced by the oscillation of the electron in the ground state \cite{madsen_strongfield_2021} via $\mathcal{D}[\boldsymbol{\chi}_{\vb{k}\mu}]$ acting until the ionization time $t_1$, when the electron transitions from the ground to the continuum state.
\textcolor{black}{These bound state oscillations have a natural formulation in the Kramers-Henneberger frame, where their contribution to HHG has been estimated \cite{madsen_strongfield_2021}.}
The transition matrix element $\bra{\vb{v}_\alpha} \vb{r}_\alpha(t_1) \ket{g}$, which includes the phase of the semi-classical action, induces a change of the field due to its coupling to the electric field operator at the ionization time $t_1$. Once the electron is ionized, it propagates in the continuum under the influence of the field. The backaction on the EM field due to the electron's dynamics in the continuum is reflected in the displacement $\mathcal{D}[\boldsymbol{\delta}_\alpha(t,t_1,\omega_k,\vb{v}_\alpha)]$. This shift of the individual EM field modes is proportional to the respective Fourier component of the electron displacement in the continuum \eqref{eq:FT_electron_displacement}. 
In the latter contribution, in contrast to the bound state oscillation, the electron does not only oscillate, but also has a drift since it can appear in the continuum with a non-vanishing momentum $\vb{p}$. \textcolor{black}{The fact that this induces a displacement in the optical field can intuitively be understood when recalling that a propagating electron in the continuum is the same as a classical charge current, and thus, when coupled to the field operator, induces a coherent displacement.}
In Fig.~\ref{Fig:delta:xi} we show the behaviour of the absolute value of the different displacements for different modes of the EM field ($n_\text{harm}$) at the end of the pulse $t=T$, for varying the ionization time $t_1$. We can see that the largest contribution to the shift on the initial field states is due to the electron displacement in the continuum via $\delta(T,t_1,\omega_k,\vb{p})$ (see Fig.~\ref{Fig:delta:xi}~(b) and (c)), as would be expected for the process of ATI. The contribution from the bound state oscillation prior to ionization is, at least, two orders of magnitude smaller than the continuum contribution (see Fig.~\ref{Fig:delta:xi}~(a)). 
Moreover, we observe that the bound state oscillation prior ionization $\chi_{\vb{k},\mu}(t_1)$ increases for later ionization times (increasing ionization time $t_1$), while the contribution from the continuum propagation $\delta(T, t_1,\omega_k,\vb{p})$ decreases. This is consistent with the fact that, for later ionization times, the more time the electron is bound, so the bigger the contribution coming from $\chi_{\vb{k},\mu}(t_1)$, and accordingly the electron spends less time propagating in the continuum, leading to a smaller $\delta(T, t_1,\omega_k,\vb{p})$. 
In Fig.~\ref{Fig:delta:xi}~(b) and (c), we show the contribution from the continuum displacement $\delta(T, t_1,\omega_k,\vb{p})$ for two different canonical momenta $\vb{p}\in \{ 0,\, 0.93\sqrt{U_p} \}$. We observe that for large canonical momentum, the shift is in general larger, and shows pronounced oscillations. 
\textcolor{black}{The different oscillation periods in the displacement $\delta(T, t_1,\omega_k,\vb{p})$ for different harmonics, which is most pronounced in Fig. \ref{Fig:delta:xi}~(c), originates from the non-linear motion of the electron in the continuum. Since the displacement of each field mode is proportional to the respective Fourier component of the electron displacement in the continuum, the different modes are shifted depending on this non-linear electron motion after ionization.}

In Fig.~\eqref{Fig:behavior:conditioning} (a) and (b) we show, respectively, the behaviour of the real and imaginary parts of the backaction on the field due to the continuum propagation of the electron $\delta_\alpha(T, t_1,\omega_k,\vb{p})$ for the fundamental mode for positive (blue solid curve) and negative (orange dashed curve) values of the initial canonical momentum. The real part shows similar behaviour in both cases with a phase difference of $\pi$ between the oscillations. The imaginary part shows the same phase difference of $\pi$, but is positive (negative) for positive (negative) initial momenta $\vb{p}$.
The different sign of the initial electron momentum dictates the propagation direction of the electron in the continuum. Since the sign of the displacement amplitude $\delta_\alpha(T, t_1,\omega_k,\vb{p})$ determines the in or out of phase shift of the fundamental mode in phase space, will lead to an increased or decreased coherent state amplitude, respectively. This interference then either leads to an enhancement or depletion of the fundamental driving laser amplitude (see discussion below).
Note that the influence of the positive and negative momentum depends on the carrier-envelope-phase (CEP) of the driving laser field. A $\pi$ phase change in the CEP leads to the same effect as interchanging the positive and negative electron momentum \cite{paulus_measurement_2003}.

However, so far we have only considered a conditioning on a single final electron momentum. In order to obtain the EM field state when including all possible final momenta of the electron, we integrate \eqref{eq:DI:term} over $\vb{v}_\alpha$ and obtain the corresponding mixed state density matrix
\begin{align}
\label{eq:rhoATI_mixed_definition}
    \rho_{ATI}(t) & = \int \dd^3v_\alpha \dyad{\phi_\alpha(\vb{v}_\alpha,t)}. 
\end{align}

The states in \eqref{eq:DI:term} (pure state for single final electron momentum), and \eqref{eq:rhoATI_mixed_definition} (mixed state for all possible final electron momenta) are the final state of the EM field in intense laser-atom interaction conditioned on the process of ATI. All relevant quantities of the optical field can be obtained from here on.

\subsubsection{\label{sec:observables}Field observables}

To obtain further insights into the dynamical behavior of the optical field during the process of ATI we compute the corresponding photon number distribution of the fundamental mode which drives the ionization process. 
We shall consider the case where the ionization is conditioned on a particular electron momentum such that we can use the pure state \eqref{eq:DI:term}, and evaluate the action of the first displacement operation 
\begin{equation}\label{eq:ATI:singlemomentum}
\begin{aligned}
    \ket{\phi_{ATI} (t,\vb{p})} &= \\ & \hspace{-1.2cm}
    \frac{-ie}{\hbar}  \int_{t_0}^t \dd t^\prime  \mathcal{D}[\boldsymbol{\delta}(t,t_1,\vb{v}_\alpha)] \vb{E}_Q(t^\prime,\vb{R}_\alpha) \cdot \bra{\vb{v}_\alpha} \vb{r}_\alpha(t^\prime) \ket{g}
    \\ &\bigotimes_{\vb{k}, \mu \simeq \vb{k}_{L,\mu}} \ket{ \chi_{\vb{k},\mu}(t^\prime) }\bigotimes_{\vb{k}, \mu \gg \vb{k}_{L,\mu}} \ket{\chi_{\vb{k},\mu}(t^\prime) }.
\end{aligned}
\end{equation}

To compute the observables of the EM field we first need to transform back into the original laboratory frame 
\begin{align}
\label{eq:ATI:singlemomentum_original_frame}
    \ket{ \Phi_{ATI} (t,\vb{p})} = e^{- i H_f t} D(\alpha_{\vb{k}_L}) \ket{\phi_{ATI}(t,\vb{p})}.
\end{align}

In the following, we are interested in the backaction on the driving field due to the process of ATI. In Fig.~\ref{Fig:delta:xi} we have seen, that the oscillation of the electron in the continuum gives rise to non-negligible contributions to the shift in the harmonic modes. 
Thus, in order to eliminate possible contribution from HHG, we project the state \eqref{eq:ATI:singlemomentum_original_frame} on the vacuum of the harmonic modes $\ket{\{0\}_{HH}} = \otimes_{\vb{k},\mu \gg \vb{k}_L \mu} \ket{0_{\vb{k}\mu}}$, which takes into account only the cases in which no harmonic photon was emitted.
The state of the fundamental mode, after the conditioning on a final electron momentum $\vb{p}$, and on zero harmonic photons $\ket{\{0\}_{HH}}$, is then given by (see Appendix \ref{sec:observable_ATI_direct} for details)
\begin{align} \label{eq:Paris_cat_ATI}
        \ket{\Phi_{ATI}^{\vb{k}_L}(t, \vb{p})} = \bra{\{ 0\}_{HH}} \ket{\Phi_{ATI} (t,\vb{p})}.
\end{align}

\begin{figure}
    \centering
    \includegraphics[width = 1.\columnwidth]{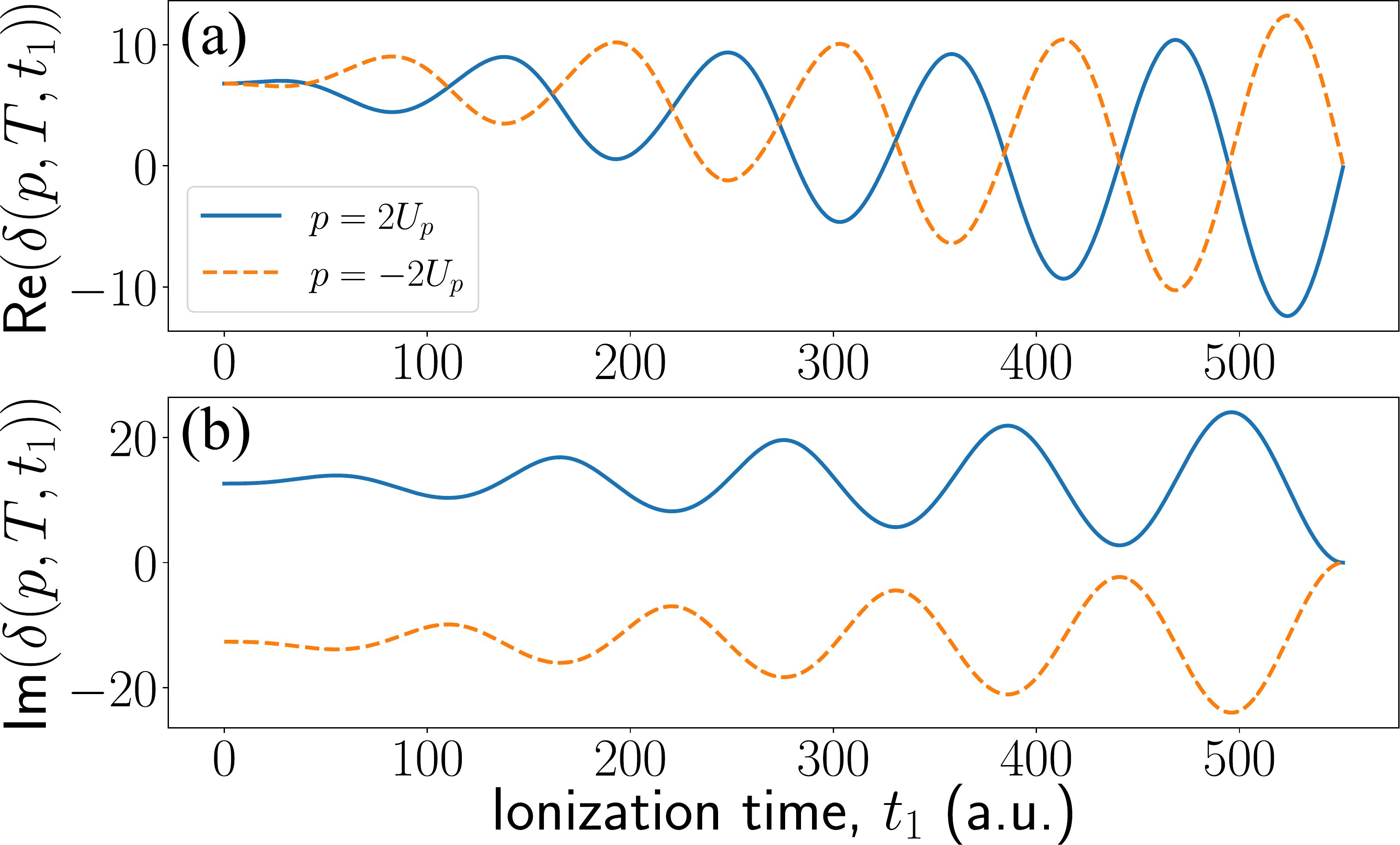}
    \caption{Real and imaginary part, respectively in (a) and (b), of the shift of the fundamental mode $\delta(T, t_1,\omega_k,\vb{p})$ for values of the canonical momentum \textcolor{black}{$p= 0.93\sqrt{U_p}$} (blue solid curve) and \textcolor{black}{$p=-0.93\sqrt{U_p}$} (orange dashed curve). The different behaviour obtained for positive and negative momentum is a consequence of the carrier--envelope--phase (CEP) (which is the phase between the carrier wave and the pulse envelope) of the employed laser field, such that a change in $\pi$ of this phase interchanges the behaviour obtained for positive and negative momentum. }
    \label{Fig:behavior:conditioning}
\end{figure}

We shall first compute the photon number distribution of the driving field (see Appendix \ref{sec:observable_ATI_direct})
\begin{align}
\label{eq:photon_prob_distr}
    P_{n_{\vb{k}_L}} (t,\vb{p}) = \abs{\bra{n_{\vb{k}_L}} \ket{\tilde \Phi_{ATI}(t,\vb{p})} }^2.
\end{align}

Having calculated the photon number distribution $P_{n_{\vb{k}_l}} (t,\vb{p})$, we further compute the photon number expectation value in the fundamental mode
\begin{align}\label{eq:mean:photon}
    \expval{n_{\vb{k}_L} (t,\vb{p})} = \sum_{n_{\vb{k}_L}} n_{\vb{k}_L} P_{n_{\vb{k}_l}} (t,\vb{p}).
\end{align}

\begin{figure}
    \centering
    \includegraphics[width = 1.\columnwidth]{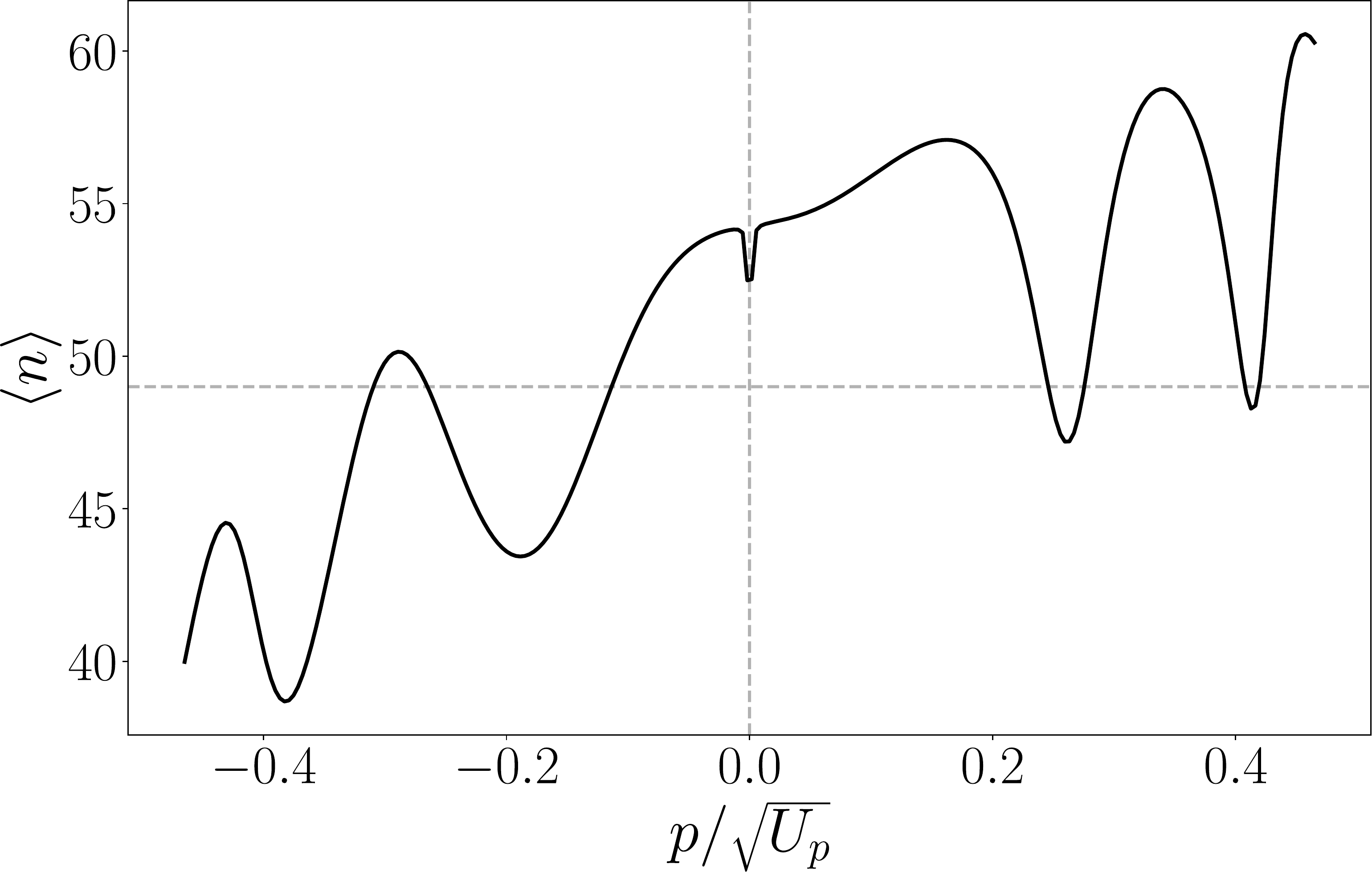}
    \caption{The mean value of the photon number \eqref{eq:mean:photon} of the fundamental mode at the end of the pulse. Here, we use $\alpha = 7i$, to reproduce the form of the considered field. In order to avoid reaching the numerical limit of the machines used for the computation, we have multiplied the field constant by 0.2 (see Appendix \ref{App:numerical:implementation} for details). We consider a field with a sinusoidal squared envelope with $\omega = 0.057$ a.u., 5 cycles of duration and field amplitude $E_0 = 0.053$ a.u.}
    \label{Fig:Mean:photon}
\end{figure}

\begin{figure}
    \centering
    \includegraphics[width = 1.\columnwidth]{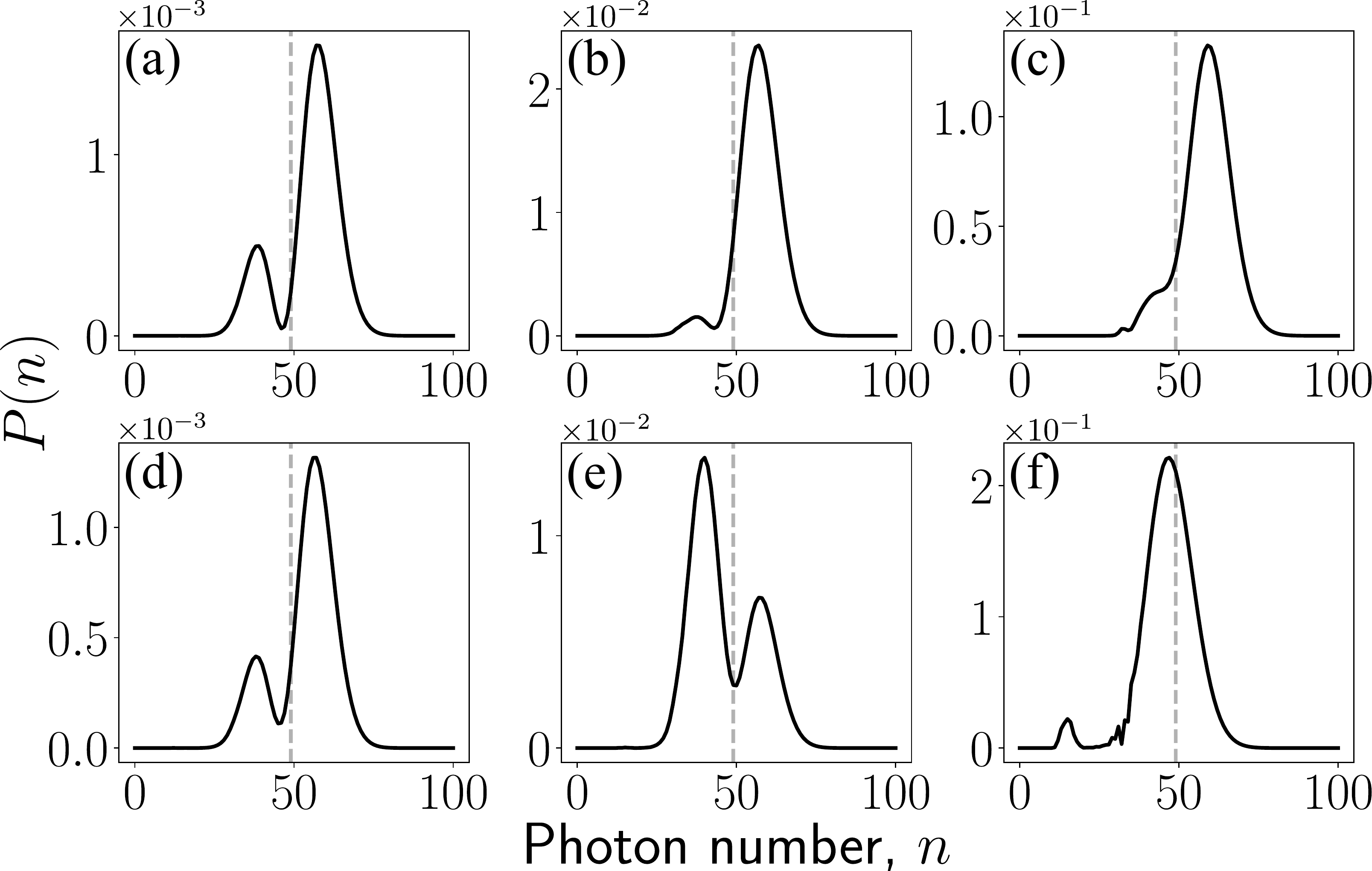}
    \caption{Photon number probability distribution \eqref{eq:photon_prob_distr} for positive (upper row) and negative (lower row) values of the canonical momentum. Specifically, we used \textcolor{black}{(a) $p=0.05\sqrt{U_p}$, (b) $p=0.14 \sqrt{U_p}$, (c) $p = 0.32\sqrt{U_p}$, (d) $p=-0.05\sqrt{U_p}$, (e) $p=-0.14\sqrt{U_p}$ and (f) $p=-0.32\sqrt{U_p}$}. We consider a field with a sinusoidal squared envelope with $\omega = 0.057$ a.u., 5 cycles of duration and field amplitude $E_0 = 0.053$ a.u. The vertical grey dashed line shows the location of the maximum photon number probability for the initial coherent state used in the numerical analysis, i.e., $\alpha = 7i$.}
    \label{Fig:Probs}
\end{figure}

In Fig.~\ref{Fig:Mean:photon}, we show the mean photon number from Eq.~\eqref{eq:mean:photon}, for values of the canonical momentum ranging from \textcolor{black}{$-0.46\sqrt{U_p}$ to $0.46\sqrt{U_p}$}, i.e. within a regime where we can neglect the rescattering effects \cite{milosevic_above-threshold_2006}. On the other hand, in Fig.~\ref{Fig:Probs} we show the photon number probability distribution for different values of the canonical momentum. In both cases, we used $\alpha = 7i$, in agreement with the form considered for the vector potential (see Appendix \ref{App:numerical:implementation} for more details about the numerical implementation). The main feature that one can observe in these plots is the different behaviour that is obtained for the conditioning over positive and negative momentum. 
As we see in Fig.~\ref{Fig:Mean:photon}, for negative values of the canonical momentum we find that the actual mean photon number of the input field gets reduced, in contrast to positive momenta which can increase the mean value of the photon number in the fundamental mode. 

Since the shift of the amplitude in the fundamental mode is mostly determined by the displacement due to the electron propagation in the continuum via $\delta(T, t_1,\omega_k,\vb{p})$ (see Fig.~\ref{Fig:delta:xi}), the different behavior for positive and negative momenta crucially depends on the phase of $\delta(T, t_1,\omega_k,\vb{p})$ (see Fig.\ref{Fig:behavior:conditioning}).
Thus, the depletion (enhancement) of the average photon number in the driving laser mode is due to the phase of the displacement for the continuum propagation for negative (positive) initial electron momenta.
The opposite sign comes from the opposite propagation direction of the electron for opposite initial momenta, such that the Fourier transform of this displacement leads to the $\pi$ phase difference in the shift of the field.
This observed behavior of the displacement for opposite momenta can be summarized as follows: for negative values of $p$, the scattered radiation by the electron interferes out-of-phase with the input field, leading to a decreasing in the overall mean photon number; for positive values of $p$, the interference takes place in-phase, and an overall enhancement in the mean photon number of the input field is obtained. 
For increasing momentum the enhancement/depletion generally increases. This is an expected feature as the bigger the canonical momentum, the bigger is the electron's excursion and therefore the value of the displacement $\abs{\delta(T, t_1,\omega_k,\vb{p})}$.

Finally, we show the photon number probability distribution in Fig.~\ref{Fig:Probs}, which witness that the fundamental mode is in a coherent state superposition when the interaction is conditioned on the ATI processes (see \eqref{eq:ATI:singlemomentum}). This is reflected in the multi-peak structure of the photon number probability distribution which can be seen in Fig.~\ref{Fig:Probs} (compare to section \ref{intro_coherent_state}). However, specially for large momenta as in Figs.~\ref{Fig:Probs}~(c) and (f) with canonical momentum \textcolor{black}{$p = 0.32\sqrt{U_p}$ and $p=-0.32\sqrt{U_p}$} respectively, we get a dominant peak. Even though in this case, and according to Fig.~\ref{Fig:delta:xi}, more peaks are expected as the value of $\delta(T, t_1,\omega_k,\vb{p})$ has more pronounced oscillations, the conditioning measurement over the harmonic modes we considered selects some specific values of $\delta(T, t_1,\omega_k,\vb{p})$, leading to a Poissonian-like behaviour for the final probability.

\subsection{Rescattering in ATI}

We are now interested in incorporating the re-scattering events in the ATI description given in the previous subsection. For that purpose, we start from \eqref{eq:ManyAtoms:ATI:Sch_identity}, and consider only a single atom within the single active electron approximation (we therefore drop the index $\alpha$)
\begin{equation}
	i\hbar \dv{}{t}\ket{\psi(\vb{v},t)}
		= e  \vb{E}_Q(t,\vb{R}) \cdot\bra{\vb{v}}\vb{r}(t) \ket{\Psi(t)},
\end{equation}
which after introducing the SFA version of the identity \eqref{eq:idendity} leads to
\begin{equation}\label{eq:SingleAtom:ATI:Sch}
\begin{aligned}
	i\hbar \dv{}{t}\ket{\phi(\vb{v},t)} &=  e\vb{E}_Q(t,\vb{R})\cdot \mel{\vb{v}}{\vb{r}(t)}{g} \ket{\phi(t)} \\
		&\quad
		+e\vb{E}_Q(t,\vb{R})\cdot \int \dd^3 v' \mel{\vb{v}}{\vb{r}(t)}{\vb{v}'} \ket{\phi(\vb{v}',t)}.
\end{aligned}
\end{equation}

In the direct ionization analysis we neglected the effect of the rescattering transition terms, i.e. the $\vb{g}(\vb{v},\vb{v}')$ term in \eqref{eq:dipCC:el}. The reason for this is that we treat them as a first order perturbation term \cite{amini2019symphony}. However, in order to describe the rescattering process, these terms need to be taken into account, such that our differential equation now reads
\begin{equation}
\label{eq:TDSE_rescattering}
    \begin{aligned}
    i \hbar \dv{}{t} \ket{\phi (\vb{v}, t)} 
        &=  e \vb{E}_Q(t,\vb{R}) \cdot \mel{\vb{v}}{\vb{r}(t)}{g}        
            \ket{\phi(t)} \\
           &\hspace{-1.cm} +e \vb{E}_Q(t,\vb{R}) \cdot \Delta\vb{r}(t,\vb{v})
           \ket{\phi(\vb{v},t)}
           \\
           &\hspace{-1.cm} -\dfrac{\hbar}{e}\int \dd^3v' 
                \exp[\dfrac{i}{\hbar} S(\vb{v},t,t_0)]
                \vb{E}_Q(t,\vb{R}) \cdot\vb{g}(\vb{v},\vb{v}') \\
            &\hspace{0.5cm}
            \times \exp[-\dfrac{i}{\hbar} S(\vb{v}',t,t_0)]
                \ket{\phi_\alpha (\vb{v}',t)},
    \end{aligned}
\end{equation}

We now perform a perturbative expansion of the quantum optical state when conditioned to ATI up to first order in perturbation theory, such that we include now the effect of the rescattering terms
\begin{equation}
\label{eq:perturbation_ansatz}
	\ket{\phi(\vb{v},t)}
		\approx \ket{\phi^{(0)}(\vb{v},t)} + \ket{\phi^{(1)}(\vb{v},t)}.
\end{equation}
Introducing these terms into the Schrödinger equation \eqref{eq:TDSE_rescattering} we get for the first order perturbation theory term (see Appendix \ref{app:rescattering} for details)
\begin{equation}
    \begin{aligned}
    \ket{\phi^{(1)}(\vb{p},t)}
        &= -\dfrac{e}{\hbar}\int^t_{t_0} \dd t_2 \int \dd^3 p' \int^{t_2}_{t_0} \dd t_1
        \\
        &\quad
            \times
            \mathcal{D}[\boldsymbol{\delta}(t,t_2,\vb{p}_\alpha)]
            \exp[\frac{i}{\hbar} S(\vb{p},t_2,t_0)] \\
        &\quad   \times\vb{E}_Q(t_2,\vb{R}) \cdot \vb{g}\Big(\vb{p}+\dfrac{e}{c}\vb{A}(t),\vb{p}'+\dfrac{e}{c}\vb{A}(t)\Big)
        \\&\quad\times
            \mathcal{D}[\boldsymbol{\delta}(t_2,t_1,\vb{p}'_\alpha)]
                \big) \exp[\frac{i}{\hbar} S(\vb{p}',t_2,t_1)] \\
        &\quad  
            \times\vb{E}_Q(t_1,\vb{R}) \cdot
            \mel{\vb{p}'-\dfrac{e}{c}\vb{A}(t)}{\vb{r}(t_1)}{g} \\
        &\quad
        \times\mathcal{D}[\boldsymbol{\chi}(t_1)]
                \ket{\phi(t_0)},
    \end{aligned}
\end{equation}
where we have expressed the final state in terms of the canonical momentum $\vb{p}$ and $\vb{p}^\prime$.

The expression above gives the complete quantum electrodynamics of the rescattering process: first the electron gets ionized at $t_1$, which has associated a displacement in the photonic quadratures; afterwards, the electron propagates in the continuum until $t_2$, acquiring the usual semi-classical phase while the electromagnetic field gets displaced by $\delta(t_2,t_1,\omega_k,\vb{p}^\prime)$, a quantity that depends on the electron displacement from $t_1$ to $t_2$; finally, at time $t_2$, the electron rescatters with the core potential such that its velocity changes from $\vb{p}'$ to $\vb{p}$, and it propagates in the continuum until it is measured. Again during this last process, the field gets displaced by $\delta(t,t_2,\omega_k,\vb{p})$, i.e. a quantity that depends on the electron displacement since the rescattering takes place at time $t_2$ until it is finally measured at time $t$. From here, and analogously to the direct ionization case, we can introduce the single-mode approximations and then proceed to compute the quantum optical observables, which leads to expressions where the rescattering terms add incoherently to the ones we already obtained in direct ATI. While for values of \textcolor{black}{$p<2\sqrt{U_p}$} we expect these terms not to play a very important role similarly to what happens in the semi-classical analysis \cite{milosevic_above-threshold_2006}, the same cannot be said for the regime \textcolor{black}{$2\sqrt{U_p} < p < 5\sqrt{U_p}$}, and will be part of future investigations.

\section{\label{sec:experiment}Experimental approach for quantum state engineering}
In this section we provide an approach where the aforementioned theoretical findings can be experimentally investigated. Specifically, we describe the operation principles of an experimental scheme that allows: a) generation of the non-classical light states by implementing conditioning approaches on the field modes after the interaction, b) the control of the quantum features of the generated non-classical light states, and c) the characterization of the quantum states of light. 

\subsection{General description}
\begin{figure*}
    \centering
    \includegraphics[width=0.95 \textwidth]{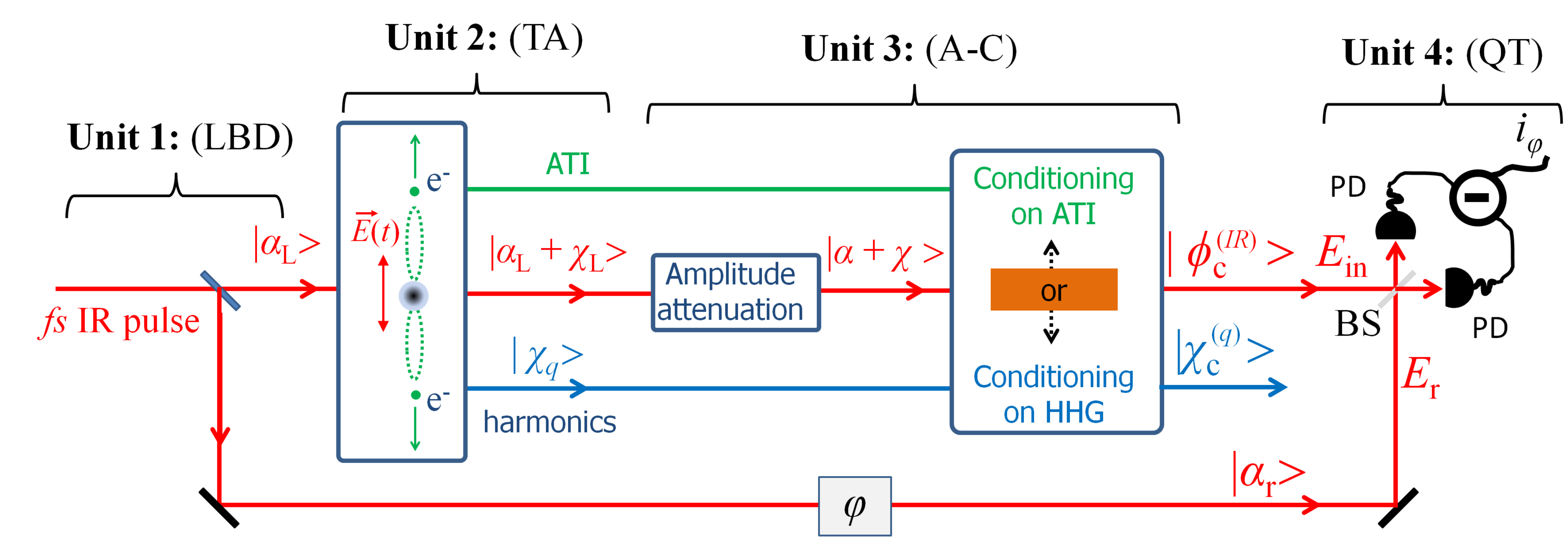}
    \caption{Schematic illustration of the operation principle of the experimental approach. Unit 1: Laser beam deliver system (LBD). Unit 2: Target area (TA). The intense laser-atom interaction is shown in the context of the electron recollision picture. The ATI photoelectrons are emitted in the direction of the polarization of the driving IR field while the generated high harmonic copropagate with the IR field.  Unit 3: IR attenuation and conditioning (A--C) on HHG and/or ATI processes. $\ket{\alpha_{L}}$ is the initial coherent state of the driving field, and $\ket{\alpha_{L}+\chi_{L}}$, $\ket{\alpha + \chi}$ are the states of the IR field after the interaction and attenuation, respectively. $\ket{\chi_{q}}$ correspond to the coherent states of the harmonic modes. $\ket{\phi_{c}^{(IR)}}$ and $\ket{\chi_{c}^{q}}$ are the field states conditioned on  HHG or ATI processes, for the IR and high--harmonic field states, respectively, which correspond to a coherent state superposition. We note that the detailed expressions of the light states are given in Section IV. Unit 4: A typical scheme of the QT method for the characterization of the state $\ket{\phi_c}$, with the state $\ket{\alpha_r}$ of the local oscillator reference field, with a controllable phase shift $\phi$. BS is a beam splitter, PD are identical IR photodiodes used from the balanced detector, and $i_{\phi}$ is the $\phi$ dependent output photocurrent difference used for the measurement of the electric field operator and the light state characterization via the reconstruction of the Wigner function.}
    \label{Fig:Exp1}
\end{figure*}

A schematic diagram of the experimental configuration is shown in Fig.~\ref{Fig:Exp1}, with the scheme divided in four units.  
Unit 1 concerns the laser beam delivery (LBD). It is used to control the properties of the driving laser field towards the laser-atom interaction region. It contains the laser beam steering, polarization control, beam shaping, pulse characterization and focusing optics.  
Unit 2 is the target area (TA) where the intense femtosecond (fs) infrared (IR) laser pulse interacts with the gas target leading to the generation of ions (that we omit from our present discussion), ATI photoelectrons and high harmonic photons emitted towards the extreme ultraviolet (XUV) spectral region. The atomic gas medium is placed at the focus of the IR beam, and the photon/electron detectors are used to measure the interaction products. 
Unit 3 (A--C) contains an optical arrangement for IR photon number attenuation, and a photon (electron) correlation approach is used to condition the field modes exiting the medium on the HHG (ATI) processes.  
Unit 4 deals with the quantum state characterization of the field, which can be achieved using the quantum tomography (QT) approach \cite{Lvovsky_Rev_QT,Breitenbach_Squeezed_QT}. It is noted that the degree of the IR attenuation in Unit 3 is associated with the limitations of the QT approach to characterize the optical field states \cite{skotiniotis2017macroscopic}, which is typically in the range of a few-photons, and does not originate from the conditioning approach itself, which in principle is applicable for high photon number light states.    

\subsection{Experimental procedure}
\begin{figure}
    \centering
    \includegraphics[width=1 \columnwidth]{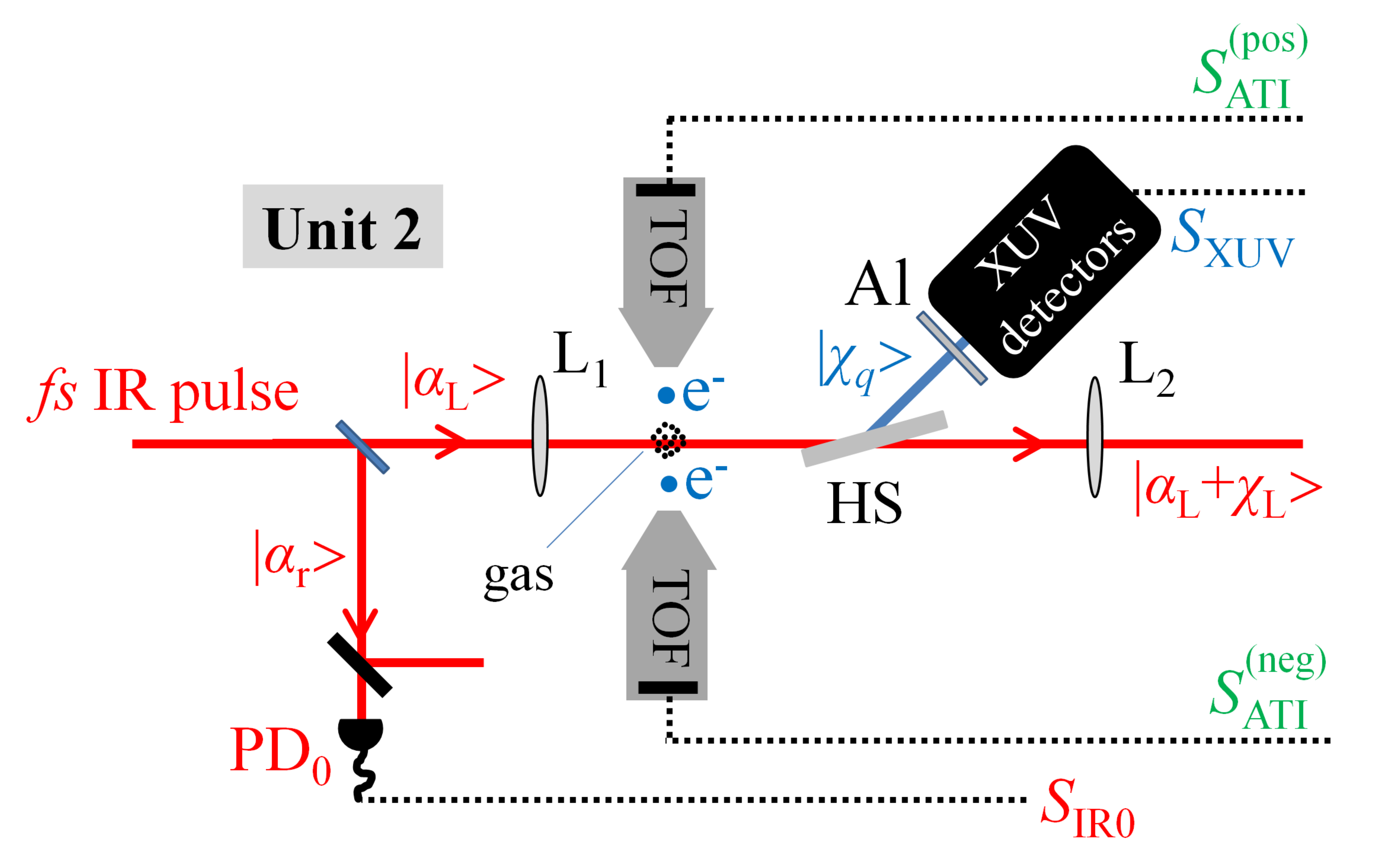}
    \caption{A more detailed scheme of Unit 2. $\textrm{L}_{1}$ is an IR focusing lens. The arrangement with the TOF spectrometers can be used for measuring the ATI spectra associated to positive and negative electron momenta. HS is a harmonic separator which transmits the IR field and reflects the high harmonics towards the XUV detectors. \emph{Al} is a thin Aluminum metal filter which transmits the harmonics with $q \geq 11$ (harmonics in the plateau and cut-off region of the spectrum). \textcolor{black}{In this way we eliminate the contribution of low order harmonics which are mainly produced by multi-photon processes. Additionally, the presence of a metal filter is an experimental requirement for blocking any residual part of the IR beam reflected by the HS.} The lens $\textrm{L}_{2}$ is used to collimate the IR beam after the interaction. With $S_\textrm{ATI}^\textrm{(pos, neg)}$ and $S_\textrm{XUV}$ we denote the integrated, over a defined region of the spectra, ATI and HHG signals, respectively. In case of using multi-cycle driving laser fields, the use of only one TOF is sufficient as $S_\textrm{ATI}^\textrm{(pos)} \approx S_\textrm{ATI}^\textrm{(neg)}$.}
    \label{Fig:Exp2}
\end{figure}

\begin{figure}
    \centering
    \includegraphics[width=1.\columnwidth]{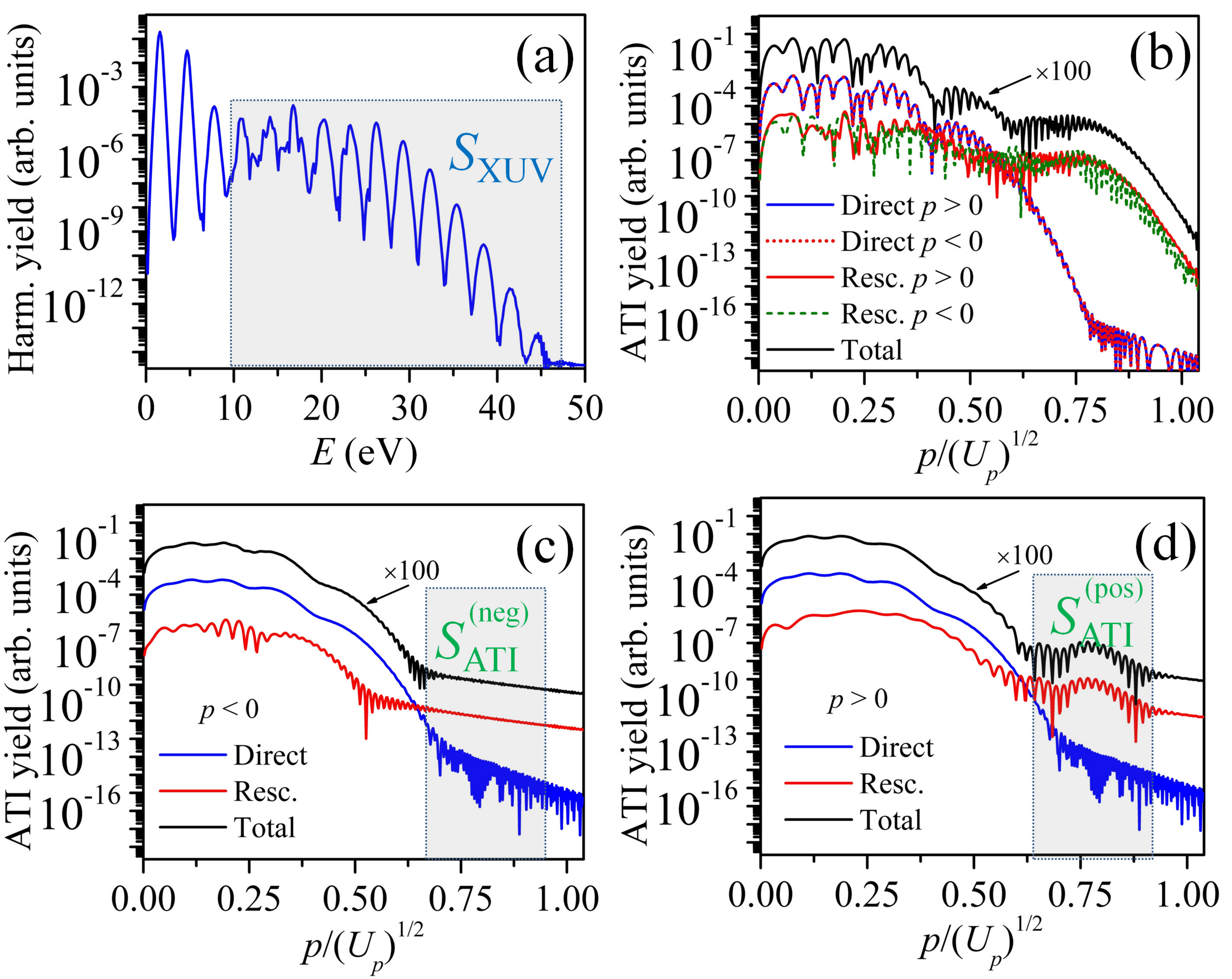}
    \caption{Calculated HHG and ATI spectra \textcolor{black}{using the \textsc{Qprop} software \cite{Qprop}}. Insets (a) and (b) respectively show the HHG and ATI spectrum produced by the the interaction of Xenon atoms with a linearly polarized \textcolor{black}{with a sinusoidal squared envelope} laser pulse of intensity $8 \times 10^{13}$ W/cm$^2$ and $\sim\!30$ fs duration. In (b) the photoelectron spectrum corresponding to direct, rescattered photoelectrons with positive ($p > 0$) and negative ($p < 0$) momenta are shown with different colors. Insets (c) and (d) respectively show the ATI spectra for negative and positive electron momenta produced by the the interaction of Xenon atoms with a linearly polarized laser pulse of intensity $8 \times 10^{13}$ W/cm$^2$ and $\sim\!5$ fs duration. The CEP effect is depicted in the high energy part of the ATI spectra of the negative and positive photoelectrons. The ATI spectra corresponding to the direct and rescattered electrons are shown with different colors. In all graphs the total ATI spectrum (black solid curve), which includes the contribution of the direct and rescattered electrons,
    has been shifted by a factor of 100 for visualization reasons. The gray shaded areas depict an example of the controllable (in the width and their position in the spectra) integrated areas of the $S_\textrm{ATI}^\textrm{(pos, neg)}$ and $S_\textrm{XUV}$.}
    \label{Fig:HHGATI}
\end{figure}

In a typical intense laser-atom interaction experiment \cite[and references therein]{Chatziathanasiou2017}, a linearly polarized IR fs laser pulse of $\expval{N_{0}}\sim 10^{14}$ photons per pulse is delivered by the Unit 1. The pulse is focused with an intensity $I_{IR} \sim 10^{14}$ W/cm$^2$ into an atomic ensemble of atomic density typically in the range of $10^{18}$ atoms/cm$^{3}$ placed in Unit 2 (see Fig.~\ref{Fig:Exp2}). The photon number and the spectrum of the generated harmonics can be measured by means of calibrated XUV photodetectors, and/or XUV--spectrometers, respectively. The charged ions (not shown in Fig.~\ref{Fig:Exp2}) and the ATI photoelectrons, can be measured by means of time--of--flight (TOF) spectrometers. The two TOF spectrometers (shown in the upper and lower part of Fig.~\ref{Fig:Exp2}) can be used for measuring the ATI spectra, and discriminating between the electrons with positive and negative momenta. This arrangement is central in case of using few--cycle laser pulses, and has been extensively used as a diagnostic of the CEP pulse to pulse stability of laser systems delivering few--cycle laser pulses \cite{Paulus2001CEP}.
As an example, in Fig.~\ref{Fig:HHGATI} we show the calculated HHG and ATI spectra produced by the interaction of Xenon atoms with a multi-- and few--cycle fs IR ($\lambda_\textrm{L} \approx$ 800 nm) laser pulse. The quantities $S_\textrm{XUV}$ and $S_\textrm{ATI}^\textrm{(pos, neg)}$ correspond to the current resulted by the integrated, over a defined region of the spectra (shown in grey shaded in Fig.~\ref{Fig:HHGATI}), harmonic intensity and ATI photoelectron signals, respectively. We note that, in case of using multi--cycle driving laser fields, the use of two TOF spectrometers is not needed as the positive and negative electron ATI spectra are almost identical, i.e. $S_\textrm{ATI}^\textrm{(pos)} \approx S_\textrm{ATI}^\textrm{(neg)}$.

\begin{figure}
    \centering
    \includegraphics[width=0.85\columnwidth]{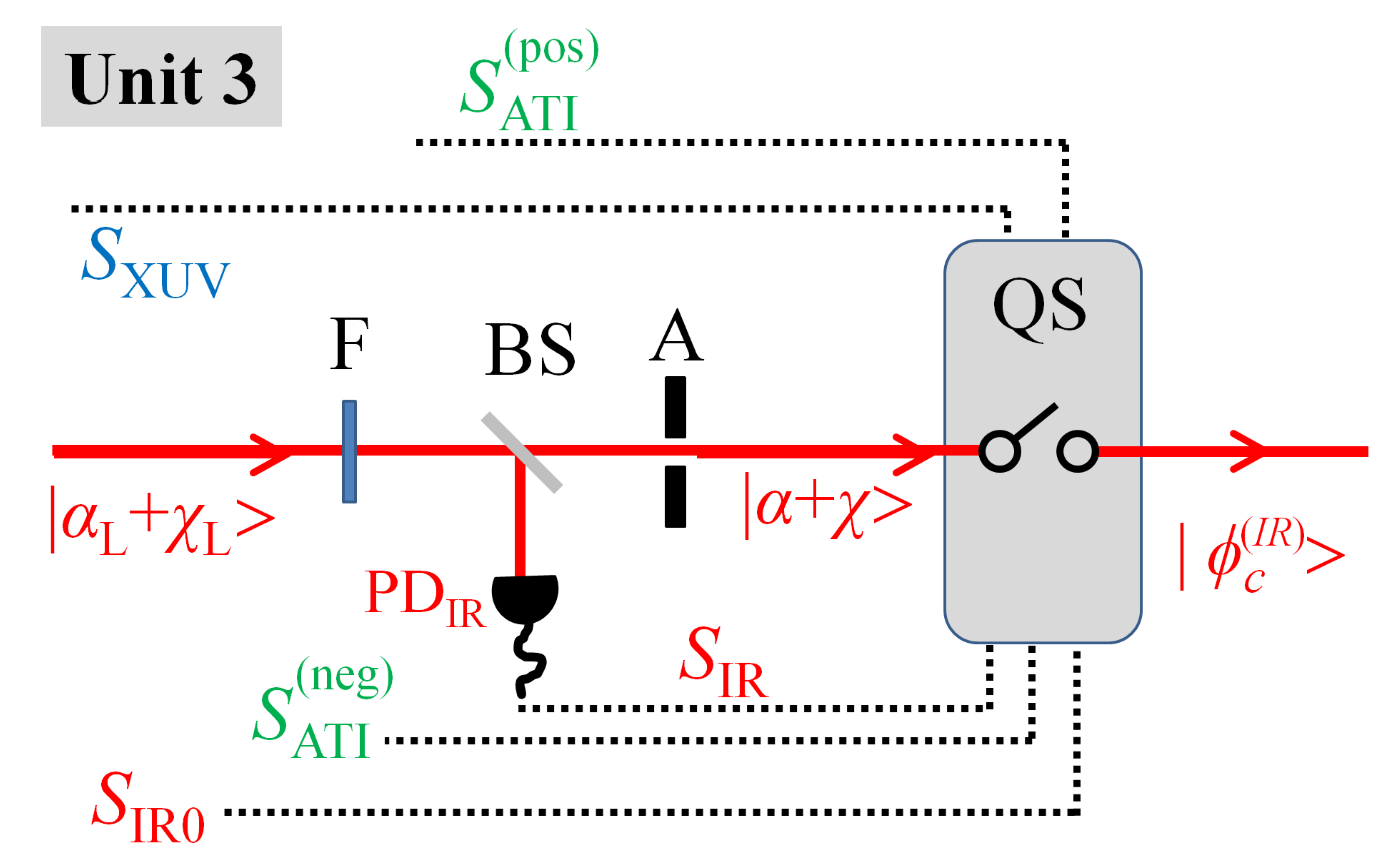}
    \caption{A schematic illustration of Unit 3. F, BS and A, are a neutral density filter, a beam splitter and an aperture, respectively. PD$_\textrm{IR}$ is an IR photodetector. $S_\textrm{XUV}$, $S_\textrm{ATI}^\textrm{(pos, neg)}$, $S_\textrm{IR}$ and $S_\textrm{IR0}$ are the signals used by the quantum spectrometer (QS) to condition the $\ket{\alpha + \chi}$ state on the HHG or ATI processes.}
    \label{Fig:Exp3}
\end{figure}

After the harmonic separator (HS), the IR beam enters in Unit 3 (Fig.~\ref{Fig:Exp3}). Here, the beam passes through an amplitude attenuation optical arrangement consisting of neutral density filters (F), a beam splitter (BS), and a spatial filter (aperture) placed in the IR beam path. In this way, the mean photon number of the IR beam after the aperture (A) (i.e. before entering the unit 4) can be reduced down to the level of a few photons per pulse. Separating some portion of the IR beam via the BS is necessary for implementing the conditioning procedure, which will be explained in more detail below. The photodiode PD$_\textrm{IR}$ is used to record the IR photon number signal $S_\textrm{IR}$ which is reflected by the BS. The $S_\textrm{XUV}$, $S_\textrm{ATI}^\textrm{(pos, neg)}$, $S_\textrm{IR}$, as well as the photon number signal of the driving laser field $S_\textrm{IR0}$ entering in Unit 2, need to be simultaneously recorded for each laser shot in order to condition the outgoing field modes on the HHG and ATI processes. The $S_\textrm{IR0}$ is used to trace the energy of the driving laser field and selects (in case that is needed) only the shots with the highest possible energy stability, typically in the level of $\lesssim 1$\%. It is noted that the electronic noise needs to be subtracted from all signals. 

\subsubsection{Conditioning in the experiment}
The quantum operations described in Section IV can be implemented by means of the Quantum Spectrometer (QS) approach \cite{Paris-ncomm,Tsatr_QS_PRL}. The QS is a shot--to--shot photon correlation--based method which provides the probability of absorbing photons from a driving laser field towards the generation of the intense laser--atom interaction products, such as HHG photons and ATI electrons. Its operation principle relies on photon statistics and the shot--to--shot correlation between the interaction products, and the energy conservation, i.e. when the signal of the interaction products increases, the IR signal $S_\textrm{IR}$ decreases. The method has been described in Refs. \cite{Paris-ncomm,lewenstein2021generation, Tsatr_QS_PRL, Lamprou_QS_Photonics2021}, and was used for the generation of optical Schrödinger ``cat-like'' and ``kitten-like'' states \cite{lewenstein2021generation,rivera2022strong} by conditioning the IR field state on the HHG process. 

\begin{figure}
    \centering
    \includegraphics[width=0.85\columnwidth]{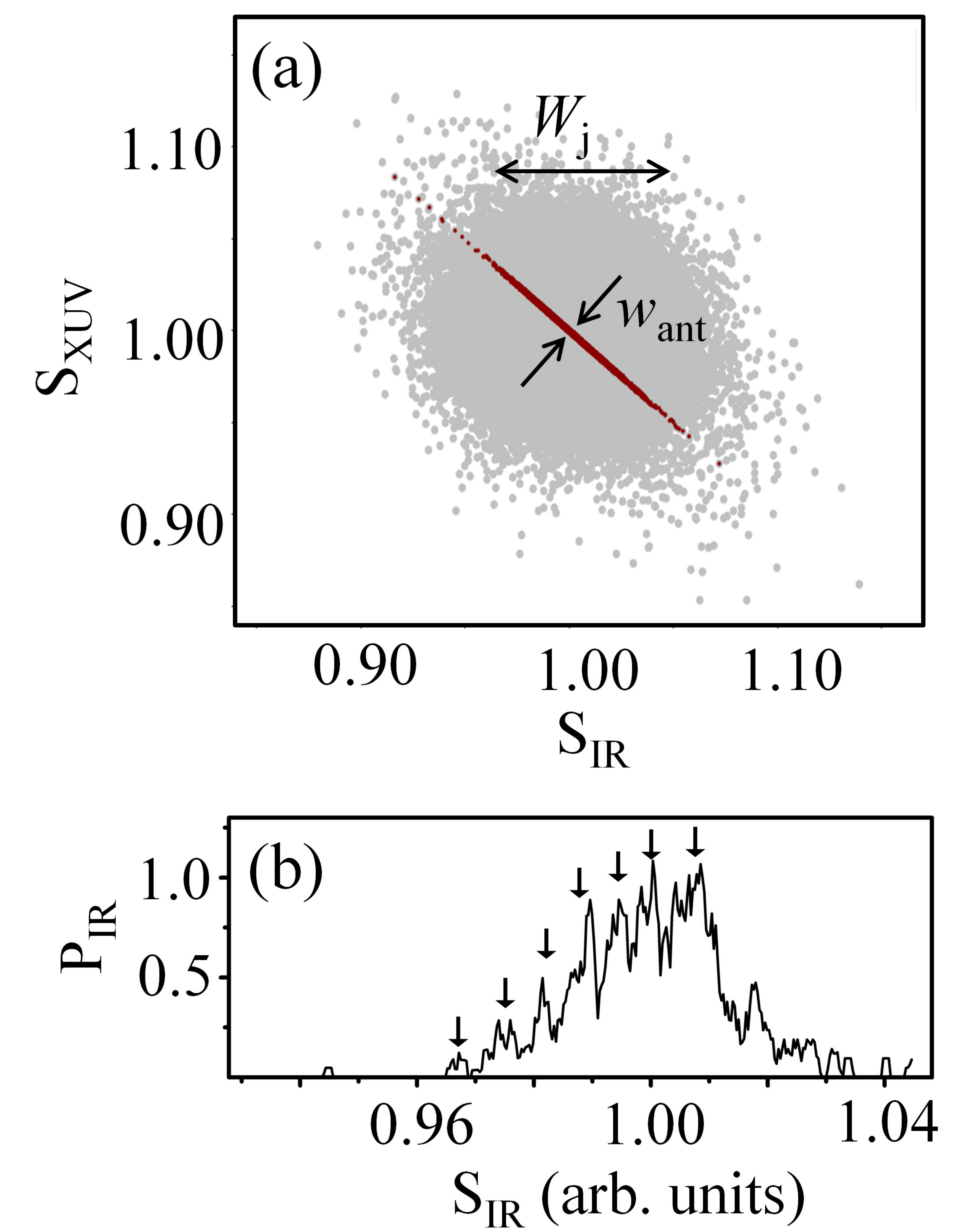}
    \caption{(a) $(S_\textrm{XUV}, S_\textrm{IR})$ photon number distribution (gray points). The mean values of $(S_\textrm{XUV}$ and $S_\textrm{IR})$ are normalized to 1. $W_{j}$ shows the width of the distribution. The red points show the selected points along the anticorrelation diagonal of a width $w_\textrm{ant}$. The distribution was created by keeping the energy stability of the driving field at the level of $\approx1$\%, and after subtracting the electronic noise from each laser shot. (b) Probability of absorbing IR photons ($\textrm{P}_{IR}$) towards HHG. The arrows depict the positions of the peaks in the multi--peak structure of $\textrm{P}_{IR}$.  Figure reproduced from Ref.~\cite{rivera2022strong}.}
    \label{Fig:correlation}
\end{figure}

Here, as an example, we briefly discuss the QS method using the HHG process. After the light--matter interaction taking place in Unit 2, the IR and XUV photon numbers are respectively, $N_\textrm{IR}$ and $N_\textrm{XUV}$. The $N_\textrm{IR}$ is smaller than the photon number of the IR field before the interaction ($N_{0}$) due to IR photon losses associated with all processes taking place in the interaction region. The IR and XUV photon numbers reaching the PD$_\textrm{IR}$ detector in Unit 3 and the XUV detector in Unit 2 are $n_\textrm{IR}$ and $n_\textrm{XUV}$, respectively. These are related with $N_{XUV}$ and $N_{IR}$ through the equations $n_\textrm{XUV}=N_\textrm{XUV}/A_\textrm{XUV}$ and $n_\textrm{IR}=N_\textrm{IR}/B_\textrm{IR}$, where $A_\textrm{XUV}$ and $B_\textrm{IR}$ are the attenuation factors corresponding to the XUV and IR photon losses introduced by the optical elements in the beam paths. The $n_\textrm{IR}$, $n_\textrm{XUV}$ and $n_{0}$ (where $n_{0}$ is the photon number of the attenuated IR field reaching the detector PD$_{0}$ in Unit 2) signals are recorded for each laser shot by a high dynamic range boxcar integrator resulting in the photocurrent outputs $S_\textrm{IR}$, $S_\textrm{XUV}$ and $S_\textrm{IR0}$. The $S_\textrm{IR0}$ is used in order to collect the laser shots which provide energy stability typically in the level of $\lesssim 1$\%. Then, and after balancing the mean value of the $S_\textrm{XUV}$ on the mean value of the $S_\textrm{IR}$, we create the joint distribution $(S_\textrm{XUV}, S_\textrm{IR})$ shown in Fig. \ref{Fig:correlation}a. The distribution is a kind of multi--dimensional map which contains the information of all processes occurring during the laser-atom interaction, and provides access to the correlated XUV--IR signals. Also, 
taking into account that 
the generation of $N_{q}$ photons of the $q$th harmonic corresponds to $q A N_{q}$ IR photons 
lost (where $A$ is the XUV absorption factor in the HHG medium), 
information about the probabilities of absorbing IR photons towards harmonic generation can be extracted. However, the number of points at which the IR photons are correlated with the generated harmonic photons is a small fraction \textcolor{black}{(typically in the range of $\sim 0.1$\%)} compared the total number of points in the distribution. To reveal these points, we take advantage of the energy conservation, and we collect only those lying along the anti--correlation diagonal of the joint distribution. These points provide the probability of absorbing IR photons ($\textrm{P}_{IR}$) towards the harmonic emission. The $\textrm{P}_{IR}$ depicts a multi-peak structure corresponding to the generated high harmonic orders (Fig.~\ref{Fig:correlation} (b)), with the spacing between the peaks to be $(\Delta q) AN_{q}=2 A N_{q}$, with $\Delta q=2$ the distance between two harmonic peaks in the HHG spectrum. The minimum value of the width ($w_\textrm{ant}$) of the anti--correlation diagonal, which defines the resolution of the QS, can be experimentally obtained by the accuracy to find the peak of the joint distribution which is $w_\textrm{ant}=W_\textrm{j}/\sqrt{k}$, where
$W_\textrm{j}$ is the percentage of the width of the joint distribution relative to its mean value and $k$ is the number of points in the distribution. Here, it is important to note that the power of the QS to resolve the multi-peak structure of $\textrm{P}_{IR}$ and to condition the IR state on the HHG process, is associated with the dynamic range of the detection system, the number of IR photons absorbed towards the harmonic emission (associated with the conversion efficiency of the HHG process), with the number of accumulated shots and the number of XUV--IR correlated points. Due to this multi--parameter dependence of the resolution, there is large space for further improvement of the QS method. 

Additionally, by selecting the points along the anti--correlation diagonal, we condition the IR field state exiting the medium on the HHG process. This is because  we select only those shots that are relevant to the harmonic emission, and we remove the unwanted background associated with all residual processes, e.g. electronic excitation or ionization. This action corresponds to the application of the $\Pi_{\Tilde{n}\neq 0} = \mathbbm{1}-\Pi_{\Tilde{0}}$ operator onto the field state given by Eq. \eqref{eq:state:condit:ground}, and its projection onto the harmonic field modes (see Section \ref{Sec:Conditioning:Theory}, Eq.\eqref{eq:Paris_IR_Projection_HHG}) \cite{stammer2022high, stammer2022theory}. This results in the creation of an IR field coherent state superposition which is given by Eq. \eqref{eq:Paris_IR_Projection_HHG}.

Hereafter, for the sake of simplicity, we express the conditioned light states by considering only a single mode for the input IR field (we further omit the oscillation term $e^{-i\omega t}$). In this case Eq.~\eqref{eq:Paris_IR_Projection_HHG} reads 
\begin{equation} \label{eq:cat_IR}
	\ket{\phi_{c}^{(\text{IR})}}_\textrm{HHG}=\ket{\alpha + \chi}-\xi_{\text{IR}} \ket{\alpha},
\end{equation}
which is a genuine optical Schrödinger ``cat-like'' state in the IR spectral range. The approach can likewise be used for creating an optical Schrödinger ``cat-like'' state in the XUV spectral range \cite{stammer2022high}. This can be achieved if the $S_\textrm{XUV}$ contains the integrated signal of all high harmonic orders except one, lets say the $q$th. In this case Eq. \eqref{eq:Paris_XUV_Projection_HHG} reads
\begin{equation} \label{eq:cat_XUV}
	\ket{\chi_{c}^{(q)}}_\textrm{HHG}=\ket{\chi_{q}}-\xi_{q} \ket{0_q},
\end{equation}
which is a coherent state superposition in the spectral range of the $q$th harmonic. In a similar way, the method can be applied to the ATI process using the photoelectron signal $S_\textrm{ATI}^\textrm{(pos, neg)}$. After selecting only the shots that are relevant to the ATI photoelectron emission from the joint distribution $(S_\textrm{ATI}^\textrm{(pos, neg)}, S_{IR})$, we condition the IR field state on the ATI process. In the case that the signal $S_\textrm{ATI}^\textrm{(pos, neg)}$ is integrated over all possible outgoing momenta, the resulting IR state is a mixed state of coherent state superpositions given by Eq. \eqref{eq:rhoATI_mixed_definition}, and is therefore written as
\begin{equation} \label{eq:cat_ATI_cont}
	\rho_{c}^{(ATI)} = \int \dd\vb{p} \ C(\vb{p}) \dyad{\Phi_{ATI}^{\vb{k}_L}(t, \vb{p})}.
\end{equation}

In case of using a multi-cycle fundamental driving field, where the ATI spectrum consists of well defined photoelectron peaks (Fig.~\ref{Fig:HHGATI} (b)), we can approximate the state by
\begin{equation} \label{eq:cat_ATI_discr}
	\rho_{c}^{(ATI)} = \sum_{\vb{p}_{i}} C(\vb{p}_{i}) \dyad{\Phi_{ATI}^{\vb{k}_L}(t, \vb{p}_i)}.
\end{equation}

We note that in Eqs. \eqref{eq:cat_IR}--\eqref{eq:cat_ATI_discr}, the factors $\xi_{\text{IR}}$, $\xi_{q}$ and $C(\vb{p})$, are complex, and reflect the coupling of the initial coherent state with the shifted one. Their absolute values are in the range $0 < |\xi_{\text{IR}}|, |\xi_{q}|, |C(\vb{p})| < 1$ and they depend on $\chi$, $\chi_{q}$ and $\vb{p}$, respectively. 

Additionally, the approach can be used for creating entangled coherent state superpositions between different driving frequency modes \cite{stammer2022high, stammer2022theory}. This can be achieved, for instance, by generating high harmonics using a multi-mode driving field. Such experiments are typically performed using a two--frequency ($\omega_{1} - \omega_{2}$) field (in the visible--infrared spectral range), usually consisting of the fundamental and its second harmonic \cite{Kim_2005, Mauritsson_2006, Fleischer_2014}. In this case, the state of the field before the interaction is $\ket{\alpha_{{\omega_{1}}}}\ket{\alpha_{{\omega_{2}}}}$, which after the interaction and attenuation becomes $\ket{\alpha_{\omega_{1}}+\chi_{\omega_{1}}}\ket{\alpha_{\omega_{2}}+\chi_{\omega_{2}}}$. Following the correlation measurement based procedure used to obtain Eq.~\eqref{eq:cat_IR}, the state conditioned on HHG is given by
\begin{equation} \label{eq:ent_IR}
	\begin{aligned}
	\ket{\phi_{c}^{(\omega_{1}, \omega{2})}}_\textrm{HHG}=\ket{\alpha_{\omega_{1}} + \chi_{\omega_{1}}}\ket{\alpha_{\omega_{2}} + \chi_{\omega_{2}}}
	\\
	-\xi_{(\omega_{1}, \omega_{2})} \ket{\alpha_{\omega_{1}}}\ket{\alpha_{\omega_{2}}},
\end{aligned}
\end{equation}
which is an entangled coherent state in the visible--infrared spectral range. The degree of entanglement for different shifts $\chi_{\omega_{1,2}}$  of the two driving field modes has been discussed in \cite{stammer2022high} in terms of an entanglement witness based on the purity of the reduced density matrix.

\subsubsection{Control of the optical quantum state}
The control of the quantum features of the light states shown in Eqs. \eqref{eq:cat_IR}--\eqref{eq:cat_ATI_discr}, relies on the control of $\chi$, $\chi_{q}$ and $\vb{p}$. This can be achieved by changing the density of atoms in the interaction region, the intensity of the employed laser field, the field polarization, the CEP in case of using a few-cycle laser system or the measured interaction products used by the QS. Evidently, there is a large number of combinations that can be used as ``knobs'' for controlling the quantum features of the coherent state superposition. A representative example which shows the power of the approach on controlling the features of the coherent state superposition can be given using Eq. \eqref{eq:cat_IR}. It can be shown that, when the shift of the coherent state $\chi$ is reduced, the overlap between the initial coherent state and the shifted one gets higher i.e. the value of $|\xi_\textrm{IR}|$ increases, and in the extreme case where $|\xi_\textrm{IR}| \to 1$ the coherent state superposition takes the form $|\phi_{c}^{(\text{IR})}\rangle_\textrm{HHG}\approx D (\alpha)\ket{1}$, which is an optical ``kitten-like'' state \cite{lewenstein2021generation,rivera2022strong}. This has been confirmed experimentally \cite{rivera2022strong} by reducing the gas density in the TA of Unit 2. On the other hand, when the shift of the coherent state $\chi$ is increased (which can be done by increasing the gas density), the overlap between the initial coherent state and the shifted one decreases, i.e. the value of $|\xi_\textrm{IR}|$ is reduced, and in the extreme case where $|\xi_{\text{IR}}| \xrightarrow{} 0$ the coherent state superposition takes the form $|\phi_{c}^{(\text{IR})}\rangle_\textrm{HHG}\approx \ket{\alpha + \chi}$, which is a regular coherent state.

Finally, it is noted that the combination of the aforementioned control ``knobs'', together with optical arrangements consisting of passive linear optical elements (such as phase shifters, beam splitters, fibers etc.), can provide an enormous high number of combinations \cite[and references therein]{Gilchrist_2004} for the generation of large optical ``cat-like'' states, and massively entangled states with controllable quantum features. We note that the beam splitters and the phase shifters are considered as one of the most important optical elements in quantum state engineering and hence, a brief description of their action on a field state, is required. 

\subsubsection{Linear optical elements and ``cat-like'' states from HHG}

{\it Phase shifter}: A phase shifter introduces a phase shift $\varphi$ in the field state. This can be achieved by exploiting the refractive index of the materials introduced in the beam path, or via optical arrangements such as a delay stage. The unitary operator which describes the action of a phase shifter on a field state is $U(\varphi)=\exp[i \varphi b^\dag b)]$. In the case the light field before the phase shifter is in a coherent state $\ket{\beta}$ the state of the outgoing field is 
\begin{equation} \label{eq:PhaseShift}
\begin{aligned}
U(\varphi)\ket{\beta}=\ket{\beta e^{i\varphi}} 
\end{aligned}
\end{equation}
Obviously, if the incoming field is in an optical CSS state, e.g. $\ket{\gamma}+\xi \ket{\beta}$, the state of the output field will be $\ket{\gamma e^{i\varphi}}+\xi \ket{\beta e^{i\varphi}}$.

{\it Beam splitter}: The beam splitter is an optical element which mixes two incoming spatial field modes into two outgoing spatial modes, and is characterized by its transmission $T$ and reflection $R = 1-T$ coefficients. These coefficients depend on the frequency, and usually on the polarization of the light field. Considering that $b$ ($b^\dag$) and $c$ ($c^\dag$) are the annihilation (creation) operators of the two incoming spatial modes onto the BS, the unitary operator which describes the action of the beam splitter on the field state is $B(\theta)=\exp[\theta(b c^\dag-b^\dag c)]$, with $\theta=\arctan(\sqrt{T})$. 

{\it Coherent states on a beam splitter}: Considering that the incoming field modes are the coherent states $\ket{\beta}$, and $\ket{\gamma}$, the outgoing state from the beam splitter is 
\begin{equation} \label{eq:BS}
\begin{aligned}
B(\theta)\ket{\beta} \ket{\gamma}&=\ket{\beta \cos{(\theta)}+\gamma \sin{(\theta)}}_{t} 
\\&\quad \otimes
\ket{-\gamma \cos{(\theta)}+\beta \sin{(\theta)}}_{r}, 
\end{aligned}
\end{equation}
where each of the components in the product state corresponds to one of the outgoing modes of the beam splitter. The subscripts $t$ and $r$ denote the transmitted and reflected parts, respectively. Here, it is important to note, that in quantum optics, when only one field mode (lets say $\ket{\beta}$)  enters the beam splitter, the input of the other mode is described by the vacuum state $\ket{0}$. In this case, it is evident from \eqref{eq:BS} that the output field state is $\ket{\beta \cos{(\theta)}}_{t} \ket{-\beta \sin{(\theta)}}_{r}$.

\begin{figure}
    \centering
    \includegraphics[width=1\columnwidth]{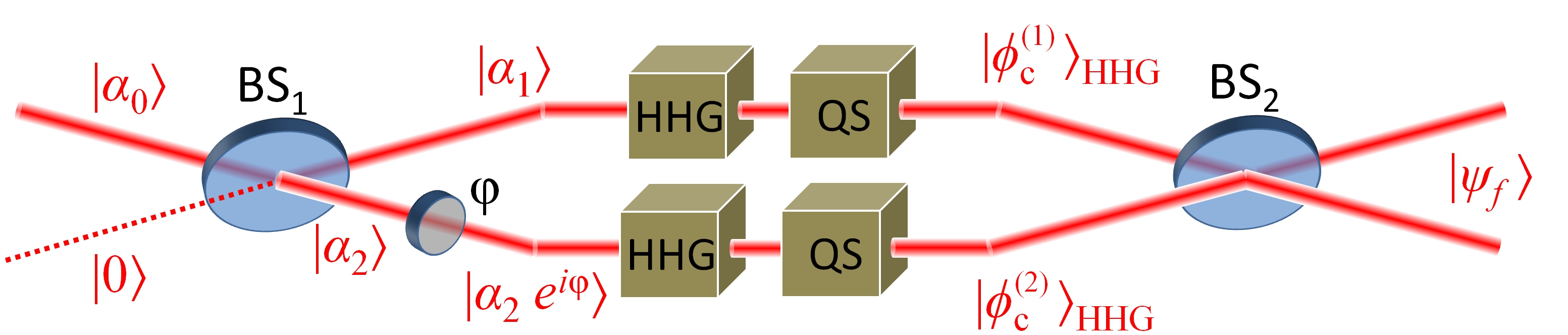}
    \caption{An optical arrangement which leads to the entangled  coherent state $\ket{\psi_{f}}$. $\ket{\alpha_{0}}$ is the initial coherent state of the driving IR field. $\ket{\alpha_{1,2}}$ are the coherent states after the beam splitter BS$_1$. $\ket{\alpha_{2}e^{i\varphi}}$ is the state of the field after the phase shifter which introduces a phase $\varphi$ in $\ket{\alpha_{2}}$. HHG and QS are the HHG areas and the QS arrangements. $\ket{\psi_{f}}$ is the final state after the last beam splitter (BS$_2$). $\ket{\phi_{c}^{(1/2)}}_\textrm{HHG}$ are the optical ``cat-like'' states created after conditioning on HHG. The state $\ket{\psi_{f}}$ can be controlled by the variables $\chi_{1}$, $\chi_{2}$, and $\varphi$.}
    \label{Fig:Exp4}
\end{figure}

{\it Optical ``cat''  states on a beam splitter}: Considering that the incoming field modes are an optical CSS state $\ket{\gamma}+\xi \ket{\beta}$, and the vacuum state $\ket{0}$, the outgoing state from the beam splitter reads
\begin{equation} \label{eq:Bell}
\begin{aligned}
B(\theta)(\ket{\gamma} + \xi \ket{\beta})\ket{0}=\ket{\gamma \cos{(\theta)}}_{t}\ket{-\gamma \sin{(\theta)}}_{r}+ 
\\
\xi \ket{\beta \cos{(\theta)}}_{t}\ket{-\beta \sin{(\theta)}}_{r}, 
\end{aligned}
\end{equation}
which, for a nonzero transmissivity of the beam splitter, takes the form of an entangled coherent state between the two spatial modes, and is of the form
\begin{equation} \label{eq:Ent_Bell1}
\begin{aligned}
\ket{\psi}=\frac {1}{\sqrt{N}}(\ket{\gamma_r}_r\ket{\gamma_t}_{t}+\xi \ket{\beta_r}_{r}\ket{\beta_t}_{t}), 
\end{aligned}
\end{equation}
where $N$ is the normalization factor, $\ket{\gamma_r}_r = \gamma \cos(\theta)$ and $\ket{\gamma_t}_t = -\gamma\sin(\theta)$ (a similar definition follows for $\beta$). 
The entanglement properties of this state is particularly interesting if the distance between the coherent states appearing in each of the terms in the superposition is large enough, and the states $\ket{\gamma_{r,t}}_{r,t}$ and $\ket{\beta_{r,t}}_{r,t}$ are almost orthonormal. 
However, within the HHG approach the generation of the coherent state superpositions is conditioned to relatively small values of the displacement $\chi$, so in this scenario it is of interest to look for approaches that allow us to enlarge the distance between the two coherent states \cite{laghaout_amplification_2013,sychev_enlargement_2017,rivera2021jcompelec}. 
We can go even further with the entanglement by using a multi--beam splitter arrangement and schemes consisting of multiple HHG and QS arrangements \cite{rivera2021jcompelec}. For example, let us assume that we use the arrangement shown in Fig.~\ref{Fig:Exp4}, where a single mode IR coherent state $\ket{\alpha_{0}}$ enters into the system which contains two beam splitters (BS$_1$ and BS$_2$), a phase shifter ($\varphi$), two HHG areas (Unit 2 in Fig.~\ref{Fig:Exp1}), and two QS (Unit 3 in Fig.~\ref{Fig:Exp1}). In this scheme the incoming field modes on the last beam splitter (BS$_2$) are both optical ``cat-like'' states $\ket{\phi_c^{(1)}} = \ket{\gamma_{1}}+\xi_{1} \ket{\beta_{1}}$ and $\ket{\phi_c^{(2)}} = \ket{\gamma_{2}}+\xi_{2} \ket{\beta_{2}}$, and thus the outgoing state from the beam splitter reads  
\begin{equation} \label{eq:Bell_1}
\begin{aligned}
B(\theta)(\ket{\gamma_{1}} + &\xi_{1} \ket{\beta_{1}})(\ket{\gamma_{2}} + \xi_{2} \ket{\beta_{2}})=
\\&\quad
\ket{\gamma_{1} \cos{(\theta)}+ \gamma_{2} \sin{(\theta)})}_{t}
\\&\quad \otimes\ket{-\gamma_{1} \sin{(\theta)} + \gamma_{2} \cos{(\theta)})}_{r} 
\\&\quad
+\xi_{2} \ket{\gamma_{1} \cos{(\theta)}+\beta_{2} \sin{(\theta)}}_{t}
\\&\quad\quad \otimes\ket{-\gamma_{1} \sin{(\theta)}+\beta_{2} \cos{(\theta)}}_{r}
\\&\quad
+\xi_{1} \ket{\beta_{1} \cos{(\theta)}+\gamma_{2} \sin{(\theta)}}_{t}
\\&\quad\quad \otimes\ket{-\beta_{1} \sin{(\theta)}+\gamma_{2} \cos{(\theta)}}_{r}
\\&\quad
+\xi_{1} \xi_{2} \ket{\beta_{1} \cos{(\theta)}+\beta_{2} \sin{(\theta)}}_{t}
\\&\quad\quad \otimes\ket{-\beta_{1} \sin{(\theta)}+\beta_{2} \cos{(\theta)}}_{r}.
\end{aligned}
\end{equation}

Using the Eqs. \eqref{eq:cat_IR}, \eqref{eq:PhaseShift} and \eqref{eq:Bell_1}  in the optical arrangement shown in Fig.~\ref{Fig:Exp4}, and considering for reasons of simplicity a 50:50 beam splitter (i.e. $\ket{\alpha_{1}}=\ket{\alpha_{2}}$ for $\theta = \pi /4$), the output field state $\ket{\psi_{f}}$ reads 
\begin{equation} \label{eq:psi_Fig13}
\begin{aligned}
\ket{\psi_{f}}&=
\ket{\frac {1}{\sqrt{2}}[\alpha_{1}(1+e^{i\varphi})+\chi_{1} + \chi_{2} e^{i\varphi}]}_{t}\\ &\quad
\otimes\ket{\frac {1}{\sqrt{2}}[-\alpha_{1}(1-e^{i\varphi})+\chi_{1} + \chi_{2} e^{i\varphi}]}_{r}\\&\quad
+\xi_{2} \ket{\frac {1}{\sqrt{2}}[\alpha_{1} (1+e^{i\varphi})+\chi_{1}]}_{t}\\&\quad\quad
\otimes\ket{\frac {1}{\sqrt{2}}[-\alpha_{1} (1-e^{i\varphi}) +\chi_{1}] }_{r}\\&\quad
+\xi_{1} \ket{\frac {1}{\sqrt{2}}[\alpha_{1} (1+e^{i\varphi}) +\chi_{2} e^{i\varphi}] }_{t}\\&\quad\quad
\otimes\ket{\frac {1}{\sqrt{2}}[-\alpha_{1} (1-e^{i\varphi}) +\chi_{2} e^{i\varphi}]}_{r}\\&\quad
+\xi_{1} \xi_{2} \ket{\frac {1}{\sqrt{2}}[\alpha_{1} (1 + e^{i\varphi})]}_{t}\\&\quad\quad\quad
\otimes\ket{\frac {1}{\sqrt{2}}[-\alpha_{1}(1 - e^{i\varphi})]}_{r}.
\end{aligned}
\end{equation}
The above expression is an entangled coherent state which can be controlled by the variables $\chi_{1}$, $\chi_{2}$ (and consequently $\xi_{1}$, $\xi_{2}$), and $\varphi$. We note that, by performing measurements over one of the modes, one can generate coherent state superpositions involving more than two coherent states.

\subsubsection{Optical quantum state characterization}\label{Sec:Quantum:characterization}
The characterization of optical light state is a large chapter in quantum optics and is practically impossible to address it completely in a single section of a manuscript. For this reason, here we focus in one of the most commonly used methods named quantum tomography (QT) \cite{Breitenbach_Squeezed_QT,Lvovsky_Rev_QT}, which relies on the use of a homodyne detection technique \cite{Bachor_book_2019} (Unit 4 in Fig.~\ref{Fig:Exp1}).  

According to classical electrodynamics, the mode of E$_{in}$ can be decomposed into two quadrature components $x$ and $p$ oscillating with a $\pi / 2$ phase difference. This is because $E_{in}(t)=(E_{0}e^{i\varphi}e^{-i\omega t}+E^*_0e^{-i\varphi}e^{i\omega t})=\frac {1}{2}(\alpha e^{-i\omega t}+\alpha^*e^{i\omega t})= \cos(\omega t) X+ \sin(\omega t) P$, where $\alpha=E_{0}e^{i\varphi}=X+iP$, $\alpha^*=E^*_{0}e^{-i\varphi}=X-iP$, with $X=\textrm{Re}[\alpha]=\frac {1}{2}(\alpha + \alpha^{*})$, $P=\textrm{Im}[\alpha]=\frac {1}{2i}(\alpha - \alpha^{*})$, and $E_{0}$ is the field amplitude. Following the canonical quantization procedure, the electric field operator $\hat{E}_{in} (t) = E_{0} (\hat{a}e^{-i\omega t}+ \hat{a}^\dagger e^{i\omega t})$ reads
\begin{equation} \label{eq:Electricfield}
\begin{aligned}
\hat{E}_{in} (t) =\sqrt{2} E_{0} ( \cos(\omega t) \hat{x}+ \sin(\omega t) \hat{p}), 
\end{aligned}
\end{equation}
where $\hat{x}=(\hat{a}+\hat{a}^\dagger)/\sqrt{2}$, and $\hat{p}=(\hat{a}-\hat{a}^\dagger)/i\sqrt{2}$ are the non-commuting quadrature field operators, and $\hat{a}$, $\hat{a}^\dagger$ are the photon annihilation and creation operators, respectively. The operators $\hat{x}$ and $\hat{p}$ are the analogues to the position and momentum operators of a particle in a harmonic potential, and satisfy the commutation and uncertainty relations $[\hat{x}, \hat{p}]=i$ and $\Delta x \Delta p \geq \frac {1}{2}$, respectively. 

In Unit 4 of Fig.~\ref{Fig:Exp1}, E$_{in}$ is the field of the state to be characterized. The E$_{in}$ is spatiotemporally overlapped on a beam splitter (BS) with the field of a local oscillator (E$_{r}$) coming from the 2$^\text{nd}$ branch of the interferometer which introduces a controllable delay $\Delta\tau$ (phase shift $\varphi$) between the E$_{r}$ and E$_{in}$ fields. An important requirement for the applicability of the method is that the photon number of the local oscillator field should be much larger than that of E$_{in}$ to be characterized (for details see Ref.~\cite{Yuen_1983}). The fields after the BS are detected by a balanced differential photodetection system consisting of two identical photodiodes (PD), which provides at each value of $\varphi$ the photocurrent difference $i_{\varphi}$. Setting the delay stage around $\Delta\tau\approx 0$, the characterization of the quantum state of light can be achieved by recording for each shot the value of $i_{\varphi}$ as a function of $\varphi$. The beauty of the homodyne detection scheme \cite{Bachor_book_2019} is that it provides access to the measurement of the values of the field quadrature. The values of the photocurrent difference $i_{\varphi}$ are directly proportional to the measurement of the electric field operator i.e., 
 \begin{equation} \label{eq:Electricfield_homo}
\begin{aligned}
 \hat{E}_{in} (\varphi) \propto \hat{x}_{\varphi}=\cos(\varphi) \hat{x}+ \sin(\varphi) \hat{p}. 
 \end{aligned}
\end{equation}
 The unknown proportionality factor between the measured photocurrent and the quadrature values in Eqs. \eqref{eq:Electricfield_homo} can be obtained by using the condition that the variance of the vacuum state is $1/2$, as for the vacuum state $\left\langle \hat{x}_{\varphi} \right\rangle=0$, and $(\Delta x_{\varphi})^{2}=1/2$. In other words, the homodyne data are scaled according to the measured vacuum state quadrature noise. Experimentally, that is all that someone has to do in order to characterize the light state. This is because repeated measurements of $\hat{x}_{\varphi}$ at each $\varphi$ provides the probability distribution $P_{\varphi}(x_{\varphi})=\langle{x_{\varphi}}|{\hat{\rho}}|{x_{\varphi}}\rangle$ of its eigenvalues $x_{\varphi}$, where $\hat\rho\equiv\dyad{\phi}$ is the density operator of the light state to be characterized and $|{x_{\varphi}}\rangle$ the eigenstate with eigenvalue $x_{\varphi}$. The density matrix $\hat\rho$, which provides complete information about the light state, can be obtained in the Fock basis by calculating the matrix elements $\rho_{nm}$ using an iterative {\it Maximum--Likelihood} procedure beautifully described in Ref.~\cite{Lvovsky_MaxLik_alg}. Having these values, the mean photon number of the light state can be obtained by the diagonal elements $\rho_{nn}$ of the density matrix $\hat \rho$, and the relation $\langle{n}\rangle=\sum n\rho_{nn}$. 
 
One of the most complete and commonly used ways to visualize the quantum character of light states is via the quasi--probability distribution in phase space $(x,p)$, namely, the Wigner function $W(x,p)$, which has been extensively addressed over the years by an enormous number of research articles and books (see for instance refs. \cite{Bachor_book_2019, Schleich_Book_2001, Gerry__Book_2001, Leonhardt__Book_2001, Leonhardt__Book_2010}). What is important to be mentioned here is that: a) the marginal distributions $P(x)=\langle{x}|{\hat{\rho}}|{x}\rangle = \int^{+\infty}_{-\infty} \dd p\, W(x,p)$ and $P(p)=\langle{p}|{\hat{\rho}}|{p}\rangle = \int^{+\infty}_{-\infty} \dd x\, W(x,p)$ yield the position and momentum distributions, respectively, and b) if we introduce a phase shift $\varphi$, the $x$ and $p$ components rotate by $\varphi$ in phase--space via the operator ${\hat{U}}(\varphi)$ shown in Eq. \eqref{eq:PhaseShift}. Taking this into account, the equation which connects the $P_{\varphi}(x_{\varphi})$ with $W(x,p)$ for each $\varphi$ is    
\begin{equation} \label{eq:Wigner}
\begin{aligned}
&P_{\varphi}(x_{\varphi})=\langle{x}|{\hat{U}}(\varphi){\hat{\rho}}{\hat{U}^\dagger}(\varphi)|{x}\rangle\\&=
\int^{+\infty}_{-\infty}\dd p_{\varphi} W(x_{\varphi} \cos{\varphi}-p_{\varphi} \sin{\varphi}, x_{\varphi} \sin{\varphi}+p_{\varphi} \cos{\varphi}),
 \end{aligned}
\end{equation}
where $x_{\varphi}=x\cos{\varphi}+p\sin{\varphi}$ and $p_{\varphi}=-x\sin{\varphi}+p\cos{\varphi}$. The above equation is called {\it Radon transformation} and has been extensively studied in the theory of tomographic imaging \cite{Herman__Book_1980}, while the inverse {\it Radon transformation}, implemented via the standard filtered back-projection algorithm, is the key tool in quantum state tomography and quantum state reconstruction \cite[and references therein]{Lvovsky_Rev_QT, Leonhardt__Book_2001}. 
Hence, the Wigner function can be reconstructed by applying the algorithm directly to the quadrature values $x_{\varphi,k}$, using the formula $W(x,p)\simeq \frac{1}{2\pi^{2}N} \sum_{k=1}^{N}K(x\cdot \cos(\varphi_{k})+p\cdot \sin(\varphi_{k})-x_{\varphi, k})$ \cite{Lvovsky_Rev_QT}, where $N$ is the number of the recorded quadrature--phase pairs $(x_{\varphi,k},\varphi_{k})$. In the previous expression, $k$ is the index of each value and $K(z)=\frac{1}{2} \int_{-\infty}^{\infty}\dd\xi\,|\xi|\exp(i\xi z)$ is called integration kernel with $z=x\cdot \cos(\varphi_{k})+p\cdot \sin(\varphi_{k})-x_{\varphi,k}$. The numerical implementation of the integration kernel requires the replacement of the infinite integration limits with a finite cutoff $k_{c}$. In order to reduce the numerical artifacts (rapid oscillations) and allow the details of the Wigner function to be resolved, the value of $k_{c}$ is typically set to be around $k_{c} \sim\! 4$. Alternatively, the $W(x,p)$ can be also obtained using the density matrix formalism \cite{Lvovsky_MaxLik_alg,Banaszek_Wig_rec_dm}.

\begin{figure}
    \centering
    \includegraphics[width=1\columnwidth]{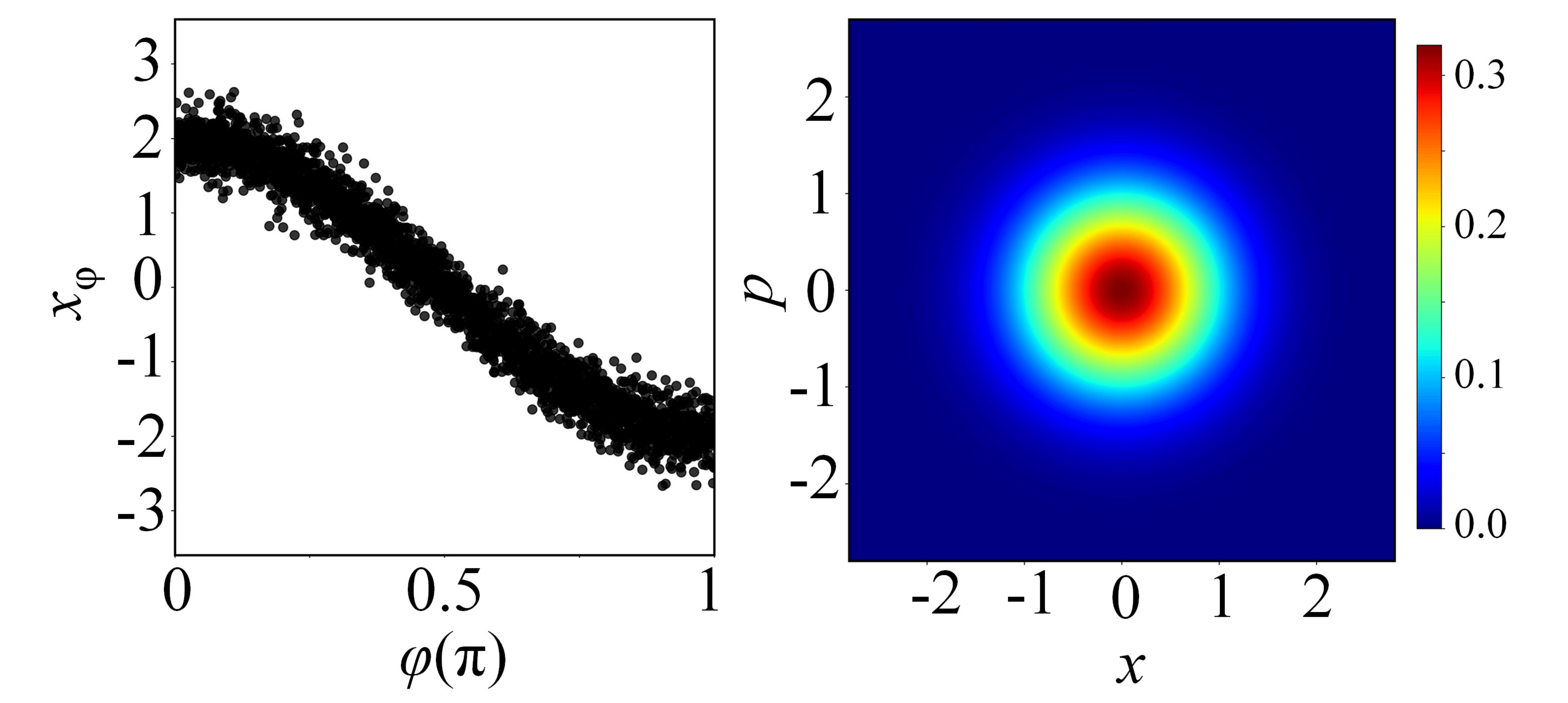}
    \caption{Quantum tomography of an optical coherent state $\ket{\alpha}$. In the left panel we show the calculated homodyne detection signal $x_{\phi}$, while the corresponding Wigner function centered at $|\alpha|=2$ is shown in the right panel.}
    \label{Fig:WignerCoherent}
\end{figure}

In the following, we provide three characteristic examples of the homodyne traces, and their corresponding Wigner functions as have been calculated for an optical coherent, ``cat'', and ``kitten-like'' states. These cases have been experimentally measured in \cite{lewenstein2021generation,rivera2022strong} by means of the arrangement shown in Fig.~\ref{Fig:Exp1}. 
In Fig.~\ref{Fig:WignerCoherent} we show the calculated homodyne trace (left panel) and the Wigner function (right panel) of a coherent state. This corresponds to the case where the HHG process is switched off. 
The calculation has been performed using a coherent state $\ket{\alpha}$ of amplitude $|\alpha|=2$. In this case, $\hat\rho\equiv\dyad{\alpha}$ and therefore the mean value $\left\langle \hat{x}_{\varphi} \right\rangle=\langle{\alpha}|\hat{x}_{\varphi}|{\alpha}\rangle$ exhibits an oscillation with $\left\langle \hat{x}_{\varphi} \right\rangle=|\alpha|\cos{\varphi}$. Furthermore, the $P_{\varphi}$ defined in Eq.~\eqref{eq:Wigner} and the quantum fluctuation (which is the same as the vacuum state $(\Delta x_{\varphi})^{2}=1/2$) remains constant along the cycle. All these characteristic features are shown by the Wigner function which depicts a Gaussian distribution given by
\begin{equation} \label{eq:WignerPlot_coherent_state}
\begin{aligned}
W(\beta) = \frac{2}{\pi} e^{-2|\beta - \alpha|^{2}}.
 \end{aligned}
\end{equation}
In the above equation we have used the transformation $\it x \equiv \Re[\beta-\alpha]$ and $\it p \equiv \Im[\beta-\alpha]$ (where $\beta$ is a variable) which centers the Wigner function at the origin as for $\beta=\alpha$, $x=0$ and $p=0$. This is a convenient way for plotting a Wigner function, as well as writing its expression in a more compact form.

\begin{figure}
    \centering
    \includegraphics[width=1\columnwidth]{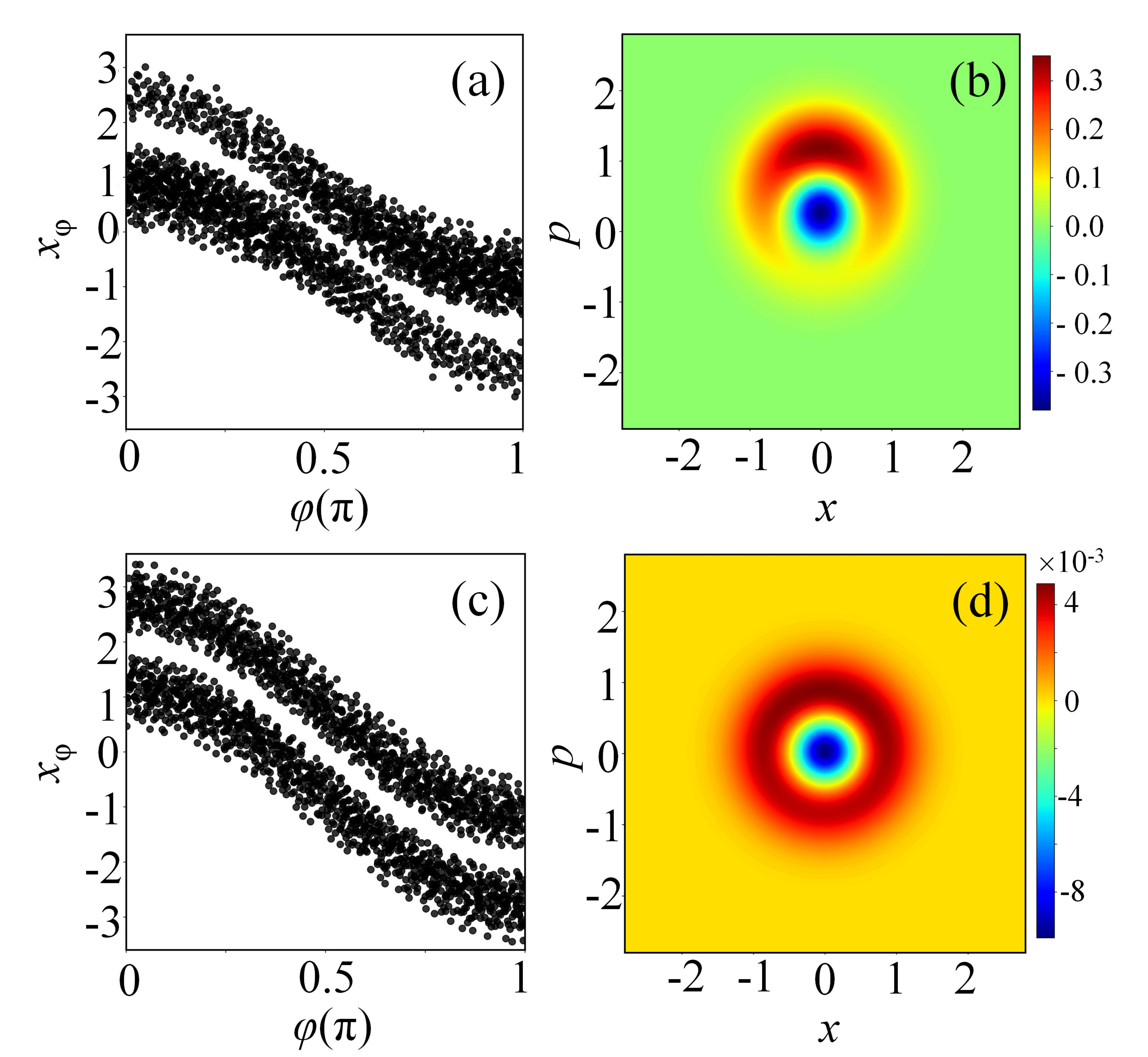}
    \caption{Quantum tomography of optical ``cat-like'' \textcolor{black}{CSS} ((a) and (b)) and ``kitten-like'' \textcolor{black}{CSS} ((c) and (d)) states created when the HHG and the QS are switched on. The left panels show the calculated homodyne detection signal $x_{\phi}$ and the right panels the corresponding Wigner functions centered at $|\alpha|$. The calculations have been preformed using the state  $\ket{\phi_{c}}=\ket{\alpha + \chi}-\xi_{(IR)} \ket{\alpha}$ (with $|\alpha|=2$) created by the conditioning on HHG process. For the optical ``cat-like'' state ((a) and (b)) the $|\chi|$ was set at 0.8, while for the ``kitten-like'' state ((c) and (d)) is reduced to 0.1.}
    \label{Fig:Wigner_cat_kitten}
\end{figure}

The situation changes dramatically when the HHG process and the QS are switched on. In Fig.~\ref{Fig:Wigner_cat_kitten} we show the calculated homodyne traces (left panels) and the corresponding Wigner functions (right panels) of an optical ``cat-like'' (Figs.~\ref{Fig:Wigner_cat_kitten} (a) and (b)) and a ``kitten-like'' (Figs.~\ref{Fig:Wigner_cat_kitten} (c) and (d)) state created when we introduce the conditioning on HHG.
In Figs.~\ref{Fig:Wigner_cat_kitten} (a) and (b), the calculation has been performed using the optical ``cat'' state $\ket{\phi_{c}}=\ket{\alpha + \chi}-\xi_{(IR)} \ket{\alpha}$ shown in Eq. \eqref{eq:cat_IR} with $|\alpha|=2$, and $|\chi|=0.8$. In Figs.~\ref{Fig:Wigner_cat_kitten} (c) and (d), the value of $|\chi|$ has been reduced to 0.1. In both cases, $\hat\rho=\dyad{{\phi_{c}}}$ and the mean value $\left\langle \hat{x}_{\varphi} \right\rangle=\bra{\phi_{c}} \hat{x}_{\varphi}\ket{\phi_{c}}$ exhibits an oscillation with $\left\langle \hat{x}_{\varphi} \right\rangle=\sqrt{\langle{n}\rangle}\cos{\varphi}$. On the other hand, and contrarily to what we saw in the case of a coherent state, $P_{\varphi}$ changes along the cycle and depicts a minimum around the center of the quantum fluctuation which is larger than the vacuum state. All these characteristic features are shown in the equation of Wigner function which reads,
\begin{equation} \label{eq:WignerPlot}
\begin{aligned}
W(\beta)
			=\ &\dfrac{2}{\pi N}
			 \Big[ e^{-2\lvert\beta - \alpha - \chi\rvert^2}
			 + e^{-\lvert\chi\rvert^2}e^{-2\lvert\beta - \alpha\rvert^2}\\
			 &- \big(
			 		e^{2(\beta - \alpha)\chi^*}
			 		+ e^{2(\beta - \alpha)^*\chi}
			 	\big)
			 	e^{-\lvert\chi\rvert^2}e^{-2\lvert\beta - \alpha\rvert^2}
			 	\Big],
 \end{aligned}
\end{equation}
where $N = 1 - e^{-\lvert\chi\rvert^2}$ is the normalization factor for Eq.~\eqref{eq:cat_IR}. The Wigner function depicts the features of a genuine optical ``cat'' (``kitten'' for $\chi=0.1$) state as it has a ring-like shape with a minimum around the center, and shows pronounced non-classical signatures by means of its negative values.

\subsubsection{Analysis of an experimentally reconstructed Wigner function}\label{Sec:Wigner analysis}
 
Obtaining quantitative information about the non-classicality of the optical coherent state superposition measured by a quantum tomography method, requires the optimization $k_{c}$ used to reconstruct the Wigner function and an error analysis. Both are associated with the number of points of the measured homodyne detection signal $x_{\phi}$, and the error introduced by the balanced detector of the homodyne detection system in the measurement of the photon number. 

\begin{figure}
    \centering
    \includegraphics[width=1 \columnwidth]{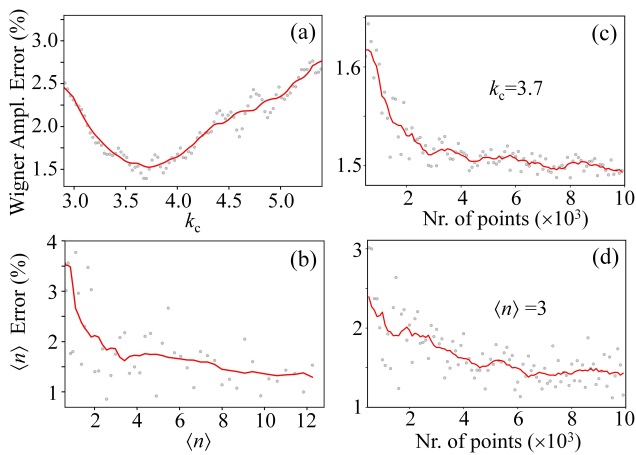}
    \caption{(a) Dependence of the error amplitude of a reconstructed Winger function of a coherent state on $k_{c}$, using a homodyne trace with $\sim 10^{4}$ points for $0\leq\phi\leq\pi$. (b) Dependence of the accuracy of measuring the mean photon number on the mean photon number of the light state. (c) Dependence of the Wigner function amplitude error on the number of points of the measured homodyne detection signal $x_{\phi}$, for $k_{c} \approx 3.7$. (d) Dependence of the accuracy of measuring the mean photon number of the light state on the number of points of the measured homodyne detection signal $x_{\phi}$, for $\langle n \rangle = 3$. In all graphs, the red solid line is a 15-points running average of the data (gray points). Figs (a) and (b) reproduced from Ref.~\cite{rivera2022strong}.}
    \label{Fig:Wigner_Error}
\end{figure}

A convenient way to estimate the error is to use the Wigner function of a coherent state \cite{rivera2022strong}. In this case the error can be obtained by comparing (subtracting) the ideal Wigner function of a coherent state with the Wigner function of the state reconstructed from the experimental data. Figure~\ref{Fig:Wigner_Error}~(a) shows how this procedure can be used to find the optimum value of $k_{c}$ and minimize the error of the amplitude of the reconstructed Wigner function $W(x,p)$. The same procedure can be used for obtaining the error in the mean photon number. This can be done using the density matrix elements $\rho_{nm}$ in the Fock basis $\ket{n}$, where the mean photon number is obtained from the diagonal elements $\rho_{nn}$, and the relation $\langle{n}\rangle=\sum n\rho_{nn}$. Figure~\ref{Fig:Wigner_Error}~(b) shows the dependence of the accuracy of measuring the photon number $\langle n \rangle$ on the mean photon number $\langle n \rangle$ of the light state (Error (\%) $= |\langle n_{rec} \rangle-\langle n_{th} \rangle| / \langle n_{th} \rangle$). Where $\langle n_{rec} \rangle$ is the mean photon number of a coherent state calculated using the number of data points of the measured homodyne detection signal $x_{\phi}$, and $\langle n_{th} \rangle$ is the mean photon number value of an ideal coherent state. As 
shown in Fig.~\ref{Fig:Wigner_Error}~(b), this procedure can been repeated for coherent states of different photon number values. The dependence of the Wigner function amplitude and mean photon number errors on the number of points of the measured homodyne detection signal $x_{\phi}$ is shown in Figs.~\ref{Fig:Wigner_Error}~(c) and (d) for $k_{c} \approx 3.7$ and $\langle n \rangle = 3$, respectively. A homodyne trace which contains $\geq 5 \times 10^3$ measured $x_{\phi}$ values (for $0\leq\phi\leq\pi$) results in a Wigner function amplitude and mean photon number error $\leq 1.5$ \%. 

After the error analysis, the non-classicality of the measured light state can be obtained by quantitatively evaluating its quantum features (such as its negativity) using the reconstructed Wigner function. The negativity is associated with the contrast of the fringe pattern created by quantum interference effects between the coherent states participating in the coherent state superposition. One of the reasons why the negativity can be reduced compared to the ideal case is the presence of decoherence effects on the coherent state superposition after its creation. Although the optical coherent state superpositions can be considered as (almost) decoherence free compared to the particle state superposition, the presence of optical losses during the propagation in optical elements affects the non-classicality of these light states. This leads to a Wigner function with reduced negativity or reduced fringe contrast. Also, the influence of the decoherence effects on the non-classicality of the state depends on the difference between the coherent states in the superposition $\chi$. Hence, large traditional cat-states ($\ket{\alpha} \pm \ket{- \alpha}$) are more sensitive to decoherence effects. The decoherence can be quantitatively studied using a noise model that introduces photon losses due to the interaction with a Gaussian reservoir \cite{rivera2022strong, Leonhardt1993}, via the von Neumann entropy $S$, or using the fidelity $F$ (see e.g. Ref.~\cite{MZhang2021}).

\section{Discussion}

The interaction of intense laser fields with matter leads to a highly non-linear electronic response, and simultaneously influences the electromagnetic field by generating radiation in the previously unpopulated field modes, and leads to a backaction on the fundamental mode which drives the process. Furthermore, the interaction induces correlations among the modes participating in the process. 
With the full quantum electrodynamical description of intense laser-matter interaction introduced in this tutorial, we were able to explicitly show how the two most common processes of high harmonic generation and above-threshold ionization, influence the quantum state of the electromagnetic field modes. Utilizing conditioning procedures on the different processes allowed us to perform quantum state engineering protocols for generating non-classical optical field states with novel properties, such as high photon number, extending towards extreme wavelength regime and using ultrashort light pulses.  

The fully quantized descriptions of HHG and ATI process allow to obtain further insights into the dynamics of intense laser-matter interaction, from a novel perspective by taking into account the quantum nature of the electromagnetic field. 
This allows to answer the questions of the non-classical properties of the electromagnetic field, and to investigate the backaction of the interaction on the driving laser field in intense laser-matter interaction. In particular, we uncovered dynamical changes of the electromagnetic field which could not be addressed by the previous semi-classical theory.

We introduced the quantum electrodynamical description of the process of HHG and ATI, considering the natural continuum of electromagnetic field modes. Considering a conditioning procedure on the emission of harmonic radiation leads to entangled coherent states, and coherent state superposition with non-classical signatures.

We obtained the quantum state of the EM field for the conditioning procedure on ATI, which allowed to obtain the backaction on the fundamental mode. We emphasize that the backaction on the EM field due to the electron dynamics in the ATI process can be decomposed in 3 steps. First, the bound electron wavefunction oscillates in the field, leading to a shift of the EM field modes itself. In the second step, the electron transitions from the ground state to the continuum, which influences the EM field state due to the coupling of the dipole transition matrix element with the electric field operator. In the last step, after ionization, the electron is driven by the field leading to a shift of the EM field state due to the charge current associated with the electron's motion in the continuum.

With the ability to compute the EM field state conditioned on ATI, we obtain field observables of the driving laser mode (photon number distribution and its mean value) to understand the underlying dynamics, and the backaction associated to the process. Interestingly, the direction of the electron propagation (determined by the initial electron momentum after ionization) determines if the shift of the field state amplitude is in or out of phase, leading to an enhancement or depletion of the fundamental mode, respectively. 

This quantum electrodynamical description, together with the conditioning procedures introduced in theory and experiment, gives rise to the opportunity of generating non-classical field states. The applicability and use of those states towards a novel platform for quantum information science will be outlined in the next section. 

\section{Outlook}

This paper provides a tutorial to a new emerging interdisciplinary area of science, in which ultrafast laser physics and attoscience merge with quantum optics and quantum information science. In this sense it serves as an invitation to researchers to join the efforts in this area, and for this reason we devote special attention to the discussion of the outlook and possible future investigations. We decided to organize this section using the different levels: {\bf operational} (concerning ``day-to-day'' generalization and continuations of the research discussed in this tutorial), {\bf tactical} (connecting the present state-of-art status to future merge with quantum technologies, focused on detection, characterization and applications of massive superposition and entangled states generated with the present framework),  and {\bf strategic} (relating to long term goals and aims - the development of the new interdisciplinary platform for quantum technologies that merges quantum information, attoscience and quantum optics).

\subsection{Outlook - Operational level}

Typically, the operational plan describes the day to day running of the operation, in our case the new area of science. The operational plan charts out ``small steps'' or milestones to achieve the higher level (tactical) goals within a realistic time-frame. The present paper sets the QED framework for attoscience, and focuses on some applications related to conditional measurements of HHG and ATI in atomic media that lead to the generation of the massive Schrödinger cat states (massive superposition and entanglement states). There is an obvious ocean of possible generalizations, and applications of the present theory that indeed can be achieved on a ``day-to-day'' basis. Here we list a few:

\begin{itemize}
    \item Even in the presently investigated schemes, there is plenty of place for further studies, such as for instance the role of the CEP, and there is a whole plethora of possibilities associated with using more complex polarization states (elliptic, time-dependent elliptic, locally chiral, etc.),  combining polarization with two-colour fields and structuring polarization in space (cf. \cite{pisanty2019knotting,pisanty2019conservation,rego2019generation,neufeld_floquet_2019, ayuso_synthetic_2019,stammer2020evidence,  ayuso_enantio-sensitive_2021, katsoulis_momentum_2021, mayer_imprinting_2021}). This might lead to creation of novel types of topological Schrödinger cats, very different from those discussed so far in the literature (cf. \cite{dziarmaga2012non}).
    \textcolor{black}{\item 
    Another important consideration is the spin of the electron, which was overlooked until recently \cite{barth_spinpolarized_2013, Hartung2016,eckart_ultrafast_2018,kaushal_looking_2018a,trabert_spin_2018}, where it has been shown that circular strong-field ionization produces spin-polarized electrons. As such, this process may be exploited to produce entanglement between the spin and  the laser field. The electronic spin provides an additional degree of freedom useful in heading towards delivering quantum technologies, as the discrete spin can be more easily measured and manipulated than the continuous electronic degrees of freedom typically considered.
    }
    \item The same QED theory, and conditioning methods, can be applied to multielectron processes in atoms (ionization in general, NSDI by the interplay between Electron Impact Ionization (EII) and Resonant Excitation with Subsequent Ionization (RESI)). These can be at least analysed using the SFA approach.  
    \item The present framework can be 
    extended to more complex targets such as molecules, where e.g. chirality can lead to otherwise symmetry-forbidden electronic responses, like electronic motion perpendicular to the polarization plane \cite{ayuso_ultrafast_2022}, and electronic orbital angular momentum in ionization with linearly polarized light \cite{planas2022strong}, thus expanding the spectrum of possible conditioning setups.
\item A completely new avenue is to apply the present framework to solid state and condensed matter systems. In the first place, this could be done for weakly correlated systems that can be described by semiconductor Bloch equations \cite{vampa_theoretical_2014,vampa_merge_2017}, or a little more flexible Wannier-Bloch approach \cite{osika_wannier-bloch_2017}.
The next step would be to apply the present framework to imaging of topological, chiral and/or strongly correlated systems, such as  topological insulators \cite{Ivanov,Chacon}  
or high $T_c$ superconductors (cf. \cite{Jensandus}).

\end{itemize}

\subsection{Outlook - Tactical level}

In order to seek for the creation of a new feasible platform for quantum technologies, one needs clearly to relate ``what we have'' to the basic aspects of quantum information science: detection, characterization, and finally, applications of generated quantum correlations and quantum entanglement. At this stage, experiments generate in a simplified view the massive superposition of the fundamental laser mode \cite{lewenstein2021generation}, or massively entangled states of several modes \cite{stammer2022high, stammer2022theory}.

The next directions of research should then focus on: 
\begin{itemize}
    \item Detection and characterization methods of massive Schrödinger cat states. This is a formidable challenge \cite{cirac1998quantum}, which for large cats is extremely demanding \cite{skotiniotis2017macroscopic}.
    \item These kind of states are in the first place useful for quantum metrology in general \cite{giovannetti2004quantum, toth2014quantum, pezze2018quantum}, and on photonic platforms \cite{polino2020photonic, gobel2015quantum, barbieri2022optical} in particular, and perhaps in quantum communications. It will thus be important to design concrete protocols for their applications.
\end{itemize}

\subsection{Outlook - Strategic level}

On the strategic level we aim at developing a novel {\bf quantum technology platform combining attoscience, quantum optics and quantum information}. Each of these areas are already at a step where proof-of-principle experiments have been developed. The point is to create the new platform that is totally and intrinsically interdisciplinary. The research presented here, focused in theory and experiments on conditioning, is only a part of this general vision aimed at realising a universal and firmly established tool to offer completely unknown solutions and developments, i.e. a set of stable and reproducible methods to generate massive entangled states and massive quantum superposition. Most, if not all of these methods, involve quantum electrodynamics of intense laser--matter interactions, i.e. the framework described in the present tutorial. Here is the more complete list of the future  research lines, related to these strategic objectives, aims and goals 

\begin{itemize}
    \item {\it Objective 1}:  Generation, characterization and concrete applications in QT of entangled/quantum correlated states using conditioning methods.
    \item {\it Objective 2}: Generation, characterization and concrete applications in QT of entangled/quantum correlated states in strong--field physics and attosecond science driven by quantum light; this includes driving by strongly squeezed light, light with orbital angular momentum (cf.  \cite{bhattacharya2022fermionic}), effects on photoelectron orbital angular momentum \cite{maxwell2021manipulating}, and more.  
    \item {\it Objective 3}:  Generation, characterization and concrete applications in QT of entangled/quantum correlated states in {\it Zerfall} processes, such as single \cite{vrakking2021control, koll2022experimental, vrakking2022ion} or double ionization (cf. \cite{maxwell2021entanglement}).
    \item {\it Objective 4}: Studying quantitative and measurable effects of decoherence in attoscience, by studying reduced density matrices of {\it Zerfall}--products in ionization processes (cf. \cite{bourassin2020quantifying,laurell2022continuous,busto2021probing}).
\end{itemize}
The results of the present tutorial  set up the basis for a road map toward this novel platform of attoscience, quantum optics and quantum information science  for quantum technologies.

\section*{Acknowledgements}

We thank Emilio Pisanty for discussion regarding this work.
ICFO group acknowledges support from: ERC AdG NOQIA; Agencia Estatal de Investigación (R$\&$D project CEX2019-000910-S, funded by MCIN/ AEI/10.13039/501100011033, Plan National FIDEUA PID2019-106901GB-I00, FPI, QUANTERA MAQS PCI2019-111828-2, Proyectos de I+D+I “Retos Colaboración” QUSPIN RTC2019-007196-7); Fundació Cellex; Fundació Mir-Puig; Generalitat de Catalunya through the European Social Fund FEDER and CERCA program (AGAUR Grant No. 2017 SGR 134, QuantumCAT \ U16-011424, co-funded by ERDF Operational Program of Catalonia 2014-2020); EU Horizon 2020 FET-OPEN OPTOlogic (Grant No 899794); National Science Centre, Poland (Symfonia Grant No. 2016/20/W/ST4/00314); European Union’s Horizon 2020 research and innovation programme under the Marie-Skłodowska-Curie grant agreement No 101029393 (STREDCH) and No 847648 (“La Caixa” Junior Leaders fellowships ID100010434: LCF/BQ/PI19/11690013, LCF/BQ/PI20/11760031, LCF/BQ/PR20/11770012, LCF/BQ/PR21/11840013).
P. Tzallas group at FORTH acknowledges LASERLABEUROPE V (H2020-EU.1.4.1.2 grant no.871124), FORTH Synergy Grant AgiIDA (grand no. 00133), the H2020 framework program for research and innovation under the NEP-Europe-Pilot project (no. 101007417). ELI-ALPS is supported by the European Union and co-financed by the European Regional Development Fund (GINOP Grant No. 2.3.6-15-2015-00001).
P.S. acknowledges funding from the European Union’s Horizon 2020
research and innovation programme under the Marie Skłodowska-Curie grant agreement No 847517. 
J.R-D. acknowledges support from the Secretaria d'Universitats i Recerca del Departament d'Empresa i Coneixement de la Generalitat de Catalunya, as well as the European Social Fund (L'FSE inverteix en el teu futur)--FEDER. 
A.S.M. acknowledges funding support from the European Union’s Horizon 2020 research and innovation programme under the Marie Sk\l odowska-Curie grant agreement, SSFI No.\ 887153.
M.F.C. acknowledges financial support from the Guangdong Province Science and Technology Major Project (Future functional materials under extreme conditions - 2021B0301030005).

\bibliography{references}

\begin{thebibliography}{184}%
\makeatletter
\providecommand \@ifxundefined [1]{%
 \@ifx{#1\undefined}
}%
\providecommand \@ifnum [1]{%
 \ifnum #1\expandafter \@firstoftwo
 \else \expandafter \@secondoftwo
 \fi
}%
\providecommand \@ifx [1]{%
 \ifx #1\expandafter \@firstoftwo
 \else \expandafter \@secondoftwo
 \fi
}%
\providecommand \natexlab [1]{#1}%
\providecommand \enquote  [1]{``#1''}%
\providecommand \bibnamefont  [1]{#1}%
\providecommand \bibfnamefont [1]{#1}%
\providecommand \citenamefont [1]{#1}%
\providecommand \href@noop [0]{\@secondoftwo}%
\providecommand \href [0]{\begingroup \@sanitize@url \@href}%
\providecommand \@href[1]{\@@startlink{#1}\@@href}%
\providecommand \@@href[1]{\endgroup#1\@@endlink}%
\providecommand \@sanitize@url [0]{\catcode `\\12\catcode `\$12\catcode
  `\&12\catcode `\#12\catcode `\^12\catcode `\_12\catcode `\%12\relax}%
\providecommand \@@startlink[1]{}%
\providecommand \@@endlink[0]{}%
\providecommand \url  [0]{\begingroup\@sanitize@url \@url }%
\providecommand \@url [1]{\endgroup\@href {#1}{\urlprefix }}%
\providecommand \urlprefix  [0]{URL }%
\providecommand \Eprint [0]{\href }%
\providecommand \doibase [0]{https://doi.org/}%
\providecommand \selectlanguage [0]{\@gobble}%
\providecommand \bibinfo  [0]{\@secondoftwo}%
\providecommand \bibfield  [0]{\@secondoftwo}%
\providecommand \translation [1]{[#1]}%
\providecommand \BibitemOpen [0]{}%
\providecommand \bibitemStop [0]{}%
\providecommand \bibitemNoStop [0]{.\EOS\space}%
\providecommand \EOS [0]{\spacefactor3000\relax}%
\providecommand \BibitemShut  [1]{\csname bibitem#1\endcsname}%
\let\auto@bib@innerbib\@empty
\bibitem [{\citenamefont {Siegman}(1986)}]{Siegman-book}%
  \BibitemOpen
  \bibfield  {author} {\bibinfo {author} {\bibfnamefont {A.}~\bibnamefont
  {Siegman}},\ }\href@noop {} {\emph {\bibinfo {title} {Lasers}}}\ (\bibinfo
  {publisher} {University Science Books, California},\ \bibinfo {year}
  {1986})\BibitemShut {NoStop}%
\bibitem [{\citenamefont {Glauber}(1963{\natexlab{a}})}]{glauber1963coherent}%
  \BibitemOpen
  \bibfield  {author} {\bibinfo {author} {\bibfnamefont {R.~J.}\ \bibnamefont
  {Glauber}},\ }\bibfield  {title} {\bibinfo {title} {Coherent and incoherent
  states of the radiation field},\ }\href@noop {} {\bibfield  {journal}
  {\bibinfo  {journal} {Phys. Rev.}\ }\textbf {\bibinfo {volume} {131}},\
  \bibinfo {pages} {2766} (\bibinfo {year} {1963}{\natexlab{a}})}\BibitemShut
  {NoStop}%
\bibitem [{\citenamefont {Glauber}(1963{\natexlab{b}})}]{glauber1963quantum}%
  \BibitemOpen
  \bibfield  {author} {\bibinfo {author} {\bibfnamefont {R.~J.}\ \bibnamefont
  {Glauber}},\ }\bibfield  {title} {\bibinfo {title} {The quantum theory of
  optical coherence},\ }\href@noop {} {\bibfield  {journal} {\bibinfo
  {journal} {Phys. Rev.}\ }\textbf {\bibinfo {volume} {130}},\ \bibinfo {pages}
  {2529} (\bibinfo {year} {1963}{\natexlab{b}})}\BibitemShut {NoStop}%
\bibitem [{\citenamefont {Sudarshan}(1963)}]{sudarshan1963equivalence}%
  \BibitemOpen
  \bibfield  {author} {\bibinfo {author} {\bibfnamefont {E.}~\bibnamefont
  {Sudarshan}},\ }\bibfield  {title} {\bibinfo {title} {Equivalence of
  semiclassical and quantum mechanical descriptions of statistical light
  beams},\ }\href@noop {} {\bibfield  {journal} {\bibinfo  {journal} {Phys.
  Rev. Lett.}\ }\textbf {\bibinfo {volume} {10}},\ \bibinfo {pages} {277}
  (\bibinfo {year} {1963})}\BibitemShut {NoStop}%
\bibitem [{\citenamefont {Eberly}\ and\ \citenamefont
  {Allen}(1987)}]{Eberlybook}%
  \BibitemOpen
  \bibfield  {author} {\bibinfo {author} {\bibfnamefont {J.~H.}\ \bibnamefont
  {Eberly}}\ and\ \bibinfo {author} {\bibfnamefont {L.}~\bibnamefont {Allen}},\
  }\href@noop {} {\emph {\bibinfo {title} {Optical Resonance and Two Level
  Atoms}}}\ (\bibinfo  {publisher} {Dover Publication, Unitated States of
  America},\ \bibinfo {year} {1987})\BibitemShut {NoStop}%
\bibitem [{\citenamefont {Piszczatowski}\ \emph {et~al.}(2009)\citenamefont
  {Piszczatowski}, \citenamefont {Łach}, \citenamefont {Przybytek},
  \citenamefont {Komasa}, \citenamefont {Pachucki},\ and\ \citenamefont
  {Jeziorski}}]{Pachucki1}%
  \BibitemOpen
  \bibfield  {author} {\bibinfo {author} {\bibfnamefont {S.}~\bibnamefont
  {Piszczatowski}}, \bibinfo {author} {\bibfnamefont {G.}~\bibnamefont
  {Łach}}, \bibinfo {author} {\bibfnamefont {M.}~\bibnamefont {Przybytek}},
  \bibinfo {author} {\bibfnamefont {J.}~\bibnamefont {Komasa}}, \bibinfo
  {author} {\bibfnamefont {K.}~\bibnamefont {Pachucki}},\ and\ \bibinfo
  {author} {\bibfnamefont {B.}~\bibnamefont {Jeziorski}},\ }\bibfield  {title}
  {\bibinfo {title} {Theoretical determination of the dissociation energy of
  molecular hydrogen},\ }\href@noop {} {\bibfield  {journal} {\bibinfo
  {journal} {Journal of Chemical Theory and Computation}\ }\textbf {\bibinfo
  {volume} {5}},\ \bibinfo {pages} {3039} (\bibinfo {year} {2009})}\BibitemShut
  {NoStop}%
\bibitem [{\citenamefont {Pachucki}\ \emph {et~al.}(1996)\citenamefont
  {Pachucki}, \citenamefont {Leibfried}, \citenamefont {Weitz}, \citenamefont
  {Huber}, \citenamefont {König},\ and\ \citenamefont {Hänsch}}]{Pachucki2}%
  \BibitemOpen
  \bibfield  {author} {\bibinfo {author} {\bibfnamefont {K.}~\bibnamefont
  {Pachucki}}, \bibinfo {author} {\bibfnamefont {D.}~\bibnamefont {Leibfried}},
  \bibinfo {author} {\bibfnamefont {M.}~\bibnamefont {Weitz}}, \bibinfo
  {author} {\bibfnamefont {A.}~\bibnamefont {Huber}}, \bibinfo {author}
  {\bibfnamefont {W.}~\bibnamefont {König}},\ and\ \bibinfo {author}
  {\bibfnamefont {T.}~\bibnamefont {Hänsch}},\ }\bibfield  {title} {\bibinfo
  {title} {Theory of the energy levels and precise two-photon spectroscopy of
  atomic hydrogen and deuterium},\ }\href@noop {} {\bibfield  {journal}
  {\bibinfo  {journal} {Journal of Physics B: Atomic, Molecular and Optical
  Physics}\ }\textbf {\bibinfo {volume} {29}},\ \bibinfo {pages} {177}
  (\bibinfo {year} {1996})}\BibitemShut {NoStop}%
\bibitem [{\citenamefont {Drake}(2020)}]{Drake}%
  \BibitemOpen
  \bibfield  {author} {\bibinfo {author} {\bibfnamefont {G.}~\bibnamefont
  {Drake}},\ }\bibfield  {title} {\bibinfo {title} {Accuracy in atomic and
  molecular data},\ }\href@noop {} {\bibfield  {journal} {\bibinfo  {journal}
  {Journal of Physics B: Atomic, Molecular and Optical Physics}\ }\textbf
  {\bibinfo {volume} {53}},\ \bibinfo {pages} {223001} (\bibinfo {year}
  {2020})}\BibitemShut {NoStop}%
\bibitem [{\citenamefont {Larson}\ and\ \citenamefont
  {Mavrogordatos}(2022)}]{Booklarson}%
  \BibitemOpen
  \bibfield  {author} {\bibinfo {author} {\bibfnamefont {J.}~\bibnamefont
  {Larson}}\ and\ \bibinfo {author} {\bibfnamefont {T.}~\bibnamefont
  {Mavrogordatos}},\ }\href@noop {} {\emph {\bibinfo {title} {Jaynes-Cummings
  Model and Its Descendants: Modern research directions (IOP Series in Quantum
  Technology)}}}\ (\bibinfo  {publisher} {IOP Publishing Ltd, Bristol},\
  \bibinfo {year} {2022})\BibitemShut {NoStop}%
\bibitem [{\citenamefont {Drake}(2022)}]{Drake-handbook}%
  \BibitemOpen
  \bibfield  {author} {\bibinfo {author} {\bibfnamefont {G.~W.}\ \bibnamefont
  {Drake}},\ }\href@noop {} {\emph {\bibinfo {title} {Springer Handbook of
  Atomic, Molecular, and Optical Physics}}}\ (\bibinfo  {publisher} {Springer
  Handbooks, Würzburg},\ \bibinfo {year} {2022})\BibitemShut {NoStop}%
\bibitem [{\citenamefont {Amini}\ \emph {et~al.}(2019)\citenamefont {Amini},
  \citenamefont {Biegert}, \citenamefont {Calegari}, \citenamefont
  {Chac{\'o}n}, \citenamefont {Ciappina}, \citenamefont {Dauphin},
  \citenamefont {Efimov}, \citenamefont {de~Morisson~Faria}, \citenamefont
  {Giergiel}, \citenamefont {Gniewek} \emph {et~al.}}]{amini2019symphony}%
  \BibitemOpen
  \bibfield  {author} {\bibinfo {author} {\bibfnamefont {K.}~\bibnamefont
  {Amini}}, \bibinfo {author} {\bibfnamefont {J.}~\bibnamefont {Biegert}},
  \bibinfo {author} {\bibfnamefont {F.}~\bibnamefont {Calegari}}, \bibinfo
  {author} {\bibfnamefont {A.}~\bibnamefont {Chac{\'o}n}}, \bibinfo {author}
  {\bibfnamefont {M.~F.}\ \bibnamefont {Ciappina}}, \bibinfo {author}
  {\bibfnamefont {A.}~\bibnamefont {Dauphin}}, \bibinfo {author} {\bibfnamefont
  {D.~K.}\ \bibnamefont {Efimov}}, \bibinfo {author} {\bibfnamefont {C.~F.}\
  \bibnamefont {de~Morisson~Faria}}, \bibinfo {author} {\bibfnamefont
  {K.}~\bibnamefont {Giergiel}}, \bibinfo {author} {\bibfnamefont
  {P.}~\bibnamefont {Gniewek}}, \emph {et~al.},\ }\bibfield  {title} {\bibinfo
  {title} {Symphony on strong field approximation},\ }\href@noop {} {\bibfield
  {journal} {\bibinfo  {journal} {Rep. Prog. Phys.}\ }\textbf {\bibinfo
  {volume} {82}},\ \bibinfo {pages} {116001} (\bibinfo {year}
  {2019})}\BibitemShut {NoStop}%
\bibitem [{\citenamefont {Agostini}\ \emph {et~al.}(1979)\citenamefont
  {Agostini}, \citenamefont {Fabre}, \citenamefont {Mainfray}, \citenamefont
  {Petite},\ and\ \citenamefont {Rahman}}]{Agostini1979}%
  \BibitemOpen
  \bibfield  {author} {\bibinfo {author} {\bibfnamefont {P.}~\bibnamefont
  {Agostini}}, \bibinfo {author} {\bibfnamefont {F.}~\bibnamefont {Fabre}},
  \bibinfo {author} {\bibfnamefont {G.}~\bibnamefont {Mainfray}}, \bibinfo
  {author} {\bibfnamefont {G.}~\bibnamefont {Petite}},\ and\ \bibinfo {author}
  {\bibfnamefont {N.~K.}\ \bibnamefont {Rahman}},\ }\bibfield  {title}
  {\bibinfo {title} {Free-free transitions following six-photon ionization of
  xenon atoms},\ }\href {https://doi.org/10.1103/PhysRevLett.42.1127}
  {\bibfield  {journal} {\bibinfo  {journal} {Phys. Rev. Lett.}\ }\textbf
  {\bibinfo {volume} {42}},\ \bibinfo {pages} {1127} (\bibinfo {year}
  {1979})}\BibitemShut {NoStop}%
\bibitem [{\citenamefont {Kruit}\ \emph {et~al.}(1983)\citenamefont {Kruit},
  \citenamefont {Kimman}, \citenamefont {Muller},\ and\ \citenamefont {van~der
  Wiel}}]{Kruit1983}%
  \BibitemOpen
  \bibfield  {author} {\bibinfo {author} {\bibfnamefont {P.}~\bibnamefont
  {Kruit}}, \bibinfo {author} {\bibfnamefont {J.}~\bibnamefont {Kimman}},
  \bibinfo {author} {\bibfnamefont {H.~G.}\ \bibnamefont {Muller}},\ and\
  \bibinfo {author} {\bibfnamefont {M.~J.}\ \bibnamefont {van~der Wiel}},\
  }\bibfield  {title} {\bibinfo {title} {Electron spectra from multiphoton
  ionization of xenon at 1064, 532, and 355 nm},\ }\href
  {https://doi.org/10.1103/PhysRevA.28.248} {\bibfield  {journal} {\bibinfo
  {journal} {Phys. Rev. A}\ }\textbf {\bibinfo {volume} {28}},\ \bibinfo
  {pages} {248} (\bibinfo {year} {1983})}\BibitemShut {NoStop}%
\bibitem [{\citenamefont {Becker}\ \emph {et~al.}(2002)\citenamefont {Becker},
  \citenamefont {Grasbon}, \citenamefont {Kopold}, \citenamefont
  {Milo\v{s}evi\'c}, \citenamefont {Paulus},\ and\ \citenamefont
  {Walther}}]{Becker2002}%
  \BibitemOpen
  \bibfield  {author} {\bibinfo {author} {\bibfnamefont {W.}~\bibnamefont
  {Becker}}, \bibinfo {author} {\bibfnamefont {F.}~\bibnamefont {Grasbon}},
  \bibinfo {author} {\bibfnamefont {R.}~\bibnamefont {Kopold}}, \bibinfo
  {author} {\bibfnamefont {D.~B.}\ \bibnamefont {Milo\v{s}evi\'c}}, \bibinfo
  {author} {\bibfnamefont {G.~G.}\ \bibnamefont {Paulus}},\ and\ \bibinfo
  {author} {\bibfnamefont {H.}~\bibnamefont {Walther}},\ }\bibfield  {title}
  {\bibinfo {title} {Above-threshold ionization: From classical features to
  quantum effects},\ }\href {https://doi.org/10.1016/S1049-250X(02)80006-4}
  {\bibfield  {journal} {\bibinfo  {journal} {Adv. At. Mol. Opt. Phys.}\
  }\textbf {\bibinfo {volume} {48}},\ \bibinfo {pages} {35} (\bibinfo {year}
  {2002})}\BibitemShut {NoStop}%
\bibitem [{\citenamefont {McPherson}\ \emph {et~al.}(1987)\citenamefont
  {McPherson}, \citenamefont {Gibson}, \citenamefont {Jara}, \citenamefont
  {Johann}, \citenamefont {Luk}, \citenamefont {McIntyre}, \citenamefont
  {Boyer},\ and\ \citenamefont {Rhodes}}]{McPherson1987}%
  \BibitemOpen
  \bibfield  {author} {\bibinfo {author} {\bibfnamefont {A.}~\bibnamefont
  {McPherson}}, \bibinfo {author} {\bibfnamefont {G.}~\bibnamefont {Gibson}},
  \bibinfo {author} {\bibfnamefont {H.}~\bibnamefont {Jara}}, \bibinfo {author}
  {\bibfnamefont {U.}~\bibnamefont {Johann}}, \bibinfo {author} {\bibfnamefont
  {T.~S.}\ \bibnamefont {Luk}}, \bibinfo {author} {\bibfnamefont {I.~A.}\
  \bibnamefont {McIntyre}}, \bibinfo {author} {\bibfnamefont {K.}~\bibnamefont
  {Boyer}},\ and\ \bibinfo {author} {\bibfnamefont {C.~K.}\ \bibnamefont
  {Rhodes}},\ }\bibfield  {title} {\bibinfo {title} {Studies of multiphoton
  production of vacuum-ultraviolet radiation in the rare gases},\ }\href
  {https://doi.org/10.1364/JOSAB.4.000595} {\bibfield  {journal} {\bibinfo
  {journal} {J. Opt. Soc. Am. B}\ }\textbf {\bibinfo {volume} {4}},\ \bibinfo
  {pages} {595} (\bibinfo {year} {1987})}\BibitemShut {NoStop}%
\bibitem [{\citenamefont {Ferray}\ \emph {et~al.}(1988)\citenamefont {Ferray},
  \citenamefont {L'Huillier}, \citenamefont {Li}, \citenamefont {Lompre},
  \citenamefont {Mainfray},\ and\ \citenamefont {C}}]{Ferray1988}%
  \BibitemOpen
  \bibfield  {author} {\bibinfo {author} {\bibfnamefont {M.}~\bibnamefont
  {Ferray}}, \bibinfo {author} {\bibfnamefont {A.}~\bibnamefont {L'Huillier}},
  \bibinfo {author} {\bibfnamefont {X.}~\bibnamefont {Li}}, \bibinfo {author}
  {\bibfnamefont {L.}~\bibnamefont {Lompre}}, \bibinfo {author} {\bibfnamefont
  {G.}~\bibnamefont {Mainfray}},\ and\ \bibinfo {author} {\bibfnamefont
  {M.}~\bibnamefont {C}},\ }\bibfield  {title} {\bibinfo {title}
  {Multiple-harmonic conversion of 1064 nm radiation in rare gases},\
  }\href@noop {} {\bibfield  {journal} {\bibinfo  {journal} {Journal of Physics
  B: Atomic, Molecular and Optical Physics}\ }\textbf {\bibinfo {volume}
  {21}},\ \bibinfo {pages} {L31} (\bibinfo {year} {1988})}\BibitemShut
  {NoStop}%
\bibitem [{\citenamefont {Lewenstein}\ and\ \citenamefont
  {L'Huillier}(2009)}]{AnneML2}%
  \BibitemOpen
  \bibfield  {author} {\bibinfo {author} {\bibfnamefont {M.}~\bibnamefont
  {Lewenstein}}\ and\ \bibinfo {author} {\bibfnamefont {A.}~\bibnamefont
  {L'Huillier}},\ }\href@noop {} {\emph {\bibinfo {title} {Strong Field Laser
  Physics}}},\ edited by\ \bibinfo {editor} {\bibfnamefont {T.}~\bibnamefont
  {Brabec}}\ (\bibinfo  {publisher} {Springer, New York},\ \bibinfo {year}
  {2009})\ Chap.\ \bibinfo {chapter} {Principles of Single Atom Physics:
  High-Order Harmonic Generation, Above-Threshold Ionization and Non-Sequential
  Ionization}, pp.\ \bibinfo {pages} {147--183}\BibitemShut {NoStop}%
\bibitem [{\citenamefont {l'Huillier}\ \emph {et~al.}(1983)\citenamefont
  {l'Huillier}, \citenamefont {Lompre}, \citenamefont {Mainfray},\ and\
  \citenamefont {Manus}}]{Anne1983a}%
  \BibitemOpen
  \bibfield  {author} {\bibinfo {author} {\bibfnamefont {A.}~\bibnamefont
  {l'Huillier}}, \bibinfo {author} {\bibfnamefont {L.~A.}\ \bibnamefont
  {Lompre}}, \bibinfo {author} {\bibfnamefont {G.}~\bibnamefont {Mainfray}},\
  and\ \bibinfo {author} {\bibfnamefont {C.}~\bibnamefont {Manus}},\ }\bibfield
   {title} {\bibinfo {title} {Multiply charged ions induced by multiphoton
  absorption in rare gases at 0.53 \ensuremath{\mu}m},\ }\href
  {https://doi.org/10.1103/PhysRevA.27.2503} {\bibfield  {journal} {\bibinfo
  {journal} {Phys. Rev. A}\ }\textbf {\bibinfo {volume} {27}},\ \bibinfo
  {pages} {2503} (\bibinfo {year} {1983})}\BibitemShut {NoStop}%
\bibitem [{\citenamefont {Corkum}(1993)}]{Corkum1993}%
  \BibitemOpen
  \bibfield  {author} {\bibinfo {author} {\bibfnamefont {P.~B.}\ \bibnamefont
  {Corkum}},\ }\bibfield  {title} {\bibinfo {title} {Plasma perspective on
  strong field multiphoton ionization},\ }\href
  {https://doi.org/10.1103/PhysRevLett.71.1994} {\bibfield  {journal} {\bibinfo
   {journal} {Phys. Rev. Lett.}\ }\textbf {\bibinfo {volume} {71}},\ \bibinfo
  {pages} {1994} (\bibinfo {year} {1993})}\BibitemShut {NoStop}%
\bibitem [{\citenamefont {Walker}\ \emph {et~al.}(1994)\citenamefont {Walker},
  \citenamefont {Sheehy}, \citenamefont {DiMauro}, \citenamefont {Agostini},
  \citenamefont {Schafer},\ and\ \citenamefont {Kulander}}]{Walker1994}%
  \BibitemOpen
  \bibfield  {author} {\bibinfo {author} {\bibfnamefont {B.}~\bibnamefont
  {Walker}}, \bibinfo {author} {\bibfnamefont {B.}~\bibnamefont {Sheehy}},
  \bibinfo {author} {\bibfnamefont {L.~F.}\ \bibnamefont {DiMauro}}, \bibinfo
  {author} {\bibfnamefont {P.}~\bibnamefont {Agostini}}, \bibinfo {author}
  {\bibfnamefont {K.~J.}\ \bibnamefont {Schafer}},\ and\ \bibinfo {author}
  {\bibfnamefont {K.~C.}\ \bibnamefont {Kulander}},\ }\bibfield  {title}
  {\bibinfo {title} {Precision measurement of strong field double ionization of
  helium},\ }\href {https://doi.org/10.1103/PhysRevLett.73.1227} {\bibfield
  {journal} {\bibinfo  {journal} {Phys. Rev. Lett.}\ }\textbf {\bibinfo
  {volume} {73}},\ \bibinfo {pages} {1227} (\bibinfo {year}
  {1994})}\BibitemShut {NoStop}%
\bibitem [{\citenamefont {Feuerstein}\ \emph {et~al.}(2001)\citenamefont
  {Feuerstein}, \citenamefont {Moshammer}, \citenamefont {Fischer},
  \citenamefont {Dorn}, \citenamefont {Schr\"oter}, \citenamefont
  {Deipenwisch}, \citenamefont {Crespo Lopez-Urrutia}, \citenamefont {H\"ohr},
  \citenamefont {Neumayer}, \citenamefont {Ullrich}, \citenamefont {Rottke},
  \citenamefont {Trump}, \citenamefont {Wittmann}, \citenamefont {Korn},\ and\
  \citenamefont {Sandner}}]{Feuerstein2001}%
  \BibitemOpen
  \bibfield  {author} {\bibinfo {author} {\bibfnamefont {B.}~\bibnamefont
  {Feuerstein}}, \bibinfo {author} {\bibfnamefont {R.}~\bibnamefont
  {Moshammer}}, \bibinfo {author} {\bibfnamefont {D.}~\bibnamefont {Fischer}},
  \bibinfo {author} {\bibfnamefont {A.}~\bibnamefont {Dorn}}, \bibinfo {author}
  {\bibfnamefont {C.~D.}\ \bibnamefont {Schr\"oter}}, \bibinfo {author}
  {\bibfnamefont {J.}~\bibnamefont {Deipenwisch}}, \bibinfo {author}
  {\bibfnamefont {J.~R.}\ \bibnamefont {Crespo Lopez-Urrutia}}, \bibinfo
  {author} {\bibfnamefont {C.}~\bibnamefont {H\"ohr}}, \bibinfo {author}
  {\bibfnamefont {P.}~\bibnamefont {Neumayer}}, \bibinfo {author}
  {\bibfnamefont {J.}~\bibnamefont {Ullrich}}, \bibinfo {author} {\bibfnamefont
  {H.}~\bibnamefont {Rottke}}, \bibinfo {author} {\bibfnamefont
  {C.}~\bibnamefont {Trump}}, \bibinfo {author} {\bibfnamefont
  {M.}~\bibnamefont {Wittmann}}, \bibinfo {author} {\bibfnamefont
  {G.}~\bibnamefont {Korn}},\ and\ \bibinfo {author} {\bibfnamefont
  {W.}~\bibnamefont {Sandner}},\ }\bibfield  {title} {\bibinfo {title}
  {Separation of recollision mechanisms in nonsequential strong field double
  ionization of {Ar}: the role of excitation tunneling},\ }\href
  {https://doi.org/10.1103/PhysRevLett.87.043003} {\bibfield  {journal}
  {\bibinfo  {journal} {Phys. Rev. Lett.}\ }\textbf {\bibinfo {volume} {87}},\
  \bibinfo {pages} {043003} (\bibinfo {year} {2001})}\BibitemShut {NoStop}%
\bibitem [{\citenamefont {Becker}\ \emph
  {et~al.}(2012{\natexlab{a}})\citenamefont {Becker}, \citenamefont {Liu},
  \citenamefont {Ho},\ and\ \citenamefont {Eberly}}]{Eberly2012}%
  \BibitemOpen
  \bibfield  {author} {\bibinfo {author} {\bibfnamefont {W.}~\bibnamefont
  {Becker}}, \bibinfo {author} {\bibfnamefont {X.}~\bibnamefont {Liu}},
  \bibinfo {author} {\bibfnamefont {P.}~\bibnamefont {Ho}},\ and\ \bibinfo
  {author} {\bibfnamefont {J.}~\bibnamefont {Eberly}},\ }\bibfield  {title}
  {\bibinfo {title} {Theories of photoelectron correlation in laser-driven
  multiple atomic ionization},\ }\href@noop {} {\bibfield  {journal} {\bibinfo
  {journal} {Review of Modern Physics}\ }\textbf {\bibinfo {volume} {84}},\
  \bibinfo {pages} {1011} (\bibinfo {year} {2012}{\natexlab{a}})}\BibitemShut
  {NoStop}%
\bibitem [{\citenamefont {Bandrauk}\ and\ \citenamefont
  {Kono}(2006)}]{Bandraukbook}%
  \BibitemOpen
  \bibfield  {author} {\bibinfo {author} {\bibfnamefont {A.}~\bibnamefont
  {Bandrauk}}\ and\ \bibinfo {author} {\bibfnamefont {H.}~\bibnamefont
  {Kono}},\ }\href@noop {} {\emph {\bibinfo {title} {Advances in Multi-Photon
  Processes and Spectroscopy: Molecules in intense laser fields: Nonlinea
  {Multiphoton} {Spectrocopy} and {Near-femtosecond} to {Sub-femtosecond}
  (attosecond) {Dynamics}}}},\ Vol.~\bibinfo {volume} {15}\ (\bibinfo
  {publisher} {World Scientific, Singapore},\ \bibinfo {year} {2006})\ pp.\
  \bibinfo {pages} {149--214}\BibitemShut {NoStop}%
\bibitem [{\citenamefont {Calegari}\ \emph {et~al.}(2016)\citenamefont
  {Calegari}, \citenamefont {Sansone}, \citenamefont {Stagira}, \citenamefont
  {Vozzi},\ and\ \citenamefont {Nisoli}}]{Calegari1}%
  \BibitemOpen
  \bibfield  {author} {\bibinfo {author} {\bibfnamefont {F.}~\bibnamefont
  {Calegari}}, \bibinfo {author} {\bibfnamefont {G.}~\bibnamefont {Sansone}},
  \bibinfo {author} {\bibfnamefont {S.}~\bibnamefont {Stagira}}, \bibinfo
  {author} {\bibfnamefont {C.}~\bibnamefont {Vozzi}},\ and\ \bibinfo {author}
  {\bibfnamefont {M.}~\bibnamefont {Nisoli}},\ }\bibfield  {title} {\bibinfo
  {title} {Advances in attosecond science},\ }\href@noop {} {\bibfield
  {journal} {\bibinfo  {journal} {Journal of Physics B: Atomic, Molecular and
  Optical Physics}\ }\textbf {\bibinfo {volume} {49}},\ \bibinfo {pages}
  {062001} (\bibinfo {year} {2016})}\BibitemShut {NoStop}%
\bibitem [{\citenamefont {Nisoli}\ \emph {et~al.}(2017)\citenamefont {Nisoli},
  \citenamefont {Decleva}, \citenamefont {Calegari}, \citenamefont {Palacios},\
  and\ \citenamefont {Martín}}]{Calegari2}%
  \BibitemOpen
  \bibfield  {author} {\bibinfo {author} {\bibfnamefont {M.}~\bibnamefont
  {Nisoli}}, \bibinfo {author} {\bibfnamefont {P.}~\bibnamefont {Decleva}},
  \bibinfo {author} {\bibfnamefont {F.}~\bibnamefont {Calegari}}, \bibinfo
  {author} {\bibfnamefont {A.}~\bibnamefont {Palacios}},\ and\ \bibinfo
  {author} {\bibfnamefont {F.}~\bibnamefont {Martín}},\ }\bibfield  {title}
  {\bibinfo {title} {Attosecond electron dynamics in molecules},\ }\href@noop
  {} {\bibfield  {journal} {\bibinfo  {journal} {Chem. Rev.}\ }\textbf
  {\bibinfo {volume} {117}},\ \bibinfo {pages} {10760} (\bibinfo {year}
  {2017})}\BibitemShut {NoStop}%
\bibitem [{\citenamefont {Liu}(2015)}]{Chinesebook}%
  \BibitemOpen
  \bibinfo {editor} {\bibfnamefont {Y.}~\bibnamefont {Liu}},\ ed.,\ \href@noop
  {} {\emph {\bibinfo {title} {Advances of Atoms and Molecules in Strong Laser
  Fields}}}\ (\bibinfo  {publisher} {World Scientific, Singapore},\ \bibinfo
  {year} {2015})\BibitemShut {NoStop}%
\bibitem [{\citenamefont {Lezius}\ \emph {et~al.}(1998)\citenamefont {Lezius},
  \citenamefont {Dobosz}, \citenamefont {Normand},\ and\ \citenamefont
  {Schmidt}}]{Lezius}%
  \BibitemOpen
  \bibfield  {author} {\bibinfo {author} {\bibfnamefont {M.}~\bibnamefont
  {Lezius}}, \bibinfo {author} {\bibfnamefont {S.}~\bibnamefont {Dobosz}},
  \bibinfo {author} {\bibfnamefont {D.}~\bibnamefont {Normand}},\ and\ \bibinfo
  {author} {\bibfnamefont {M.}~\bibnamefont {Schmidt}},\ }\bibfield  {title}
  {\bibinfo {title} {Explosion dynamics of rare gas clusters in strong laser
  fields},\ }\href@noop {} {\bibfield  {journal} {\bibinfo  {journal} {Phys.
  Rev. Lett.}\ }\textbf {\bibinfo {volume} {80}},\ \bibinfo {pages} {261}
  (\bibinfo {year} {1998})}\BibitemShut {NoStop}%
\bibitem [{\citenamefont {Jungreuthmayer}\ \emph {et~al.}(2004)\citenamefont
  {Jungreuthmayer}, \citenamefont {Geissler}, \citenamefont {Zanghellini},\
  and\ \citenamefont {Brabec}}]{Brabec}%
  \BibitemOpen
  \bibfield  {author} {\bibinfo {author} {\bibfnamefont {C.}~\bibnamefont
  {Jungreuthmayer}}, \bibinfo {author} {\bibfnamefont {M.}~\bibnamefont
  {Geissler}}, \bibinfo {author} {\bibfnamefont {J.}~\bibnamefont
  {Zanghellini}},\ and\ \bibinfo {author} {\bibfnamefont {T.}~\bibnamefont
  {Brabec}},\ }\bibfield  {title} {\bibinfo {title} {Microscopic analysis of
  large-cluster explosion in intense laser fields},\ }\href@noop {} {\bibfield
  {journal} {\bibinfo  {journal} {Phys. Rev. Lett.}\ }\textbf {\bibinfo
  {volume} {92}},\ \bibinfo {pages} {133401} (\bibinfo {year}
  {2004})}\BibitemShut {NoStop}%
\bibitem [{\citenamefont {Siedschlag}\ \emph {et~al.}(2004)\citenamefont
  {Siedschlag}, \citenamefont {Saalmann},\ and\ \citenamefont {Rost}}]{Rost}%
  \BibitemOpen
  \bibfield  {author} {\bibinfo {author} {\bibfnamefont {C.}~\bibnamefont
  {Siedschlag}}, \bibinfo {author} {\bibfnamefont {U.}~\bibnamefont
  {Saalmann}},\ and\ \bibinfo {author} {\bibfnamefont {J.}~\bibnamefont
  {Rost}},\ }\href@noop {} {\emph {\bibinfo {title} {Latest Advances in Atomic
  Cluster Collisions: CLUSTERS IN INTENSE LASER FIELDS}}},\ Vol.~\bibinfo
  {volume} {15}\ (\bibinfo  {publisher} {World Scientific},\ \bibinfo {year}
  {2004})\ pp.\ \bibinfo {pages} {271--278}\BibitemShut {NoStop}%
\bibitem [{\citenamefont {Vampa}\ and\ \citenamefont
  {Brabec}(2017)}]{vampa_merge_2017}%
  \BibitemOpen
  \bibfield  {author} {\bibinfo {author} {\bibfnamefont {G.}~\bibnamefont
  {Vampa}}\ and\ \bibinfo {author} {\bibfnamefont {T.}~\bibnamefont {Brabec}},\
  }\bibfield  {title} {\bibinfo {title} {Merge of high harmonic generation from
  gases and solids and its implications for attosecond science},\ }\href
  {https://doi.org/10.1088/1361-6455/aa528d} {\bibfield  {journal} {\bibinfo
  {journal} {J. Phys. B}\ }\textbf {\bibinfo {volume} {50}},\ \bibinfo {pages}
  {083001} (\bibinfo {year} {2017})}\BibitemShut {NoStop}%
\bibitem [{\citenamefont {Jiménez-Galán}\ \emph {et~al.}(2020)\citenamefont
  {Jiménez-Galán}, \citenamefont {Silva}, \citenamefont {Smirnova},\ and\
  \citenamefont {Ivanov}}]{Ivanov}%
  \BibitemOpen
  \bibfield  {author} {\bibinfo {author} {\bibfnamefont {A.}~\bibnamefont
  {Jiménez-Galán}}, \bibinfo {author} {\bibfnamefont {R.~E.~F.}\ \bibnamefont
  {Silva}}, \bibinfo {author} {\bibfnamefont {O.}~\bibnamefont {Smirnova}},\
  and\ \bibinfo {author} {\bibfnamefont {M.}~\bibnamefont {Ivanov}},\
  }\bibfield  {title} {\bibinfo {title} {Lightwave control of topological
  properties in 2d materials for sub-cycle and non-resonant valley
  manipulation},\ }\href@noop {} {\bibfield  {journal} {\bibinfo  {journal}
  {Nat. Phot.}\ }\textbf {\bibinfo {volume} {14}},\ \bibinfo {pages} {728}
  (\bibinfo {year} {2020})}\BibitemShut {NoStop}%
\bibitem [{\citenamefont {Chacón}\ \emph {et~al.}(2020)\citenamefont
  {Chacón}, \citenamefont {Kim}, \citenamefont {Zhu}, \citenamefont {Kelly},
  \citenamefont {Dauphin}, \citenamefont {Pisanty}, \citenamefont {Maxwell},
  \citenamefont {Picón}, \citenamefont {Ciappina}, \citenamefont {Kim},
  \citenamefont {Ticknor}, \citenamefont {Saxena},\ and\ \citenamefont
  {Lewenstein}}]{Chacon}%
  \BibitemOpen
  \bibfield  {author} {\bibinfo {author} {\bibfnamefont {A.}~\bibnamefont
  {Chacón}}, \bibinfo {author} {\bibfnamefont {D.}~\bibnamefont {Kim}},
  \bibinfo {author} {\bibfnamefont {W.}~\bibnamefont {Zhu}}, \bibinfo {author}
  {\bibfnamefont {S.}~\bibnamefont {Kelly}}, \bibinfo {author} {\bibfnamefont
  {A.}~\bibnamefont {Dauphin}}, \bibinfo {author} {\bibfnamefont
  {E.}~\bibnamefont {Pisanty}}, \bibinfo {author} {\bibfnamefont
  {A.}~\bibnamefont {Maxwell}}, \bibinfo {author} {\bibfnamefont
  {A.}~\bibnamefont {Picón}}, \bibinfo {author} {\bibfnamefont
  {M.}~\bibnamefont {Ciappina}}, \bibinfo {author} {\bibfnamefont
  {D.}~\bibnamefont {Kim}}, \bibinfo {author} {\bibfnamefont {C.}~\bibnamefont
  {Ticknor}}, \bibinfo {author} {\bibfnamefont {A.}~\bibnamefont {Saxena}},\
  and\ \bibinfo {author} {\bibfnamefont {M.}~\bibnamefont {Lewenstein}},\
  }\bibfield  {title} {\bibinfo {title} {Circular dichroism in higher-order
  harmonic generation: Heralding topological phases and transitions in chern
  insulators},\ }\href@noop {} {\bibfield  {journal} {\bibinfo  {journal}
  {Phys. Rev. B}\ }\textbf {\bibinfo {volume} {102}},\ \bibinfo {pages}
  {134115} (\bibinfo {year} {2020})}\BibitemShut {NoStop}%
\bibitem [{\citenamefont {Eberly}\ \emph {et~al.}(1984)\citenamefont {Eberly},
  \citenamefont {Wódkiewicz},\ and\ \citenamefont {Shore}}]{Wodkiewicz1}%
  \BibitemOpen
  \bibfield  {author} {\bibinfo {author} {\bibfnamefont {J.}~\bibnamefont
  {Eberly}}, \bibinfo {author} {\bibfnamefont {K.}~\bibnamefont
  {Wódkiewicz}},\ and\ \bibinfo {author} {\bibfnamefont {B.}~\bibnamefont
  {Shore}},\ }\bibfield  {title} {\bibinfo {title} {Noise in strong laser-atom
  interactions: Phase telegraph noise},\ }\href
  {https://doi.org/10.1103/PhysRevA.30.2381} {\bibfield  {journal} {\bibinfo
  {journal} {Phys. Rev. A}\ }\textbf {\bibinfo {volume} {30}},\ \bibinfo
  {pages} {2381} (\bibinfo {year} {1984})}\BibitemShut {NoStop}%
\bibitem [{\citenamefont {Wódkiewicz}\ and\ \citenamefont
  {Eberly}(1986)}]{Wodkiewicz2}%
  \BibitemOpen
  \bibfield  {author} {\bibinfo {author} {\bibfnamefont {K.}~\bibnamefont
  {Wódkiewicz}}\ and\ \bibinfo {author} {\bibfnamefont {J.}~\bibnamefont
  {Eberly}},\ }\bibfield  {title} {\bibinfo {title} {Effects of hidden
  transient noise processes on multiphoton absorption},\ }\href
  {https://doi.org/https://doi.org/10.1364/JOSAB.3.000628} {\bibfield
  {journal} {\bibinfo  {journal} {Journal of the Optical Society of America B}\
  }\textbf {\bibinfo {volume} {3}},\ \bibinfo {pages} {628} (\bibinfo {year}
  {1986})}\BibitemShut {NoStop}%
\bibitem [{\citenamefont {Maxwell}\ \emph {et~al.}(2022)\citenamefont
  {Maxwell}, \citenamefont {Madsen},\ and\ \citenamefont
  {Lewenstein}}]{maxwell2021entanglement}%
  \BibitemOpen
  \bibfield  {author} {\bibinfo {author} {\bibfnamefont {A.~S.}\ \bibnamefont
  {Maxwell}}, \bibinfo {author} {\bibfnamefont {L.~B.}\ \bibnamefont
  {Madsen}},\ and\ \bibinfo {author} {\bibfnamefont {M.}~\bibnamefont
  {Lewenstein}},\ }\bibfield  {title} {\bibinfo {title} {Entanglement of
  {{Orbital Angular Momentum}} in {{Non}}-{{Sequential Double Ionization}}},\
  }\href@noop {} {\bibfield  {journal} {\bibinfo  {journal} {Nat. Commun. [in
  press]}\ } (\bibinfo {year} {2022})},\ \Eprint
  {https://arxiv.org/abs/2111.10148} {arXiv:2111.10148} \BibitemShut {NoStop}%
\bibitem [{\citenamefont {Artemyev}\ \emph
  {et~al.}(2017{\natexlab{a}})\citenamefont {Artemyev}, \citenamefont
  {Cederbaum},\ and\ \citenamefont {Demekhin}}]{Cederbaum1}%
  \BibitemOpen
  \bibfield  {author} {\bibinfo {author} {\bibfnamefont {A.~N.}\ \bibnamefont
  {Artemyev}}, \bibinfo {author} {\bibfnamefont {L.~S.}\ \bibnamefont
  {Cederbaum}},\ and\ \bibinfo {author} {\bibfnamefont {P.~V.}\ \bibnamefont
  {Demekhin}},\ }\bibfield  {title} {\bibinfo {title} {Impact of intense laser
  pulses on the autoionization dynamics of the \$2s2p\$ doubly excited state of
  {He}},\ }\href {https://doi.org/10.1103/PhysRevA.96.033410} {\bibfield
  {journal} {\bibinfo  {journal} {Phys. Rev. A}\ }\textbf {\bibinfo {volume}
  {96}},\ \bibinfo {pages} {033410} (\bibinfo {year}
  {2017}{\natexlab{a}})}\BibitemShut {NoStop}%
\bibitem [{\citenamefont {Artemyev}\ \emph
  {et~al.}(2017{\natexlab{b}})\citenamefont {Artemyev}, \citenamefont
  {Cederbaum},\ and\ \citenamefont {Demekhin}}]{Cederbaum2}%
  \BibitemOpen
  \bibfield  {author} {\bibinfo {author} {\bibfnamefont {A.~N.}\ \bibnamefont
  {Artemyev}}, \bibinfo {author} {\bibfnamefont {L.~S.}\ \bibnamefont
  {Cederbaum}},\ and\ \bibinfo {author} {\bibfnamefont {P.~V.}\ \bibnamefont
  {Demekhin}},\ }\bibfield  {title} {\bibinfo {title} {Impact of two-electron
  dynamics and correlations on high-order-harmonic generation in {He}},\ }\href
  {https://doi.org/10.1103/PhysRevA.95.033402} {\bibfield  {journal} {\bibinfo
  {journal} {Phys. Rev. A}\ }\textbf {\bibinfo {volume} {95}},\ \bibinfo
  {pages} {033402} (\bibinfo {year} {2017}{\natexlab{b}})}\BibitemShut
  {NoStop}%
\bibitem [{\citenamefont {Runge}\ and\ \citenamefont {Gross}(1984)}]{Gross1}%
  \BibitemOpen
  \bibfield  {author} {\bibinfo {author} {\bibfnamefont {E.}~\bibnamefont
  {Runge}}\ and\ \bibinfo {author} {\bibfnamefont {E.~K.~U.}\ \bibnamefont
  {Gross}},\ }\bibfield  {title} {\bibinfo {title} {Density-{Functional}
  {Theory} for {Time}-{Dependent} {Systems}},\ }\href
  {https://doi.org/10.1103/PhysRevLett.52.997} {\bibfield  {journal} {\bibinfo
  {journal} {Phys. Rev. Lett.}\ }\textbf {\bibinfo {volume} {52}},\ \bibinfo
  {pages} {997} (\bibinfo {year} {1984})}\BibitemShut {NoStop}%
\bibitem [{\citenamefont {Castro}\ \emph {et~al.}(2015)\citenamefont {Castro},
  \citenamefont {Rubio},\ and\ \citenamefont {Gross}}]{Gross2}%
  \BibitemOpen
  \bibfield  {author} {\bibinfo {author} {\bibfnamefont {A.}~\bibnamefont
  {Castro}}, \bibinfo {author} {\bibfnamefont {A.}~\bibnamefont {Rubio}},\ and\
  \bibinfo {author} {\bibfnamefont {E.~K.~U.}\ \bibnamefont {Gross}},\
  }\bibfield  {title} {\bibinfo {title} {Enhancing and controlling single-atom
  high-harmonic generation spectra: a time-dependent density-functional
  scheme},\ }\href {https://doi.org/10.1140/epjb/e2015-50889-7} {\bibfield
  {journal} {\bibinfo  {journal} {The European Physical Journal B}\ }\textbf
  {\bibinfo {volume} {88}},\ \bibinfo {pages} {191} (\bibinfo {year}
  {2015})}\BibitemShut {NoStop}%
\bibitem [{\citenamefont {Woźniak}\ \emph {et~al.}(2021)\citenamefont
  {Woźniak}, \citenamefont {Lesiuk}, \citenamefont {Przybytek}, \citenamefont
  {Efimov}, \citenamefont {Prauzner-Bechcicki}, \citenamefont {Mandrysz},
  \citenamefont {Ciappina}, \citenamefont {Pisanty}, \citenamefont
  {Zakrzewski}, \citenamefont {Lewenstein},\ and\ \citenamefont
  {Moszyński}}]{Moszynski1}%
  \BibitemOpen
  \bibfield  {author} {\bibinfo {author} {\bibfnamefont {A.~P.}\ \bibnamefont
  {Woźniak}}, \bibinfo {author} {\bibfnamefont {M.}~\bibnamefont {Lesiuk}},
  \bibinfo {author} {\bibfnamefont {M.}~\bibnamefont {Przybytek}}, \bibinfo
  {author} {\bibfnamefont {D.~K.}\ \bibnamefont {Efimov}}, \bibinfo {author}
  {\bibfnamefont {J.~S.}\ \bibnamefont {Prauzner-Bechcicki}}, \bibinfo {author}
  {\bibfnamefont {M.}~\bibnamefont {Mandrysz}}, \bibinfo {author}
  {\bibfnamefont {M.}~\bibnamefont {Ciappina}}, \bibinfo {author}
  {\bibfnamefont {E.}~\bibnamefont {Pisanty}}, \bibinfo {author} {\bibfnamefont
  {J.}~\bibnamefont {Zakrzewski}}, \bibinfo {author} {\bibfnamefont
  {M.}~\bibnamefont {Lewenstein}},\ and\ \bibinfo {author} {\bibfnamefont
  {R.}~\bibnamefont {Moszyński}},\ }\bibfield  {title} {\bibinfo {title} {A
  systematic construction of {Gaussian} basis sets for the description of laser
  field ionization and high-harmonic generation},\ }\href
  {https://doi.org/10.1063/5.0040879} {\bibfield  {journal} {\bibinfo
  {journal} {The Journal of Chemical Physics}\ }\textbf {\bibinfo {volume}
  {154}},\ \bibinfo {pages} {094111} (\bibinfo {year} {2021})}\BibitemShut
  {NoStop}%
\bibitem [{\citenamefont {Woźniak}\ \emph {et~al.}(2022)\citenamefont
  {Woźniak}, \citenamefont {Przybytek}, \citenamefont {Lewenstein},\ and\
  \citenamefont {Moszyński}}]{Moszynski2}%
  \BibitemOpen
  \bibfield  {author} {\bibinfo {author} {\bibfnamefont {A.~P.}\ \bibnamefont
  {Woźniak}}, \bibinfo {author} {\bibfnamefont {M.}~\bibnamefont {Przybytek}},
  \bibinfo {author} {\bibfnamefont {M.}~\bibnamefont {Lewenstein}},\ and\
  \bibinfo {author} {\bibfnamefont {R.}~\bibnamefont {Moszyński}},\ }\bibfield
   {title} {\bibinfo {title} {Effects of electronic correlation on the high
  harmonic generation in helium: a time-dependent configuration interaction
  singles vs time-dependent full configuration interaction study},\ }\href
  {http://arxiv.org/abs/2201.09339} {\bibfield  {journal} {\bibinfo  {journal}
  {arXiv:2201.09339 [physics]}\ } (\bibinfo {year} {2022})}\BibitemShut
  {NoStop}%
\bibitem [{\citenamefont {Joachain}\ \emph {et~al.}(2012)\citenamefont
  {Joachain}, \citenamefont {Kylstra},\ and\ \citenamefont
  {Potvliege}}]{Joachain-book}%
  \BibitemOpen
  \bibfield  {author} {\bibinfo {author} {\bibfnamefont {C.~J.}\ \bibnamefont
  {Joachain}}, \bibinfo {author} {\bibfnamefont {N.~J.}\ \bibnamefont
  {Kylstra}},\ and\ \bibinfo {author} {\bibfnamefont {R.}~\bibnamefont
  {Potvliege}},\ }\href@noop {} {\emph {\bibinfo {title} {Atoms in {Intense}
  {Laser} {Fields}}}}\ (\bibinfo  {publisher} {Cambridge University Press,
  Cambridge},\ \bibinfo {address} {Cambridge},\ \bibinfo {year}
  {2012})\BibitemShut {NoStop}%
\bibitem [{\citenamefont {Leopold}\ and\ \citenamefont
  {Percival}(1978)}]{Percival1}%
  \BibitemOpen
  \bibfield  {author} {\bibinfo {author} {\bibfnamefont {J.~G.}\ \bibnamefont
  {Leopold}}\ and\ \bibinfo {author} {\bibfnamefont {I.~C.}\ \bibnamefont
  {Percival}},\ }\bibfield  {title} {\bibinfo {title} {Microwave {Ionization}
  and {Excitation} of {Rydberg} {Atoms}},\ }\href
  {https://doi.org/10.1103/PhysRevLett.41.944} {\bibfield  {journal} {\bibinfo
  {journal} {Phys. Rev. Lett.}\ }\textbf {\bibinfo {volume} {41}},\ \bibinfo
  {pages} {944} (\bibinfo {year} {1978})}\BibitemShut {NoStop}%
\bibitem [{\citenamefont {Percival}(1977)}]{Percival2}%
  \BibitemOpen
  \bibfield  {author} {\bibinfo {author} {\bibfnamefont {I.~C.}\ \bibnamefont
  {Percival}},\ }\bibfield  {title} {\bibinfo {title} {Semiclassical theory of
  {Bound} {States}},\ }in\ \href {https://doi.org/10.1002/9780470142554.ch1}
  {\emph {\bibinfo {booktitle} {Advances in {Chemical} {Physics}}}}\ (\bibinfo
  {publisher} {John Wiley \& Sons, Ltd},\ \bibinfo {year} {1977})\ pp.\
  \bibinfo {pages} {1--61}\BibitemShut {NoStop}%
\bibitem [{\citenamefont {Grochmalicki}\ \emph {et~al.}(1991)\citenamefont
  {Grochmalicki}, \citenamefont {Lewenstein},\ and\ \citenamefont
  {Rz\c~a\.zewski}}]{Grochmalicki}%
  \BibitemOpen
  \bibfield  {author} {\bibinfo {author} {\bibfnamefont {J.}~\bibnamefont
  {Grochmalicki}}, \bibinfo {author} {\bibfnamefont {M.}~\bibnamefont
  {Lewenstein}},\ and\ \bibinfo {author} {\bibfnamefont {K.}~\bibnamefont
  {Rz\c~a\.zewski}},\ }\bibfield  {title} {\bibinfo {title} {Stabilization of
  atoms in superintense laser fields: {Is} it real?},\ }\href
  {https://doi.org/10.1103/PhysRevLett.66.1038} {\bibfield  {journal} {\bibinfo
   {journal} {Phys. Rev. Lett.}\ }\textbf {\bibinfo {volume} {66}},\ \bibinfo
  {pages} {1038} (\bibinfo {year} {1991})}\BibitemShut {NoStop}%
\bibitem [{\citenamefont {Vampa}\ \emph {et~al.}(2014)\citenamefont {Vampa},
  \citenamefont {McDonald}, \citenamefont {Orlando}, \citenamefont {Klug},
  \citenamefont {Corkum},\ and\ \citenamefont
  {Brabec}}]{vampa_theoretical_2014}%
  \BibitemOpen
  \bibfield  {author} {\bibinfo {author} {\bibfnamefont {G.}~\bibnamefont
  {Vampa}}, \bibinfo {author} {\bibfnamefont {C.~R.}\ \bibnamefont {McDonald}},
  \bibinfo {author} {\bibfnamefont {G.}~\bibnamefont {Orlando}}, \bibinfo
  {author} {\bibfnamefont {D.~D.}\ \bibnamefont {Klug}}, \bibinfo {author}
  {\bibfnamefont {P.~B.}\ \bibnamefont {Corkum}},\ and\ \bibinfo {author}
  {\bibfnamefont {T.}~\bibnamefont {Brabec}},\ }\bibfield  {title} {\bibinfo
  {title} {Theoretical {Analysis} of {High}-{Harmonic} {Generation} in
  {Solids}},\ }\href {https://doi.org/10.1103/PhysRevLett.113.073901}
  {\bibfield  {journal} {\bibinfo  {journal} {Phys. Rev. Lett.}\ }\textbf
  {\bibinfo {volume} {113}},\ \bibinfo {pages} {073901} (\bibinfo {year}
  {2014})}\BibitemShut {NoStop}%
\bibitem [{\citenamefont {Osika}\ \emph {et~al.}(2017)\citenamefont {Osika},
  \citenamefont {Chacón}, \citenamefont {Ortmann}, \citenamefont {Suárez},
  \citenamefont {Pérez-Hernández}, \citenamefont {Szafran}, \citenamefont
  {Ciappina}, \citenamefont {Sols}, \citenamefont {Landsman},\ and\
  \citenamefont {Lewenstein}}]{osika_wannier-bloch_2017}%
  \BibitemOpen
  \bibfield  {author} {\bibinfo {author} {\bibfnamefont {E.~N.}\ \bibnamefont
  {Osika}}, \bibinfo {author} {\bibfnamefont {A.}~\bibnamefont {Chacón}},
  \bibinfo {author} {\bibfnamefont {L.}~\bibnamefont {Ortmann}}, \bibinfo
  {author} {\bibfnamefont {N.}~\bibnamefont {Suárez}}, \bibinfo {author}
  {\bibfnamefont {J.~A.}\ \bibnamefont {Pérez-Hernández}}, \bibinfo {author}
  {\bibfnamefont {B.}~\bibnamefont {Szafran}}, \bibinfo {author} {\bibfnamefont
  {M.~F.}\ \bibnamefont {Ciappina}}, \bibinfo {author} {\bibfnamefont
  {F.}~\bibnamefont {Sols}}, \bibinfo {author} {\bibfnamefont {A.~S.}\
  \bibnamefont {Landsman}},\ and\ \bibinfo {author} {\bibfnamefont
  {M.}~\bibnamefont {Lewenstein}},\ }\bibfield  {title} {\bibinfo {title}
  {Wannier-{Bloch} {Approach} to {Localization} in {High}-{Harmonics}
  {Generation} in {Solids}},\ }\href
  {https://doi.org/10.1103/PhysRevX.7.021017} {\bibfield  {journal} {\bibinfo
  {journal} {Phys. Rev. X}\ }\textbf {\bibinfo {volume} {7}},\ \bibinfo {pages}
  {021017} (\bibinfo {year} {2017})}\BibitemShut {NoStop}%
\bibitem [{\citenamefont {Bauer}\ and\ \citenamefont {Hansen}(2018)}]{Bauer1}%
  \BibitemOpen
  \bibfield  {author} {\bibinfo {author} {\bibfnamefont {D.}~\bibnamefont
  {Bauer}}\ and\ \bibinfo {author} {\bibfnamefont {K.~K.}\ \bibnamefont
  {Hansen}},\ }\bibfield  {title} {\bibinfo {title} {High-{Harmonic}
  {Generation} in {Solids} with and without {Topological} {Edge} {States}},\
  }\href {https://doi.org/10.1103/PhysRevLett.120.177401} {\bibfield  {journal}
  {\bibinfo  {journal} {Phys. Rev. Lett.}\ }\textbf {\bibinfo {volume} {120}},\
  \bibinfo {pages} {177401} (\bibinfo {year} {2018})}\BibitemShut {NoStop}%
\bibitem [{\citenamefont {Jürß}\ and\ \citenamefont {Bauer}(2019)}]{Bauer2}%
  \BibitemOpen
  \bibfield  {author} {\bibinfo {author} {\bibfnamefont {C.}~\bibnamefont
  {Jürß}}\ and\ \bibinfo {author} {\bibfnamefont {D.}~\bibnamefont {Bauer}},\
  }\bibfield  {title} {\bibinfo {title} {High-harmonic generation in
  {Su}-{Schrieffer}-{Heeger} chains},\ }\href
  {https://doi.org/10.1103/PhysRevB.99.195428} {\bibfield  {journal} {\bibinfo
  {journal} {Phys. Rev. B}\ }\textbf {\bibinfo {volume} {99}},\ \bibinfo
  {pages} {195428} (\bibinfo {year} {2019})}\BibitemShut {NoStop}%
\bibitem [{\citenamefont {Baldelli}\ \emph {et~al.}(2022)\citenamefont
  {Baldelli}, \citenamefont {Bhattacharya}, \citenamefont {González-Cuadra},
  \citenamefont {Lewenstein},\ and\ \citenamefont {Graß}}]{Baldelli}%
  \BibitemOpen
  \bibfield  {author} {\bibinfo {author} {\bibfnamefont {N.}~\bibnamefont
  {Baldelli}}, \bibinfo {author} {\bibfnamefont {U.}~\bibnamefont
  {Bhattacharya}}, \bibinfo {author} {\bibfnamefont {D.}~\bibnamefont
  {González-Cuadra}}, \bibinfo {author} {\bibfnamefont {M.}~\bibnamefont
  {Lewenstein}},\ and\ \bibinfo {author} {\bibfnamefont {T.}~\bibnamefont
  {Graß}},\ }\bibfield  {title} {\bibinfo {title} {Detecting {Majorana} {Zero}
  {Modes} via {Strong} {Field} {Dynamics}},\ }\href
  {http://arxiv.org/abs/2202.03547} {\bibfield  {journal} {\bibinfo  {journal}
  {arXiv:2202.03547 [cond-mat]}\ } (\bibinfo {year} {2022})}\BibitemShut
  {NoStop}%
\bibitem [{\citenamefont {Alcalà}\ \emph {et~al.}(2022)\citenamefont
  {Alcalà}, \citenamefont {Bhattacharya}, \citenamefont {Biegert},
  \citenamefont {Ciappina}, \citenamefont {Elu}, \citenamefont {Graß},
  \citenamefont {Grochowski}, \citenamefont {Lewenstein}, \citenamefont
  {Palau}, \citenamefont {Sidiropoulos}, \citenamefont {Steinle},\ and\
  \citenamefont {Tyulnev}}]{Jensandus}%
  \BibitemOpen
  \bibfield  {author} {\bibinfo {author} {\bibfnamefont {J.}~\bibnamefont
  {Alcalà}}, \bibinfo {author} {\bibfnamefont {U.}~\bibnamefont
  {Bhattacharya}}, \bibinfo {author} {\bibfnamefont {J.}~\bibnamefont
  {Biegert}}, \bibinfo {author} {\bibfnamefont {M.}~\bibnamefont {Ciappina}},
  \bibinfo {author} {\bibfnamefont {U.}~\bibnamefont {Elu}}, \bibinfo {author}
  {\bibfnamefont {T.}~\bibnamefont {Graß}}, \bibinfo {author} {\bibfnamefont
  {P.~T.}\ \bibnamefont {Grochowski}}, \bibinfo {author} {\bibfnamefont
  {M.}~\bibnamefont {Lewenstein}}, \bibinfo {author} {\bibfnamefont
  {A.}~\bibnamefont {Palau}}, \bibinfo {author} {\bibfnamefont {T.~P.~H.}\
  \bibnamefont {Sidiropoulos}}, \bibinfo {author} {\bibfnamefont
  {T.}~\bibnamefont {Steinle}},\ and\ \bibinfo {author} {\bibfnamefont
  {I.}~\bibnamefont {Tyulnev}},\ }\bibfield  {title} {\bibinfo {title} {High
  harmonic spectroscopy of quantum phase transitions in a high-${T}_c$
  superconductor},\ }\href {http://arxiv.org/abs/2201.09515} {\bibfield
  {journal} {\bibinfo  {journal} {arXiv:2201.09515 [cond-mat,
  physics:physics]}\ } (\bibinfo {year} {2022})}\BibitemShut {NoStop}%
\bibitem [{\citenamefont {Georges}\ \emph {et~al.}(1996)\citenamefont
  {Georges}, \citenamefont {Kotliar}, \citenamefont {Krauth},\ and\
  \citenamefont {Rozenberg}}]{DMFT1}%
  \BibitemOpen
  \bibfield  {author} {\bibinfo {author} {\bibfnamefont {A.}~\bibnamefont
  {Georges}}, \bibinfo {author} {\bibfnamefont {G.}~\bibnamefont {Kotliar}},
  \bibinfo {author} {\bibfnamefont {W.}~\bibnamefont {Krauth}},\ and\ \bibinfo
  {author} {\bibfnamefont {M.~J.}\ \bibnamefont {Rozenberg}},\ }\bibfield
  {title} {\bibinfo {title} {Dynamical mean-field theory of strongly correlated
  fermion systems and the limit of infinite dimensions},\ }\href
  {https://doi.org/10.1103/RevModPhys.68.13} {\bibfield  {journal} {\bibinfo
  {journal} {Rev. Mod. Phys.}\ }\textbf {\bibinfo {volume} {68}},\ \bibinfo
  {pages} {13} (\bibinfo {year} {1996})}\BibitemShut {NoStop}%
\bibitem [{\citenamefont {Kotliar}\ \emph {et~al.}(2006)\citenamefont
  {Kotliar}, \citenamefont {Savrasov}, \citenamefont {Haule}, \citenamefont
  {Oudovenko}, \citenamefont {Parcollet},\ and\ \citenamefont
  {Marianetti}}]{DMFT2}%
  \BibitemOpen
  \bibfield  {author} {\bibinfo {author} {\bibfnamefont {G.}~\bibnamefont
  {Kotliar}}, \bibinfo {author} {\bibfnamefont {S.~Y.}\ \bibnamefont
  {Savrasov}}, \bibinfo {author} {\bibfnamefont {K.}~\bibnamefont {Haule}},
  \bibinfo {author} {\bibfnamefont {V.~S.}\ \bibnamefont {Oudovenko}}, \bibinfo
  {author} {\bibfnamefont {O.}~\bibnamefont {Parcollet}},\ and\ \bibinfo
  {author} {\bibfnamefont {C.~A.}\ \bibnamefont {Marianetti}},\ }\bibfield
  {title} {\bibinfo {title} {Electronic structure calculations with dynamical
  mean-field theory},\ }\href {https://doi.org/10.1103/RevModPhys.78.865}
  {\bibfield  {journal} {\bibinfo  {journal} {Rev. Mod. Phys.}\ }\textbf
  {\bibinfo {volume} {78}},\ \bibinfo {pages} {865} (\bibinfo {year}
  {2006})}\BibitemShut {NoStop}%
\bibitem [{\citenamefont {Vollhardt}(2010)}]{Vollhardt}%
  \BibitemOpen
  \bibfield  {author} {\bibinfo {author} {\bibfnamefont {D.}~\bibnamefont
  {Vollhardt}},\ }\bibfield  {title} {\bibinfo {title} {Dynamical
  {Mean}‐{Field} {Theory} of {Electronic} {Correlations} in {Models} and
  {Materials}},\ }\href {https://doi.org/10.1063/1.3518901} {\bibfield
  {journal} {\bibinfo  {journal} {AIP Conference Proceedings}\ }\textbf
  {\bibinfo {volume} {1297}},\ \bibinfo {pages} {339} (\bibinfo {year}
  {2010})}\BibitemShut {NoStop}%
\bibitem [{\citenamefont {Lee}(2007)}]{PALee}%
  \BibitemOpen
  \bibfield  {author} {\bibinfo {author} {\bibfnamefont {P.~A.}\ \bibnamefont
  {Lee}},\ }\bibfield  {title} {\bibinfo {title} {From high temperature
  superconductivity to quantum spin liquid: progress in strong correlation
  physics},\ }\href {https://doi.org/10.1088/0034-4885/71/1/012501} {\bibfield
  {journal} {\bibinfo  {journal} {Rep. Prog. Phys.}\ }\textbf {\bibinfo
  {volume} {71}},\ \bibinfo {pages} {012501} (\bibinfo {year}
  {2007})}\BibitemShut {NoStop}%
\bibitem [{\citenamefont {Eberly}\ and\ \citenamefont
  {Javanainen}(1988)}]{ATI-review1}%
  \BibitemOpen
  \bibfield  {author} {\bibinfo {author} {\bibfnamefont {J.~H.}\ \bibnamefont
  {Eberly}}\ and\ \bibinfo {author} {\bibfnamefont {J.}~\bibnamefont
  {Javanainen}},\ }\bibfield  {title} {\bibinfo {title} {Above-threshold
  ionisation},\ }\href {https://doi.org/10.1088/0143-0807/9/4/004} {\bibfield
  {journal} {\bibinfo  {journal} {Eur J. Phys.}\ }\textbf {\bibinfo {volume}
  {9}},\ \bibinfo {pages} {265} (\bibinfo {year} {1988})}\BibitemShut {NoStop}%
\bibitem [{\citenamefont {Eberly}\ \emph {et~al.}(1991)\citenamefont {Eberly},
  \citenamefont {Javanainen},\ and\ \citenamefont
  {Rz\c~a\.zewski}}]{ATI-review2}%
  \BibitemOpen
  \bibfield  {author} {\bibinfo {author} {\bibfnamefont {J.~H.}\ \bibnamefont
  {Eberly}}, \bibinfo {author} {\bibfnamefont {J.}~\bibnamefont {Javanainen}},\
  and\ \bibinfo {author} {\bibfnamefont {K.}~\bibnamefont {Rz\c~a\.zewski}},\
  }\bibfield  {title} {\bibinfo {title} {Above-threshold ionization},\
  }\href@noop {} {\bibfield  {journal} {\bibinfo  {journal} {Physics Reports}\
  }\textbf {\bibinfo {volume} {204}},\ \bibinfo {pages} {331} (\bibinfo {year}
  {1991})}\BibitemShut {NoStop}%
\bibitem [{\citenamefont {Suárez}\ \emph {et~al.}(2015)\citenamefont
  {Suárez}, \citenamefont {Chacón}, \citenamefont {Ciappina}, \citenamefont
  {Biegert},\ and\ \citenamefont {Lewenstein}}]{Noslen1}%
  \BibitemOpen
  \bibfield  {author} {\bibinfo {author} {\bibfnamefont {N.}~\bibnamefont
  {Suárez}}, \bibinfo {author} {\bibfnamefont {A.}~\bibnamefont {Chacón}},
  \bibinfo {author} {\bibfnamefont {M.~F.}\ \bibnamefont {Ciappina}}, \bibinfo
  {author} {\bibfnamefont {J.}~\bibnamefont {Biegert}},\ and\ \bibinfo {author}
  {\bibfnamefont {M.}~\bibnamefont {Lewenstein}},\ }\bibfield  {title}
  {\bibinfo {title} {Above-threshold ionization and photoelectron spectra in
  atomic systems driven by strong laser fields},\ }\href
  {https://doi.org/10.1103/PhysRevA.92.063421} {\bibfield  {journal} {\bibinfo
  {journal} {Phys. Rev. A}\ }\textbf {\bibinfo {volume} {92}},\ \bibinfo
  {pages} {063421} (\bibinfo {year} {2015})}\BibitemShut {NoStop}%
\bibitem [{\citenamefont {Suárez}\ \emph {et~al.}(2016)\citenamefont
  {Suárez}, \citenamefont {Chacón}, \citenamefont {Ciappina}, \citenamefont
  {Wolter}, \citenamefont {Biegert},\ and\ \citenamefont
  {Lewenstein}}]{Noslen2}%
  \BibitemOpen
  \bibfield  {author} {\bibinfo {author} {\bibfnamefont {N.}~\bibnamefont
  {Suárez}}, \bibinfo {author} {\bibfnamefont {A.}~\bibnamefont {Chacón}},
  \bibinfo {author} {\bibfnamefont {M.~F.}\ \bibnamefont {Ciappina}}, \bibinfo
  {author} {\bibfnamefont {B.}~\bibnamefont {Wolter}}, \bibinfo {author}
  {\bibfnamefont {J.}~\bibnamefont {Biegert}},\ and\ \bibinfo {author}
  {\bibfnamefont {M.}~\bibnamefont {Lewenstein}},\ }\bibfield  {title}
  {\bibinfo {title} {Above-threshold ionization and laser-induced electron
  diffraction in diatomic molecules},\ }\href
  {https://doi.org/10.1103/PhysRevA.94.043423} {\bibfield  {journal} {\bibinfo
  {journal} {Phys. Rev. A}\ }\textbf {\bibinfo {volume} {94}},\ \bibinfo
  {pages} {043423} (\bibinfo {year} {2016})}\BibitemShut {NoStop}%
\bibitem [{\citenamefont {Suárez}\ \emph {et~al.}(2017)\citenamefont
  {Suárez}, \citenamefont {Chacón}, \citenamefont {Pérez-Hernández},
  \citenamefont {Biegert}, \citenamefont {Lewenstein},\ and\ \citenamefont
  {Ciappina}}]{Noslen3}%
  \BibitemOpen
  \bibfield  {author} {\bibinfo {author} {\bibfnamefont {N.}~\bibnamefont
  {Suárez}}, \bibinfo {author} {\bibfnamefont {A.}~\bibnamefont {Chacón}},
  \bibinfo {author} {\bibfnamefont {J.~A.}\ \bibnamefont {Pérez-Hernández}},
  \bibinfo {author} {\bibfnamefont {J.}~\bibnamefont {Biegert}}, \bibinfo
  {author} {\bibfnamefont {M.}~\bibnamefont {Lewenstein}},\ and\ \bibinfo
  {author} {\bibfnamefont {M.~F.}\ \bibnamefont {Ciappina}},\ }\bibfield
  {title} {\bibinfo {title} {High-order-harmonic generation in atomic and
  molecular systems},\ }\href {https://doi.org/10.1103/PhysRevA.95.033415}
  {\bibfield  {journal} {\bibinfo  {journal} {Phys. Rev. A}\ }\textbf {\bibinfo
  {volume} {95}},\ \bibinfo {pages} {033415} (\bibinfo {year}
  {2017})}\BibitemShut {NoStop}%
\bibitem [{\citenamefont {Suárez}\ \emph {et~al.}(2018)\citenamefont
  {Suárez}, \citenamefont {Chacón}, \citenamefont {Pisanty}, \citenamefont
  {Ortmann}, \citenamefont {Landsman}, \citenamefont {Picón}, \citenamefont
  {Biegert}, \citenamefont {Lewenstein},\ and\ \citenamefont
  {Ciappina}}]{NOslen4}%
  \BibitemOpen
  \bibfield  {author} {\bibinfo {author} {\bibfnamefont {N.}~\bibnamefont
  {Suárez}}, \bibinfo {author} {\bibfnamefont {A.}~\bibnamefont {Chacón}},
  \bibinfo {author} {\bibfnamefont {E.}~\bibnamefont {Pisanty}}, \bibinfo
  {author} {\bibfnamefont {L.}~\bibnamefont {Ortmann}}, \bibinfo {author}
  {\bibfnamefont {A.~S.}\ \bibnamefont {Landsman}}, \bibinfo {author}
  {\bibfnamefont {A.}~\bibnamefont {Picón}}, \bibinfo {author} {\bibfnamefont
  {J.}~\bibnamefont {Biegert}}, \bibinfo {author} {\bibfnamefont
  {M.}~\bibnamefont {Lewenstein}},\ and\ \bibinfo {author} {\bibfnamefont
  {M.~F.}\ \bibnamefont {Ciappina}},\ }\bibfield  {title} {\bibinfo {title}
  {Above-threshold ionization in multicenter molecules: {The} role of the
  initial state},\ }\href {https://doi.org/10.1103/PhysRevA.97.033415}
  {\bibfield  {journal} {\bibinfo  {journal} {Phys. Rev. A}\ }\textbf {\bibinfo
  {volume} {97}},\ \bibinfo {pages} {033415} (\bibinfo {year}
  {2018})}\BibitemShut {NoStop}%
\bibitem [{\citenamefont {Suárez~Rojas}(2018)}]{Noslen5}%
  \BibitemOpen
  \bibfield  {author} {\bibinfo {author} {\bibfnamefont {N.}~\bibnamefont
  {Suárez~Rojas}},\ }\bibfield  {title} {\bibinfo {title} {Strong-field
  processes in atoms and polyatomic molecules},\ }\href
  {https://upcommons.upc.edu/handle/2117/114001} {\bibfield  {journal}
  {\bibinfo  {journal} {TDX (Tesis Doctorals en Xarxa)}\ } (\bibinfo {year}
  {2018})}\BibitemShut {NoStop}%
\bibitem [{\citenamefont {Schuricke}\ \emph {et~al.}(2011)\citenamefont
  {Schuricke}, \citenamefont {Zhu}, \citenamefont {Steinmann}, \citenamefont
  {Simeonidis}, \citenamefont {Ivanov}, \citenamefont {Kheifets}, \citenamefont
  {Grum-Grzhimailo}, \citenamefont {Bartschat}, \citenamefont {Dorn},\ and\
  \citenamefont {Ullrich}}]{Schuricke2011}%
  \BibitemOpen
  \bibfield  {author} {\bibinfo {author} {\bibfnamefont {M.}~\bibnamefont
  {Schuricke}}, \bibinfo {author} {\bibfnamefont {G.}~\bibnamefont {Zhu}},
  \bibinfo {author} {\bibfnamefont {J.}~\bibnamefont {Steinmann}}, \bibinfo
  {author} {\bibfnamefont {K.}~\bibnamefont {Simeonidis}}, \bibinfo {author}
  {\bibfnamefont {I.}~\bibnamefont {Ivanov}}, \bibinfo {author} {\bibfnamefont
  {A.}~\bibnamefont {Kheifets}}, \bibinfo {author} {\bibfnamefont {A.~N.}\
  \bibnamefont {Grum-Grzhimailo}}, \bibinfo {author} {\bibfnamefont
  {K.}~\bibnamefont {Bartschat}}, \bibinfo {author} {\bibfnamefont
  {A.}~\bibnamefont {Dorn}},\ and\ \bibinfo {author} {\bibfnamefont
  {J.}~\bibnamefont {Ullrich}},\ }\bibfield  {title} {\bibinfo {title}
  {Strong-field ionization of lithium},\ }\href
  {https://doi.org/10.1103/PhysRevA.83.023413} {\bibfield  {journal} {\bibinfo
  {journal} {Phys. Rev. A}\ }\textbf {\bibinfo {volume} {83}},\ \bibinfo
  {pages} {023413} (\bibinfo {year} {2011})}\BibitemShut {NoStop}%
\bibitem [{\citenamefont {Taylor}\ \emph {et~al.}(2003)\citenamefont {Taylor},
  \citenamefont {Parker}, \citenamefont {Meharg},\ and\ \citenamefont
  {Dundas}}]{Taylor1}%
  \BibitemOpen
  \bibfield  {author} {\bibinfo {author} {\bibfnamefont {K.~T.}\ \bibnamefont
  {Taylor}}, \bibinfo {author} {\bibfnamefont {J.~S.}\ \bibnamefont {Parker}},
  \bibinfo {author} {\bibfnamefont {K.~J.}\ \bibnamefont {Meharg}},\ and\
  \bibinfo {author} {\bibfnamefont {D.}~\bibnamefont {Dundas}},\ }\bibfield
  {title} {\bibinfo {title} {Laser-driven helium at 780 nm},\ }\href
  {https://doi.org/10.1140/epjd/e2003-00074-0} {\bibfield  {journal} {\bibinfo
  {journal} {The European Physical Journal D - Atomic, Molecular, Optical and
  Plasma Physics}\ }\textbf {\bibinfo {volume} {26}},\ \bibinfo {pages} {67}
  (\bibinfo {year} {2003})}\BibitemShut {NoStop}%
\bibitem [{\citenamefont {Parker}\ \emph {et~al.}(2000)\citenamefont {Parker},
  \citenamefont {Moore}, \citenamefont {Dundas},\ and\ \citenamefont
  {Taylor}}]{Taylor2}%
  \BibitemOpen
  \bibfield  {author} {\bibinfo {author} {\bibfnamefont {J.~S.}\ \bibnamefont
  {Parker}}, \bibinfo {author} {\bibfnamefont {L.~R.}\ \bibnamefont {Moore}},
  \bibinfo {author} {\bibfnamefont {D.}~\bibnamefont {Dundas}},\ and\ \bibinfo
  {author} {\bibfnamefont {K.~T.}\ \bibnamefont {Taylor}},\ }\bibfield  {title}
  {\bibinfo {title} {Double ionization of helium at 390 nm},\ }\href
  {https://doi.org/10.1088/0953-4075/33/20/106} {\bibfield  {journal} {\bibinfo
   {journal} {Journal of Physics B: Atomic, Molecular and Optical Physics}\
  }\textbf {\bibinfo {volume} {33}},\ \bibinfo {pages} {L691} (\bibinfo {year}
  {2000})}\BibitemShut {NoStop}%
\bibitem [{\citenamefont {Moore}\ \emph {et~al.}(2001)\citenamefont {Moore},
  \citenamefont {Parker}, \citenamefont {Dundas}, \citenamefont {Cairns},\ and\
  \citenamefont {Taylor}}]{moore_two_2001}%
  \BibitemOpen
  \bibfield  {author} {\bibinfo {author} {\bibfnamefont {L.}~\bibnamefont
  {Moore}}, \bibinfo {author} {\bibfnamefont {J.}~\bibnamefont {Parker}},
  \bibinfo {author} {\bibfnamefont {D.}~\bibnamefont {Dundas}}, \bibinfo
  {author} {\bibfnamefont {K.}~\bibnamefont {Cairns}},\ and\ \bibinfo {author}
  {\bibfnamefont {K.}~\bibnamefont {Taylor}},\ }\bibfield  {title} {\bibinfo
  {title} {The two electron response in laser driven helium: {NATO} {Advanced}
  {Research} {Workshop} on {Super}-{Intense} {Laser}-{Atom} {Physics}}\
  }(\bibinfo {year} {2001})\ pp.\ \bibinfo {pages} {107--116}\BibitemShut
  {NoStop}%
\bibitem [{\citenamefont {Muller}(2001)}]{Muller1}%
  \BibitemOpen
  \bibfield  {author} {\bibinfo {author} {\bibfnamefont {H.~G.}\ \bibnamefont
  {Muller}},\ }\bibfield  {title} {\bibinfo {title} {Non-sequential double
  ionization of helium and related wave-function dynamics obtained from a
  five-dimensional grid calculation.},\ }\href
  {https://doi.org/10.1364/OE.8.000417} {\bibfield  {journal} {\bibinfo
  {journal} {Optics Express}\ }\textbf {\bibinfo {volume} {8}},\ \bibinfo
  {pages} {417} (\bibinfo {year} {2001})}\BibitemShut {NoStop}%
\bibitem [{\citenamefont {Grobe}\ and\ \citenamefont
  {Eberly}(1992)}]{Ebery-Grobe}%
  \BibitemOpen
  \bibfield  {author} {\bibinfo {author} {\bibfnamefont {R.}~\bibnamefont
  {Grobe}}\ and\ \bibinfo {author} {\bibfnamefont {J.~H.}\ \bibnamefont
  {Eberly}},\ }\bibfield  {title} {\bibinfo {title} {Photoelectron spectra for
  a two-electron system in a strong laser field},\ }\bibfield  {journal}
  {\bibinfo  {journal} {Phys. Rev. Lett.}\ }\textbf {\bibinfo {volume} {68}},\
  \href {https://doi.org/10.1103/PhysRevLett.68.2905}
  {10.1103/PhysRevLett.68.2905} (\bibinfo {year} {1992})\BibitemShut {NoStop}%
\bibitem [{\citenamefont {Liu}\ \emph {et~al.}(1999)\citenamefont {Liu},
  \citenamefont {Eberly}, \citenamefont {Haan},\ and\ \citenamefont
  {Grobe}}]{eberly-corr}%
  \BibitemOpen
  \bibfield  {author} {\bibinfo {author} {\bibfnamefont {W.-C.}\ \bibnamefont
  {Liu}}, \bibinfo {author} {\bibfnamefont {J.~H.}\ \bibnamefont {Eberly}},
  \bibinfo {author} {\bibfnamefont {S.~L.}\ \bibnamefont {Haan}},\ and\
  \bibinfo {author} {\bibfnamefont {R.}~\bibnamefont {Grobe}},\ }\bibfield
  {title} {\bibinfo {title} {Correlation {Effects} in {Two}-{Electron} {Model}
  {Atoms} in {Intense} {Laser} {Fields}},\ }\href
  {https://doi.org/10.1103/PhysRevLett.83.520} {\bibfield  {journal} {\bibinfo
  {journal} {Phys. Rev. Lett.}\ }\textbf {\bibinfo {volume} {83}},\ \bibinfo
  {pages} {520} (\bibinfo {year} {1999})}\BibitemShut {NoStop}%
\bibitem [{\citenamefont {Becker}\ \emph
  {et~al.}(2012{\natexlab{b}})\citenamefont {Becker}, \citenamefont {Liu},
  \citenamefont {Ho},\ and\ \citenamefont {Eberly}}]{Becker-rmp}%
  \BibitemOpen
  \bibfield  {author} {\bibinfo {author} {\bibfnamefont {W.}~\bibnamefont
  {Becker}}, \bibinfo {author} {\bibfnamefont {X.}~\bibnamefont {Liu}},
  \bibinfo {author} {\bibfnamefont {P.~J.}\ \bibnamefont {Ho}},\ and\ \bibinfo
  {author} {\bibfnamefont {J.~H.}\ \bibnamefont {Eberly}},\ }\bibfield  {title}
  {\bibinfo {title} {Theories of photoelectron correlation in laser-driven
  multiple atomic ionization},\ }\href
  {https://doi.org/10.1103/RevModPhys.84.1011} {\bibfield  {journal} {\bibinfo
  {journal} {Reviews of Modern Physics}\ }\textbf {\bibinfo {volume} {84}},\
  \bibinfo {pages} {1011} (\bibinfo {year} {2012}{\natexlab{b}})}\BibitemShut
  {NoStop}%
\bibitem [{\citenamefont {{de Morisson Faria}}\ and\ \citenamefont
  {Liu}(2011)}]{demorissonfaria_electron_2011}%
  \BibitemOpen
  \bibfield  {author} {\bibinfo {author} {\bibfnamefont {C.~F.}\ \bibnamefont
  {{de Morisson Faria}}}\ and\ \bibinfo {author} {\bibfnamefont
  {X.}~\bibnamefont {Liu}},\ }\bibfield  {title} {\bibinfo {title}
  {Electron\textendash electron correlation in strong laser fields},\ }\href
  {https://doi.org/10.1080/09500340.2010.543958} {\bibfield  {journal}
  {\bibinfo  {journal} {Journal of Modern Optics}\ }\textbf {\bibinfo {volume}
  {58}},\ \bibinfo {pages} {1076} (\bibinfo {year} {2011})}\BibitemShut
  {NoStop}%
\bibitem [{\citenamefont {Sacha}\ and\ \citenamefont
  {Eckhardt}(2001)}]{sacha1}%
  \BibitemOpen
  \bibfield  {author} {\bibinfo {author} {\bibfnamefont {K.}~\bibnamefont
  {Sacha}}\ and\ \bibinfo {author} {\bibfnamefont {B.}~\bibnamefont
  {Eckhardt}},\ }\bibfield  {title} {\bibinfo {title} {Pathways to double
  ionization of atoms in strong fields},\ }\href
  {https://doi.org/10.1103/PhysRevA.63.043414} {\bibfield  {journal} {\bibinfo
  {journal} {Phys. Rev. A}\ }\textbf {\bibinfo {volume} {63}},\ \bibinfo
  {pages} {043414} (\bibinfo {year} {2001})}\BibitemShut {NoStop}%
\bibitem [{\citenamefont {Prauzner-Bechcicki}\ \emph
  {et~al.}(2007)\citenamefont {Prauzner-Bechcicki}, \citenamefont {Sacha},
  \citenamefont {Eckhardt},\ and\ \citenamefont {Zakrzewski}}]{sacha2}%
  \BibitemOpen
  \bibfield  {author} {\bibinfo {author} {\bibfnamefont {J.~S.}\ \bibnamefont
  {Prauzner-Bechcicki}}, \bibinfo {author} {\bibfnamefont {K.}~\bibnamefont
  {Sacha}}, \bibinfo {author} {\bibfnamefont {B.}~\bibnamefont {Eckhardt}},\
  and\ \bibinfo {author} {\bibfnamefont {J.}~\bibnamefont {Zakrzewski}},\
  }\bibfield  {title} {\bibinfo {title} {Time-{Resolved} {Quantum} {Dynamics}
  of {Double} {Ionization} in {Strong} {Laser} {Fields}},\ }\href
  {https://doi.org/10.1103/PhysRevLett.98.203002} {\bibfield  {journal}
  {\bibinfo  {journal} {Phys. Rev. Lett.}\ }\textbf {\bibinfo {volume} {98}},\
  \bibinfo {pages} {203002} (\bibinfo {year} {2007})}\BibitemShut {NoStop}%
\bibitem [{\citenamefont {Efimov}\ \emph {et~al.}(2021)\citenamefont {Efimov},
  \citenamefont {Maksymov}, \citenamefont {Ciappina}, \citenamefont
  {Prauzner-Bechcicki}, \citenamefont {Lewenstein},\ and\ \citenamefont
  {Zakrzewski}}]{Dmitry}%
  \BibitemOpen
  \bibfield  {author} {\bibinfo {author} {\bibfnamefont {D.~K.}\ \bibnamefont
  {Efimov}}, \bibinfo {author} {\bibfnamefont {A.}~\bibnamefont {Maksymov}},
  \bibinfo {author} {\bibfnamefont {M.}~\bibnamefont {Ciappina}}, \bibinfo
  {author} {\bibfnamefont {J.~S.}\ \bibnamefont {Prauzner-Bechcicki}}, \bibinfo
  {author} {\bibfnamefont {M.}~\bibnamefont {Lewenstein}},\ and\ \bibinfo
  {author} {\bibfnamefont {J.}~\bibnamefont {Zakrzewski}},\ }\bibfield  {title}
  {\bibinfo {title} {Three-electron correlations in strong laser field
  ionization},\ }\href {https://doi.org/10.1364/OE.431572} {\bibfield
  {journal} {\bibinfo  {journal} {Optics Express}\ }\textbf {\bibinfo {volume}
  {29}},\ \bibinfo {pages} {26526} (\bibinfo {year} {2021})}\BibitemShut
  {NoStop}%
\bibitem [{\citenamefont {Maxwell}\ and\ \citenamefont {{Figueira de Morisson
  Faria}}(2016)}]{maxwell_controlling_2016}%
  \BibitemOpen
  \bibfield  {author} {\bibinfo {author} {\bibfnamefont {A.~S.}\ \bibnamefont
  {Maxwell}}\ and\ \bibinfo {author} {\bibfnamefont {C.}~\bibnamefont
  {{Figueira de Morisson Faria}}},\ }\bibfield  {title} {\bibinfo {title}
  {Controlling {{Below-Threshold Nonsequential Double Ionization}} via
  {{Quantum Interference}}},\ }\href
  {https://doi.org/10.1103/PhysRevLett.116.143001} {\bibfield  {journal}
  {\bibinfo  {journal} {Phys. Rev. Lett.}\ }\textbf {\bibinfo {volume} {116}},\
  \bibinfo {pages} {143001} (\bibinfo {year} {2016})}\BibitemShut {NoStop}%
\bibitem [{\citenamefont {Guo}\ and\ \citenamefont
  {{\AA}berg}(1988)}]{guo_quantum_1988}%
  \BibitemOpen
  \bibfield  {author} {\bibinfo {author} {\bibfnamefont {D.~S.}\ \bibnamefont
  {Guo}}\ and\ \bibinfo {author} {\bibfnamefont {T.}~\bibnamefont
  {{\AA}berg}},\ }\bibfield  {title} {\bibinfo {title} {Quantum
  electrodynamical approach to multiphoton ionisation in the high-intensity
  {{H}} field},\ }\href {https://doi.org/10.1088/0305-4470/21/24/013}
  {\bibfield  {journal} {\bibinfo  {journal} {Journal of Physics A:
  Mathematical and General}\ }\textbf {\bibinfo {volume} {21}},\ \bibinfo
  {pages} {4577} (\bibinfo {year} {1988})}\BibitemShut {NoStop}%
\bibitem [{\citenamefont {Fu}\ \emph {et~al.}(2001)\citenamefont {Fu},
  \citenamefont {Wang}, \citenamefont {Li},\ and\ \citenamefont
  {Gao}}]{fu_interrelation_2001}%
  \BibitemOpen
  \bibfield  {author} {\bibinfo {author} {\bibfnamefont {P.}~\bibnamefont
  {Fu}}, \bibinfo {author} {\bibfnamefont {B.}~\bibnamefont {Wang}}, \bibinfo
  {author} {\bibfnamefont {X.}~\bibnamefont {Li}},\ and\ \bibinfo {author}
  {\bibfnamefont {L.}~\bibnamefont {Gao}},\ }\bibfield  {title} {\bibinfo
  {title} {Interrelation between high-order harmonic generation and
  above-threshold ionization},\ }\href
  {https://doi.org/10.1103/PhysRevA.64.063401} {\bibfield  {journal} {\bibinfo
  {journal} {Physical Review A - Atomic, Molecular, and Optical Physics}\
  }\textbf {\bibinfo {volume} {64}},\ \bibinfo {pages} {063401} (\bibinfo
  {year} {2001})}\BibitemShut {NoStop}%
\bibitem [{\citenamefont {Wang}\ \emph {et~al.}(2007)\citenamefont {Wang},
  \citenamefont {Gao}, \citenamefont {Li}, \citenamefont {Guo},\ and\
  \citenamefont {Fu}}]{wang_frequencydomain_2007}%
  \BibitemOpen
  \bibfield  {author} {\bibinfo {author} {\bibfnamefont {B.}~\bibnamefont
  {Wang}}, \bibinfo {author} {\bibfnamefont {L.}~\bibnamefont {Gao}}, \bibinfo
  {author} {\bibfnamefont {X.}~\bibnamefont {Li}}, \bibinfo {author}
  {\bibfnamefont {D.~S.}\ \bibnamefont {Guo}},\ and\ \bibinfo {author}
  {\bibfnamefont {P.}~\bibnamefont {Fu}},\ }\bibfield  {title} {\bibinfo
  {title} {Frequency-domain theory of high-order above-threshold ionization
  based on nonperturbative quantum electrodynamics},\ }\href
  {https://doi.org/10.1103/PhysRevA.75.063419} {\bibfield  {journal} {\bibinfo
  {journal} {Physical Review A - Atomic, Molecular, and Optical Physics}\
  }\textbf {\bibinfo {volume} {75}},\ \bibinfo {pages} {063419} (\bibinfo
  {year} {2007})}\BibitemShut {NoStop}%
\bibitem [{\citenamefont {Wang}\ \emph {et~al.}(2012)\citenamefont {Wang},
  \citenamefont {Guo}, \citenamefont {Chen}, \citenamefont {Yan},\ and\
  \citenamefont {Fu}}]{wang_frequencydomain_2012}%
  \BibitemOpen
  \bibfield  {author} {\bibinfo {author} {\bibfnamefont {B.}~\bibnamefont
  {Wang}}, \bibinfo {author} {\bibfnamefont {Y.}~\bibnamefont {Guo}}, \bibinfo
  {author} {\bibfnamefont {J.}~\bibnamefont {Chen}}, \bibinfo {author}
  {\bibfnamefont {Z.-C.}\ \bibnamefont {Yan}},\ and\ \bibinfo {author}
  {\bibfnamefont {P.}~\bibnamefont {Fu}},\ }\bibfield  {title} {\bibinfo
  {title} {Frequency-domain theory of nonsequential double ionization in
  intense laser fields based on nonperturbative {{QED}}},\ }\href
  {https://doi.org/10.1103/PhysRevA.85.023402} {\bibfield  {journal} {\bibinfo
  {journal} {Phys. Rev. A}\ }\textbf {\bibinfo {volume} {85}},\ \bibinfo
  {pages} {023402} (\bibinfo {year} {2012})}\BibitemShut {NoStop}%
\bibitem [{\citenamefont {Gonoskov}\ \emph {et~al.}(2016)\citenamefont
  {Gonoskov}, \citenamefont {Tsatrafyllis}, \citenamefont {Kominis},\ and\
  \citenamefont {Tzallas}}]{Paris-srep}%
  \BibitemOpen
  \bibfield  {author} {\bibinfo {author} {\bibfnamefont {I.~A.}\ \bibnamefont
  {Gonoskov}}, \bibinfo {author} {\bibfnamefont {N.}~\bibnamefont
  {Tsatrafyllis}}, \bibinfo {author} {\bibfnamefont {I.~K.}\ \bibnamefont
  {Kominis}},\ and\ \bibinfo {author} {\bibfnamefont {P.}~\bibnamefont
  {Tzallas}},\ }\bibfield  {title} {\bibinfo {title} {Quantum optical
  signatures in strong-field laser physics: {Infrared} photon counting in
  high-order-harmonic generation},\ }\href {https://doi.org/10.1038/srep32821}
  {\bibfield  {journal} {\bibinfo  {journal} {Scientific Reports}\ }\textbf
  {\bibinfo {volume} {6}},\ \bibinfo {pages} {32821} (\bibinfo {year}
  {2016})}\BibitemShut {NoStop}%
\bibitem [{\citenamefont {Tsatrafyllis}\ \emph {et~al.}(2017)\citenamefont
  {Tsatrafyllis}, \citenamefont {Kominis}, \citenamefont {Gonoskov},\ and\
  \citenamefont {Tzallas}}]{Paris-ncomm}%
  \BibitemOpen
  \bibfield  {author} {\bibinfo {author} {\bibfnamefont {N.}~\bibnamefont
  {Tsatrafyllis}}, \bibinfo {author} {\bibfnamefont {I.~K.}\ \bibnamefont
  {Kominis}}, \bibinfo {author} {\bibfnamefont {I.~A.}\ \bibnamefont
  {Gonoskov}},\ and\ \bibinfo {author} {\bibfnamefont {P.}~\bibnamefont
  {Tzallas}},\ }\bibfield  {title} {\bibinfo {title} {High-order harmonics
  measured by the photon statistics of the infrared driving-field exiting the
  atomic medium},\ }\href {https://doi.org/10.1038/ncomms15170} {\bibfield
  {journal} {\bibinfo  {journal} {Nature Communications}\ }\textbf {\bibinfo
  {volume} {8}},\ \bibinfo {pages} {15170} (\bibinfo {year}
  {2017})}\BibitemShut {NoStop}%
\bibitem [{\citenamefont {Gorlach}\ \emph {et~al.}(2020)\citenamefont
  {Gorlach}, \citenamefont {Neufeld}, \citenamefont {Rivera}, \citenamefont
  {Cohen},\ and\ \citenamefont {Kaminer}}]{Ido-ncomm}%
  \BibitemOpen
  \bibfield  {author} {\bibinfo {author} {\bibfnamefont {A.}~\bibnamefont
  {Gorlach}}, \bibinfo {author} {\bibfnamefont {O.}~\bibnamefont {Neufeld}},
  \bibinfo {author} {\bibfnamefont {N.}~\bibnamefont {Rivera}}, \bibinfo
  {author} {\bibfnamefont {O.}~\bibnamefont {Cohen}},\ and\ \bibinfo {author}
  {\bibfnamefont {I.}~\bibnamefont {Kaminer}},\ }\bibfield  {title} {\bibinfo
  {title} {The quantum-optical nature of high harmonic generation},\ }\href
  {https://doi.org/10.1038/s41467-020-18218-w} {\bibfield  {journal} {\bibinfo
  {journal} {Nature Communications}\ }\textbf {\bibinfo {volume} {11}},\
  \bibinfo {pages} {4598} (\bibinfo {year} {2020})}\BibitemShut {NoStop}%
\bibitem [{\citenamefont {Rivera}\ and\ \citenamefont
  {Kaminer}(2020)}]{Ido-nphysrev}%
  \BibitemOpen
  \bibfield  {author} {\bibinfo {author} {\bibfnamefont {N.}~\bibnamefont
  {Rivera}}\ and\ \bibinfo {author} {\bibfnamefont {I.}~\bibnamefont
  {Kaminer}},\ }\bibfield  {title} {\bibinfo {title} {Light–matter
  interactions with photonic quasiparticles},\ }\href
  {https://doi.org/10.1038/s42254-020-0224-2} {\bibfield  {journal} {\bibinfo
  {journal} {Nature Reviews Physics}\ }\textbf {\bibinfo {volume} {2}},\
  \bibinfo {pages} {538} (\bibinfo {year} {2020})}\BibitemShut {NoStop}%
\bibitem [{\citenamefont {Fuchs}\ \emph {et~al.}(2022)\citenamefont {Fuchs},
  \citenamefont {Abel}, \citenamefont {Nathanael}, \citenamefont {Reinhard},
  \citenamefont {Wiesner}, \citenamefont {Wünsche}, \citenamefont
  {Skruszewicz}, \citenamefont {Rödel}, \citenamefont {Born}, \citenamefont
  {Schmidt},\ and\ \citenamefont {Paulus}}]{Gerhard-aphysB}%
  \BibitemOpen
  \bibfield  {author} {\bibinfo {author} {\bibfnamefont {S.}~\bibnamefont
  {Fuchs}}, \bibinfo {author} {\bibfnamefont {J.~J.}\ \bibnamefont {Abel}},
  \bibinfo {author} {\bibfnamefont {J.}~\bibnamefont {Nathanael}}, \bibinfo
  {author} {\bibfnamefont {J.}~\bibnamefont {Reinhard}}, \bibinfo {author}
  {\bibfnamefont {F.}~\bibnamefont {Wiesner}}, \bibinfo {author} {\bibfnamefont
  {M.}~\bibnamefont {Wünsche}}, \bibinfo {author} {\bibfnamefont
  {S.}~\bibnamefont {Skruszewicz}}, \bibinfo {author} {\bibfnamefont
  {C.}~\bibnamefont {Rödel}}, \bibinfo {author} {\bibfnamefont
  {D.}~\bibnamefont {Born}}, \bibinfo {author} {\bibfnamefont {H.}~\bibnamefont
  {Schmidt}},\ and\ \bibinfo {author} {\bibfnamefont {G.~G.}\ \bibnamefont
  {Paulus}},\ }\bibfield  {title} {\bibinfo {title} {Photon counting of extreme
  ultraviolet high harmonics using a superconducting nanowire single-photon
  detector},\ }\href {https://doi.org/10.1007/s00340-022-07754-6} {\bibfield
  {journal} {\bibinfo  {journal} {Applied Physics B}\ }\textbf {\bibinfo
  {volume} {128}},\ \bibinfo {pages} {26} (\bibinfo {year} {2022})}\BibitemShut
  {NoStop}%
\bibitem [{\citenamefont {Spasibko}\ \emph {et~al.}(2017)\citenamefont
  {Spasibko}, \citenamefont {Kopylov}, \citenamefont {Krutyanskiy},
  \citenamefont {Murzina}, \citenamefont {Leuchs},\ and\ \citenamefont
  {Chekhova}}]{spasibko2017multiphoton}%
  \BibitemOpen
  \bibfield  {author} {\bibinfo {author} {\bibfnamefont {K.~Y.}\ \bibnamefont
  {Spasibko}}, \bibinfo {author} {\bibfnamefont {D.~A.}\ \bibnamefont
  {Kopylov}}, \bibinfo {author} {\bibfnamefont {V.~L.}\ \bibnamefont
  {Krutyanskiy}}, \bibinfo {author} {\bibfnamefont {T.~V.}\ \bibnamefont
  {Murzina}}, \bibinfo {author} {\bibfnamefont {G.}~\bibnamefont {Leuchs}},\
  and\ \bibinfo {author} {\bibfnamefont {M.~V.}\ \bibnamefont {Chekhova}},\
  }\bibfield  {title} {\bibinfo {title} {Multiphoton effects enhanced due to
  ultrafast photon-number fluctuations},\ }\href@noop {} {\bibfield  {journal}
  {\bibinfo  {journal} {Physical Review Letters}\ }\textbf {\bibinfo {volume}
  {119}},\ \bibinfo {pages} {223603} (\bibinfo {year} {2017})}\BibitemShut
  {NoStop}%
\bibitem [{\citenamefont {Manceau}\ \emph {et~al.}(2019)\citenamefont
  {Manceau}, \citenamefont {Spasibko}, \citenamefont {Leuchs}, \citenamefont
  {Filip},\ and\ \citenamefont {Chekhova}}]{manceau2019indefinite}%
  \BibitemOpen
  \bibfield  {author} {\bibinfo {author} {\bibfnamefont {M.}~\bibnamefont
  {Manceau}}, \bibinfo {author} {\bibfnamefont {K.~Y.}\ \bibnamefont
  {Spasibko}}, \bibinfo {author} {\bibfnamefont {G.}~\bibnamefont {Leuchs}},
  \bibinfo {author} {\bibfnamefont {R.}~\bibnamefont {Filip}},\ and\ \bibinfo
  {author} {\bibfnamefont {M.~V.}\ \bibnamefont {Chekhova}},\ }\bibfield
  {title} {\bibinfo {title} {Indefinite-mean pareto photon distribution from
  amplified quantum noise},\ }\href@noop {} {\bibfield  {journal} {\bibinfo
  {journal} {Physical review letters}\ }\textbf {\bibinfo {volume} {123}},\
  \bibinfo {pages} {123606} (\bibinfo {year} {2019})}\BibitemShut {NoStop}%
\bibitem [{\citenamefont {Gorlach}\ \emph {et~al.}(2022)\citenamefont
  {Gorlach}, \citenamefont {Tzur}, \citenamefont {Birk}, \citenamefont
  {Kr{\"u}ger}, \citenamefont {Rivera}, \citenamefont {Cohen},\ and\
  \citenamefont {Kaminer}}]{gorlach2022high}%
  \BibitemOpen
  \bibfield  {author} {\bibinfo {author} {\bibfnamefont {A.}~\bibnamefont
  {Gorlach}}, \bibinfo {author} {\bibfnamefont {M.~E.}\ \bibnamefont {Tzur}},
  \bibinfo {author} {\bibfnamefont {M.}~\bibnamefont {Birk}}, \bibinfo {author}
  {\bibfnamefont {M.}~\bibnamefont {Kr{\"u}ger}}, \bibinfo {author}
  {\bibfnamefont {N.}~\bibnamefont {Rivera}}, \bibinfo {author} {\bibfnamefont
  {O.}~\bibnamefont {Cohen}},\ and\ \bibinfo {author} {\bibfnamefont
  {I.}~\bibnamefont {Kaminer}},\ }\bibfield  {title} {\bibinfo {title} {High
  harmonic generation driven by quantum light},\ }\href@noop {} {\bibfield
  {journal} {\bibinfo  {journal} {arXiv preprint arXiv:2211.03188}\ } (\bibinfo
  {year} {2022})}\BibitemShut {NoStop}%
\bibitem [{\citenamefont {Varr\'o}(2021)}]{Sandor-phot1}%
  \BibitemOpen
  \bibfield  {author} {\bibinfo {author} {\bibfnamefont {S.}~\bibnamefont
  {Varr\'o}},\ }\bibfield  {title} {\bibinfo {title} {Quantum {Optical}
  {Aspects} of {High}-{Harmonic} {Generation}},\ }\href
  {https://doi.org/10.3390/photonics8070269} {\bibfield  {journal} {\bibinfo
  {journal} {Photonics}\ }\textbf {\bibinfo {volume} {8}},\ \bibinfo {pages}
  {269} (\bibinfo {year} {2021})}\BibitemShut {NoStop}%
\bibitem [{\citenamefont {F\"oldi}\ \emph {et~al.}(2021)\citenamefont
  {F\"oldi}, \citenamefont {Magashegyi}, \citenamefont {Gombk\"oto},\ and\
  \citenamefont {Varr\'o}}]{Sandor-phot2}%
  \BibitemOpen
  \bibfield  {author} {\bibinfo {author} {\bibfnamefont {P.}~\bibnamefont
  {F\"oldi}}, \bibinfo {author} {\bibfnamefont {I.}~\bibnamefont {Magashegyi}},
  \bibinfo {author} {\bibfnamefont {A.}~\bibnamefont {Gombk\"oto}},\ and\
  \bibinfo {author} {\bibfnamefont {S.}~\bibnamefont {Varr\'o}},\ }\bibfield
  {title} {\bibinfo {title} {Describing {High}-{Order} {Harmonic} {Generation}
  {Using} {Quantum} {Optical} {Models}},\ }\href
  {https://doi.org/10.3390/photonics8070263} {\bibfield  {journal} {\bibinfo
  {journal} {Photonics}\ }\textbf {\bibinfo {volume} {8}},\ \bibinfo {pages}
  {263} (\bibinfo {year} {2021})}\BibitemShut {NoStop}%
\bibitem [{\citenamefont {Gombk\"oto}\ \emph {et~al.}(2021)\citenamefont
  {Gombk\"oto}, \citenamefont {F\"oldi},\ and\ \citenamefont
  {Varr\'o}}]{Sandor-pra}%
  \BibitemOpen
  \bibfield  {author} {\bibinfo {author} {\bibfnamefont {A.}~\bibnamefont
  {Gombk\"oto}}, \bibinfo {author} {\bibfnamefont {P.}~\bibnamefont
  {F\"oldi}},\ and\ \bibinfo {author} {\bibfnamefont {S.}~\bibnamefont
  {Varr\'o}},\ }\bibfield  {title} {\bibinfo {title} {Quantum-optical
  description of photon statistics and cross correlations in high-order
  harmonic generation},\ }\href {https://doi.org/10.1103/PhysRevA.104.033703}
  {\bibfield  {journal} {\bibinfo  {journal} {Phys. Rev. A}\ }\textbf {\bibinfo
  {volume} {104}},\ \bibinfo {pages} {033703} (\bibinfo {year}
  {2021})}\BibitemShut {NoStop}%
\bibitem [{\citenamefont {Lewenstein}\ \emph {et~al.}(2021)\citenamefont
  {Lewenstein}, \citenamefont {Ciappina}, \citenamefont {Pisanty},
  \citenamefont {Rivera-Dean}, \citenamefont {Stammer}, \citenamefont {{Th.
  Lamprou}},\ and\ \citenamefont {Tzallas}}]{lewenstein2021generation}%
  \BibitemOpen
  \bibfield  {author} {\bibinfo {author} {\bibfnamefont {M.}~\bibnamefont
  {Lewenstein}}, \bibinfo {author} {\bibfnamefont {M.}~\bibnamefont
  {Ciappina}}, \bibinfo {author} {\bibfnamefont {E.}~\bibnamefont {Pisanty}},
  \bibinfo {author} {\bibfnamefont {J.}~\bibnamefont {Rivera-Dean}}, \bibinfo
  {author} {\bibfnamefont {P.}~\bibnamefont {Stammer}}, \bibinfo {author}
  {\bibnamefont {{Th. Lamprou}}},\ and\ \bibinfo {author} {\bibfnamefont
  {P.}~\bibnamefont {Tzallas}},\ }\bibfield  {title} {\bibinfo {title}
  {Generation of optical schr{\"o}dinger cat states in intense laser--matter
  interactions},\ }\href@noop {} {\bibfield  {journal} {\bibinfo  {journal}
  {Nature Physics}\ }\textbf {\bibinfo {volume} {17}},\ \bibinfo {pages} {1104}
  (\bibinfo {year} {2021})}\BibitemShut {NoStop}%
\bibitem [{\citenamefont {Stammer}\ \emph
  {et~al.}(2022{\natexlab{a}})\citenamefont {Stammer}, \citenamefont
  {Rivera-Dean}, \citenamefont {{Th. Lamprou}}, \citenamefont {Pisanty},
  \citenamefont {Ciappina}, \citenamefont {Tzallas},\ and\ \citenamefont
  {Lewenstein}}]{stammer2022high}%
  \BibitemOpen
  \bibfield  {author} {\bibinfo {author} {\bibfnamefont {P.}~\bibnamefont
  {Stammer}}, \bibinfo {author} {\bibfnamefont {J.}~\bibnamefont
  {Rivera-Dean}}, \bibinfo {author} {\bibnamefont {{Th. Lamprou}}}, \bibinfo
  {author} {\bibfnamefont {E.}~\bibnamefont {Pisanty}}, \bibinfo {author}
  {\bibfnamefont {M.~F.}\ \bibnamefont {Ciappina}}, \bibinfo {author}
  {\bibfnamefont {P.}~\bibnamefont {Tzallas}},\ and\ \bibinfo {author}
  {\bibfnamefont {M.}~\bibnamefont {Lewenstein}},\ }\bibfield  {title}
  {\bibinfo {title} {High photon number entangled states and coherent state
  superposition from the extreme ultraviolet to the far infrared},\ }\href@noop
  {} {\bibfield  {journal} {\bibinfo  {journal} {Phys. Rev. Lett.}\ }\textbf
  {\bibinfo {volume} {128}},\ \bibinfo {pages} {123603} (\bibinfo {year}
  {2022}{\natexlab{a}})}\BibitemShut {NoStop}%
\bibitem [{\citenamefont {Rivera-Dean}\ \emph {et~al.}(2022)\citenamefont
  {Rivera-Dean}, \citenamefont {{Th. Lamprou}}, \citenamefont {Pisanty},
  \citenamefont {Stammer}, \citenamefont {Ord{\'o}{\~n}ez}, \citenamefont
  {Maxwell}, \citenamefont {Ciappina}, \citenamefont {Lewenstein},\ and\
  \citenamefont {Tzallas}}]{rivera2022strong}%
  \BibitemOpen
  \bibfield  {author} {\bibinfo {author} {\bibfnamefont {J.}~\bibnamefont
  {Rivera-Dean}}, \bibinfo {author} {\bibnamefont {{Th. Lamprou}}}, \bibinfo
  {author} {\bibfnamefont {E.}~\bibnamefont {Pisanty}}, \bibinfo {author}
  {\bibfnamefont {P.}~\bibnamefont {Stammer}}, \bibinfo {author} {\bibfnamefont
  {A.}~\bibnamefont {Ord{\'o}{\~n}ez}}, \bibinfo {author} {\bibfnamefont
  {A.}~\bibnamefont {Maxwell}}, \bibinfo {author} {\bibfnamefont
  {M.}~\bibnamefont {Ciappina}}, \bibinfo {author} {\bibfnamefont
  {M.}~\bibnamefont {Lewenstein}},\ and\ \bibinfo {author} {\bibfnamefont
  {P.}~\bibnamefont {Tzallas}},\ }\bibfield  {title} {\bibinfo {title} {Strong
  laser fields and their power to generate controllable high-photon-number
  coherent-state superpositions},\ }\href@noop {} {\bibfield  {journal}
  {\bibinfo  {journal} {Phys. Rev. A}\ }\textbf {\bibinfo {volume} {105}},\
  \bibinfo {pages} {033714} (\bibinfo {year} {2022})}\BibitemShut {NoStop}%
\bibitem [{\citenamefont {Stammer}(2022)}]{stammer2022theory}%
  \BibitemOpen
  \bibfield  {author} {\bibinfo {author} {\bibfnamefont {P.}~\bibnamefont
  {Stammer}},\ }\bibfield  {title} {\bibinfo {title} {Theory of entanglement
  and measurement in high harmonic generation},\ }\href@noop {} {\bibfield
  {journal} {\bibinfo  {journal} {arXiv preprint arXiv:2203.04354}\ } (\bibinfo
  {year} {2022})}\BibitemShut {NoStop}%
\bibitem [{\citenamefont {Lamprou}\ \emph {et~al.}(2021)\citenamefont
  {Lamprou}, \citenamefont {Lopez-Martens}, \citenamefont {Haessler},
  \citenamefont {Liontos}, \citenamefont {Kahaly}, \citenamefont {Rivera-Dean},
  \citenamefont {Stammer}, \citenamefont {Pisanty}, \citenamefont {Ciappina},
  \citenamefont {Lewenstein},\ and\ \citenamefont
  {Tzallas}}]{Lamprou_QS_Photonics2021}%
  \BibitemOpen
  \bibfield  {author} {\bibinfo {author} {\bibfnamefont {T.}~\bibnamefont
  {Lamprou}}, \bibinfo {author} {\bibfnamefont {R.}~\bibnamefont
  {Lopez-Martens}}, \bibinfo {author} {\bibfnamefont {S.}~\bibnamefont
  {Haessler}}, \bibinfo {author} {\bibfnamefont {I.}~\bibnamefont {Liontos}},
  \bibinfo {author} {\bibfnamefont {S.}~\bibnamefont {Kahaly}}, \bibinfo
  {author} {\bibfnamefont {J.}~\bibnamefont {Rivera-Dean}}, \bibinfo {author}
  {\bibfnamefont {P.}~\bibnamefont {Stammer}}, \bibinfo {author} {\bibfnamefont
  {E.}~\bibnamefont {Pisanty}}, \bibinfo {author} {\bibfnamefont {M.~F.}\
  \bibnamefont {Ciappina}}, \bibinfo {author} {\bibfnamefont {M.}~\bibnamefont
  {Lewenstein}},\ and\ \bibinfo {author} {\bibfnamefont {P.}~\bibnamefont
  {Tzallas}},\ }\bibfield  {title} {\bibinfo {title} {Quantum--optical
  spectrometry in relativistic laser--plasma interactions using the
  high-harmonic generation process: A proposal},\ }\href@noop {} {\bibfield
  {journal} {\bibinfo  {journal} {Photonics}\ }\textbf {\bibinfo {volume}
  {8}},\ \bibinfo {pages} {192} (\bibinfo {year} {2021})}\BibitemShut {NoStop}%
\bibitem [{\citenamefont {Rivera-Dean}\ \emph {et~al.}(2021)\citenamefont
  {Rivera-Dean}, \citenamefont {Stammer}, \citenamefont {Pisanty},
  \citenamefont {{Th. Lamprou}}, \citenamefont {Tzallas}, \citenamefont
  {Lewenstein},\ and\ \citenamefont {Ciappina}}]{rivera2021jcompelec}%
  \BibitemOpen
  \bibfield  {author} {\bibinfo {author} {\bibfnamefont {J.}~\bibnamefont
  {Rivera-Dean}}, \bibinfo {author} {\bibfnamefont {P.}~\bibnamefont
  {Stammer}}, \bibinfo {author} {\bibfnamefont {E.}~\bibnamefont {Pisanty}},
  \bibinfo {author} {\bibnamefont {{Th. Lamprou}}}, \bibinfo {author}
  {\bibfnamefont {P.}~\bibnamefont {Tzallas}}, \bibinfo {author} {\bibfnamefont
  {M.}~\bibnamefont {Lewenstein}},\ and\ \bibinfo {author} {\bibfnamefont
  {M.~F.}\ \bibnamefont {Ciappina}},\ }\bibfield  {title} {\bibinfo {title}
  {New schemes for creating large optical {Schrödinger} cat states using
  strong laser fields},\ }\href {https://doi.org/10.1007/s10825-021-01789-2}
  {\bibfield  {journal} {\bibinfo  {journal} {Journal of Computational
  Electronics}\ }\textbf {\bibinfo {volume} {20}},\ \bibinfo {pages} {2111}
  (\bibinfo {year} {2021})}\BibitemShut {NoStop}%
\bibitem [{\citenamefont {Tannoudji}\ \emph {et~al.}(1992)\citenamefont
  {Tannoudji}, \citenamefont {Grynberg},\ and\ \citenamefont
  {Dupont-Roe}}]{tannoudji1992atom}%
  \BibitemOpen
  \bibfield  {author} {\bibinfo {author} {\bibfnamefont {C.~C.}\ \bibnamefont
  {Tannoudji}}, \bibinfo {author} {\bibfnamefont {G.}~\bibnamefont
  {Grynberg}},\ and\ \bibinfo {author} {\bibfnamefont {J.}~\bibnamefont
  {Dupont-Roe}},\ }\href@noop {} {\emph {\bibinfo {title} {Atom-photon
  interactions}}}\ (\bibinfo  {publisher} {New York, NY (United States); John
  Wiley and Sons Inc.},\ \bibinfo {year} {1992})\BibitemShut {NoStop}%
\bibitem [{\citenamefont {Cohen-Tannoudji}\ \emph {et~al.}(1997)\citenamefont
  {Cohen-Tannoudji}, \citenamefont {Dupont-Roc},\ and\ \citenamefont
  {Grynberg}}]{cohen1997photons}%
  \BibitemOpen
  \bibfield  {author} {\bibinfo {author} {\bibfnamefont {C.}~\bibnamefont
  {Cohen-Tannoudji}}, \bibinfo {author} {\bibfnamefont {J.}~\bibnamefont
  {Dupont-Roc}},\ and\ \bibinfo {author} {\bibfnamefont {G.}~\bibnamefont
  {Grynberg}},\ }\href@noop {} {\emph {\bibinfo {title} {Photons and
  Atoms-Introduction to Quantum Electrodynamics}}}\ (\bibinfo  {publisher}
  {Wiley-VCH, Weinheim},\ \bibinfo {year} {1997})\BibitemShut {NoStop}%
\bibitem [{\citenamefont {Glauber}\ and\ \citenamefont
  {Lewenstein}(1991)}]{Roy1991pra}%
  \BibitemOpen
  \bibfield  {author} {\bibinfo {author} {\bibfnamefont {R.~J.}\ \bibnamefont
  {Glauber}}\ and\ \bibinfo {author} {\bibfnamefont {M.}~\bibnamefont
  {Lewenstein}},\ }\bibfield  {title} {\bibinfo {title} {Quantum optics of
  dielectric media},\ }\href {https://doi.org/10.1103/PhysRevA.43.467}
  {\bibfield  {journal} {\bibinfo  {journal} {Phys. Rev. A}\ }\textbf {\bibinfo
  {volume} {43}},\ \bibinfo {pages} {467} (\bibinfo {year} {1991})}\BibitemShut
  {NoStop}%
\bibitem [{\citenamefont {Vogel}\ and\ \citenamefont
  {Welsch}(2006)}]{vogel2006quantum}%
  \BibitemOpen
  \bibfield  {author} {\bibinfo {author} {\bibfnamefont {W.}~\bibnamefont
  {Vogel}}\ and\ \bibinfo {author} {\bibfnamefont {D.-G.}\ \bibnamefont
  {Welsch}},\ }\href@noop {} {\emph {\bibinfo {title} {Quantum optics}}}\
  (\bibinfo  {publisher} {John Wiley \& Sons, Weinheim},\ \bibinfo {year}
  {2006})\BibitemShut {NoStop}%
\bibitem [{\citenamefont {Scully}\ and\ \citenamefont
  {Zubairy}(2001)}]{ScullyBook}%
  \BibitemOpen
  \bibfield  {author} {\bibinfo {author} {\bibfnamefont {M.~O.}\ \bibnamefont
  {Scully}}\ and\ \bibinfo {author} {\bibfnamefont {M.~S.}\ \bibnamefont
  {Zubairy}},\ }\href@noop {} {\emph {\bibinfo {title} {Quantum optics}}}\
  (\bibinfo  {publisher} {Cambridge University Press, Cambridge},\ \bibinfo
  {year} {2001})\BibitemShut {NoStop}%
\bibitem [{\citenamefont {Haroche}\ and\ \citenamefont
  {Raimond}(2006)}]{Haroche-book}%
  \BibitemOpen
  \bibfield  {author} {\bibinfo {author} {\bibfnamefont {S.}~\bibnamefont
  {Haroche}}\ and\ \bibinfo {author} {\bibfnamefont {J.-M.}\ \bibnamefont
  {Raimond}},\ }\href@noop {} {\emph {\bibinfo {title} {Exploring the quantum:
  atoms, cavities and photons}}}\ (\bibinfo  {publisher} {Oxford Graduate
  Texts, Oxford},\ \bibinfo {year} {2006})\BibitemShut {NoStop}%
\bibitem [{\citenamefont {Hudson}(1974)}]{hudson_when_1974}%
  \BibitemOpen
  \bibfield  {author} {\bibinfo {author} {\bibfnamefont {R.~L.}\ \bibnamefont
  {Hudson}},\ }\bibfield  {title} {\bibinfo {title} {When is the wigner
  quasi-probability density non-negative?},\ }\href
  {https://doi.org/10.1016/0034-4877(74)90007-X} {\bibfield  {journal}
  {\bibinfo  {journal} {Reports on Mathematical Physics}\ }\textbf {\bibinfo
  {volume} {6}},\ \bibinfo {pages} {249} (\bibinfo {year} {1974})}\BibitemShut
  {NoStop}%
\bibitem [{\citenamefont {Kenfack}\ and\ \citenamefont
  {\.Zyczkowski}(2004)}]{kenfack_negativity_2004}%
  \BibitemOpen
  \bibfield  {author} {\bibinfo {author} {\bibfnamefont {A.}~\bibnamefont
  {Kenfack}}\ and\ \bibinfo {author} {\bibfnamefont {K.}~\bibnamefont
  {\.Zyczkowski}},\ }\bibfield  {title} {\bibinfo {title} {Negativity of the
  {Wigner} function as an indicator of non-classicality},\ }\href
  {https://doi.org/10.1088/1464-4266/6/10/003} {\bibfield  {journal} {\bibinfo
  {journal} {Journal of Optics B: Quantum and Semiclassical Optics}\ }\textbf
  {\bibinfo {volume} {6}},\ \bibinfo {pages} {396} (\bibinfo {year}
  {2004})}\BibitemShut {NoStop}%
\bibitem [{\citenamefont {Asbóth}\ \emph {et~al.}(2004)\citenamefont
  {Asbóth}, \citenamefont {Adam}, \citenamefont {Koniorczyk},\ and\
  \citenamefont {Janszky}}]{asboth_coherent-state_2004}%
  \BibitemOpen
  \bibfield  {author} {\bibinfo {author} {\bibfnamefont {J.~K.}\ \bibnamefont
  {Asbóth}}, \bibinfo {author} {\bibfnamefont {P.}~\bibnamefont {Adam}},
  \bibinfo {author} {\bibfnamefont {M.}~\bibnamefont {Koniorczyk}},\ and\
  \bibinfo {author} {\bibfnamefont {J.}~\bibnamefont {Janszky}},\ }\bibfield
  {title} {\bibinfo {title} {Coherent-state qubits: entanglement and
  decoherence},\ }\href {https://doi.org/10.1140/epjd/e2004-00094-2} {\bibfield
   {journal} {\bibinfo  {journal} {The European Physical Journal D - Atomic,
  Molecular, Optical and Plasma Physics}\ }\textbf {\bibinfo {volume} {30}},\
  \bibinfo {pages} {403} (\bibinfo {year} {2004})}\BibitemShut {NoStop}%
\bibitem [{\citenamefont {Sanders}(1992)}]{sanders_entangled_1992}%
  \BibitemOpen
  \bibfield  {author} {\bibinfo {author} {\bibfnamefont {B.~C.}\ \bibnamefont
  {Sanders}},\ }\bibfield  {title} {\bibinfo {title} {Entangled coherent
  states},\ }\href {https://doi.org/10.1103/PhysRevA.45.6811} {\bibfield
  {journal} {\bibinfo  {journal} {Phys. Rev. A}\ }\textbf {\bibinfo {volume}
  {45}},\ \bibinfo {pages} {6811} (\bibinfo {year} {1992})}\BibitemShut
  {NoStop}%
\bibitem [{\citenamefont {Gilchrist}\ \emph
  {et~al.}(2004{\natexlab{a}})\citenamefont {Gilchrist}, \citenamefont
  {Nemoto}, \citenamefont {Munro}, \citenamefont {Ralph}, \citenamefont
  {Glancy}, \citenamefont {Braunstein},\ and\ \citenamefont
  {Milburn}}]{gilchrist_schrodinger_2004}%
  \BibitemOpen
  \bibfield  {author} {\bibinfo {author} {\bibfnamefont {A.}~\bibnamefont
  {Gilchrist}}, \bibinfo {author} {\bibfnamefont {K.}~\bibnamefont {Nemoto}},
  \bibinfo {author} {\bibfnamefont {W.~J.}\ \bibnamefont {Munro}}, \bibinfo
  {author} {\bibfnamefont {T.~C.}\ \bibnamefont {Ralph}}, \bibinfo {author}
  {\bibfnamefont {S.}~\bibnamefont {Glancy}}, \bibinfo {author} {\bibfnamefont
  {S.~L.}\ \bibnamefont {Braunstein}},\ and\ \bibinfo {author} {\bibfnamefont
  {G.~J.}\ \bibnamefont {Milburn}},\ }\bibfield  {title} {\bibinfo {title}
  {Schrödinger cats and their power for quantum information processing},\
  }\href {https://doi.org/10.1088/1464-4266/6/8/032} {\bibfield  {journal}
  {\bibinfo  {journal} {Journal of Optics B: Quantum and Semiclassical Optics}\
  }\textbf {\bibinfo {volume} {6}},\ \bibinfo {pages} {S828} (\bibinfo {year}
  {2004}{\natexlab{a}})}\BibitemShut {NoStop}%
\bibitem [{\citenamefont {Jouguet}\ \emph {et~al.}(2013)\citenamefont
  {Jouguet}, \citenamefont {Kunz-Jacques}, \citenamefont {Leverrier},
  \citenamefont {Grangier},\ and\ \citenamefont
  {Diamanti}}]{jouguet_experimental_2013}%
  \BibitemOpen
  \bibfield  {author} {\bibinfo {author} {\bibfnamefont {P.}~\bibnamefont
  {Jouguet}}, \bibinfo {author} {\bibfnamefont {S.}~\bibnamefont
  {Kunz-Jacques}}, \bibinfo {author} {\bibfnamefont {A.}~\bibnamefont
  {Leverrier}}, \bibinfo {author} {\bibfnamefont {P.}~\bibnamefont
  {Grangier}},\ and\ \bibinfo {author} {\bibfnamefont {E.}~\bibnamefont
  {Diamanti}},\ }\bibfield  {title} {\bibinfo {title} {Experimental
  demonstration of long-distance continuous-variable quantum key
  distribution},\ }\href {https://doi.org/10.1038/nphoton.2013.63} {\bibfield
  {journal} {\bibinfo  {journal} {Nature Photonics}\ }\textbf {\bibinfo
  {volume} {7}},\ \bibinfo {pages} {378} (\bibinfo {year} {2013})}\BibitemShut
  {NoStop}%
\bibitem [{\citenamefont {Lloyd}\ and\ \citenamefont
  {Braunstein}(1999)}]{lloyd_quantum_1999}%
  \BibitemOpen
  \bibfield  {author} {\bibinfo {author} {\bibfnamefont {S.}~\bibnamefont
  {Lloyd}}\ and\ \bibinfo {author} {\bibfnamefont {S.~L.}\ \bibnamefont
  {Braunstein}},\ }\bibfield  {title} {\bibinfo {title} {Quantum {Computation}
  over {Continuous} {Variables}},\ }\href
  {https://doi.org/10.1103/PhysRevLett.82.1784} {\bibfield  {journal} {\bibinfo
   {journal} {Phys. Rev. Lett.}\ }\textbf {\bibinfo {volume} {82}},\ \bibinfo
  {pages} {1784} (\bibinfo {year} {1999})}\BibitemShut {NoStop}%
\bibitem [{\citenamefont {Ralph}\ \emph {et~al.}(2003)\citenamefont {Ralph},
  \citenamefont {Gilchrist}, \citenamefont {Milburn}, \citenamefont {Munro},\
  and\ \citenamefont {Glancy}}]{ralph_quantum_2003}%
  \BibitemOpen
  \bibfield  {author} {\bibinfo {author} {\bibfnamefont {T.~C.}\ \bibnamefont
  {Ralph}}, \bibinfo {author} {\bibfnamefont {A.}~\bibnamefont {Gilchrist}},
  \bibinfo {author} {\bibfnamefont {G.~J.}\ \bibnamefont {Milburn}}, \bibinfo
  {author} {\bibfnamefont {W.~J.}\ \bibnamefont {Munro}},\ and\ \bibinfo
  {author} {\bibfnamefont {S.}~\bibnamefont {Glancy}},\ }\bibfield  {title}
  {\bibinfo {title} {Quantum computation with optical coherent states},\ }\href
  {https://doi.org/10.1103/PhysRevA.68.042319} {\bibfield  {journal} {\bibinfo
  {journal} {Phys. Rev. A}\ }\textbf {\bibinfo {volume} {68}},\ \bibinfo
  {pages} {042319} (\bibinfo {year} {2003})}\BibitemShut {NoStop}%
\bibitem [{\citenamefont {Joo}\ \emph {et~al.}(2011)\citenamefont {Joo},
  \citenamefont {Munro},\ and\ \citenamefont {Spiller}}]{joo_quantum_2011}%
  \BibitemOpen
  \bibfield  {author} {\bibinfo {author} {\bibfnamefont {J.}~\bibnamefont
  {Joo}}, \bibinfo {author} {\bibfnamefont {W.~J.}\ \bibnamefont {Munro}},\
  and\ \bibinfo {author} {\bibfnamefont {T.~P.}\ \bibnamefont {Spiller}},\
  }\bibfield  {title} {\bibinfo {title} {Quantum {Metrology} with {Entangled}
  {Coherent} {States}},\ }\href
  {https://doi.org/10.1103/PhysRevLett.107.083601} {\bibfield  {journal}
  {\bibinfo  {journal} {Phys. Rev. Lett.}\ }\textbf {\bibinfo {volume} {107}},\
  \bibinfo {pages} {083601} (\bibinfo {year} {2011})}\BibitemShut {NoStop}%
\bibitem [{\citenamefont {Aichelburg}\ and\ \citenamefont
  {Grosse}(1977)}]{aichelburg_exactly_1977}%
  \BibitemOpen
  \bibfield  {author} {\bibinfo {author} {\bibfnamefont {P.~C.}\ \bibnamefont
  {Aichelburg}}\ and\ \bibinfo {author} {\bibfnamefont {H.}~\bibnamefont
  {Grosse}},\ }\bibfield  {title} {\bibinfo {title} {Exactly soluble system of
  relativistic two-body interaction},\ }\href
  {https://doi.org/10.1103/PhysRevD.16.1900} {\bibfield  {journal} {\bibinfo
  {journal} {Physical Review D}\ }\textbf {\bibinfo {volume} {16}},\ \bibinfo
  {pages} {1900} (\bibinfo {year} {1977})}\BibitemShut {NoStop}%
\bibitem [{\citenamefont {Aichelburg}(1977)}]{aichelburg_model_1977}%
  \BibitemOpen
  \bibfield  {author} {\bibinfo {author} {\bibfnamefont {P.~C.}\ \bibnamefont
  {Aichelburg}},\ }\bibfield  {title} {\bibinfo {title} {Model for
  {Relativistic} {Two}-body {Interaction} with {Radiation} {Reaction}},\ }\href
  {https://doi.org/10.1103/PhysRevLett.38.451} {\bibfield  {journal} {\bibinfo
  {journal} {Phys. Rev. Lett.}\ }\textbf {\bibinfo {volume} {38}},\ \bibinfo
  {pages} {451} (\bibinfo {year} {1977})}\BibitemShut {NoStop}%
\bibitem [{\citenamefont {R\c{z}a\.zewski}\ and\ \citenamefont
  {\.Zakowicz}(1976)}]{rzazewski_initial_1976}%
  \BibitemOpen
  \bibfield  {author} {\bibinfo {author} {\bibfnamefont {K.}~\bibnamefont
  {R\c{z}a\.zewski}}\ and\ \bibinfo {author} {\bibfnamefont {W.}~\bibnamefont
  {\.Zakowicz}},\ }\bibfield  {title} {\bibinfo {title} {Initial value problem
  and causality of radiating oscillator},\ }\href
  {https://doi.org/10.1088/0305-4470/9/7/018} {\bibfield  {journal} {\bibinfo
  {journal} {Journal of Physics A: Mathematical and General}\ }\textbf
  {\bibinfo {volume} {9}},\ \bibinfo {pages} {1159} (\bibinfo {year}
  {1976})}\BibitemShut {NoStop}%
\bibitem [{\citenamefont {R\c{z}a\.zewski}\ and\ \citenamefont
  {\.Zakowicz}(1980)}]{rzazewski_initial_1980}%
  \BibitemOpen
  \bibfield  {author} {\bibinfo {author} {\bibfnamefont {K.}~\bibnamefont
  {R\c{z}a\.zewski}}\ and\ \bibinfo {author} {\bibfnamefont {W.}~\bibnamefont
  {\.Zakowicz}},\ }\bibfield  {title} {\bibinfo {title} {Initial value problem
  for two oscillators interacting with electromagnetic field},\ }\href
  {https://doi.org/10.1063/1.524426} {\bibfield  {journal} {\bibinfo  {journal}
  {Journal of Mathematical Physics}\ }\textbf {\bibinfo {volume} {21}},\
  \bibinfo {pages} {378} (\bibinfo {year} {1980})}\BibitemShut {NoStop}%
\bibitem [{\citenamefont {Tong}\ and\ \citenamefont
  {Chu}(1997)}]{tong_density-functional_1997}%
  \BibitemOpen
  \bibfield  {author} {\bibinfo {author} {\bibfnamefont {X.-M.}\ \bibnamefont
  {Tong}}\ and\ \bibinfo {author} {\bibfnamefont {S.-I.}\ \bibnamefont {Chu}},\
  }\bibfield  {title} {\bibinfo {title} {Density-functional theory with
  optimized effective potential and self-interaction correction for ground
  states and autoionizing resonances},\ }\href
  {https://doi.org/10.1103/PhysRevA.55.3406} {\bibfield  {journal} {\bibinfo
  {journal} {Phys. Rev. A}\ }\textbf {\bibinfo {volume} {55}},\ \bibinfo
  {pages} {3406} (\bibinfo {year} {1997})}\BibitemShut {NoStop}%
\bibitem [{\citenamefont {Tong}\ and\ \citenamefont
  {Lin}(2005)}]{tong_empirical_2005}%
  \BibitemOpen
  \bibfield  {author} {\bibinfo {author} {\bibfnamefont {X.~M.}\ \bibnamefont
  {Tong}}\ and\ \bibinfo {author} {\bibfnamefont {C.~D.}\ \bibnamefont {Lin}},\
  }\bibfield  {title} {\bibinfo {title} {Empirical formula for static field
  ionization rates of atoms and molecules by lasers in the barrier-suppression
  regime},\ }\href {https://doi.org/10.1088/0953-4075/38/15/001} {\bibfield
  {journal} {\bibinfo  {journal} {Journal of Physics B: Atomic, Molecular and
  Optical Physics}\ }\textbf {\bibinfo {volume} {38}},\ \bibinfo {pages} {2593}
  (\bibinfo {year} {2005})}\BibitemShut {NoStop}%
\bibitem [{\citenamefont {Lewenstein}\ \emph {et~al.}(1994)\citenamefont
  {Lewenstein}, \citenamefont {Balcou}, \citenamefont {Ivanov}, \citenamefont
  {L’Huillier},\ and\ \citenamefont {Corkum}}]{lewenstein1994theory}%
  \BibitemOpen
  \bibfield  {author} {\bibinfo {author} {\bibfnamefont {M.}~\bibnamefont
  {Lewenstein}}, \bibinfo {author} {\bibfnamefont {P.}~\bibnamefont {Balcou}},
  \bibinfo {author} {\bibfnamefont {M.~Y.}\ \bibnamefont {Ivanov}}, \bibinfo
  {author} {\bibfnamefont {A.}~\bibnamefont {L’Huillier}},\ and\ \bibinfo
  {author} {\bibfnamefont {P.~B.}\ \bibnamefont {Corkum}},\ }\bibfield  {title}
  {\bibinfo {title} {Theory of high-harmonic generation by low-frequency laser
  fields},\ }\href@noop {} {\bibfield  {journal} {\bibinfo  {journal} {Phys.
  Rev. A}\ }\textbf {\bibinfo {volume} {49}},\ \bibinfo {pages} {2117}
  (\bibinfo {year} {1994})}\BibitemShut {NoStop}%
\bibitem [{\citenamefont {Sundaram}\ and\ \citenamefont
  {Milonni}(1990)}]{sundaram1990high}%
  \BibitemOpen
  \bibfield  {author} {\bibinfo {author} {\bibfnamefont {B.}~\bibnamefont
  {Sundaram}}\ and\ \bibinfo {author} {\bibfnamefont {P.~W.}\ \bibnamefont
  {Milonni}},\ }\bibfield  {title} {\bibinfo {title} {High-order harmonic
  generation: simplified model and relevance of single-atom theories to
  experiment},\ }\href@noop {} {\bibfield  {journal} {\bibinfo  {journal}
  {Phys. Rev. A}\ }\textbf {\bibinfo {volume} {41}},\ \bibinfo {pages} {6571}
  (\bibinfo {year} {1990})}\BibitemShut {NoStop}%
\bibitem [{\citenamefont {Bauer}\ and\ \citenamefont {Koval}(2006)}]{Qprop}%
  \BibitemOpen
  \bibfield  {author} {\bibinfo {author} {\bibfnamefont {D.}~\bibnamefont
  {Bauer}}\ and\ \bibinfo {author} {\bibfnamefont {P.}~\bibnamefont {Koval}},\
  }\bibfield  {title} {\bibinfo {title} {Qprop: {A} {Schrödinger}-solver for
  intense laser–atom interaction},\ }\href
  {https://doi.org/10.1016/j.cpc.2005.11.001} {\bibfield  {journal} {\bibinfo
  {journal} {Computer Physics Communications}\ }\textbf {\bibinfo {volume}
  {174}},\ \bibinfo {pages} {396} (\bibinfo {year} {2006})}\BibitemShut
  {NoStop}%
\bibitem [{\citenamefont {Nielsen}\ and\ \citenamefont
  {Chuang}(2000)}]{NielsenandChuang}%
  \BibitemOpen
  \bibfield  {author} {\bibinfo {author} {\bibfnamefont {M.~A.}\ \bibnamefont
  {Nielsen}}\ and\ \bibinfo {author} {\bibfnamefont {I.~L.}\ \bibnamefont
  {Chuang}},\ }\href@noop {} {\emph {\bibinfo {title} {Quantum {Computation}
  and {Quantum} {Information}}}}\ (\bibinfo  {publisher} {Cambridge University
  Press, Cambridge},\ \bibinfo {year} {2000})\BibitemShut {NoStop}%
\bibitem [{\citenamefont {Krause}\ \emph {et~al.}(1992)\citenamefont {Krause},
  \citenamefont {Schafer},\ and\ \citenamefont
  {Kulander}}]{krause_high-order_1992}%
  \BibitemOpen
  \bibfield  {author} {\bibinfo {author} {\bibfnamefont {J.~L.}\ \bibnamefont
  {Krause}}, \bibinfo {author} {\bibfnamefont {K.~J.}\ \bibnamefont
  {Schafer}},\ and\ \bibinfo {author} {\bibfnamefont {K.~C.}\ \bibnamefont
  {Kulander}},\ }\bibfield  {title} {\bibinfo {title} {High-order harmonic
  generation from atoms and ions in the high intensity regime},\ }\href
  {https://doi.org/10.1103/PhysRevLett.68.3535} {\bibfield  {journal} {\bibinfo
   {journal} {Phys. Rev. Lett.}\ }\textbf {\bibinfo {volume} {68}},\ \bibinfo
  {pages} {3535} (\bibinfo {year} {1992})}\BibitemShut {NoStop}%
\bibitem [{\citenamefont {van Enk}\ and\ \citenamefont
  {Hirota}(2001)}]{van2001entangled}%
  \BibitemOpen
  \bibfield  {author} {\bibinfo {author} {\bibfnamefont {S.~J.}\ \bibnamefont
  {van Enk}}\ and\ \bibinfo {author} {\bibfnamefont {O.}~\bibnamefont
  {Hirota}},\ }\bibfield  {title} {\bibinfo {title} {Entangled coherent states:
  Teleportation and decoherence},\ }\href@noop {} {\bibfield  {journal}
  {\bibinfo  {journal} {Physical Review A}\ }\textbf {\bibinfo {volume} {64}},\
  \bibinfo {pages} {022313} (\bibinfo {year} {2001})}\BibitemShut {NoStop}%
\bibitem [{\citenamefont {Othman}\ and\ \citenamefont
  {Yevick}(2018)}]{othman2018quantum}%
  \BibitemOpen
  \bibfield  {author} {\bibinfo {author} {\bibfnamefont {A.}~\bibnamefont
  {Othman}}\ and\ \bibinfo {author} {\bibfnamefont {D.}~\bibnamefont
  {Yevick}},\ }\bibfield  {title} {\bibinfo {title} {Quantum properties of the
  superposition of two nearly identical coherent states},\ }\href@noop {}
  {\bibfield  {journal} {\bibinfo  {journal} {International Journal of
  Theoretical Physics}\ }\textbf {\bibinfo {volume} {57}},\ \bibinfo {pages}
  {2293} (\bibinfo {year} {2018})}\BibitemShut {NoStop}%
\bibitem [{\citenamefont {Madsen}(2021)}]{madsen_strongfield_2021}%
  \BibitemOpen
  \bibfield  {author} {\bibinfo {author} {\bibfnamefont {L.~B.}\ \bibnamefont
  {Madsen}},\ }\bibfield  {title} {\bibinfo {title} {Strong-field approximation
  for high-order harmonic generation in infrared laser pulses in the
  accelerated {{Kramers-Henneberger}} frame},\ }\href
  {https://doi.org/10.1103/PhysRevA.104.033117} {\bibfield  {journal} {\bibinfo
   {journal} {Physical Review A}\ }\textbf {\bibinfo {volume} {104}},\ \bibinfo
  {pages} {033117} (\bibinfo {year} {2021})}\BibitemShut {NoStop}%
\bibitem [{\citenamefont {Paulus}\ \emph {et~al.}(2003)\citenamefont {Paulus},
  \citenamefont {Lindner}, \citenamefont {Walther}, \citenamefont {Baltuška},
  \citenamefont {Goulielmakis}, \citenamefont {Lezius},\ and\ \citenamefont
  {Krausz}}]{paulus_measurement_2003}%
  \BibitemOpen
  \bibfield  {author} {\bibinfo {author} {\bibfnamefont {G.~G.}\ \bibnamefont
  {Paulus}}, \bibinfo {author} {\bibfnamefont {F.}~\bibnamefont {Lindner}},
  \bibinfo {author} {\bibfnamefont {H.}~\bibnamefont {Walther}}, \bibinfo
  {author} {\bibfnamefont {A.}~\bibnamefont {Baltuška}}, \bibinfo {author}
  {\bibfnamefont {E.}~\bibnamefont {Goulielmakis}}, \bibinfo {author}
  {\bibfnamefont {M.}~\bibnamefont {Lezius}},\ and\ \bibinfo {author}
  {\bibfnamefont {F.}~\bibnamefont {Krausz}},\ }\bibfield  {title} {\bibinfo
  {title} {Measurement of the {Phase} of {Few}-{Cycle} {Laser} {Pulses}},\
  }\href {https://doi.org/10.1103/PhysRevLett.91.253004} {\bibfield  {journal}
  {\bibinfo  {journal} {Phys. Rev. Lett.}\ }\textbf {\bibinfo {volume} {91}},\
  \bibinfo {pages} {253004} (\bibinfo {year} {2003})}\BibitemShut {NoStop}%
\bibitem [{\citenamefont {Milošević}\ \emph {et~al.}(2006)\citenamefont
  {Milošević}, \citenamefont {Paulus}, \citenamefont {Bauer},\ and\
  \citenamefont {Becker}}]{milosevic_above-threshold_2006}%
  \BibitemOpen
  \bibfield  {author} {\bibinfo {author} {\bibfnamefont {D.~B.}\ \bibnamefont
  {Milošević}}, \bibinfo {author} {\bibfnamefont {G.~G.}\ \bibnamefont
  {Paulus}}, \bibinfo {author} {\bibfnamefont {D.}~\bibnamefont {Bauer}},\ and\
  \bibinfo {author} {\bibfnamefont {W.}~\bibnamefont {Becker}},\ }\bibfield
  {title} {\bibinfo {title} {Above-threshold ionization by few-cycle pulses},\
  }\href {https://doi.org/10.1088/0953-4075/39/14/R01} {\bibfield  {journal}
  {\bibinfo  {journal} {Journal of Physics B: Atomic, Molecular and Optical
  Physics}\ }\textbf {\bibinfo {volume} {39}},\ \bibinfo {pages} {R203}
  (\bibinfo {year} {2006})}\BibitemShut {NoStop}%
\bibitem [{\citenamefont {Lvovsky}\ and\ \citenamefont
  {Raymer}(2009)}]{Lvovsky_Rev_QT}%
  \BibitemOpen
  \bibfield  {author} {\bibinfo {author} {\bibfnamefont {A.~I.}\ \bibnamefont
  {Lvovsky}}\ and\ \bibinfo {author} {\bibfnamefont {M.~G.}\ \bibnamefont
  {Raymer}},\ }\bibfield  {title} {\bibinfo {title} {Continuous-variable
  optical quantum-state tomography},\ }\href
  {https://doi.org/10.1103/RevModPhys.81.299} {\bibfield  {journal} {\bibinfo
  {journal} {Rev. Mod. Phys.}\ }\textbf {\bibinfo {volume} {81}},\ \bibinfo
  {pages} {299} (\bibinfo {year} {2009})}\BibitemShut {NoStop}%
\bibitem [{\citenamefont {Breitenbach}\ \emph {et~al.}(1997)\citenamefont
  {Breitenbach}, \citenamefont {Schiller},\ and\ \citenamefont
  {Mlynek}}]{Breitenbach_Squeezed_QT}%
  \BibitemOpen
  \bibfield  {author} {\bibinfo {author} {\bibfnamefont {G.}~\bibnamefont
  {Breitenbach}}, \bibinfo {author} {\bibfnamefont {S.}~\bibnamefont
  {Schiller}},\ and\ \bibinfo {author} {\bibfnamefont {J.}~\bibnamefont
  {Mlynek}},\ }\bibfield  {title} {\bibinfo {title} {Measurement of the quantum
  states of squeezed light},\ }\href {https://doi.org/10.1038/387471a0}
  {\bibfield  {journal} {\bibinfo  {journal} {Nature}\ }\textbf {\bibinfo
  {volume} {387}},\ \bibinfo {pages} {471} (\bibinfo {year}
  {1997})}\BibitemShut {NoStop}%
\bibitem [{\citenamefont {Skotiniotis}\ \emph {et~al.}(2017)\citenamefont
  {Skotiniotis}, \citenamefont {D{\"u}r},\ and\ \citenamefont
  {Sekatski}}]{skotiniotis2017macroscopic}%
  \BibitemOpen
  \bibfield  {author} {\bibinfo {author} {\bibfnamefont {M.}~\bibnamefont
  {Skotiniotis}}, \bibinfo {author} {\bibfnamefont {W.}~\bibnamefont
  {D{\"u}r}},\ and\ \bibinfo {author} {\bibfnamefont {P.}~\bibnamefont
  {Sekatski}},\ }\bibfield  {title} {\bibinfo {title} {Macroscopic
  superpositions require tremendous measurement devices},\ }\href@noop {}
  {\bibfield  {journal} {\bibinfo  {journal} {Quantum}\ }\textbf {\bibinfo
  {volume} {1}},\ \bibinfo {pages} {34} (\bibinfo {year} {2017})}\BibitemShut
  {NoStop}%
\bibitem [{\citenamefont {Chatziathanasiou}\ \emph {et~al.}(2006)\citenamefont
  {Chatziathanasiou}, \citenamefont {Kahaly}, \citenamefont {Skantzakis},
  \citenamefont {Sansone}, \citenamefont {Lopez-Martens}, \citenamefont
  {Haessler}, \citenamefont {Varju}, \citenamefont {Tsakiris}, \citenamefont
  {Charalambidis},\ and\ \citenamefont {Tzallas}}]{Chatziathanasiou2017}%
  \BibitemOpen
  \bibfield  {author} {\bibinfo {author} {\bibfnamefont {S.}~\bibnamefont
  {Chatziathanasiou}}, \bibinfo {author} {\bibfnamefont {S.}~\bibnamefont
  {Kahaly}}, \bibinfo {author} {\bibfnamefont {E.}~\bibnamefont {Skantzakis}},
  \bibinfo {author} {\bibfnamefont {G.}~\bibnamefont {Sansone}}, \bibinfo
  {author} {\bibfnamefont {R.}~\bibnamefont {Lopez-Martens}}, \bibinfo {author}
  {\bibfnamefont {S.}~\bibnamefont {Haessler}}, \bibinfo {author}
  {\bibfnamefont {K.}~\bibnamefont {Varju}}, \bibinfo {author} {\bibfnamefont
  {G.~D.}\ \bibnamefont {Tsakiris}}, \bibinfo {author} {\bibfnamefont
  {D.}~\bibnamefont {Charalambidis}},\ and\ \bibinfo {author} {\bibfnamefont
  {P.}~\bibnamefont {Tzallas}},\ }\bibfield  {title} {\bibinfo {title}
  {Generation of attosecond light pulses from gas and solid state media},\
  }\href@noop {} {\bibfield  {journal} {\bibinfo  {journal} {Photonics}\
  }\textbf {\bibinfo {volume} {4}},\ \bibinfo {pages} {26} (\bibinfo {year}
  {2006})}\BibitemShut {NoStop}%
\bibitem [{\citenamefont {Paulus}\ \emph {et~al.}(2001)\citenamefont {Paulus},
  \citenamefont {Grasbon}, \citenamefont {Walther}, \citenamefont {Villoresi},
  \citenamefont {Nisoli}, \citenamefont {Stagira}, \citenamefont {Priori},\
  and\ \citenamefont {Silvestri}}]{Paulus2001CEP}%
  \BibitemOpen
  \bibfield  {author} {\bibinfo {author} {\bibfnamefont {G.~G.}\ \bibnamefont
  {Paulus}}, \bibinfo {author} {\bibfnamefont {F.}~\bibnamefont {Grasbon}},
  \bibinfo {author} {\bibfnamefont {H.}~\bibnamefont {Walther}}, \bibinfo
  {author} {\bibfnamefont {P.}~\bibnamefont {Villoresi}}, \bibinfo {author}
  {\bibfnamefont {M.}~\bibnamefont {Nisoli}}, \bibinfo {author} {\bibfnamefont
  {S.}~\bibnamefont {Stagira}}, \bibinfo {author} {\bibfnamefont
  {E.}~\bibnamefont {Priori}},\ and\ \bibinfo {author} {\bibfnamefont {S.~D.}\
  \bibnamefont {Silvestri}},\ }\bibfield  {title} {\bibinfo {title}
  {Absolute-phase phenomena in photoionization with few--cycle laser pulses},\
  }\href@noop {} {\bibfield  {journal} {\bibinfo  {journal} {Nature}\ }\textbf
  {\bibinfo {volume} {414}},\ \bibinfo {pages} {182} (\bibinfo {year}
  {2001})}\BibitemShut {NoStop}%
\bibitem [{\citenamefont {Tsatrafyllis}\ \emph {et~al.}(2019)\citenamefont
  {Tsatrafyllis}, \citenamefont {K\"uhn}, \citenamefont {Dumergue},
  \citenamefont {Foldi}, \citenamefont {Kahaly}, \citenamefont {Cormier},
  \citenamefont {Gonoskov}, \citenamefont {Kiss}, \citenamefont {Varju},
  \citenamefont {Varro},\ and\ \citenamefont {Tzallas}}]{Tsatr_QS_PRL}%
  \BibitemOpen
  \bibfield  {author} {\bibinfo {author} {\bibfnamefont {N.}~\bibnamefont
  {Tsatrafyllis}}, \bibinfo {author} {\bibfnamefont {S.}~\bibnamefont
  {K\"uhn}}, \bibinfo {author} {\bibfnamefont {M.}~\bibnamefont {Dumergue}},
  \bibinfo {author} {\bibfnamefont {P.}~\bibnamefont {Foldi}}, \bibinfo
  {author} {\bibfnamefont {S.}~\bibnamefont {Kahaly}}, \bibinfo {author}
  {\bibfnamefont {E.}~\bibnamefont {Cormier}}, \bibinfo {author} {\bibfnamefont
  {I.~A.}\ \bibnamefont {Gonoskov}}, \bibinfo {author} {\bibfnamefont
  {B.}~\bibnamefont {Kiss}}, \bibinfo {author} {\bibfnamefont {K.}~\bibnamefont
  {Varju}}, \bibinfo {author} {\bibfnamefont {S.}~\bibnamefont {Varro}},\ and\
  \bibinfo {author} {\bibfnamefont {P.}~\bibnamefont {Tzallas}},\ }\bibfield
  {title} {\bibinfo {title} {Quantum optical signatures in a strong laser pulse
  after interaction with semiconductors},\ }\href
  {https://doi.org/10.1103/PhysRevLett.122.193602} {\bibfield  {journal}
  {\bibinfo  {journal} {Phys. Rev. Lett.}\ }\textbf {\bibinfo {volume} {122}},\
  \bibinfo {pages} {193602} (\bibinfo {year} {2019})}\BibitemShut {NoStop}%
\bibitem [{\citenamefont {Kim}\ \emph {et~al.}(2005)\citenamefont {Kim},
  \citenamefont {Kim}, \citenamefont {Kim}, \citenamefont {Lee}, \citenamefont
  {Lee}, \citenamefont {Park}, \citenamefont {Cho},\ and\ \citenamefont
  {Nam}}]{Kim_2005}%
  \BibitemOpen
  \bibfield  {author} {\bibinfo {author} {\bibfnamefont {I.~J.}\ \bibnamefont
  {Kim}}, \bibinfo {author} {\bibfnamefont {C.~M.}\ \bibnamefont {Kim}},
  \bibinfo {author} {\bibfnamefont {H.~T.}\ \bibnamefont {Kim}}, \bibinfo
  {author} {\bibfnamefont {G.~H.}\ \bibnamefont {Lee}}, \bibinfo {author}
  {\bibfnamefont {Y.~S.}\ \bibnamefont {Lee}}, \bibinfo {author} {\bibfnamefont
  {J.~Y.}\ \bibnamefont {Park}}, \bibinfo {author} {\bibfnamefont {D.~J.}\
  \bibnamefont {Cho}},\ and\ \bibinfo {author} {\bibfnamefont {C.~H.}\
  \bibnamefont {Nam}},\ }\bibfield  {title} {\bibinfo {title} {Highly efficient
  high--harmonic generation in an orthogonally polarized two--color laser
  field},\ }\href@noop {} {\bibfield  {journal} {\bibinfo  {journal} {Phys.
  Rev. Lett.}\ }\textbf {\bibinfo {volume} {94}},\ \bibinfo {pages} {243901}
  (\bibinfo {year} {2005})}\BibitemShut {NoStop}%
\bibitem [{\citenamefont {Mauritsson}\ \emph {et~al.}(2006)\citenamefont
  {Mauritsson}, \citenamefont {Johnsson}, \citenamefont {Gustafsson},
  \citenamefont {L’Huillier}, \citenamefont {Schafer},\ and\ \citenamefont
  {Gaarde}}]{Mauritsson_2006}%
  \BibitemOpen
  \bibfield  {author} {\bibinfo {author} {\bibfnamefont {J.}~\bibnamefont
  {Mauritsson}}, \bibinfo {author} {\bibfnamefont {P.}~\bibnamefont
  {Johnsson}}, \bibinfo {author} {\bibfnamefont {E.}~\bibnamefont
  {Gustafsson}}, \bibinfo {author} {\bibfnamefont {A.}~\bibnamefont
  {L’Huillier}}, \bibinfo {author} {\bibfnamefont {K.~J.}\ \bibnamefont
  {Schafer}},\ and\ \bibinfo {author} {\bibfnamefont {M.~B.}\ \bibnamefont
  {Gaarde}},\ }\bibfield  {title} {\bibinfo {title} {Attosecond pulse trains
  generated using two color laser fields},\ }\href@noop {} {\bibfield
  {journal} {\bibinfo  {journal} {Phys. Rev. Lett.}\ }\textbf {\bibinfo
  {volume} {97}},\ \bibinfo {pages} {013001} (\bibinfo {year}
  {2006})}\BibitemShut {NoStop}%
\bibitem [{\citenamefont {Fleischer}\ \emph {et~al.}(2014)\citenamefont
  {Fleischer}, \citenamefont {Kfir}, \citenamefont {Diskin}, \citenamefont
  {Sidorenko}, ,\ and\ \citenamefont {Cohen}}]{Fleischer_2014}%
  \BibitemOpen
  \bibfield  {author} {\bibinfo {author} {\bibfnamefont {A.}~\bibnamefont
  {Fleischer}}, \bibinfo {author} {\bibfnamefont {O.}~\bibnamefont {Kfir}},
  \bibinfo {author} {\bibfnamefont {T.}~\bibnamefont {Diskin}}, \bibinfo
  {author} {\bibfnamefont {P.}~\bibnamefont {Sidorenko}}, ,\ and\ \bibinfo
  {author} {\bibfnamefont {O.}~\bibnamefont {Cohen}},\ }\bibfield  {title}
  {\bibinfo {title} {Spin angular momentum and tunable polarization in
  highharmonic generation},\ }\href@noop {} {\bibfield  {journal} {\bibinfo
  {journal} {Nat. Photonics}\ }\textbf {\bibinfo {volume} {8}},\ \bibinfo
  {pages} {543} (\bibinfo {year} {2014})}\BibitemShut {NoStop}%
\bibitem [{\citenamefont {Gilchrist}\ \emph
  {et~al.}(2004{\natexlab{b}})\citenamefont {Gilchrist}, \citenamefont
  {Nemoto}, \citenamefont {Munro}, \citenamefont {Ralph}, \citenamefont
  {Glancy}, \citenamefont {Braunstein},\ and\ \citenamefont
  {Milburn}}]{Gilchrist_2004}%
  \BibitemOpen
  \bibfield  {author} {\bibinfo {author} {\bibfnamefont {A.}~\bibnamefont
  {Gilchrist}}, \bibinfo {author} {\bibfnamefont {K.}~\bibnamefont {Nemoto}},
  \bibinfo {author} {\bibfnamefont {W.~J.}\ \bibnamefont {Munro}}, \bibinfo
  {author} {\bibfnamefont {T.~C.}\ \bibnamefont {Ralph}}, \bibinfo {author}
  {\bibfnamefont {S.}~\bibnamefont {Glancy}}, \bibinfo {author} {\bibfnamefont
  {S.~L.}\ \bibnamefont {Braunstein}},\ and\ \bibinfo {author} {\bibfnamefont
  {G.~J.}\ \bibnamefont {Milburn}},\ }\bibfield  {title} {\bibinfo {title}
  {Schrödinger cats and their power for quantum information processing},\
  }\href@noop {} {\bibfield  {journal} {\bibinfo  {journal} {J. Opt B: Quantum
  Semiclass.}\ }\textbf {\bibinfo {volume} {6}},\ \bibinfo {pages} {5828}
  (\bibinfo {year} {2004}{\natexlab{b}})}\BibitemShut {NoStop}%
\bibitem [{\citenamefont {Laghaout}\ \emph {et~al.}(2013)\citenamefont
  {Laghaout}, \citenamefont {Neergaard-Nielsen}, \citenamefont {Rigas},
  \citenamefont {Kragh}, \citenamefont {Tipsmark},\ and\ \citenamefont
  {Andersen}}]{laghaout_amplification_2013}%
  \BibitemOpen
  \bibfield  {author} {\bibinfo {author} {\bibfnamefont {A.}~\bibnamefont
  {Laghaout}}, \bibinfo {author} {\bibfnamefont {J.~S.}\ \bibnamefont
  {Neergaard-Nielsen}}, \bibinfo {author} {\bibfnamefont {I.}~\bibnamefont
  {Rigas}}, \bibinfo {author} {\bibfnamefont {C.}~\bibnamefont {Kragh}},
  \bibinfo {author} {\bibfnamefont {A.}~\bibnamefont {Tipsmark}},\ and\
  \bibinfo {author} {\bibfnamefont {U.~L.}\ \bibnamefont {Andersen}},\
  }\bibfield  {title} {\bibinfo {title} {Amplification of realistic
  {Schr}{\textbackslash}"odinger-cat-state-like states by homodyne heralding},\
  }\href {https://doi.org/10.1103/PhysRevA.87.043826} {\bibfield  {journal}
  {\bibinfo  {journal} {Phys. Rev. A}\ }\textbf {\bibinfo {volume} {87}},\
  \bibinfo {pages} {043826} (\bibinfo {year} {2013})}\BibitemShut {NoStop}%
\bibitem [{\citenamefont {Sychev}\ \emph {et~al.}(2017)\citenamefont {Sychev},
  \citenamefont {Ulanov}, \citenamefont {Pushkina}, \citenamefont {Richards},
  \citenamefont {Fedorov},\ and\ \citenamefont
  {Lvovsky}}]{sychev_enlargement_2017}%
  \BibitemOpen
  \bibfield  {author} {\bibinfo {author} {\bibfnamefont {D.~V.}\ \bibnamefont
  {Sychev}}, \bibinfo {author} {\bibfnamefont {A.~E.}\ \bibnamefont {Ulanov}},
  \bibinfo {author} {\bibfnamefont {A.~A.}\ \bibnamefont {Pushkina}}, \bibinfo
  {author} {\bibfnamefont {M.~W.}\ \bibnamefont {Richards}}, \bibinfo {author}
  {\bibfnamefont {I.~A.}\ \bibnamefont {Fedorov}},\ and\ \bibinfo {author}
  {\bibfnamefont {A.~I.}\ \bibnamefont {Lvovsky}},\ }\bibfield  {title}
  {\bibinfo {title} {Enlargement of optical {Schrödinger}'s cat states},\
  }\href {https://doi.org/10.1038/nphoton.2017.57} {\bibfield  {journal}
  {\bibinfo  {journal} {Nat. Phot.}\ }\textbf {\bibinfo {volume} {11}},\
  \bibinfo {pages} {379} (\bibinfo {year} {2017})}\BibitemShut {NoStop}%
\bibitem [{\citenamefont {Bachor}\ and\ \citenamefont
  {Ralph}(2019)}]{Bachor_book_2019}%
  \BibitemOpen
  \bibfield  {author} {\bibinfo {author} {\bibfnamefont {H.}~\bibnamefont
  {Bachor}}\ and\ \bibinfo {author} {\bibfnamefont {T.}~\bibnamefont {Ralph}},\
  }\href@noop {} {\emph {\bibinfo {title} {A Guide to Experiments in Quantum
  Optics}}}\ (\bibinfo  {publisher} {Wiley‐VCH Verlag},\ \bibinfo {address}
  {Weinheim, Germany},\ \bibinfo {year} {2019})\BibitemShut {NoStop}%
\bibitem [{\citenamefont {Yuen}\ and\ \citenamefont {Chan}(1983)}]{Yuen_1983}%
  \BibitemOpen
  \bibfield  {author} {\bibinfo {author} {\bibfnamefont {H.~P.}\ \bibnamefont
  {Yuen}}\ and\ \bibinfo {author} {\bibfnamefont {V.~W.~S.}\ \bibnamefont
  {Chan}},\ }\bibfield  {title} {\bibinfo {title} {Noise in homodyne and
  heterodyne detection},\ }\href@noop {} {\bibfield  {journal} {\bibinfo
  {journal} {Opt. Lett.}\ }\textbf {\bibinfo {volume} {8}},\ \bibinfo {pages}
  {177} (\bibinfo {year} {1983})}\BibitemShut {NoStop}%
\bibitem [{\citenamefont {Lvovsky}(2004)}]{Lvovsky_MaxLik_alg}%
  \BibitemOpen
  \bibfield  {author} {\bibinfo {author} {\bibfnamefont {A.~I.}\ \bibnamefont
  {Lvovsky}},\ }\bibfield  {title} {\bibinfo {title} {Iterative
  maximum-likelihood reconstruction in quantum homodyne tomography},\ }\href
  {https://doi.org/10.1088/1464-4266/6/6/014} {\bibfield  {journal} {\bibinfo
  {journal} {Journal of Optics B: Quantum and Semiclassical Optics}\ }\textbf
  {\bibinfo {volume} {6}},\ \bibinfo {pages} {S556} (\bibinfo {year}
  {2004})}\BibitemShut {NoStop}%
\bibitem [{\citenamefont {Schleich}(2001)}]{Schleich_Book_2001}%
  \BibitemOpen
  \bibfield  {author} {\bibinfo {author} {\bibfnamefont {W.~P.}\ \bibnamefont
  {Schleich}},\ }\href@noop {} {\emph {\bibinfo {title} {Quantum Optics in
  Phase Space}}}\ (\bibinfo  {publisher} {Wiley-VHC Verlag},\ \bibinfo
  {address} {Weinheim, Germany},\ \bibinfo {year} {2001})\BibitemShut {NoStop}%
\bibitem [{\citenamefont {Gerry}\ and\ \citenamefont
  {Knight}(2005)}]{Gerry__Book_2001}%
  \BibitemOpen
  \bibfield  {author} {\bibinfo {author} {\bibfnamefont {C.}~\bibnamefont
  {Gerry}}\ and\ \bibinfo {author} {\bibfnamefont {P.}~\bibnamefont {Knight}},\
  }\href@noop {} {\emph {\bibinfo {title} {Introductory Quantum Optics}}}\
  (\bibinfo  {publisher} {Cambridge University Press},\ \bibinfo {address}
  {Cambridge, UK},\ \bibinfo {year} {2005})\BibitemShut {NoStop}%
\bibitem [{\citenamefont {Leonhardt}({\natexlab{a}})}]{Leonhardt__Book_2001}%
  \BibitemOpen
  \bibfield  {author} {\bibinfo {author} {\bibfnamefont {U.}~\bibnamefont
  {Leonhardt}},\ }\bibfield  {title} {\bibinfo {title} {Measuring the quantum
  state of light},\ }\href@noop {} {\bibfield  {journal} {\bibinfo  {journal}
  {(Eds., P. Knight and A. Miller, Cambridge University Press, UK, 1997)}\ }
  ({\natexlab{a}})}\BibitemShut {NoStop}%
\bibitem [{\citenamefont {Leonhardt}({\natexlab{b}})}]{Leonhardt__Book_2010}%
  \BibitemOpen
  \bibfield  {author} {\bibinfo {author} {\bibfnamefont {U.}~\bibnamefont
  {Leonhardt}},\ }\bibfield  {title} {\bibinfo {title} {Essential quantum
  optics},\ }\href@noop {} {\bibfield  {journal} {\bibinfo  {journal}
  {(Cambridge University Press, UK, 2010)}\ } ({\natexlab{b}})}\BibitemShut
  {NoStop}%
\bibitem [{\citenamefont {Herman}(1980)}]{Herman__Book_1980}%
  \BibitemOpen
  \bibfield  {author} {\bibinfo {author} {\bibfnamefont {G.~T.}\ \bibnamefont
  {Herman}},\ }\href@noop {} {\emph {\bibinfo {title} {Image Reconstruction
  from projections: The fundamentals of Computerized tomography}}}\ (\bibinfo
  {publisher} {Academic Press},\ \bibinfo {address} {New York},\ \bibinfo
  {year} {1980})\BibitemShut {NoStop}%
\bibitem [{\citenamefont {Banaszek}(1999)}]{Banaszek_Wig_rec_dm}%
  \BibitemOpen
  \bibfield  {author} {\bibinfo {author} {\bibfnamefont {K.}~\bibnamefont
  {Banaszek}},\ }\bibfield  {title} {\bibinfo {title} {Quantum homodyne
  tomography with a priori constraints},\ }\href
  {https://doi.org/10.1103/PhysRevA.59.4797} {\bibfield  {journal} {\bibinfo
  {journal} {Phys. Rev. A}\ }\textbf {\bibinfo {volume} {59}},\ \bibinfo
  {pages} {4797} (\bibinfo {year} {1999})}\BibitemShut {NoStop}%
\bibitem [{\citenamefont {M.Leonhardt}(1993)}]{Leonhardt1993}%
  \BibitemOpen
  \bibfield  {author} {\bibinfo {author} {\bibnamefont {M.Leonhardt}},\
  }\bibfield  {title} {\bibinfo {title} {Quantum statistics of a lossless beam
  splitter: Su(2) symmetry in phase space},\ }\href
  {http://doi.org/10.1103/PhysRevA.48.3265} {\bibfield  {journal} {\bibinfo
  {journal} {Phys. Rev. A}\ }\textbf {\bibinfo {volume} {48}},\ \bibinfo
  {pages} {3265} (\bibinfo {year} {1993})}\BibitemShut {NoStop}%
\bibitem [{\citenamefont {Zhang}\ \emph {et~al.}(2021)\citenamefont {Zhang},
  \citenamefont {Kang}, \citenamefont {Wang}, \citenamefont {Xu}, \citenamefont
  {Su},\ and\ \citenamefont {Peng}}]{MZhang2021}%
  \BibitemOpen
  \bibfield  {author} {\bibinfo {author} {\bibfnamefont {M.}~\bibnamefont
  {Zhang}}, \bibinfo {author} {\bibfnamefont {H.}~\bibnamefont {Kang}},
  \bibinfo {author} {\bibfnamefont {M.}~\bibnamefont {Wang}}, \bibinfo {author}
  {\bibfnamefont {F.}~\bibnamefont {Xu}}, \bibinfo {author} {\bibfnamefont
  {X.}~\bibnamefont {Su}},\ and\ \bibinfo {author} {\bibfnamefont
  {K.}~\bibnamefont {Peng}},\ }\bibfield  {title} {\bibinfo {title}
  {Quantifying quantum coherence of optical cat states},\ }\href
  {http://doi.org/10.1364/PRJ.418417} {\bibfield  {journal} {\bibinfo
  {journal} {Photonics Research}\ }\textbf {\bibinfo {volume} {9}},\ \bibinfo
  {pages} {887} (\bibinfo {year} {2021})}\BibitemShut {NoStop}%
\bibitem [{\citenamefont {Pisanty}\ \emph
  {et~al.}(2019{\natexlab{a}})\citenamefont {Pisanty}, \citenamefont {Machado},
  \citenamefont {Vicu{\~n}a-Hern{\'a}ndez}, \citenamefont {Pic{\'o}n},
  \citenamefont {Celi}, \citenamefont {Torres},\ and\ \citenamefont
  {Lewenstein}}]{pisanty2019knotting}%
  \BibitemOpen
  \bibfield  {author} {\bibinfo {author} {\bibfnamefont {E.}~\bibnamefont
  {Pisanty}}, \bibinfo {author} {\bibfnamefont {G.~J.}\ \bibnamefont
  {Machado}}, \bibinfo {author} {\bibfnamefont {V.}~\bibnamefont
  {Vicu{\~n}a-Hern{\'a}ndez}}, \bibinfo {author} {\bibfnamefont
  {A.}~\bibnamefont {Pic{\'o}n}}, \bibinfo {author} {\bibfnamefont
  {A.}~\bibnamefont {Celi}}, \bibinfo {author} {\bibfnamefont {J.~P.}\
  \bibnamefont {Torres}},\ and\ \bibinfo {author} {\bibfnamefont
  {M.}~\bibnamefont {Lewenstein}},\ }\bibfield  {title} {\bibinfo {title}
  {Knotting fractional-order knots with the polarization state of light},\
  }\href@noop {} {\bibfield  {journal} {\bibinfo  {journal} {Nature Photonics}\
  }\textbf {\bibinfo {volume} {13}},\ \bibinfo {pages} {569} (\bibinfo {year}
  {2019}{\natexlab{a}})}\BibitemShut {NoStop}%
\bibitem [{\citenamefont {Pisanty}\ \emph
  {et~al.}(2019{\natexlab{b}})\citenamefont {Pisanty}, \citenamefont {Rego},
  \citenamefont {San~Rom{\'a}n}, \citenamefont {Pic{\'o}n}, \citenamefont
  {Dorney}, \citenamefont {Kapteyn}, \citenamefont {Murnane}, \citenamefont
  {Plaja}, \citenamefont {Lewenstein},\ and\ \citenamefont
  {Hern{\'a}ndez-Garc{\'\i}a}}]{pisanty2019conservation}%
  \BibitemOpen
  \bibfield  {author} {\bibinfo {author} {\bibfnamefont {E.}~\bibnamefont
  {Pisanty}}, \bibinfo {author} {\bibfnamefont {L.}~\bibnamefont {Rego}},
  \bibinfo {author} {\bibfnamefont {J.}~\bibnamefont {San~Rom{\'a}n}}, \bibinfo
  {author} {\bibfnamefont {A.}~\bibnamefont {Pic{\'o}n}}, \bibinfo {author}
  {\bibfnamefont {K.~M.}\ \bibnamefont {Dorney}}, \bibinfo {author}
  {\bibfnamefont {H.~C.}\ \bibnamefont {Kapteyn}}, \bibinfo {author}
  {\bibfnamefont {M.~M.}\ \bibnamefont {Murnane}}, \bibinfo {author}
  {\bibfnamefont {L.}~\bibnamefont {Plaja}}, \bibinfo {author} {\bibfnamefont
  {M.}~\bibnamefont {Lewenstein}},\ and\ \bibinfo {author} {\bibfnamefont
  {C.}~\bibnamefont {Hern{\'a}ndez-Garc{\'\i}a}},\ }\bibfield  {title}
  {\bibinfo {title} {Conservation of torus-knot angular momentum in high-order
  harmonic generation},\ }\href@noop {} {\bibfield  {journal} {\bibinfo
  {journal} {Phys. Rev. Lett.}\ }\textbf {\bibinfo {volume} {122}},\ \bibinfo
  {pages} {203201} (\bibinfo {year} {2019}{\natexlab{b}})}\BibitemShut
  {NoStop}%
\bibitem [{\citenamefont {Rego}\ \emph {et~al.}(2019)\citenamefont {Rego},
  \citenamefont {Dorney}, \citenamefont {Brooks}, \citenamefont {Nguyen},
  \citenamefont {Liao}, \citenamefont {San~Rom{\'a}n}, \citenamefont {Couch},
  \citenamefont {Liu}, \citenamefont {Pisanty}, \citenamefont {Lewenstein}
  \emph {et~al.}}]{rego2019generation}%
  \BibitemOpen
  \bibfield  {author} {\bibinfo {author} {\bibfnamefont {L.}~\bibnamefont
  {Rego}}, \bibinfo {author} {\bibfnamefont {K.~M.}\ \bibnamefont {Dorney}},
  \bibinfo {author} {\bibfnamefont {N.~J.}\ \bibnamefont {Brooks}}, \bibinfo
  {author} {\bibfnamefont {Q.~L.}\ \bibnamefont {Nguyen}}, \bibinfo {author}
  {\bibfnamefont {C.-T.}\ \bibnamefont {Liao}}, \bibinfo {author}
  {\bibfnamefont {J.}~\bibnamefont {San~Rom{\'a}n}}, \bibinfo {author}
  {\bibfnamefont {D.~E.}\ \bibnamefont {Couch}}, \bibinfo {author}
  {\bibfnamefont {A.}~\bibnamefont {Liu}}, \bibinfo {author} {\bibfnamefont
  {E.}~\bibnamefont {Pisanty}}, \bibinfo {author} {\bibfnamefont
  {M.}~\bibnamefont {Lewenstein}}, \emph {et~al.},\ }\bibfield  {title}
  {\bibinfo {title} {Generation of extreme-ultraviolet beams with time-varying
  orbital angular momentum},\ }\href@noop {} {\bibfield  {journal} {\bibinfo
  {journal} {Science}\ }\textbf {\bibinfo {volume} {364}},\ \bibinfo {pages}
  {eaaw9486} (\bibinfo {year} {2019})}\BibitemShut {NoStop}%
\bibitem [{\citenamefont {Neufeld}\ \emph {et~al.}(2019)\citenamefont
  {Neufeld}, \citenamefont {Podolsky},\ and\ \citenamefont
  {Cohen}}]{neufeld_floquet_2019}%
  \BibitemOpen
  \bibfield  {author} {\bibinfo {author} {\bibfnamefont {O.}~\bibnamefont
  {Neufeld}}, \bibinfo {author} {\bibfnamefont {D.}~\bibnamefont {Podolsky}},\
  and\ \bibinfo {author} {\bibfnamefont {O.}~\bibnamefont {Cohen}},\ }\bibfield
   {title} {\bibinfo {title} {Floquet group theory and its application to
  selection rules in harmonic generation},\ }\href
  {https://doi.org/10.1038/s41467-018-07935-y} {\bibfield  {journal} {\bibinfo
  {journal} {Nature Communications}\ }\textbf {\bibinfo {volume} {10}},\
  \bibinfo {pages} {405} (\bibinfo {year} {2019})}\BibitemShut {NoStop}%
\bibitem [{\citenamefont {Ayuso}\ \emph {et~al.}(2019)\citenamefont {Ayuso},
  \citenamefont {Neufeld}, \citenamefont {Ordonez}, \citenamefont {Decleva},
  \citenamefont {Lerner}, \citenamefont {Cohen}, \citenamefont {Ivanov},\ and\
  \citenamefont {Smirnova}}]{ayuso_synthetic_2019}%
  \BibitemOpen
  \bibfield  {author} {\bibinfo {author} {\bibfnamefont {D.}~\bibnamefont
  {Ayuso}}, \bibinfo {author} {\bibfnamefont {O.}~\bibnamefont {Neufeld}},
  \bibinfo {author} {\bibfnamefont {A.~F.}\ \bibnamefont {Ordonez}}, \bibinfo
  {author} {\bibfnamefont {P.}~\bibnamefont {Decleva}}, \bibinfo {author}
  {\bibfnamefont {G.}~\bibnamefont {Lerner}}, \bibinfo {author} {\bibfnamefont
  {O.}~\bibnamefont {Cohen}}, \bibinfo {author} {\bibfnamefont
  {M.}~\bibnamefont {Ivanov}},\ and\ \bibinfo {author} {\bibfnamefont
  {O.}~\bibnamefont {Smirnova}},\ }\bibfield  {title} {\bibinfo {title}
  {Synthetic chiral light for efficient control of chiral light–matter
  interaction},\ }\href {https://doi.org/10.1038/s41566-019-0531-2} {\bibfield
  {journal} {\bibinfo  {journal} {Nature Photonics}\ }\textbf {\bibinfo
  {volume} {13}},\ \bibinfo {pages} {866} (\bibinfo {year} {2019})}\BibitemShut
  {NoStop}%
\bibitem [{\citenamefont {Stammer}\ \emph {et~al.}(2020)\citenamefont
  {Stammer}, \citenamefont {Patchkovskii},\ and\ \citenamefont
  {Morales}}]{stammer2020evidence}%
  \BibitemOpen
  \bibfield  {author} {\bibinfo {author} {\bibfnamefont {P.}~\bibnamefont
  {Stammer}}, \bibinfo {author} {\bibfnamefont {S.}~\bibnamefont
  {Patchkovskii}},\ and\ \bibinfo {author} {\bibfnamefont {F.}~\bibnamefont
  {Morales}},\ }\bibfield  {title} {\bibinfo {title} {Evidence of
  ac-stark-shifted resonances in intense two-color circularly polarized laser
  fields},\ }\href@noop {} {\bibfield  {journal} {\bibinfo  {journal} {Physical
  Review A}\ }\textbf {\bibinfo {volume} {101}},\ \bibinfo {pages} {033405}
  (\bibinfo {year} {2020})}\BibitemShut {NoStop}%
\bibitem [{\citenamefont {Ayuso}\ \emph {et~al.}(2021)\citenamefont {Ayuso},
  \citenamefont {Ordonez}, \citenamefont {Decleva}, \citenamefont {Ivanov},\
  and\ \citenamefont {Smirnova}}]{ayuso_enantio-sensitive_2021}%
  \BibitemOpen
  \bibfield  {author} {\bibinfo {author} {\bibfnamefont {D.}~\bibnamefont
  {Ayuso}}, \bibinfo {author} {\bibfnamefont {A.~F.}\ \bibnamefont {Ordonez}},
  \bibinfo {author} {\bibfnamefont {P.}~\bibnamefont {Decleva}}, \bibinfo
  {author} {\bibfnamefont {M.}~\bibnamefont {Ivanov}},\ and\ \bibinfo {author}
  {\bibfnamefont {O.}~\bibnamefont {Smirnova}},\ }\bibfield  {title} {\bibinfo
  {title} {Enantio-sensitive unidirectional light bending},\ }\href
  {https://doi.org/10.1038/s41467-021-24118-4} {\bibfield  {journal} {\bibinfo
  {journal} {Nature Communications}\ }\textbf {\bibinfo {volume} {12}},\
  \bibinfo {pages} {3951} (\bibinfo {year} {2021})}\BibitemShut {NoStop}%
\bibitem [{\citenamefont {Katsoulis}\ \emph {et~al.}(2021)\citenamefont
  {Katsoulis}, \citenamefont {Dube}, \citenamefont {Corkum}, \citenamefont
  {Staudte},\ and\ \citenamefont {Emmanouilidou}}]{katsoulis_momentum_2021}%
  \BibitemOpen
  \bibfield  {author} {\bibinfo {author} {\bibfnamefont {G.~P.}\ \bibnamefont
  {Katsoulis}}, \bibinfo {author} {\bibfnamefont {Z.}~\bibnamefont {Dube}},
  \bibinfo {author} {\bibfnamefont {P.}~\bibnamefont {Corkum}}, \bibinfo
  {author} {\bibfnamefont {A.}~\bibnamefont {Staudte}},\ and\ \bibinfo {author}
  {\bibfnamefont {A.}~\bibnamefont {Emmanouilidou}},\ }\bibfield  {title}
  {\bibinfo {title} {Momentum scalar triple product as a measure of chirality
  in electron ionization dynamics of strongly-driven atoms},\ }\href
  {http://arxiv.org/abs/2112.02670} {\bibfield  {journal} {\bibinfo  {journal}
  {arXiv:2112.02670 [physics]}\ } (\bibinfo {year} {2021})},\ \bibinfo {note}
  {arXiv: 2112.02670}\BibitemShut {NoStop}%
\bibitem [{\citenamefont {Mayer}\ \emph {et~al.}(2021)\citenamefont {Mayer},
  \citenamefont {Ivanov},\ and\ \citenamefont
  {Smirnova}}]{mayer_imprinting_2021}%
  \BibitemOpen
  \bibfield  {author} {\bibinfo {author} {\bibfnamefont {N.}~\bibnamefont
  {Mayer}}, \bibinfo {author} {\bibfnamefont {M.}~\bibnamefont {Ivanov}},\ and\
  \bibinfo {author} {\bibfnamefont {O.}~\bibnamefont {Smirnova}},\ }\bibfield
  {title} {\bibinfo {title} {Imprinting chirality on atoms using synthetic
  chiral light},\ }\href {http://arxiv.org/abs/2112.02658} {\bibfield
  {journal} {\bibinfo  {journal} {arXiv:2112.02658 [physics]}\ } (\bibinfo
  {year} {2021})},\ \bibinfo {note} {arXiv: 2112.02658}\BibitemShut {NoStop}%
\bibitem [{\citenamefont {Dziarmaga}\ \emph {et~al.}(2012)\citenamefont
  {Dziarmaga}, \citenamefont {Zurek},\ and\ \citenamefont
  {Zwolak}}]{dziarmaga2012non}%
  \BibitemOpen
  \bibfield  {author} {\bibinfo {author} {\bibfnamefont {J.}~\bibnamefont
  {Dziarmaga}}, \bibinfo {author} {\bibfnamefont {W.~H.}\ \bibnamefont
  {Zurek}},\ and\ \bibinfo {author} {\bibfnamefont {M.}~\bibnamefont
  {Zwolak}},\ }\bibfield  {title} {\bibinfo {title} {Non-local quantum
  superpositions of topological defects},\ }\href@noop {} {\bibfield  {journal}
  {\bibinfo  {journal} {Nature Physics}\ }\textbf {\bibinfo {volume} {8}},\
  \bibinfo {pages} {49} (\bibinfo {year} {2012})}\BibitemShut {NoStop}%
\bibitem [{\citenamefont {Barth}\ and\ \citenamefont
  {Smirnova}(2013)}]{barth_spinpolarized_2013}%
  \BibitemOpen
  \bibfield  {author} {\bibinfo {author} {\bibfnamefont {I.}~\bibnamefont
  {Barth}}\ and\ \bibinfo {author} {\bibfnamefont {O.}~\bibnamefont
  {Smirnova}},\ }\bibfield  {title} {\bibinfo {title} {Spin-polarized electrons
  produced by strong-field ionization},\ }\href
  {https://doi.org/10.1103/PhysRevA.88.013401} {\bibfield  {journal} {\bibinfo
  {journal} {Physical Review A}\ }\textbf {\bibinfo {volume} {88}},\ \bibinfo
  {pages} {013401} (\bibinfo {year} {2013})}\BibitemShut {NoStop}%
\bibitem [{\citenamefont {Hartung}\ \emph {et~al.}(2016)\citenamefont
  {Hartung}, \citenamefont {Morales}, \citenamefont {Kunitski}, \citenamefont
  {Henrichs}, \citenamefont {Laucke}, \citenamefont {Richter}, \citenamefont
  {Jahnke}, \citenamefont {Kalinin}, \citenamefont {Sch{\"o}ffler},
  \citenamefont {Schmidt}, \citenamefont {Ivanov}, \citenamefont {Smirnova},\
  and\ \citenamefont {D{\"o}rner}}]{Hartung2016}%
  \BibitemOpen
  \bibfield  {author} {\bibinfo {author} {\bibfnamefont {A.}~\bibnamefont
  {Hartung}}, \bibinfo {author} {\bibfnamefont {F.}~\bibnamefont {Morales}},
  \bibinfo {author} {\bibfnamefont {M.}~\bibnamefont {Kunitski}}, \bibinfo
  {author} {\bibfnamefont {K.}~\bibnamefont {Henrichs}}, \bibinfo {author}
  {\bibfnamefont {A.}~\bibnamefont {Laucke}}, \bibinfo {author} {\bibfnamefont
  {M.}~\bibnamefont {Richter}}, \bibinfo {author} {\bibfnamefont
  {T.}~\bibnamefont {Jahnke}}, \bibinfo {author} {\bibfnamefont
  {A.}~\bibnamefont {Kalinin}}, \bibinfo {author} {\bibfnamefont
  {M.}~\bibnamefont {Sch{\"o}ffler}}, \bibinfo {author} {\bibfnamefont
  {L.~P.~H.}\ \bibnamefont {Schmidt}}, \bibinfo {author} {\bibfnamefont
  {M.}~\bibnamefont {Ivanov}}, \bibinfo {author} {\bibfnamefont
  {O.}~\bibnamefont {Smirnova}},\ and\ \bibinfo {author} {\bibfnamefont
  {R.}~\bibnamefont {D{\"o}rner}},\ }\bibfield  {title} {\bibinfo {title}
  {Electron spin polarization in strong-field ionization of xenon atoms},\
  }\href {https://doi.org/10.1038/nphoton.2016.109} {\bibfield  {journal}
  {\bibinfo  {journal} {Nature Photonics}\ }\textbf {\bibinfo {volume} {10}},\
  \bibinfo {pages} {526} (\bibinfo {year} {2016})}\BibitemShut {NoStop}%
\bibitem [{\citenamefont {Eckart}\ \emph {et~al.}(2018)\citenamefont {Eckart},
  \citenamefont {Kunitski}, \citenamefont {Richter}, \citenamefont {Hartung},
  \citenamefont {Rist}, \citenamefont {Trinter}, \citenamefont {Fehre},
  \citenamefont {Schlott}, \citenamefont {Henrichs}, \citenamefont {Schmidt},
  \citenamefont {Jahnke}, \citenamefont {Sch{\"o}ffler}, \citenamefont {Liu},
  \citenamefont {Barth}, \citenamefont {Kaushal}, \citenamefont {Morales},
  \citenamefont {Ivanov}, \citenamefont {Smirnova},\ and\ \citenamefont
  {D{\"o}rner}}]{eckart_ultrafast_2018}%
  \BibitemOpen
  \bibfield  {author} {\bibinfo {author} {\bibfnamefont {S.}~\bibnamefont
  {Eckart}}, \bibinfo {author} {\bibfnamefont {M.}~\bibnamefont {Kunitski}},
  \bibinfo {author} {\bibfnamefont {M.}~\bibnamefont {Richter}}, \bibinfo
  {author} {\bibfnamefont {A.}~\bibnamefont {Hartung}}, \bibinfo {author}
  {\bibfnamefont {J.}~\bibnamefont {Rist}}, \bibinfo {author} {\bibfnamefont
  {F.}~\bibnamefont {Trinter}}, \bibinfo {author} {\bibfnamefont
  {K.}~\bibnamefont {Fehre}}, \bibinfo {author} {\bibfnamefont
  {N.}~\bibnamefont {Schlott}}, \bibinfo {author} {\bibfnamefont
  {K.}~\bibnamefont {Henrichs}}, \bibinfo {author} {\bibfnamefont {L.~P.~H.}\
  \bibnamefont {Schmidt}}, \bibinfo {author} {\bibfnamefont {T.}~\bibnamefont
  {Jahnke}}, \bibinfo {author} {\bibfnamefont {M.}~\bibnamefont
  {Sch{\"o}ffler}}, \bibinfo {author} {\bibfnamefont {K.}~\bibnamefont {Liu}},
  \bibinfo {author} {\bibfnamefont {I.}~\bibnamefont {Barth}}, \bibinfo
  {author} {\bibfnamefont {J.}~\bibnamefont {Kaushal}}, \bibinfo {author}
  {\bibfnamefont {F.}~\bibnamefont {Morales}}, \bibinfo {author} {\bibfnamefont
  {M.}~\bibnamefont {Ivanov}}, \bibinfo {author} {\bibfnamefont
  {O.}~\bibnamefont {Smirnova}},\ and\ \bibinfo {author} {\bibfnamefont
  {R.}~\bibnamefont {D{\"o}rner}},\ }\bibfield  {title} {\bibinfo {title}
  {Ultrafast preparation and detection of ring currents in single atoms},\
  }\href {https://doi.org/10.1038/s41567-018-0080-5} {\bibfield  {journal}
  {\bibinfo  {journal} {Nature Physics}\ }\textbf {\bibinfo {volume} {14}},\
  \bibinfo {pages} {701} (\bibinfo {year} {2018})}\BibitemShut {NoStop}%
\bibitem [{\citenamefont {Kaushal}\ and\ \citenamefont
  {Smirnova}(2018)}]{kaushal_looking_2018a}%
  \BibitemOpen
  \bibfield  {author} {\bibinfo {author} {\bibfnamefont {J.}~\bibnamefont
  {Kaushal}}\ and\ \bibinfo {author} {\bibfnamefont {O.}~\bibnamefont
  {Smirnova}},\ }\bibfield  {title} {\bibinfo {title} {Looking inside the
  tunnelling barrier {{III}}: {{Spin}} polarisation in strong field ionisation
  from orbitals with high angular momentum},\ }\href
  {https://doi.org/10.1088/1361-6455/aad133} {\bibfield  {journal} {\bibinfo
  {journal} {Journal of Physics B: Atomic, Molecular and Optical Physics}\
  }\textbf {\bibinfo {volume} {51}},\ \bibinfo {pages} {174003} (\bibinfo
  {year} {2018})}\BibitemShut {NoStop}%
\bibitem [{\citenamefont {Trabert}\ \emph {et~al.}(2018)\citenamefont
  {Trabert}, \citenamefont {Hartung}, \citenamefont {Eckart}, \citenamefont
  {Trinter}, \citenamefont {Kalinin}, \citenamefont {Sch{\"o}ffler},
  \citenamefont {Schmidt}, \citenamefont {Jahnke}, \citenamefont {Kunitski},\
  and\ \citenamefont {D{\"o}rner}}]{trabert_spin_2018}%
  \BibitemOpen
  \bibfield  {author} {\bibinfo {author} {\bibfnamefont {D.}~\bibnamefont
  {Trabert}}, \bibinfo {author} {\bibfnamefont {A.}~\bibnamefont {Hartung}},
  \bibinfo {author} {\bibfnamefont {S.}~\bibnamefont {Eckart}}, \bibinfo
  {author} {\bibfnamefont {F.}~\bibnamefont {Trinter}}, \bibinfo {author}
  {\bibfnamefont {A.}~\bibnamefont {Kalinin}}, \bibinfo {author} {\bibfnamefont
  {M.}~\bibnamefont {Sch{\"o}ffler}}, \bibinfo {author} {\bibfnamefont
  {L.~P.~H.}\ \bibnamefont {Schmidt}}, \bibinfo {author} {\bibfnamefont
  {T.}~\bibnamefont {Jahnke}}, \bibinfo {author} {\bibfnamefont
  {M.}~\bibnamefont {Kunitski}},\ and\ \bibinfo {author} {\bibfnamefont
  {R.}~\bibnamefont {D{\"o}rner}},\ }\bibfield  {title} {\bibinfo {title} {Spin
  and {{Angular Momentum}} in {{Strong-Field Ionization}}},\ }\href
  {https://doi.org/10.1103/PhysRevLett.120.043202} {\bibfield  {journal}
  {\bibinfo  {journal} {Physical Review Letters}\ }\textbf {\bibinfo {volume}
  {120}},\ \bibinfo {pages} {43202} (\bibinfo {year} {2018})},\ \Eprint
  {https://arxiv.org/abs/1711.03935} {arXiv:1711.03935} \BibitemShut {NoStop}%
\bibitem [{\citenamefont {Ayuso}\ \emph {et~al.}(2022)\citenamefont {Ayuso},
  \citenamefont {Ordonez},\ and\ \citenamefont
  {Smirnova}}]{ayuso_ultrafast_2022}%
  \BibitemOpen
  \bibfield  {author} {\bibinfo {author} {\bibfnamefont {D.}~\bibnamefont
  {Ayuso}}, \bibinfo {author} {\bibfnamefont {A.~F.}\ \bibnamefont {Ordonez}},\
  and\ \bibinfo {author} {\bibfnamefont {O.}~\bibnamefont {Smirnova}},\
  }\bibfield  {title} {\bibinfo {title} {Ultrafast chirality: the road to
  efficient chiral measurements},\ }\href {http://arxiv.org/abs/2203.00580}
  {\bibfield  {journal} {\bibinfo  {journal} {arXiv:2203.00580 [physics]}\ }
  (\bibinfo {year} {2022})},\ \bibinfo {note} {arXiv: 2203.00580}\BibitemShut
  {NoStop}%
\bibitem [{\citenamefont {Planas}\ \emph {et~al.}(2022)\citenamefont {Planas},
  \citenamefont {Ord{\'o}{\~n}ez}, \citenamefont {Lewenstein},\ and\
  \citenamefont {Maxwell}}]{planas2022strong}%
  \BibitemOpen
  \bibfield  {author} {\bibinfo {author} {\bibfnamefont {X.~B.}\ \bibnamefont
  {Planas}}, \bibinfo {author} {\bibfnamefont {A.}~\bibnamefont
  {Ord{\'o}{\~n}ez}}, \bibinfo {author} {\bibfnamefont {M.}~\bibnamefont
  {Lewenstein}},\ and\ \bibinfo {author} {\bibfnamefont {A.~S.}\ \bibnamefont
  {Maxwell}},\ }\bibfield  {title} {\bibinfo {title} {Strong-field chiral
  imaging with twisted photoelectrons},\ }\href@noop {} {\bibfield  {journal}
  {\bibinfo  {journal} {arXiv preprint arXiv:2202.07289}\ } (\bibinfo {year}
  {2022})}\BibitemShut {NoStop}%
\bibitem [{\citenamefont {Cirac}\ \emph {et~al.}(1998)\citenamefont {Cirac},
  \citenamefont {Lewenstein}, \citenamefont {M{\o}lmer},\ and\ \citenamefont
  {Zoller}}]{cirac1998quantum}%
  \BibitemOpen
  \bibfield  {author} {\bibinfo {author} {\bibfnamefont {J.~I.}\ \bibnamefont
  {Cirac}}, \bibinfo {author} {\bibfnamefont {M.}~\bibnamefont {Lewenstein}},
  \bibinfo {author} {\bibfnamefont {K.}~\bibnamefont {M{\o}lmer}},\ and\
  \bibinfo {author} {\bibfnamefont {P.}~\bibnamefont {Zoller}},\ }\bibfield
  {title} {\bibinfo {title} {Quantum superposition states of bose-einstein
  condensates},\ }\href@noop {} {\bibfield  {journal} {\bibinfo  {journal}
  {Phys. Rev. A}\ }\textbf {\bibinfo {volume} {57}},\ \bibinfo {pages} {1208}
  (\bibinfo {year} {1998})}\BibitemShut {NoStop}%
\bibitem [{\citenamefont {Giovannetti}\ \emph {et~al.}(2004)\citenamefont
  {Giovannetti}, \citenamefont {Lloyd},\ and\ \citenamefont
  {Maccone}}]{giovannetti2004quantum}%
  \BibitemOpen
  \bibfield  {author} {\bibinfo {author} {\bibfnamefont {V.}~\bibnamefont
  {Giovannetti}}, \bibinfo {author} {\bibfnamefont {S.}~\bibnamefont {Lloyd}},\
  and\ \bibinfo {author} {\bibfnamefont {L.}~\bibnamefont {Maccone}},\
  }\bibfield  {title} {\bibinfo {title} {Quantum-enhanced measurements: beating
  the standard quantum limit},\ }\href@noop {} {\bibfield  {journal} {\bibinfo
  {journal} {Science}\ }\textbf {\bibinfo {volume} {306}},\ \bibinfo {pages}
  {1330} (\bibinfo {year} {2004})}\BibitemShut {NoStop}%
\bibitem [{\citenamefont {T{\'o}th}\ and\ \citenamefont
  {Apellaniz}(2014)}]{toth2014quantum}%
  \BibitemOpen
  \bibfield  {author} {\bibinfo {author} {\bibfnamefont {G.}~\bibnamefont
  {T{\'o}th}}\ and\ \bibinfo {author} {\bibfnamefont {I.}~\bibnamefont
  {Apellaniz}},\ }\bibfield  {title} {\bibinfo {title} {Quantum metrology from
  a quantum information science perspective},\ }\href@noop {} {\bibfield
  {journal} {\bibinfo  {journal} {Journal of Physics A: Mathematical and
  Theoretical}\ }\textbf {\bibinfo {volume} {47}},\ \bibinfo {pages} {424006}
  (\bibinfo {year} {2014})}\BibitemShut {NoStop}%
\bibitem [{\citenamefont {Pezze}\ \emph {et~al.}(2018)\citenamefont {Pezze},
  \citenamefont {Smerzi}, \citenamefont {Oberthaler}, \citenamefont {Schmied},\
  and\ \citenamefont {Treutlein}}]{pezze2018quantum}%
  \BibitemOpen
  \bibfield  {author} {\bibinfo {author} {\bibfnamefont {L.}~\bibnamefont
  {Pezze}}, \bibinfo {author} {\bibfnamefont {A.}~\bibnamefont {Smerzi}},
  \bibinfo {author} {\bibfnamefont {M.~K.}\ \bibnamefont {Oberthaler}},
  \bibinfo {author} {\bibfnamefont {R.}~\bibnamefont {Schmied}},\ and\ \bibinfo
  {author} {\bibfnamefont {P.}~\bibnamefont {Treutlein}},\ }\bibfield  {title}
  {\bibinfo {title} {Quantum metrology with nonclassical states of atomic
  ensembles},\ }\href@noop {} {\bibfield  {journal} {\bibinfo  {journal}
  {Reviews of Modern Physics}\ }\textbf {\bibinfo {volume} {90}},\ \bibinfo
  {pages} {035005} (\bibinfo {year} {2018})}\BibitemShut {NoStop}%
\bibitem [{\citenamefont {Polino}\ \emph {et~al.}(2020)\citenamefont {Polino},
  \citenamefont {Valeri}, \citenamefont {Spagnolo},\ and\ \citenamefont
  {Sciarrino}}]{polino2020photonic}%
  \BibitemOpen
  \bibfield  {author} {\bibinfo {author} {\bibfnamefont {E.}~\bibnamefont
  {Polino}}, \bibinfo {author} {\bibfnamefont {M.}~\bibnamefont {Valeri}},
  \bibinfo {author} {\bibfnamefont {N.}~\bibnamefont {Spagnolo}},\ and\
  \bibinfo {author} {\bibfnamefont {F.}~\bibnamefont {Sciarrino}},\ }\bibfield
  {title} {\bibinfo {title} {Photonic quantum metrology},\ }\href@noop {}
  {\bibfield  {journal} {\bibinfo  {journal} {AVS Quantum Science}\ }\textbf
  {\bibinfo {volume} {2}},\ \bibinfo {pages} {024703} (\bibinfo {year}
  {2020})}\BibitemShut {NoStop}%
\bibitem [{\citenamefont {G{\"o}bel}\ and\ \citenamefont
  {Siegner}(2015)}]{gobel2015quantum}%
  \BibitemOpen
  \bibfield  {author} {\bibinfo {author} {\bibfnamefont {E.~O.}\ \bibnamefont
  {G{\"o}bel}}\ and\ \bibinfo {author} {\bibfnamefont {U.}~\bibnamefont
  {Siegner}},\ }\href@noop {} {\emph {\bibinfo {title} {Quantum Metrology:
  Foundation of Units and Measurements}}}\ (\bibinfo  {publisher} {John Wiley
  \& Sons},\ \bibinfo {year} {2015})\BibitemShut {NoStop}%
\bibitem [{\citenamefont {Barbieri}(2022)}]{barbieri2022optical}%
  \BibitemOpen
  \bibfield  {author} {\bibinfo {author} {\bibfnamefont {M.}~\bibnamefont
  {Barbieri}},\ }\bibfield  {title} {\bibinfo {title} {Optical quantum
  metrology},\ }\href@noop {} {\bibfield  {journal} {\bibinfo  {journal} {PRX
  Quantum}\ }\textbf {\bibinfo {volume} {3}},\ \bibinfo {pages} {010202}
  (\bibinfo {year} {2022})}\BibitemShut {NoStop}%
\bibitem [{\citenamefont {Bhattacharya}\ \emph {et~al.}(2022)\citenamefont
  {Bhattacharya}, \citenamefont {Chaudhary}, \citenamefont {Grass},
  \citenamefont {Johnson}, \citenamefont {Wall},\ and\ \citenamefont
  {Lewenstein}}]{bhattacharya2022fermionic}%
  \BibitemOpen
  \bibfield  {author} {\bibinfo {author} {\bibfnamefont {U.}~\bibnamefont
  {Bhattacharya}}, \bibinfo {author} {\bibfnamefont {S.}~\bibnamefont
  {Chaudhary}}, \bibinfo {author} {\bibfnamefont {T.}~\bibnamefont {Grass}},
  \bibinfo {author} {\bibfnamefont {A.~S.}\ \bibnamefont {Johnson}}, \bibinfo
  {author} {\bibfnamefont {S.}~\bibnamefont {Wall}},\ and\ \bibinfo {author}
  {\bibfnamefont {M.}~\bibnamefont {Lewenstein}},\ }\bibfield  {title}
  {\bibinfo {title} {Fermionic chern insulator from twisted light with linear
  polarization},\ }\href@noop {} {\bibfield  {journal} {\bibinfo  {journal}
  {Physical Review B}\ }\textbf {\bibinfo {volume} {105}},\ \bibinfo {pages}
  {L081406} (\bibinfo {year} {2022})}\BibitemShut {NoStop}%
\bibitem [{\citenamefont {Maxwell}\ \emph {et~al.}(2021)\citenamefont
  {Maxwell}, \citenamefont {Armstrong}, \citenamefont {Ciappina}, \citenamefont
  {Pisanty}, \citenamefont {Kang}, \citenamefont {Brown}, \citenamefont
  {Lewenstein},\ and\ \citenamefont
  {de~Morisson~Faria}}]{maxwell2021manipulating}%
  \BibitemOpen
  \bibfield  {author} {\bibinfo {author} {\bibfnamefont {A.}~\bibnamefont
  {Maxwell}}, \bibinfo {author} {\bibfnamefont {G.}~\bibnamefont {Armstrong}},
  \bibinfo {author} {\bibfnamefont {M.}~\bibnamefont {Ciappina}}, \bibinfo
  {author} {\bibfnamefont {E.}~\bibnamefont {Pisanty}}, \bibinfo {author}
  {\bibfnamefont {Y.}~\bibnamefont {Kang}}, \bibinfo {author} {\bibfnamefont
  {A.}~\bibnamefont {Brown}}, \bibinfo {author} {\bibfnamefont
  {M.}~\bibnamefont {Lewenstein}},\ and\ \bibinfo {author} {\bibfnamefont
  {C.~F.}\ \bibnamefont {de~Morisson~Faria}},\ }\bibfield  {title} {\bibinfo
  {title} {Manipulating twisted electrons in strong-field ionization},\
  }\href@noop {} {\bibfield  {journal} {\bibinfo  {journal} {Faraday
  Discussions}\ }\textbf {\bibinfo {volume} {228}},\ \bibinfo {pages} {394}
  (\bibinfo {year} {2021})}\BibitemShut {NoStop}%
\bibitem [{\citenamefont {Vrakking}(2021)}]{vrakking2021control}%
  \BibitemOpen
  \bibfield  {author} {\bibinfo {author} {\bibfnamefont {M.~J.}\ \bibnamefont
  {Vrakking}},\ }\bibfield  {title} {\bibinfo {title} {Control of attosecond
  entanglement and coherence},\ }\href@noop {} {\bibfield  {journal} {\bibinfo
  {journal} {Phys. Rev. Lett.}\ }\textbf {\bibinfo {volume} {126}},\ \bibinfo
  {pages} {113203} (\bibinfo {year} {2021})}\BibitemShut {NoStop}%
\bibitem [{\citenamefont {Koll}\ \emph {et~al.}(2022)\citenamefont {Koll},
  \citenamefont {Maikowski}, \citenamefont {Drescher}, \citenamefont
  {Witting},\ and\ \citenamefont {Vrakking}}]{koll2022experimental}%
  \BibitemOpen
  \bibfield  {author} {\bibinfo {author} {\bibfnamefont {L.-M.}\ \bibnamefont
  {Koll}}, \bibinfo {author} {\bibfnamefont {L.}~\bibnamefont {Maikowski}},
  \bibinfo {author} {\bibfnamefont {L.}~\bibnamefont {Drescher}}, \bibinfo
  {author} {\bibfnamefont {T.}~\bibnamefont {Witting}},\ and\ \bibinfo {author}
  {\bibfnamefont {M.~J.}\ \bibnamefont {Vrakking}},\ }\bibfield  {title}
  {\bibinfo {title} {Experimental control of quantum-mechanical entanglement in
  an attosecond pump-probe experiment},\ }\href@noop {} {\bibfield  {journal}
  {\bibinfo  {journal} {Phys. Rev. Lett.}\ }\textbf {\bibinfo {volume} {128}},\
  \bibinfo {pages} {043201} (\bibinfo {year} {2022})}\BibitemShut {NoStop}%
\bibitem [{\citenamefont {Vrakking}(2022)}]{vrakking2022ion}%
  \BibitemOpen
  \bibfield  {author} {\bibinfo {author} {\bibfnamefont {M.~J.}\ \bibnamefont
  {Vrakking}},\ }\bibfield  {title} {\bibinfo {title} {Ion-photoelectron
  entanglement in photoionization with chirped laser pulses},\ }\href@noop {}
  {\bibfield  {journal} {\bibinfo  {journal} {Journal of Physics B: Atomic,
  Molecular and Optical Physics}\ } (\bibinfo {year} {2022})}\BibitemShut
  {NoStop}%
\bibitem [{\citenamefont {Bourassin-Bouchet}\ \emph {et~al.}(2020)\citenamefont
  {Bourassin-Bouchet}, \citenamefont {Barreau}, \citenamefont {Gruson},
  \citenamefont {Hergott}, \citenamefont {Qu{\'e}r{\'e}}, \citenamefont
  {Sali{\`e}res},\ and\ \citenamefont {Ruchon}}]{bourassin2020quantifying}%
  \BibitemOpen
  \bibfield  {author} {\bibinfo {author} {\bibfnamefont {C.}~\bibnamefont
  {Bourassin-Bouchet}}, \bibinfo {author} {\bibfnamefont {L.}~\bibnamefont
  {Barreau}}, \bibinfo {author} {\bibfnamefont {V.}~\bibnamefont {Gruson}},
  \bibinfo {author} {\bibfnamefont {J.-F.}\ \bibnamefont {Hergott}}, \bibinfo
  {author} {\bibfnamefont {F.}~\bibnamefont {Qu{\'e}r{\'e}}}, \bibinfo {author}
  {\bibfnamefont {P.}~\bibnamefont {Sali{\`e}res}},\ and\ \bibinfo {author}
  {\bibfnamefont {T.}~\bibnamefont {Ruchon}},\ }\bibfield  {title} {\bibinfo
  {title} {Quantifying decoherence in attosecond metrology},\ }\href@noop {}
  {\bibfield  {journal} {\bibinfo  {journal} {Phys. Rev. X}\ }\textbf {\bibinfo
  {volume} {10}},\ \bibinfo {pages} {031048} (\bibinfo {year}
  {2020})}\BibitemShut {NoStop}%
\bibitem [{\citenamefont {Laurell}\ \emph {et~al.}(2022)\citenamefont
  {Laurell}, \citenamefont {Finkelstein-Shapiro}, \citenamefont {Dittel},
  \citenamefont {Guo}, \citenamefont {Demjaha}, \citenamefont {Ammitzb{\"o}ll},
  \citenamefont {Weissenbilder}, \citenamefont {Neori{\v{c}}i{\'c}},
  \citenamefont {Luo}, \citenamefont {Gisselbrecht} \emph
  {et~al.}}]{laurell2022continuous}%
  \BibitemOpen
  \bibfield  {author} {\bibinfo {author} {\bibfnamefont {H.}~\bibnamefont
  {Laurell}}, \bibinfo {author} {\bibfnamefont {D.}~\bibnamefont
  {Finkelstein-Shapiro}}, \bibinfo {author} {\bibfnamefont {C.}~\bibnamefont
  {Dittel}}, \bibinfo {author} {\bibfnamefont {C.}~\bibnamefont {Guo}},
  \bibinfo {author} {\bibfnamefont {R.}~\bibnamefont {Demjaha}}, \bibinfo
  {author} {\bibfnamefont {M.}~\bibnamefont {Ammitzb{\"o}ll}}, \bibinfo
  {author} {\bibfnamefont {R.}~\bibnamefont {Weissenbilder}}, \bibinfo {author}
  {\bibfnamefont {L.}~\bibnamefont {Neori{\v{c}}i{\'c}}}, \bibinfo {author}
  {\bibfnamefont {S.}~\bibnamefont {Luo}}, \bibinfo {author} {\bibfnamefont
  {M.}~\bibnamefont {Gisselbrecht}}, \emph {et~al.},\ }\bibfield  {title}
  {\bibinfo {title} {Continuous variable quantum state tomography of
  photoelectrons},\ }\href@noop {} {\bibfield  {journal} {\bibinfo  {journal}
  {arXiv preprint arXiv:2202.06798}\ } (\bibinfo {year} {2022})}\BibitemShut
  {NoStop}%
\bibitem [{\citenamefont {Busto}\ \emph {et~al.}(2021)\citenamefont {Busto},
  \citenamefont {Laurell}, \citenamefont {Shapiro}, \citenamefont
  {Alexandridi}, \citenamefont {Isinger}, \citenamefont {Nandi}, \citenamefont
  {Squibb}, \citenamefont {Turconi}, \citenamefont {Zhong}, \citenamefont
  {Arnold} \emph {et~al.}}]{busto2021probing}%
  \BibitemOpen
  \bibfield  {author} {\bibinfo {author} {\bibfnamefont {D.}~\bibnamefont
  {Busto}}, \bibinfo {author} {\bibfnamefont {H.}~\bibnamefont {Laurell}},
  \bibinfo {author} {\bibfnamefont {D.~F.}\ \bibnamefont {Shapiro}}, \bibinfo
  {author} {\bibfnamefont {C.}~\bibnamefont {Alexandridi}}, \bibinfo {author}
  {\bibfnamefont {M.}~\bibnamefont {Isinger}}, \bibinfo {author} {\bibfnamefont
  {S.}~\bibnamefont {Nandi}}, \bibinfo {author} {\bibfnamefont
  {R.}~\bibnamefont {Squibb}}, \bibinfo {author} {\bibfnamefont
  {M.}~\bibnamefont {Turconi}}, \bibinfo {author} {\bibfnamefont
  {S.}~\bibnamefont {Zhong}}, \bibinfo {author} {\bibfnamefont
  {C.}~\bibnamefont {Arnold}}, \emph {et~al.},\ }\bibfield  {title} {\bibinfo
  {title} {Probing electronic decoherence with high-resolution attosecond
  photoelectron interferometry},\ }\href@noop {} {\bibfield  {journal}
  {\bibinfo  {journal} {arXiv preprint arXiv:2111.12037}\ } (\bibinfo {year}
  {2021})}\BibitemShut {NoStop}%
\bibitem [{\citenamefont {Stammer}\ \emph
  {et~al.}(2022{\natexlab{b}})\citenamefont {Stammer}, \citenamefont
  {Rivera-Dean}, \citenamefont {Maxwell}, \citenamefont {Lamprou},
  \citenamefont {Ordóñez}, \citenamefont {Ciappina}, \citenamefont
  {Tzallas},\ and\ \citenamefont {Lewenstein}}]{ZenodoLink}%
  \BibitemOpen
  \bibfield  {author} {\bibinfo {author} {\bibfnamefont {P.}~\bibnamefont
  {Stammer}}, \bibinfo {author} {\bibfnamefont {J.}~\bibnamefont
  {Rivera-Dean}}, \bibinfo {author} {\bibfnamefont {A.}~\bibnamefont
  {Maxwell}}, \bibinfo {author} {\bibfnamefont {T.}~\bibnamefont {Lamprou}},
  \bibinfo {author} {\bibfnamefont {A.}~\bibnamefont {Ordóñez}}, \bibinfo
  {author} {\bibfnamefont {M.~F.}\ \bibnamefont {Ciappina}}, \bibinfo {author}
  {\bibfnamefont {P.}~\bibnamefont {Tzallas}},\ and\ \bibinfo {author}
  {\bibfnamefont {M.}~\bibnamefont {Lewenstein}},\ }\href
  {https://doi.org/10.5281/zenodo.6617182} {\bibinfo {title} {Quantum
  electrodynamics of ultra-intense laser-matter interactions}},\ \bibinfo
  {howpublished} {Zenodo} (\bibinfo {year} {2022}{\natexlab{b}})\BibitemShut
  {NoStop}%
\bibitem [{\citenamefont {Virtanen}\ \emph {et~al.}(2020)\citenamefont
  {Virtanen}, \citenamefont {Gommers}, \citenamefont {Oliphant}, \citenamefont
  {Haberland}, \citenamefont {Reddy}, \citenamefont {Cournapeau}, \citenamefont
  {Burovski}, \citenamefont {Peterson}, \citenamefont {Weckesser},
  \citenamefont {Bright}, \citenamefont {{van der Walt}}, \citenamefont
  {Brett}, \citenamefont {Wilson}, \citenamefont {Millman}, \citenamefont
  {Mayorov}, \citenamefont {Nelson}, \citenamefont {Jones}, \citenamefont
  {Kern}, \citenamefont {Larson}, \citenamefont {Carey}, \citenamefont {Polat},
  \citenamefont {Feng}, \citenamefont {Moore}, \citenamefont {{VanderPlas}},
  \citenamefont {Laxalde}, \citenamefont {Perktold}, \citenamefont {Cimrman},
  \citenamefont {Henriksen}, \citenamefont {Quintero}, \citenamefont {Harris},
  \citenamefont {Archibald}, \citenamefont {Ribeiro}, \citenamefont
  {Pedregosa}, \citenamefont {{van Mulbregt}},\ and\ \citenamefont {{SciPy 1.0
  Contributors}}}]{2020SciPy-NMeth}%
  \BibitemOpen
  \bibfield  {author} {\bibinfo {author} {\bibfnamefont {P.}~\bibnamefont
  {Virtanen}}, \bibinfo {author} {\bibfnamefont {R.}~\bibnamefont {Gommers}},
  \bibinfo {author} {\bibfnamefont {T.~E.}\ \bibnamefont {Oliphant}}, \bibinfo
  {author} {\bibfnamefont {M.}~\bibnamefont {Haberland}}, \bibinfo {author}
  {\bibfnamefont {T.}~\bibnamefont {Reddy}}, \bibinfo {author} {\bibfnamefont
  {D.}~\bibnamefont {Cournapeau}}, \bibinfo {author} {\bibfnamefont
  {E.}~\bibnamefont {Burovski}}, \bibinfo {author} {\bibfnamefont
  {P.}~\bibnamefont {Peterson}}, \bibinfo {author} {\bibfnamefont
  {W.}~\bibnamefont {Weckesser}}, \bibinfo {author} {\bibfnamefont
  {J.}~\bibnamefont {Bright}}, \bibinfo {author} {\bibfnamefont {S.~J.}\
  \bibnamefont {{van der Walt}}}, \bibinfo {author} {\bibfnamefont
  {M.}~\bibnamefont {Brett}}, \bibinfo {author} {\bibfnamefont
  {J.}~\bibnamefont {Wilson}}, \bibinfo {author} {\bibfnamefont {K.~J.}\
  \bibnamefont {Millman}}, \bibinfo {author} {\bibfnamefont {N.}~\bibnamefont
  {Mayorov}}, \bibinfo {author} {\bibfnamefont {A.~R.~J.}\ \bibnamefont
  {Nelson}}, \bibinfo {author} {\bibfnamefont {E.}~\bibnamefont {Jones}},
  \bibinfo {author} {\bibfnamefont {R.}~\bibnamefont {Kern}}, \bibinfo {author}
  {\bibfnamefont {E.}~\bibnamefont {Larson}}, \bibinfo {author} {\bibfnamefont
  {C.~J.}\ \bibnamefont {Carey}}, \bibinfo {author} {\bibfnamefont
  {{\.I}.}~\bibnamefont {Polat}}, \bibinfo {author} {\bibfnamefont
  {Y.}~\bibnamefont {Feng}}, \bibinfo {author} {\bibfnamefont {E.~W.}\
  \bibnamefont {Moore}}, \bibinfo {author} {\bibfnamefont {J.}~\bibnamefont
  {{VanderPlas}}}, \bibinfo {author} {\bibfnamefont {D.}~\bibnamefont
  {Laxalde}}, \bibinfo {author} {\bibfnamefont {J.}~\bibnamefont {Perktold}},
  \bibinfo {author} {\bibfnamefont {R.}~\bibnamefont {Cimrman}}, \bibinfo
  {author} {\bibfnamefont {I.}~\bibnamefont {Henriksen}}, \bibinfo {author}
  {\bibfnamefont {E.~A.}\ \bibnamefont {Quintero}}, \bibinfo {author}
  {\bibfnamefont {C.~R.}\ \bibnamefont {Harris}}, \bibinfo {author}
  {\bibfnamefont {A.~M.}\ \bibnamefont {Archibald}}, \bibinfo {author}
  {\bibfnamefont {A.~H.}\ \bibnamefont {Ribeiro}}, \bibinfo {author}
  {\bibfnamefont {F.}~\bibnamefont {Pedregosa}}, \bibinfo {author}
  {\bibfnamefont {P.}~\bibnamefont {{van Mulbregt}}},\ and\ \bibinfo {author}
  {\bibnamefont {{SciPy 1.0 Contributors}}},\ }\bibfield  {title} {\bibinfo
  {title} {{{SciPy} 1.0: Fundamental Algorithms for Scientific Computing in
  Python}},\ }\href {https://doi.org/10.1038/s41592-019-0686-2} {\bibfield
  {journal} {\bibinfo  {journal} {Nat. Meth.}\ }\textbf {\bibinfo {volume}
  {17}},\ \bibinfo {pages} {261} (\bibinfo {year} {2020})}\BibitemShut
  {NoStop}%
\end{thebibliography}%

\newpage
\onecolumngrid
\appendix

\section{Field commutation relations}\label{App:quantization:EM}

The commutation relation between the canonical field variables $\vb{A}(\vb{r})$ and $\vb{\Pi}(\vb{r})$ defined respectively in Eqs.~\eqref{eq:vector_potential} and \eqref{eq:field:momentum} is given by
\begin{align}
\left[ A_m(\vb{r}), \Pi_n(\vb{r}^\prime) \right] = i \hbar \delta^\perp_{mn}(\vb{r} - \vb{r}^\prime),
\end{align}
with the transverse $\delta$-function 
\begin{align}
\delta_{mn}^\perp(\vb{r} - \vb{r}^\prime) = \frac{1}{(2 \pi)^3} \int \dd^3k \left[ \delta_{mn} - \frac{k_m k_n}{k^2} \right] e^{i \vb{k} \cdot (\vb{r} - \vb{r}^\prime)}.
\end{align}

These canonical conjugate field variables are related to the field operators, $\vb{B} = \nabla \times \vb{A}$, and $\vb{E}^\perp = - \vb{\Pi}/ \epsilon_0$, of the magnetic field and the transverse part of electric field, respectively. The free-field Hamiltonian $H_f$ can then be expressed in terms of the physical operators 
\begin{align}
H_f = \frac{1}{2} \int d^3r \left[ \epsilon_0 \vb{E}^2(\vb{r}) + \frac{1}{\mu_0} \vb{B}^2(\vb{r}) \right],
\end{align}
which has the same form as the energy of the classical electromagnetic field, and it is required that the observables obey the commutation relation 
\begin{align}
\left[ E_m(\vb{r}) , B_n(\vb{r}^\prime) \right] = - \epsilon_{mnl} \frac{i\hbar}{\epsilon_0} \pdv{x_l} \delta(\vb{r} - \vb{r}^\prime).
\end{align}

\section{Interaction Hamiltonian}\label{App:PWZ:transf}

The minimal-coupling Hamiltonian given in Eq.~\eqref{eq:H_pAdip} can be written in terms of the transformed dynamical variables from \eqref{eq:new_field_momentum}, \eqref{eq:new_electron_momentum}, to obtain the multipolar-coupling Hamiltonian
\begin{equation}
    \begin{aligned}
    H^\prime &=  \frac{1}{2} \int \dd^3r \left\{ \frac{1}{\epsilon_0} \left[ \vb{\Pi}^\prime(\vb{r}) + \vb{P}_\perp(\vb{r}) \right]^2 + \frac{1}{\mu_0} \left[ \nabla \times \vb{A}(\vb{r})\right]^2 \right\} 
    \\
    & \quad+ \sum_\alpha \bigg[\frac{1}{2 m} \Big\{ \vb{p}^\prime_\alpha - e \int_0^1 \dd s\ s (\vb{r}_\alpha - \vb{R}_\alpha)
    \times \left[ \nabla 
    \times \vb{A}(\vb{R}_\alpha + s(\vb{r}_\alpha - \vb{R}_\alpha)) \right]  \Big\}^2 + V_{\rm at}(\vb{r}_\alpha, \vb{R}_\alpha))\bigg],
    \end{aligned}
\end{equation}
where, as mentioned in Section \ref{Sec:las:mat:int}, $V_{\rm at}(\vb{r}_\alpha, \vb{R}_\alpha)$ is the effective potential felt by the single active electron, and $\vb{\Pi}^\prime$ and $\vb{p}'$ are given by Eqs.~\eqref{eq:new_field_momentum} and \eqref{eq:new_electron_momentum} respectively.

Using the mode expansion of the dynamical field variables of the transverse part of the electromagnetic field, the total Hamiltonian can be decomposed in the following way
\begin{align}
H^\prime = H_f^\prime + H_{\rm at}^\prime + H_{int}^\prime,
\end{align}
where the free-field Hamiltonian $H_f^\prime$ has the same form as \eqref{eq:H_free_field} but in terms of the new mode operators 
\begin{equation}
\begin{aligned}
H_f^\prime &= \frac{1}{2} \int \dd^3r \left\{ \frac{1}{\epsilon_0} {\vb{\Pi}^\prime}^2(\vb{r}) + \frac{1}{\mu_0} \left[ \nabla \times \vb{A}(\vb{r})\right]^2 \right\} = \sum_\mu \int \dd^3k \, \hbar \omega_k \left(  {a^\prime}^\dag_{\bf{k} \mu} a^\prime_{\bf{k} \mu} + \frac{1}{2} \right).
\end{aligned}
\end{equation}

We have separated the purely electronic (atomic or molecular) Hamiltonian 
\begin{equation}
\begin{aligned}
H_{\rm at}^\prime &= \sum_\alpha \left[\frac{{\mathbf{p}^\prime}^2_\alpha}{2 m} + V_{\rm at}(\vb{r}_\alpha, \vb{R}_\alpha) + \frac{1}{2\epsilon_0} \int \dd^3r \mathbf{P}_\perp^2 (\mathbf{r})\right] = \sum_\alpha \left[\frac{{\mathbf{p}^\prime}^2_\alpha}{2 m_\alpha} + \tilde V_{\rm at}(\vb{r}_\alpha, \vb{R}_\alpha) \right], 
\end{aligned}
\end{equation}
where $\tilde V_{\rm at}(\vb{r}_\alpha, \vb{R}_\alpha)$ is the final effective potential for single active electrons that includes effects of polarization. Finally, the interaction Hamiltonian reads 
\begin{equation}
\begin{aligned}
\label{eq:H_multipolar}
H_{int}^\prime = & - \frac{e}{\epsilon_0 } \sum_\alpha \int_0^1 \dd s (\vb{r}_\alpha - \vb{R}_\alpha) \cdot \vb{\Pi}^\prime (\vb{R}_\alpha + s (\vb{r}_\alpha - \vb{R}_\alpha)) \\
& + \frac{e}{2 m } \sum_\alpha \int_0^1 \dd s \ s \left\{ [ (\vb{r}_\alpha - \vb{R}_\alpha) \times \vb{p}^\prime_\alpha] [\nabla \times \vb{A}(\vb{R}_\alpha + s (\vb{r}_\alpha - \vb{R}_\alpha))] + \operatorname{h.c.} \right\} \\
& + \frac{e^2}{2 m }  \sum_\alpha \left\{ \int_0^1 \dd s\ s  (\vb{r}_\alpha - \vb{R}_\alpha) \times  [\nabla \times \vb{A}(\vb{R}_\alpha + s (\vb{r}_\alpha - \vb{R}_\alpha))]  \right\}^2.
\end{aligned}
\end{equation}

The first term is the most relevant in our analysis since it couples the transverse displacement field $\vb{\Pi}^\prime (\vb{r}) = -  [\epsilon_0 \vb{E}_\perp (\vb{r}) + \vb{P}_\perp (\vb{r})]$ with the atomic polarization $\vb{P} (\vb{r})$. The second and third term describe the interaction of the paramagnetic magnetization with the magnetic induction field $\vb{B}(\vb{r}) = \nabla \times \vb{A}(\vb{r})$, and the diamagnetic energy of the system which is quadratic in the magnetic induction field, respectively.

\section{HHG spectrum}\label{App:HHG:Spectrum}

The coherent contribution to the HHG spectrum is proportional to 
\begin{equation}\label{eq:HHG_spectrum1}
    \begin{aligned}
    S(\omega_{\vb{k},\mu}) & \propto \lim_{t \to \infty} \abs{\chi_{\vb{k},\mu}(t)}^2 \\
    & = \tilde g(k)^2 N \abs{\vb{\epsilon}_{\vb{k},\mu} \cdot \expval{\vb{d}(\omega_{\vb{k}})}}^2 + \tilde g(k)^2 \sum_{\alpha \neq \alpha^\prime} e^{- i \vb{k} (\vb{R}_\alpha - \vb{R}_{\alpha^\prime})} \vb{\epsilon}_{\vb{k},\mu} \cdot \expval{\vb{d}_\alpha(\omega_{\vb{k}})} \vb{\epsilon}_{\vb{k},\mu} \cdot \expval{\vb{d}_{\alpha^\prime}(\omega_{\vb{k}})}^*,
    \end{aligned}
\end{equation}
where we have assumed that each of the $N$ atoms contributes equally to the HHG spectrum such that 
\begin{align}
    \sum_{\alpha = \alpha^\prime} \abs{\vb{\epsilon}_{\vb{k},\mu} \cdot \expval{\vb{d}_\alpha(\omega_{\vb{k}})}}^2 = N \abs{\vb{\epsilon}_{\vb{k},\mu} \cdot \expval{\vb{d}(\omega_{\vb{k}})}}^2.    
\end{align}

\textcolor{black}{Assuming that the $N$ atoms contribute equally to the HHG spectrum of course implies the additional assumption that all atoms see the same radiation field of the driving source. In realistic experiments this is almost never the case, and focal averaging over the spatial beam profile needs to be taken into account. However, for the concepts and methods introduced in this manuscript this only plays a minor role since the post-selection scheme does not depend on the particular structure of the HHG spectra, and holds true for generic harmonic amplitudes $\chi_{\vb{k},\mu}$.}

Following the same assumptions that lead to the dipole approximation, i.e. $\vb{k} \cdot \vb{R}_\alpha \ll 1$, and again assuming that all $N$ atoms equally contribute to the HHG process, the second term in \eqref{eq:HHG_spectrum1} is given by \begin{equation}
\begin{aligned}
    \tilde g(k)^2 &\sum_{\alpha \neq \alpha^\prime} e^{- i \vb{k} (\vb{R}_\alpha - \vb{R}_{\alpha^\prime})} \vb{\epsilon}_{\vb{k},\mu} \cdot \expval{\vb{d}_\alpha(\omega_{\vb{k}})} \vb{\epsilon}_{\vb{k},\mu} \cdot \expval{\vb{d}_{\alpha^\prime}(\omega_{\vb{k}})}^* =  \tilde g(k)^2 (N^2 - N)  \abs{\vb{\epsilon}_{\vb{k},\mu} \cdot \expval{\vb{d}(\omega_{\vb{k}})}}^2,
\end{aligned}
\end{equation}
such that the spectrum is given by 
\begin{align}
    S(\omega_{\vb{k},\mu}) \propto \lim_{t \to \infty} \abs{\chi_{\vb{k},\mu}(t)}^2 = \tilde g(k)^2 N^2  \abs{\vb{\epsilon}_{\vb{k},\mu} \cdot \expval{\vb{d}(\omega_{\vb{k}})}}^2.
\end{align}

\section{Derivation of electronic transition matrix elements}

In the main part of the manuscript we need the time-dependent dipole transition matrix elements between the ground state $\ket{g}$ and a continuum state $\ket{\vb{v}}$, or between two different continuum states, see for instance \eqref{eq:transition_gs_continuum_aprox} or \eqref{eq:C-C:approx}, respectively.  
To obtain the transition matrix elements we use that $\vb{r}(t) = U_{sc}^\dagger(t) \vb{r} U_{sc}(t)$, and we will further use that the action of the propagator on the electronic state is known from the semi-classical theory \cite{lewenstein1994theory}
\begin{equation}
\begin{aligned}\label{eq:TDSE_solution_semiclassical}
   \ket{\psi_{g/\vb{v}}(t)} &=  U_{sc}(t) \ket{g/\vb{v}} = e^{i I_p t / \hbar} \left( a_{g/\vb{v}}(t) \ket{g} + \int \dd^3 v^\prime b_{g/\vb{v}}(\vb{v}^\prime, t) \ket{ \vb{v}^\prime}  \right),
\end{aligned}
\end{equation}
which are the solutions of the time-dependent Schrödinger equation with the ground state $\ket{g}$ or the continuum state $\ket{\vb{v}}$ as the initial condition, respectively.

\subsection{\label{sec:app_amplitudes}Electron ground and continuum state population amplitudes}

The amplitudes \eqref{eq:TDSE_solution_semiclassical} of the ground and continuum states obey the following coupled differential equations 
\begin{equation}\label{eq:coupled_DGL_amplitudes1}
    \dv{}{t} a_{g/\vb{v}}(t) = \frac{i}{\hbar} \int \dd^3 v^\prime \vb{E}_{cl}(t) \cdot \vb{d}^*(\vb{v}^\prime) b_{g/\vb{v}}(\vb{v}^\prime,t)
\end{equation}
\begin{equation}\label{eq:coupled_DGL_amplitudes2}
    \begin{aligned}
    \dv{}{t} b_{g/\vb{v}}(\vb{v}^\prime,t)  = & -\frac{i}{\hbar} \left( \frac{\vb{v^\prime}^2}{2m}
    + I_p \right) b_{g/\vb{v}}(\vb{v}^\prime,t) + \frac{i}{\hbar} \vb{E}_{cl}(t) \cdot \vb{d}(\vb{v}^\prime) a_{g/\vb{v}}(t) \\
    & + e \vb{E}_{cl}(t) \cdot \nabla_{\vb{v}^\prime} b_{g/\vb{v}}(\vb{v}^\prime,t) + i \vb{E}_{cl}(t) \cdot \int \dd^3 v^{\prime \prime} b_{g/\vb{v}}(\vb{v}^{\prime \prime} , t) \vb{g}(\vb{v}^\prime,\vb{v}^{\prime \prime}), 
    \end{aligned}
\end{equation}
where $\vb{d}(\vb{v}^\prime) = -e \bra{\vb{v}^\prime} \vb{r} \ket{g}$. 
However, to solve this coupled differential equations we shall perform the following approximation by neglecting re-scattering events and only consider the contribution from direct electrons, i.e. we neglect the continuum-continuum transitions in \eqref{eq:coupled_DGL_amplitudes2} (term with $\vb{g}(\vb{v}^\prime,\vb{v}^{\prime \prime})$). 
We can therefore solve the differential equation for the continuum amplitude with the initial condition, $b_g(\vb{v}^\prime,t_0) = 0$ and $a_g(t_0)= 1$, i.e. initial ground state population 
\begin{equation}\label{eq:DGL_b_approximated}
    \begin{aligned}
    \dv{}{t} b_g(\vb{v}^\prime,t) &= - \frac{i}{\hbar} \left( \frac{\vb{v^\prime}^2}{2m} + I_p \right) b_g(\vb{v}^\prime,t) + \frac{i}{\hbar} \vb{E}_{cl}(t) \cdot \vb{d}(\vb{v}^\prime) a_g(t) + e \vb{E}_{cl}(t) \cdot \nabla_{\vb{v}^\prime} b_g(\vb{v}^\prime,t), 
    \end{aligned}
\end{equation}
with the solution given by 
\begin{equation}
\label{eq:b_final_initial_g}
    \begin{aligned}
    b_g(\vb{p}^\prime,t) &= \frac{i}{\hbar } \int_{t_0}^t \dd t^\prime \vb{E}_{cl}(t^\prime) \cdot \vb{d}\left(\vb{p}^\prime + \frac{e}{c} \vb{A}_{cl}(t^\prime) \right) \exp{- \frac{i}{\hbar} S(\vb{p}',t,t')} a_g(t^\prime),
    \end{aligned}
\end{equation}
where we have substituted the canonical momentum $\vb{p}^\prime = m\vb{v}^\prime - \frac{e}{c} \vb{A}(t)$, such that the gauge invariant kinetic momentum $m \dot{\vb{r}} (t) = \vb{p} + \frac{e}{c} \vb{A} (t)$ appears in the expression. With this, we denote by $S(\vb{p},t,t')$ the semi-classical action given by
\begin{equation}
    S(\vb{p},t,t')
        = \int^{t}_{t'} \dd t''
            \bigg[
                \dfrac{1}{2m}
                \Big( 
                    \vb{p}' +\dfrac{e}{c} \vb{A}_{cl}(t'')
                \Big)^2 + I_p
            \bigg].
\end{equation}
This can now be inserted into the differential equation for the ground state amplitude \eqref{eq:coupled_DGL_amplitudes1} and we obtain 
\begin{align}
    \dv{}{t} a_g(t) = - \int_{t_0}^t \dd t^\prime \gamma(t,t^\prime) a_g(t^\prime), 
\end{align}
where 
\begin{equation}
\begin{aligned}
    \gamma(t,t^\prime) &= \frac{1 }{\hbar^2}  \int \dd^3p \vb{E}_{cl}(t) \cdot \vb{d}^*(\vb{p}+ \frac{e}{c}\vb{A}_{cl}(t)) \vb{E}_{cl}(t^\prime) \cdot \vb{d}\left(\vb{p} + \frac{e}{c} \vb{A}_{cl}(t^\prime)\right) \exp{- \frac{i}{\hbar} S(\vb{p},t,t')}.
\end{aligned}
\end{equation}

We now assume that the rate of change of $a_g(t)$ is slow compared to the field, such that we can replace $a_g(t^\prime)$ by $a_g(t)$ in the integrand, and the solution is given by
\begin{equation}\label{eq:a_final_initial_g}
    \begin{aligned}
    a_g(t) &= \exp{- \int_{t_0}^t \dd t^\prime W(t^\prime)} a_g(t_0) = \exp{- \int_{t_0}^t \dd t^\prime W(t^\prime)}, 
    \end{aligned}
\end{equation}
with the ionization rate given by 
\begin{align}
    W(t) = \int_{t_0}^t \dd t^\prime \gamma(t,t^\prime). 
\end{align}

We further need the amplitudes with the continuum state as the initial condition $b_{\vb{v}}(\vb{v}^\prime,t_0) = \delta(\vb{v} - \vb{v^\prime})$ and $a_{\vb{v}}(t_0) = 0$. The corresponding solution of \eqref{eq:DGL_b_approximated} for $b_{\vb{v}}(\vb{v}^\prime,t)$ is given by 
\begin{equation}\label{eq:b_final_initial_v}
    \begin{aligned}
    b_{\vb{v}}(\vb{v}^\prime,t) &= \frac{i}{\hbar } \int_{t_0}^t \dd t^\prime \vb{E}_{cl}(t^\prime) \cdot \vb{d}\left(\vb{p}^\prime + \frac{e}{c} \vb{A}_{cl}(t^\prime)\right) \exp{- \frac{i}{\hbar} S(\vb{p}^\prime,t,t^\prime)} a_{\vb{v}}(t^\prime) + \exp{- \frac{i}{\hbar} S(\vb{p}^\prime,t,t_0) } \delta(\vb{v} - \vb{v^\prime}),
    \end{aligned}
\end{equation}
where the last term takes into account the initial condition $b_{\vb{v}}(\vb{v}^\prime,t_0) =\delta(\vb{v} - \vb{v^\prime})$.
Under the same assumptions for the solution of the differential equation of $a_{\vb{v}}(t)$ in \eqref{eq:coupled_DGL_amplitudes1}, we find that for $a_{\vb{v}}(t_0) = 0$
\begin{equation}\label{eq:a_final_initial_v}
    \begin{aligned}
    a_{\vb{v}}(t) &= \frac{i}{\hbar} \int \dd^3 v^\prime \int_{t_0}^t \dd t^\prime \vb{E}_{cl}(t^\prime) \cdot \vb{d}^*(\vb{v}^\prime) \exp{- \frac{i}{\hbar} S(\vb{p}',t^\prime,t_0)} \exp{- \int_{t^\prime}^t \dd\tau W(\tau)} \delta(\vb{v} - \vb{v^\prime}) \\
    &= \frac{i}{\hbar} \int_{t_0}^t \dd t^\prime \vb{E}_{cl}(t^\prime) \cdot \vb{d}^*(\vb{v}) \exp{- \frac{i}{\hbar}
    S(\vb{p},t^\prime,t_0)} \exp{- \int_{t^\prime}^t \dd\tau W(\tau)}.
    \end{aligned}
\end{equation}

We shall assume that the probability of recombination to the ground state is small when the electron is initially in the continuum. This means we approximate $a_{\vb{v}}(t) \approx 0$, which simplifies the amplitude of the continuum population 
\begin{equation}\label{eq:b_final_initial_v_approx2}
    \begin{aligned}
    b_{\vb{p}}(\vb{p}^\prime,t) &= \exp{ - \frac{i}{\hbar} S(\vb{p}',t,t_0)} \delta (\vb{p} - \vb{p}^\prime).
    \end{aligned}
\end{equation}

We further assume that the depletion of the ground state can be neglected such that $a_g(t) \simeq 1$ and accordingly the continuum population \eqref{eq:b_final_initial_g} reads
\begin{equation}\label{eq:b_final_initial_g2}
    \begin{aligned}
    b_g(\vb{p}^\prime,t) &=
        \frac{i}{\hbar } \int_{t_0}^t \dd t^\prime \vb{E}_{cl}(t^\prime) \cdot \vb{d}\left(\vb{p}^\prime(t) + \frac{e}{c} \vb{A}_{cl}(t^\prime) \right) \exp{- \frac{i}{\hbar} S(\vb{p},t,t')}.
    \end{aligned}
\end{equation}

We have expressed both amplitudes in terms of the gauge invariant kinetic momentum $\vb{p} = \vb{v} - \frac{e}{c} \vb{A}(t)$.

\subsection{\label{sec:app_bc_element}Bound-continuum transition matrix element}

In \eqref{eq:transition_gs_continuum_aprox} we need the time-dependent transition matrix element between the ground state $\ket{g}$ and the continuum state $\vb{v_\alpha}$ for the process of ATI. We can use \eqref{eq:TDSE_solution_semiclassical}, and have 
\begin{align}
    \bra{\vb{v}} \vb{r}(t) \ket{g} = \bra{\psi_{\vb{v}}(t)} \vb{r} \ket{\psi_g(t)} \simeq \int \dd^3v^\prime b_{\vb{v}}^*(\vb{v}^\prime, t) \bra{\vb{v}^\prime} \vb{r} \ket{g},
\end{align}
where we have used that $a_{\vb{v}}(t) \approx 0$ and $a_g(t) \simeq 1$. And we have further neglected C-C transition matrix elements. Inserting \eqref{eq:b_final_initial_v_approx2} we have
\begin{equation}
\begin{aligned}
    \bra{\vb{v}} \vb{r}_\alpha(t) \ket{g} \simeq \exp{\frac{i}{\hbar} S(\vb{p},t,t_0)} \bra{\vb{p} + e/c \vb{A}(t) } \vb{r} \ket{g}.
\end{aligned}
\end{equation}

\subsection{\label{sec:app_cc_elements}Continuum-Continuum transition matrix element}

For the time-dependent transition matrix element between the two continuum states in \eqref{eq:C-C:approx} we have 
\begin{equation}
    \begin{aligned}
    \bra{\vb{v}} \vb{r}(t) \ket{\vb{v}^\prime} &= \bra{\psi_{\vb{v}}(t)} \vb{r} \ket{\psi_{\vb{v}^\prime}(t)} \simeq \int \dd^3 v^{\prime \prime} \int \dd^3 v^{\prime \prime\prime}
				b^*_{\vb{v}}(\vb{v}^{\prime \prime}, t) b_{\vb{v}'}(\vb{v}^{\prime \prime\prime}, t)
					\mel{\vb{v}^{\prime \prime}}{\vb{r}}{\vb{v}^{\prime \prime\prime}},
    \end{aligned}
\end{equation}
where we have again used that $a_{\vb{v}}(t) \approx 0$ and $a_g(t) \simeq 1$.

\section{\label{sec:observable_ATI_direct}Field observables for direct ATI}
The pure state of the EM field conditioned on ATI \eqref{eq:DI:term} is given by 
\begin{equation}\label{eq:ATI:singlemomentum_app}
\begin{aligned}
    \ket{\phi_{ATI} (t,\vb{p})} &= \frac{-ie}{\hbar}  \int_{t_0}^t \dd t^\prime  \prod_{\vb{k}, \mu} e^{i \varphi_{\vb{k} \mu}(t, t^\prime, \vb{v}_\alpha)}
    D[\delta_\alpha (t,t^\prime,\omega_k,\vb{v}_\alpha)] 
    \vb{E}_Q(t^\prime,\vb{R}_\alpha) \cdot \bra{\vb{v}_\alpha} \vb{r}_\alpha(t^\prime) \ket{g}
    \\& \hspace{6cm}
    \bigotimes_{\vb{k}, \mu \simeq \vb{k}_{L,\mu}} \ket{ \chi_{\vb{k},\mu}(t^\prime) } \bigotimes_{\vb{k}, \mu \gg \vb{k}_{L,\mu}} \ket{\chi_{\vb{k},\mu}(t^\prime) }.
\end{aligned}
\end{equation}

We can now use that 
\begin{equation}
\begin{aligned}
    &\prod_{\vb{k}, \mu} e^{i \varphi_{\vb{k} \mu}(t, t^\prime, \vb{v}_\alpha)} D[\delta_\alpha (t,t^\prime,\omega_k,\vb{v}_\alpha)]  \vb{E}_Q(t^\prime,\vb{R}_\alpha)
    = \left[ \vb{E}_Q(t^\prime,\vb{R}_\alpha) - \vb{E}_{\delta}(t^\prime,\vb{R}_\alpha) \right]
    \prod_{\vb{k}, \mu} e^{i \varphi_{\vb{k} \mu}(t, t^\prime, \vb{v}_\alpha)} D[\delta_\alpha (t,t^\prime,\omega_k,\vb{v}_\alpha)],
\end{aligned}
\end{equation}
where 
\begin{equation}
\begin{aligned}
    \vb{E}_{\delta}(t^\prime,\vb{R}_\alpha) &= i \sum_\mu \int \dd^3 k \tilde g(k) \epsilon_{k \mu}
    \Big[ \delta_\alpha (t,t^\prime,\omega_k,\vb{v}_\alpha) e^{-i \omega_k t +i \mathbf{k } \cdot \mathbf{R}_\alpha}  -  \text{c.c.}\Big]
\end{aligned}
\end{equation}
such that the state of the field now reads
\begin{equation}
\begin{aligned}
    \ket{\phi_{ATI} (t,\vb{p})} = & \frac{-ie}{\hbar}  \int_{t_0}^t \dd t^\prime  \left[ \vb{E}_Q(t^\prime,\vb{R}_\alpha) - \vb{E}_{\delta}(t^\prime,\vb{R}_\alpha) \right] \cdot \bra{\vb{v}_\alpha} \vb{r}_\alpha(t^\prime) \ket{g} \\
    &
    \bigg[
    \bigotimes_{\vb{k}, \mu \simeq \vb{k}_{L,\mu}} e^{i [ \varphi_{\vb{k} \mu}(t, t^\prime, \vb{v}_\alpha) + \varphi_{\vb{k}\mu}^{\chi,\delta}(t,t^\prime) ]} \ket{ \chi_{\vb{k},\mu}(t^\prime) + \delta_\alpha (t,t^\prime,\omega_k,\vb{v}_\alpha) }
     \bigg]\\
    &
    \bigg[
    \bigotimes_{\vb{k}, \mu \gg \vb{k}_{L,\mu}} e^{i [ \varphi_{\vb{k} \mu}(t, t^\prime, \vb{v}_\alpha) + \varphi_{\vb{k}\mu}^{\chi,\delta}(t,t^\prime) ]} \ket{\chi_{\vb{k},\mu}(t^\prime) + \delta_\alpha (t,t^\prime,\omega_k,\vb{v}_\alpha)}
    \bigg],
\end{aligned}
\end{equation}
where 
\begin{align}
\label{eq:BCH_prefactor}
    \varphi_{\vb{k}\mu}^{\chi,\delta}(t,t^\prime) = \operatorname{Im}\left[ \delta_\alpha (t,t^\prime,\omega_k,\vb{v}_\alpha) \chi_{\vb{k},\mu}^*(t^\prime) \right].
\end{align}

We can now do the transformation back to the original laboratory frame 
\begin{align}
    \ket{ \Phi_{ATI} (t,\vb{p})} = e^{- i H_f t} D(\alpha_{\vb{k}_L}) \ket{\phi_{ATI}(t,\vb{p})}, 
\end{align}
and we find that 
\begin{equation}\label{eq:final_ATI_state_labframe}
    \begin{aligned}
        \ket{\Phi_{ATI} (t,\vb{p})} &=  \frac{-ie}{\hbar}  \int_{t_0}^t \dd t^\prime  \left[ \vb{E}_Q(t^\prime - t,\vb{R}_\alpha) - \vb{E}_{\delta}(t^\prime,\vb{R}_\alpha) - \vb{E}_{cl}(t^\prime, \vb{R}_\alpha) \right] \cdot \bra{\vb{v}_\alpha} \vb{r}_\alpha(t^\prime) \ket{g} \\
    & \bigotimes_{\vb{k}, \mu \simeq \vb{k}_{L,\mu}} e^{i [ \varphi_{\vb{k} \mu}(t, t^\prime, \vb{v}_\alpha) + \varphi_{\vb{k}\mu}^{\chi,\delta}(t,t^\prime) + \varphi_{\vb{k}\mu}^{\chi,\delta,\alpha}(t,t^\prime) ]} \ket{ \left[ \alpha_{\vb{k}_L} + \chi_{\vb{k},\mu}(t^\prime) + \delta_\alpha (t,t^\prime,\omega_k,\vb{v}_\alpha) \right] e^{- i \omega_{\vb{k}} t} } \\
    & \bigotimes_{\vb{k}, \mu \gg \vb{k}_{L,\mu}} e^{i [ \varphi_{\vb{k} \mu}(t, t^\prime, \vb{v}_\alpha) + \varphi_{\vb{k}\mu}^{\chi,\delta}(t,t^\prime) ]} \ket{ \left[\chi_{\vb{k},\mu}(t^\prime) + \delta_\alpha (t,t^\prime,\omega_k,\vb{v}_\alpha) \right] e^{- i \omega_{\vb{k}} t} }, 
    \end{aligned}
\end{equation}
where
\begin{align}
    \varphi_{\vb{k}\mu}^{\chi,\delta,\alpha}(t,t^\prime) = \operatorname{Im}\left\{ \alpha_{\vb{k}_L} \left[  \delta_\alpha^* (t,t^\prime,\omega_k,\vb{v}_\alpha) + \chi_{\vb{k},\mu}^*(t^\prime) \right] \right\}.
\end{align}

We have further used that 
\begin{equation}
\begin{aligned}
    &e^{- i H_f t} \vb{E}_Q(t^\prime, \vb{R}_\alpha) e^{i H_f t} = \vb{E}_Q(t^\prime - t , \vb{R}_\alpha)
    = i \sum_\mu \int \dd^3k \tilde g(k) \epsilon_{k \mu} \left[ a_{k \mu} e^{-i \omega_k (t^\prime - t) +i \mathbf{k } \cdot \mathbf{R}_\alpha}  - \text{h.c.} \right]. 
\end{aligned}
\end{equation}

This state \eqref{eq:final_ATI_state_labframe} is the field state conditioned on ATI with a single electron momentum in the laboratory frame. We shall now project this state on the harmonic vacuum $\ket{\{ 0 \}_{HH}}$ in order to obtain the state of the driving field mode in the situation where no harmonic radiation is generated. 
We thus find that 
\begin{equation}
    \begin{aligned}
    \ket{\Phi_{ATI}^{\vb{k}_L}(t, \vb{p})} = &  \bra{\{ 0\}_{HH}} \ket{\Phi_{ATI} (t,\vb{p})}\\
    = & \frac{-ie}{\hbar}  \int_{t_0}^t \dd t^\prime  \left[ \vb{E}_{Q, \vb{k}_L}(t^\prime - t,\vb{R}_\alpha) + i \sum_\mu \int \dd^3k \Theta(\vb{k}_{HH}) \tilde g(k) \epsilon_{\vb{k},\mu} e^{-i \omega_k t^\prime +i \mathbf{k } \cdot \mathbf{R}_\alpha} \left[ \chi_{\vb{k},\mu}(t^\prime) + \delta_\alpha (t,t^\prime,\omega_k,\vb{v}_\alpha) \right]  \right]\\
    & \left. - \vb{E}_{\delta}(t^\prime,\vb{R}_\alpha) - \vb{E}_{cl}(t^\prime, \vb{R}_\alpha)
 \right] \cdot \bra{\vb{v}_\alpha} \vb{r}_\alpha(t^\prime) \ket{g} \prod_{\vb{k},\mu \gg \vb{k}_{L,\mu}}\Big[ e^{i [ \varphi_{\vb{k} \mu}(t, t^\prime, \vb{v}_\alpha) + \varphi_{\vb{k}\mu}^{\chi,\delta}(t,t^\prime) ]} e^{- \frac{1}{2}\abs{\chi_{\vb{k},\mu}(t^\prime) + \delta_\alpha (t,t^\prime,\omega_k,\vb{v}_\alpha)}^2}\Big] \\
    & \bigotimes_{\vb{k}, \mu \simeq \vb{k}_{L,\mu}} e^{i [ \varphi_{\vb{k} \mu}(t, t^\prime, \vb{v}_\alpha) + \varphi_{\vb{k}\mu}^{\chi,\delta}(t,t^\prime) + \varphi_{\vb{k}\mu}^{\chi,\delta,\alpha}(t,t^\prime) ]} \ket{ \left[ \alpha_{\vb{k}_L} + \chi_{\vb{k},\mu}(t^\prime) + \delta_\alpha (t,t^\prime,\omega_k,\vb{v}_\alpha) \right] e^{- i \omega_{\vb{k}} t} },
    \end{aligned}
\end{equation}
where $\Theta(\vb{k}_{HH})$ is a Heaviside function taking into account only the contributions with field momentum $\vb{k}_{HH} \gg \vb{k}_L$. Having in mind that the action of the electric field operator over a generic coherent state $\ket{\beta_{\vb{k}}}$ can be written as
\begin{equation}
    \begin{aligned}
    \vb{E}_{Q,\vb{k}}(t,\vb{R}_\alpha)\ket{\beta_{\vb{k}}}
        &= \vb{E}_{\beta. \vb{k}}(t,\vb{R}_\alpha)\ket{\beta_{\vb{k}}} - i \Tilde{g}(k) \epsilon_{\vb{k}, \mu} D(\beta_{\vb{k}}) \ket{1_{\vb{k}}},
    \end{aligned}
\end{equation}
we can then write
\begin{equation}
    \begin{aligned}
    &\vb{E}_{Q,\vb{k}_L}(t'-t,\vb{R}_\alpha)
        \ket{\left[ 
                    \alpha_{\vb{k}_L} 
                    + \chi_{\vb{k},\mu}(t^\prime)
                    + \delta_\alpha 
                        (t,t^\prime,\omega_k,\vb{v}_\alpha) \right] e^{- i \omega_{\vb{k}} t}}
    \\
    &\hspace{2cm}
    = \big[
        \vb{E}_{cl}(t',\vb{R}_\alpha)
        + \vb{E}_{\delta}(t',\vb{R}_\alpha)
        + \vb{E}_{\chi}(t',\vb{R}_\alpha)
    \big]
    \ket{\left[ 
                    \alpha_{\vb{k}_L} 
                    + \xi_{\vb{k},\mu}(t^\prime)
                    + \delta_\alpha 
                        (t,t^\prime,\omega_k,\vb{v}_\alpha) \right] e^{- i \omega_{\vb{k}} t}}
    \\
    &\hspace{2cm}\quad
    -i\tilde g(\vb{k}_L) \epsilon_{\vb{k}_L,\mu} D\big(\left[ 
                    \alpha_{\vb{k}_L} 
                    + \chi_{\vb{k},\mu}(t^\prime)
                    + \delta_\alpha 
                        (t,t^\prime,\omega_k,\vb{v}_\alpha) \right] e^{- i \omega_{\vb{k}} t}
        \big)
        \ket{1}
    \end{aligned}
\end{equation}
such that the conditioned to ATI state reads
\begin{equation}
    \begin{aligned}
    \ket{\Phi_{ATI}^{\vb{k}_L}(t, \vb{p})}
        &= \bra{\{ 0\}_{HH}} \ket{\Phi_{ATI} (t,\vb{p})}
        \\
        &= \dfrac{-ie}{\hbar}
            \int^t_{t_0}\dd t'
                \bigg[
                    \vb{E}_\chi(t',\vb{R}_\alpha)
                    + \int \dd^3k \Theta(\vb{k}_{HH}) \tilde g(k) \epsilon_{\vb{k},\mu} e^{-i \omega_k t^\prime +i \mathbf{k } \cdot \mathbf{R}_\alpha} \left[ \chi_{\vb{k},\mu}(t^\prime) + \delta_\alpha (t,t^\prime,\omega_k,\vb{v}_\alpha) \right]
                \bigg]
                \\ &\quad
                \cdot \bra{\vb{v}_\alpha} \vb{r}_\alpha(t^\prime) \ket{g}
                \mathcal{C}_\text{HH}(p,t,t')
                \bigotimes_{\vb{k}, \mu \simeq \vb{k}_{L,\mu}} e^{i\theta_{\vb{k},\mu}(t,t',\vb{p},\alpha)} \ket{ \left[ \alpha_{\vb{k}_L} + \chi_{\vb{k},\mu}(t^\prime) + \delta_\alpha (t,t^\prime,\omega_k,\vb{v}_\alpha) \right] e^{- i \omega_{\vb{k}} t} }
                \\ &\quad
            - \dfrac{e}{\hbar}\int \dd^3 k \Pi(\vb{k}_L)
                \Tilde{g}(\vb{k})
             \int^t_{t_0}\dd t'
             \boldsymbol{\epsilon}_{\vb{k_L},\mu}
                \cdot \bra{\vb{v}_\alpha} \vb{r}_\alpha(t^\prime) \ket{g}
                \mathcal{C}_\text{HH}(p,t,t')\quad
                \\
                &\quad\quad
                \times
                \prod_{\vb{k,\mu}\simeq \vb{k}_{L,\mu}}
                D\big(\left[ 
                    \alpha_{\vb{k}} 
                    + \chi_{\vb{k},\mu}(t^\prime)
                    + \delta_\alpha 
                        (t,t^\prime,\omega_k,\vb{v}_\alpha) \right] e^{- i \omega_{\vb{k}} t}
                \big)
                \bigotimes_{\vb{k}', \mu \simeq \vb{k}_{L,\mu}} e^{i\theta_{\vb{k}',\mu}(t,t',\vb{p},\alpha)} \ket{1_{\vb{k}},\{0\}_{\vb{k'}\neq\vb{k}}},
    \end{aligned}
\end{equation}
where $\Theta(\vb{k}_\text{HH})$ is a Heaviside function that takes into account contributions for which $\vb{k}_\text{HH} \gg \vb{k}_L$, $\Pi(\vb{k}_L)$ is a rectangular function that is one whenever $\vb{k} \simeq \vb{k}_L$, and zero otherwise. We have further defined
\begin{equation}
    \begin{aligned}
    \theta_{\vb{k},\mu}(t,t',\vb{p},\alpha)
        &= \phi_{\vb{k},\mu}(t,t',\vb{v}_\alpha) +\operatorname{Im}\left[ \delta_\alpha (t,t^\prime,\omega_k,\vb{v}_\alpha) \chi_{\vb{k},\mu}^*(t^\prime) \right] + \operatorname{Im}\left\{ \alpha_{\vb{k}_L} \left[  \delta_\alpha^* (t,t^\prime,\omega_k,\vb{v}_\alpha) + \chi_{\vb{k},\mu}^*(t^\prime) \right] \right\},
    \end{aligned}
\end{equation}
which is the BCH phase that appears together with other phases arising from applying the commutation relation between two displacement operators. Finally, we have defined
\begin{equation}\label{eq:conditioning:funct}
    \begin{aligned}
    C_\text{HH}(p,t,t') &=    \prod_{\vb{k},\mu\gg \vb{k}_{L,\mu}}
                \Big[
                    e^{i \big\{ \varphi_{\vb{k} \mu}(t, t^\prime, \vb{v}_\alpha) + \operatorname{Im}\left[ \delta_\alpha (t,t^\prime,\omega_k,\vb{v}_\alpha) \chi_{\vb{k},\mu}^*(t^\prime) \right] \big\}}
                    e^{- \frac{1}{2}\abs{\chi_{\vb{k},\mu}(t^\prime) + \delta_\alpha (t,t^\prime,\omega_k,\vb{v}_\alpha)}^2}
                \Big],
    \end{aligned}
\end{equation}
which is a weight function that arises as a consequence of the conditioning onto the vacuum state of the harmonics modes. We have also defined
\begin{equation}
    \begin{aligned}
    \vb{E}_{\chi}(t^\prime,\vb{R}_\alpha) &= i \sum_\mu \int \dd^3 k \tilde g(k) \epsilon_{k \mu} \big[ \chi_{\vb{k},\mu} (t^\prime) e^{-i \omega_k t +i \mathbf{k } \cdot \mathbf{R}_\alpha} - \chi_{\vb{k},\mu}^* (t^\prime) e^{i \omega_k t-i \mathbf{k }\mathbf{R}_\alpha} \big].
    \end{aligned}
\end{equation}

Finally, we project the previous expression with respect to the a Fock state basis $\ket{n}$ to get the photon number probability amplitude for the direct term. Note that, in order to get expressions that can be analyzed computationally, we consider a single fundamental mode, corresponding to the central frequency of the employed electric field
\begin{equation}
    \begin{aligned}
    \ket{\Phi_{ATI}^{\vb{k}_L}(t, \vb{p})}
        &= \bra{\{ 0\}_{HH}} \ket{\Phi_{ATI} (t,\vb{p})}
        \\
        &= \dfrac{-ie}{\hbar}
            \int^t_{t_0}\dd t'
                \bigg[
                    \vb{E}_\chi(t',\vb{R}_\alpha)
                    + \int d^3k \Theta(\vb{k}_{HH}) \tilde g(k) \epsilon_{\vb{k},\mu} e^{-i \omega_k t^\prime +i \mathbf{k } \cdot \mathbf{R}_\alpha} \left[ \chi_{\vb{k},\mu}(t^\prime) + \delta_\alpha (t,t^\prime,\omega_k,\vb{v}_\alpha) \right]
                \bigg]
                \\ &\quad
                \cdot \bra{\vb{v}_\alpha} \vb{r}_\alpha(t^\prime) \ket{g}
                \mathcal{C}_\text{HH}(p,t,t')
                e^{i\theta_{\vb{k},\mu}(t,t',\vb{p},\alpha)} e^{-\tfrac12\lvert \alpha_{\vb{k}_L} + \chi_{\vb{k},\mu}(t^\prime) + \delta_\alpha (t,t^\prime,\omega_k,\vb{v}_\alpha)\rvert^2}
                 \\
                 &\quad \times
                 \dfrac{(\left[ \alpha_{\vb{k}_L} + \chi_{\vb{k},\mu}(t^\prime) + \delta_\alpha (t,t^\prime,\omega_k,\vb{v}_\alpha) \right] e^{-i \omega_{\vb{k}} t})^n}{\sqrt{n!}}
                \\ &\quad
            - \dfrac{e}{\hbar}\int \dd^3 k \Pi(\vb{k}_L)
                \Tilde{g}(\vb{k})
             \int^t_{t_0}\dd t'
             \boldsymbol{\epsilon}_{\vb{k_L},\mu}
                \cdot \bra{\vb{v}_\alpha} \vb{r}_\alpha(t^\prime) \ket{g}
                \mathcal{C}_\text{HH}(p,t,t')\quad
                \\
                &\quad\quad
                \times
                 e^{i\theta_{\vb{k}',\mu}(t,t',\vb{p},\alpha)} e^{-\tfrac12\lvert \alpha_{\vb{k}_L} + \chi_{\vb{k},\mu}(t^\prime) + \delta_\alpha (t,t^\prime,\omega_k,\vb{v}_\alpha)\rvert^2}
                 \\&\quad
                \Big[
                    \sqrt{n}\dfrac{(\left[ \alpha_{\vb{k}_L} + \chi_{\vb{k},\mu}(t^\prime) + \delta_\alpha (t,t^\prime,\omega_k,\vb{v}_\alpha) \right] e^{-i \omega_{\vb{k}} t})^{n-1}}{\sqrt{n-1!}}
                    - \dfrac{(\left[ \alpha_{\vb{k}_L} + \chi_{\vb{k},\mu}(t^\prime) + \delta_\alpha (t,t^\prime,\omega_k,\vb{v}_\alpha) \right] e^{-i \omega_{\vb{k}} t})^n}{\sqrt{n!}}
                \Big],
    \end{aligned}
\end{equation}
such that, up to a normalization factor, the photon number probability distribution for the $\vb{k}_{L,\mu}$ mode is given by
\begin{equation}
    P_{n_{\vb{k}_L}}(t,\vb{p})
        = \Big\lvert \braket{n}{\Phi_{ATI}^{\vb{k}_L}(t, \vb{p})}\Big\rvert^2.
\end{equation}

\section{\label{app:rescattering}Perturbation theory for rescattering in ATI}

From the perturbation theory analysis considered in the main part of the manuscript \eqref{eq:perturbation_ansatz} for the introduction of the rescattering part of the state, we get the following set of equations for each of the perturbation orders
\begin{equation}
    \begin{aligned}
    &i\hbar\dv{}{t} \ket{\phi^{(0)} (\vb{v}, t)}
        = e \vb{E}_Q(t,\vb{R}) \cdot \mel{\vb{v}}{\vb{r}(t)}{g}        
            \ket{\phi(t)} +e \vb{E}_Q(t,\vb{R}) \cdot\Delta \vb{r}(t, \vb{v}_\alpha)
           \ket{\phi^{(0)} (\vb{v},t)}\\
    & i\hbar\dv{}{t} \ket{\phi^{(1)} (\vb{v}, t)}
        = e \vb{E}_Q(t,\vb{R})\Delta \vb{r}(t, \vb{v})
            \ket{\phi^{(1)} (\vb{v},t)} - \dfrac{\hbar}{e}\int \dd^3v' 
                \exp[\frac{i}{\hbar} S(\vb{v}',t,t_0)]
                \vb{E}_Q(t,\vb{R}) \cdot\vb{g}(\vb{v},\vb{v}')
                \\
                & \hspace{3.0cm}\times
                \exp[-\frac{i}{\hbar} S(\vb{v}',t',t_0)]
                \ket{\phi^{(0)} (\vb{v}',t)}.
    \end{aligned}
\end{equation}

Here, we define $\Delta \vb{r}(t, \vb{v})$ as in Eq.~\eqref{eq:electr:displac}, so that the first equation describes the dynamics of the zeroth order perturbation term which characterizes direct ionization phenomena, and which has been previously discussed. On the other hand, the second term describes the rescattering dynamics, i.e., electrons that, once ionized, and can then re-scatter at the potential of the parent ion. Similarly to what we had in the direct ionization term, the solution to this differential equation is given as the superposition of an homogeneous term similar to that in Eq.~\eqref{eq:ATI:purestate}, plus an inhomogeneous term of the form
\begin{equation}\label{eq:rescattering}
    \begin{aligned}
    \ket{\phi^{(1)}(\vb{v},t)}
        = & -i\int^t_{t_0} \dd t_2 \int \dd^3 v'
            \prod_{\vb{k},\mu}
                e^{i \varphi_{\vb{k} \mu}(t, t_2, \vb{v})}
                D\big(
                    \delta(t,t_2,\omega_k,\vb{v})
                \big)
            \exp[\frac{i}{\hbar} S(\vb{v},t,t_0)] \\
        &  \times  \vb{E}_Q(t_2,\vb{R}) \cdot\vb{g}(\vb{v},\vb{v}')
            \exp[-\frac{i}{\hbar} S(t_2,t_0,\vb{v}')]
            \ket{\phi^{(0)}(\vb{v}',t_2)}.
    \end{aligned}
\end{equation}

However, since the electron is initially in the ground state so that $\ket{\phi^{(1)}(\vb{v},t_0)} = 0$ which implies that the homogeneous solution does not participate in the final solution. Then, the rescattering part of the state is entirely described by Eq.~\eqref{eq:rescattering}.

\section{Description of the numerical implementation}\label{App:numerical:implementation}

In the following we describe the numerical procedure used to obtain the field observables in the main part of the manuscript (see Section \ref{sec:observables}). 
The analysis is implemented in Python, and the codes can be found in \cite{ZenodoLink}.

\smallskip
Our numerical implementation consists of the following steps:
\begin{enumerate}[1.]
    \item Once the vector potential $\vb{A}(t)$ of the pulse under consideration is defined, we compute the expectation value of the time-dependent dipole moment $\langle \vb{d}(t)\rangle$ under the SFA framework.
    \item Given a time interval $[t_0,T]$, where $t_0$ and $T$ corresponds to the beginning and end of the pulse respectively, we compute $\chi(t)$ and $C_\text{HH}(p,T,t)$ given by Eq. \eqref{eq:conditioning:funct}. We use a step size of $\Delta t = 1.0$. For the displacement via $\delta(t,t_0,\omega_k,p)$ we compute the analytic exact expression for a $\sin^2$ envelope pulse.    
    \item Finally, we use an interpolating scheme such that we can have a numerical function for $\chi(t)$ and $C_\text{HH}(p,T,t)$, which can adapt to the time integrals that need to be done for computing the photon-number probability distribution.
\end{enumerate}

Furthermore, along this section we will work with atomic units (a.u.), such that $\hbar=1, e^2=1,m=1$ and $\epsilon_0 = 4\pi$.

\subsection{SFA expressions for the time-dependent dipole}

The expectation value of the time-dependent dipole moment is defined as 
\begin{equation}\label{eq:time:dependent:dipole}
    \langle \vb{d}(t)\rangle
        = \mel{\psi(t)}{\vb{d}}{\psi(t)},
\end{equation}
where $\ket{\psi(t)}$ is the wavefunction of a single electron in the intense laser field. To obtain the wavefunction we introduce the SFA approximations \cite{lewenstein1994theory}, namely (i) the contribution of all bound states except the ground state are neglected; (ii) the depletion of the ground state can be neglected; and (iii) in the continuum, the electron can be treated as a free particle, that is, we can neglect the interaction with the binding potential. Thus, we can write the semi-classical quantum state describing the electron as
\begin{equation}
    \ket{\psi(t)}
        = e^{i I_p t}\ket{g}
            + \int \dd^3 p \ b(\vb{p},t) \ket{\vb{p}+\vb{A}(t)},
\end{equation}
where the probability amplitude of finding the electron in the continuum state $\ket{\vb{p} + \vb{A}(t)}$ is given by
\begin{equation}
    \begin{aligned}
    b(\vb{p},t)
        &= i \int^t_{t_0} \dd t' \vb{E}(t') \cdot 
                            \vb{d}(\vb{p}+\vb{A}(t'))
            \exp{-i \int^t_{t'} \dd \tau
                \Big[
                    \big(\vb{p} + \vb{A}(\tau)\big)^2 + I_p
                \Big]
                },
    \end{aligned}
\end{equation}
where $\vb{d}(\vb{p} + \vb{A}(t)) = \bra{\vb{p} + \vb{A}(t)} \vb{d} \ket{g}$ is the transition dipole moment between the ground and the continuum state.
Introducing these expressions in \eqref{eq:time:dependent:dipole}, and neglecting the contributions from the continuum-continuum transitions, we get the SFA expression for the dipole moment expectation value
\begin{equation}
\label{eq:dipole_exp_appendix}
    \begin{aligned}
    \langle \vb{d}(t)\rangle
        &= i\int \dd^3p \int^t_{t_0} \dd t'
            \vb{d}^*(\vb{p}+\vb{A}(t))
            \vb{E}(t') \cdot 
                            \vb{d}(\vb{p}+\vb{A}(t'))
            \exp{-i \int^t_{t'} \dd \tau
                \Big[
                    \big(\vb{p} + \vb{A}(\tau)\big)^2 + I_p
                \Big]} + \text{c.c.}
    \end{aligned}
\end{equation}

For the numerical analysis, we considered a linearly polarized field with vector potential given by
\begin{equation}\label{Eq:app:numerical:pulse}
    \vb{A}(t) 
        = \epsilon_z \dfrac{E_0}{\omega}
            \sin[2](\dfrac{\omega}{2n_\text{cyc}})
            \sin(\omega t),
\end{equation}
where $n_\text{cyc}$ is the total number of cycles of the pulse with a $\sin^2$ envelope, $\epsilon_z$ is a unitary vector along the $z$ axis describing the polarization of the state. In all numerical analysis, we use $\omega = 0.057$ a.u., and $E_0 = 0.053$ a.u. On the other hand, we consider the ground state wavefunction of the electron to be described by
\begin{equation}
    \braket{x}{g}
        = \sqrt{\dfrac{\lambda^3}{\pi}}e^{-\lambda r},
\end{equation}
such that the dipole transition matrix element connecting the continuum state with the ground state is given by
\begin{equation}
    \begin{aligned}
    \vb{d}(\vb{p}+\vb{A}(t))
        & = \bra{\vb{p} + \vb{A}(t))} \vb{d} \ket{g} =-i \sqrt{\dfrac{\lambda^3}{\pi}}
            \dfrac{\big(\vb{p}+\vb{A}(t)\big)}{(2\pi)^{3/2}}
            \dfrac{32\pi \lambda}{(\lambda^2 + (p+A(t))^2)^3}.
    \end{aligned}
\end{equation}

\begin{figure}
    \centering
    \includegraphics[width = 0.8\columnwidth]{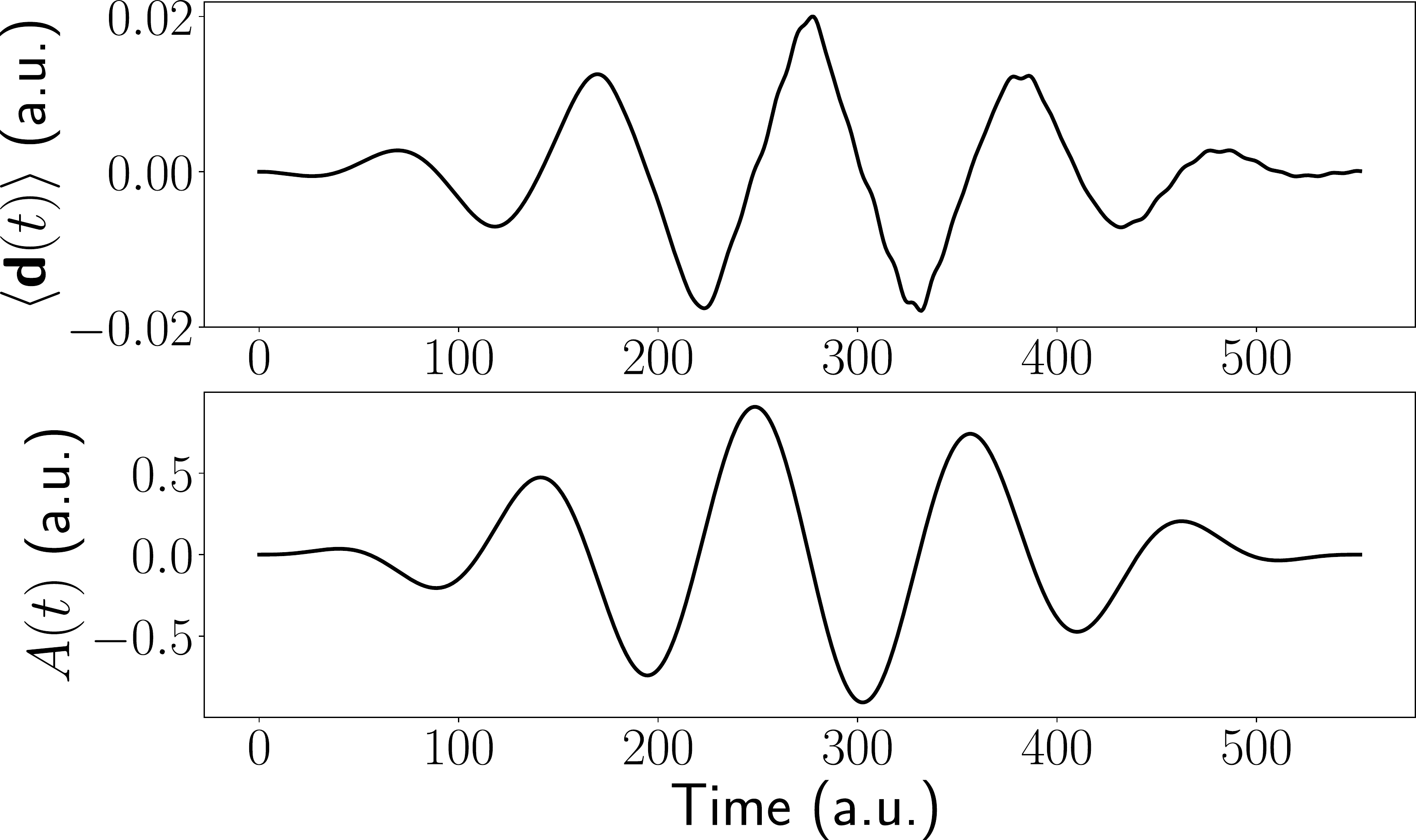}
    \caption{(a) Dipole moment expectation value obtained from the numerical calculations under the SFA approach. (b) Vector potential used to obtain the dipole moment shown in (a). The pulse shown has $n_\text{cyc}=5$ cycles.}
    \label{Fig:app:Dipole}
\end{figure}

More particularly, for a 1s state of an hydrogen atom we set $I_p=0.5$ a.u., and $\lambda = 1$. In Fig.~\ref{Fig:app:Dipole} we show in (a) the dipole moment expectation value \eqref{eq:dipole_exp_appendix} obtained, and in (b) the corresponding vector potential with $n_\text{cyc} = 5$. In order to perform the two integrals, we have first discretized the one for the momentum (which we perform in 1D), and used the \texttt{quad} function from the \texttt{SciPy} package \cite{2020SciPy-NMeth}.

\subsection{Calculation and interpolation of the quantum optical displacements}
With the dipole moment obtained in the previous subsection within the SFA approach, we can now compute the photonic displacement given by $\xi(t)$, which is defined as
\begin{equation}
    \chi(t) \propto \int^t_{t_0} \ \dd t' e^{i\omega t'}
                    \langle \vb{d}(t') \rangle.
\end{equation}

We further compute
\begin{equation}
    \begin{aligned}
    C_\text{HH}(p,T,t') &= \prod_{\vb{k},\mu\gg \vb{k}_{L,\mu}}
                \Big[
                    e^{i \big\{ \varphi_{\vb{k} \mu}(T, t^\prime, \vb{v}_\alpha) + \operatorname{Im}\left[ \delta_\alpha (T,t^\prime,\omega_k,\vb{v}_\alpha) \chi_{\vb{k},\mu}^*(t^\prime) \right] \big\}} e^{- \frac{1}{2}\abs{\chi_{\vb{k},\mu}(t^\prime) + \delta_\alpha (T,t^\prime,\omega_k,\vb{v}_\alpha)}^2}
                \Big],
    \end{aligned}
\end{equation}
where $\delta(p,T,t',\omega_k)$ is obtained analytically.
In particular, the computation of $C_\text{HH}(p,T,t')$ involves the calculation of the BCH phase $\varphi_{\vb{k} \mu}(T, t^\prime, \vb{v}_\alpha)$ from \eqref{eq:BCH_prefactor}. For this, we consider a discrete set of harmonic modes consisting of the 2nd up to the 21st harmonic, and perform an interpolation for each of the BCH phase factors. In principle we could consider more harmonic orders, but their amplitude $\abs{\delta(t,t',\omega_k,\vb{v})}$ rapidly decrease for increasing harmonic orders, and therefore their contributions are negligible. Since the effect of $\chi_{\vb{k},\mu}$ can be neglected in comparison to that introduced by $\delta(t,t',\omega_k,\vb{v})$, we neglect their effect in the previous conditioning operation. Finally, once the $C_\text{HH}(p,T,t')$ has been obtained for each value of the ionization time $t'$, we proceed to use an interpolation scheme to compute the photon number probability distribution.

\begin{figure}
    \centering
    \includegraphics[width = 0.8\columnwidth]{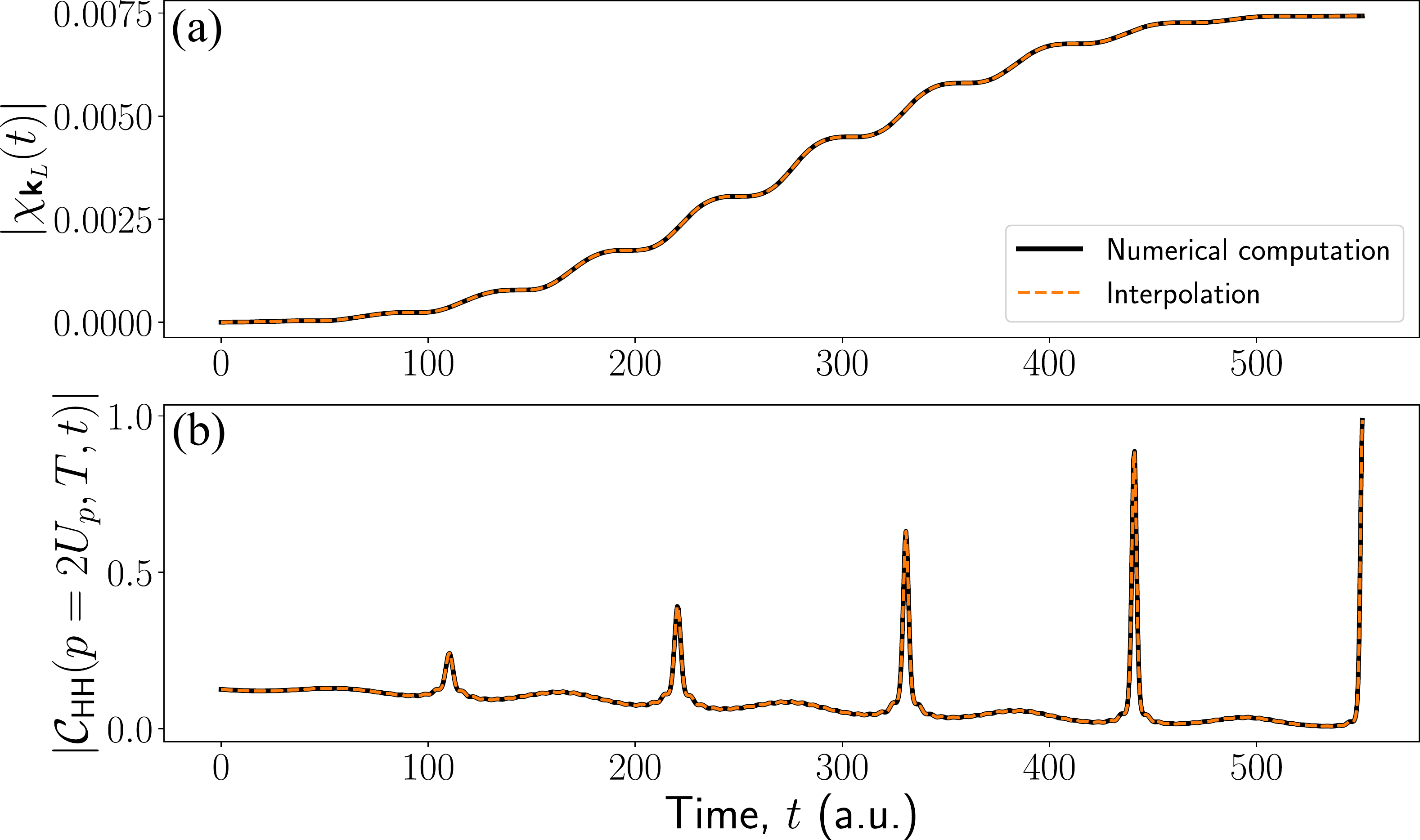}
    \caption{Behaviour of the (a) $\chi_{\vb{k}_L}(t)$ and (b) $C_\text{HH}(p=2U_p,T,t)$ with a time interval of $\Delta t = 1.0$ (solid black curve), and the function obtained from the interpolation (dashed orange curve). For the interpolation plot, we considered a time step $\Delta t = 0.01$.}
    \label{Fig:app:Interp}
\end{figure}

In both cases, similarly to what we did before, we perform this integral with the \texttt{quad} function, and is followed by performing an interpolation using the \texttt{interp1d} function provided by \texttt{SciPy}. The reason behind this is that, after the interpolation, we get two functions that exactly behave as $\chi(t)$ and $C_\text{HH}(p,T,t')$ within the $[t_0,T]$ interval. Thus, by doing this we avoid the nested integrals that appear within the definition of the photon-number probability distribution. This provides a huge numerical speed up. In particular, for doing the interpolation, we used the \texttt{cubic} option which considers polynomials of zeroth to third order, and that better fits the oscillating behaviour of both functions. The results of the interpolation are shown with the dashed orange curves in Fig.~\ref{Fig:app:Interp}.

For computing the dipole moment expectation value, we defined the time step to be equal to $\Delta t = 1.0$. In Fig.~\ref{Fig:app:Interp}, we show the behaviour of $\chi(t)$ and $\delta(p,T,t)$ (with $p= 0.1 U_p$) defined by the integral (solid black line), and after the interpolation (cyan dashed line). For the interpolation plot, we considered a time step $\Delta t = 0.01$.

\subsection{Further considerations}
As written in the main part of the manuscript, the photon number probability distribution is computed in the original reference frame, i.e. after undoing the displacement operation $D(\alpha)$. The usual regime of intense laser field physics has coherent states amplitudes of the driving laser in the order of $|\alpha| \sim 10^6$. Such high photon numbers can only be hardly implemented in the numerical analysis due to the numerical accuracy. Thus, in order to avoid this problem, we artificially reduce the value of $\alpha$ and multiply the $g(k)$ by an extra factor $\lambda < 1$ to account for this reduction in the computed values of $\delta(p,t,t')$ and $\xi(t)$. 

On the other hand, the specific value of $\alpha$ that we used in our numerical analysis is $\alpha = 7i$. This value is obtained from the definition of the vector potential, which is proportional to
\begin{equation}
    \hat{A}(t) \propto f(t)
            \big[
                \hat{a}^\dagger e^{i\omega t}
                + \hat{a}e^{-i\omega t}
            \big].
\end{equation}

Note that in the last expression we have introduced the envelope function via $f(t)$, which appears as a consequence of the multimode nature of our quantum optical description. If we now compute the average value of the vector potential operator with respect to the selected coherent state, we get
\begin{equation}
    \langle A(t) \rangle =
        \mel{7i}{\hat{A}(t)}{7i}
        \propto
            14f(t)\sin(\omega t),
\end{equation}
which has the form of the vector potential we used in the previous part of our numerical analysis. Note that one can similarly get the same phase by looking at the real and imaginary part of $\alpha_{\vb{k},\mu}$ in Fig.~\ref{Fig:behaviour:alpha} of the main text. 

\end{document}